\documentclass[11pt,a4paper,twoside]{memoir} 

\usepackage[T1]{fontenc} 
\usepackage[utf8]{inputenc}

\usepackage{cite}
\usepackage{graphicx,tikz,bbm,bm} 
\usepackage{color,xcolor} 
\usepackage{verbatim}
\usepackage{amsmath,amssymb,braket} 
\usepackage[printonlyused,withpage]{acronym} 
\usepackage{rotating} 
\usepackage[bookmarksnumbered=true,colorlinks=true, citecolor=blue, linkcolor=violet, anchorcolor=.,filecolor=.,menucolor=.,runcolor=.,urlcolor=.]{hyperref} 
\usepackage{microtype} 

\usepackage{lipsum}
\usepackage{mathrsfs}

\usepackage{appendix}
\usepackage{chngcntr}
\usepackage{etoolbox}

\makeatletter
\renewcommand{\fnum@figure}{\textsc{\figurename~\thefigure}} 
\makeatother

\setlrmarginsandblock{34.13mm}{*}{0.9} 
\setulmarginsandblock{34.13mm}{*}{*} 
\setmarginnotes{17pt}{51pt}{\onelineskip} 
\setheadfoot{\onelineskip}{2\onelineskip} 
\setheaderspaces{*}{2\onelineskip}{*} 
\checkandfixthelayout 

\makeatletter
\makechapterstyle{thesis}{

\setlength{\beforechapskip}{0pt}
\setlength{\midchapskip}{0pt}
\setlength{\afterchapskip}{0pt}

}
\makeatother

\maxsecnumdepth{subsubsection}
\maxtocdepth{subsubsection}

\newcommand{\be}{\begin{equation}}
\newcommand{\ee}{\end{equation}}
\newcommand{\mcH}{\mathcal{H}}
\newcommand{\vg}{\vec{g}}
\newcommand{\vphi}{\vec{\phi}}
\newcommand{\vchi}{\vec{\chi}}
\newcommand{\vp}{\vec{p}}
\newcommand{\valpha}{\vec{\alpha}}

\newcommand{\Tr}{\text{Tr}}
\newcommand{\mcO}{\mathcal{O}}
\newcommand{\mfh}{\mathfrak{h}}
\newcommand{\mfH}{\mathfrak{H}}
\newcommand{\mcN}{\mathcal{N}}
\newcommand{\h}{\hat}

\newcommand{\mbV}{\mathbb{V}}
\newcommand{\mbK}{\mathbb{K}}

\newcommand{\mcT}{\mathcal{T}}
\newcommand{\mcP}{\mathcal{P}}

\newcommand{\mbfD}{\mathbf{D}} 
\newcommand{\mbfF}{\mathbf{F}}  
\newcommand{\mbfT}{\mathbf{T}}  

\newcommand{\su}{\mathfrak{su}}
\newcommand{\g}{\mathfrak{g}}

\newcommand{\ex}{\text{ex}}
\DeclareMathOperator{\f}{full}
\newcommand{\phy}{\text{phy}}
\DeclareMathOperator{\can}{can}
\DeclareMathOperator{\cov}{cov}
\newcommand{\dep}{\text{dep}}

\newcommand{\bigtimes}{\mathbin{\tikz [x=1.4ex,y=1.4ex,line width=.2ex] \draw (0,0) -- (1,1) (0,1) -- (1,0);}}

\newcommand{\N}{\mathfrak{N}}
\newcommand{\n}{\mathfrak{n}}

\newcommand{\mbI}{\mathbb{I}}

\newcommand{\msK}{\mathscr{K}}
\newcommand{\msV}{\mathscr{V}}
\newcommand{\tth}{\text{th}}
\newcommand{\mbfV}{\mathbf{V}}
\newcommand{\co}{\text{co}}
\newcommand{\mfa}{\mathfrak{a}}

\newcommand{\w}{\wedge}
\newcommand{\p}{\partial}
\newcommand{\mcC}{\mathcal{C}}
\newcommand{\mcY}{\mathcal{Y}}
\newcommand{\mcX}{\mathcal{X}}

\newcommand{\mfA}{\mathfrak{A}}
\newcommand{\mcG}{\mathcal{G}}
\newcommand{\mfI}{\mathfrak{I}}

\newcommand{\tet}{
  \mathchoice
    {\includegraphics[height=1.3ex]{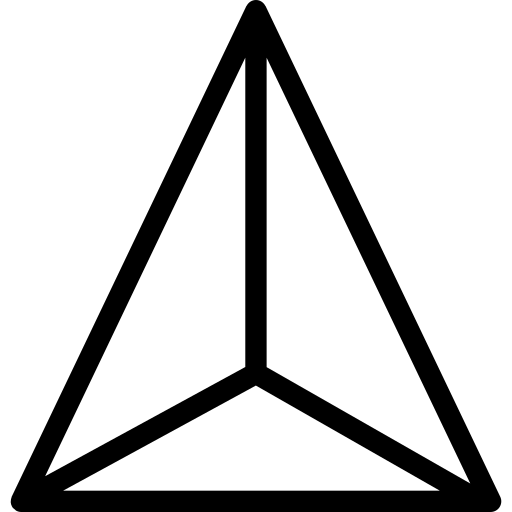}} 
    {\includegraphics[height=1.3ex]{Figures/tetrahedron3}} 
    {\includegraphics[height=1.1ex]{Figures/tetrahedron3}} 
    {\includegraphics[height=.8ex]{Figures/tetrahedron3}} 
}

\title{\textbf{\huge{On Generalised Statistical Equilibrium \\ and Discrete Quantum Gravity}}}

\date{}

\makeindex 

\newenvironment{acknowledgments}
{\abstract}
{\endabstract}

\AtBeginEnvironment{subappendices}{%
\chapter*{Appendices}
\addcontentsline{toc}{section}{Appendices}
\counterwithin{figure}{section}
\counterwithin{table}{section}
}

\AtEndEnvironment{subappendices}{%
\counterwithout{figure}{section}
\counterwithout{table}{section}
}

\begin{document}

\begin{titlingpage}
\maketitle 
\end{titlingpage}


\thispagestyle{empty} 


\begin{vplace}[1]
\begin{center} \textit{\large{for my grandparents, my guiding stars \\ and my parents, my guiding lights}} \end{center}
\end{vplace}

\cleartorecto

\newpage

\thispagestyle{empty}

\begin{vplace}[1]
 I declare that I have completed the thesis independently using only the aids and tools specified. I have not applied for a doctor's degree in the doctoral subject elsewhere and do not hold a corresponding doctor's degree. I have taken due note of the Faculty of Mathematics and Natural Sciences PhD Regulations, published in the Official Gazette of Humboldt-Universit{\"a}t zu Berlin Nr. 42/2018 on 11.07.2018. 

\vspace{0.5cm}

Ich erkl{\"a}re, dass ich die Dissertation selbst{\"a}ndig und nur unter Verwendung der von mir gem{\"a}{\ss} $\S$ 7 Abs. 3 der Promotionsordnung der Mathematisch-Naturwissenschaftlichen Fakult{\"a}t, ver{\"o}ffentlicht im Amtlichen Mitteilungsblatt der Humboldt-Universit{\"a}t zu Berlin Nr. 42/2018 am 11.07.2018 angegebenen Hilfsmittel angefertigt habe. 

\vspace{1cm}

\begin{flushright}
\begin{tabular}{@{}l@{}}
Isha Kotecha \\
June 11, 2020
\end{tabular}
\end{flushright} 

\end{vplace}

\frontmatter 

\newpage

\begin{abstract}

Statistical equilibrium configurations are known to be important in the physics of macroscopic systems with a large number of constituent degrees of freedom. They are expected to be crucial also in discrete quantum gravity, where dynamical spacetime should emerge from the collective physics of the underlying quantum gravitational degrees of freedom. However, defining statistical equilibrium in a background independent system is a challenging open issue, primarily due to the absence of absolute notions of time and energy. This is especially so in non-perturbative quantum gravity frameworks that are devoid of usual space and time structures. In this thesis, we investigate aspects of a generalisation of statistical equilibrium, specifically Gibbs states, suitable for background independent systems. We emphasise on an information theoretic characterisation for equilibrium based on the maximum entropy principle. Subsequently, we explore the resultant generalised Gibbs states in a discrete quantum gravitational system, composed of many candidate quanta of geometry of combinatorial and algebraic type (or, convex polyhedra). We utilise their field theoretic formulation of group field theory and various many-body techniques for our investigations. We construct concrete examples of quantum gravitational generalised Gibbs states, associated with different generators, e.g. geometric volume operator, momentum operators and classical closure constraint for polyhedra. We further develop inequivalent thermal representations based on entangled, two-mode squeezed, thermofield double vacua, which are induced by a class of generalised Gibbs states. In these thermal representations, we define a class of thermal condensates which encode statistical fluctuations in volume of the quantum geometry. We apply these states in the condensate cosmology programme of group field theory, where the key idea is that a macroscopic homogeneous spacetime can be approximated by a dynamical condensate phase of the underlying quantum gravity system. We study the relational effective cosmological dynamics extracted from a class of free group field theory models, for homogeneous and isotropic spacetimes. We find the correct classical limit of Friedmann equations at late times, with a bounce and accelerated expansion at early times.

\end{abstract}


\cleartorecto


\begin{acknowledgments}

Science is a journey driven by curiosity and learning. For me, this journey which began many years ago in high school, has reached a personal milestone with the completion of this thesis and my PhD. It would have been impossible without the help, encouragement and friendship of many people, to whom I am deeply grateful.

It is fortunate, as a young researcher, to be able to immerse in as stimulating an environment as that of the Albert Einstein Institute and Humboldt University of Berlin. I am indebted to Daniele Oriti, Hermann Nicolai, Deutscher Akademischer Austauschdienst (DAAD) and International Max Planck Research School for making this possible. I am especially thankful to Daniele for his mentorship and constant guidance, which helped me greatly to navigate through research topics and workings in academia. In addition to patiently guiding me through the dense literature for tackling these questions, he introduced me to important foundational problems in the field, while also encouraging me to explore topics of personal interest. 

My PhD experience was further enriched by the support of several people. Many special thanks are due to my collaborators, Goffredo Chirco and Mehdi Assanioussi, for numerous insightful discussions and friendly advice. I am also sincerely grateful to Rob Myers and Sylvain Carrozza, for the wonderful opportunity to spend some time at Perimeter Institute as a visiting graduate fellow. I am thankful to Joseph Ben Geloun, Axel Kleinschmidt, Sandra Faber, Sumati Surya and Amihay Hanany for their help, at various different stages during and before the start of my PhD. 

My time at AEI was made enjoyable by the presence of many friends and colleagues. For this, I would like to thank Alexander Kegeles, Claudio Paganini, Johannes Th\"urigen, Seungjin Lee, Mingyi Zhang, Ana Alonso-Serrano, Olof Ahl\'en, Alice Di Tucci, Caroline Jonas, Sebastian Bramberger, Lars Kreutzer, Jan Gerken and Hugo Camargo. I am particularly thankful to Alexander for a thorough reading of an earlier version of this manuscript and valuable comments; and, to Claudio for help with the German translation of the summary. Working at AEI, living in Germany and completing the PhD was made a lot easier due to the proactive support of several people, to whom I am very grateful: Anika Rast, Darya Niakhaichyk and Constance M{\"u}nchow at AEI; Bettina Wandel at DAAD; Jennifer Sabernak and Carolin Schneider at Potsdam International Community Center; and, Daniel Schaan and Milena Bauer at HU-Berlin. 

To my loving family, I am forever grateful. To my parents, Kotecha and Patel, for their open-mindedness and inspiration. To Kishan, Priyanka and Riddhi, for being my pillars of strength. Foremost, to my husband Meet, for his relentless support, unwavering faith and for being a true friend, ever since this journey began.

\end{acknowledgments}


\cleartorecto

\begin{vplace}[1]
\begin{flushright}
\begin{tabular}{@{}l@{}}
Where the voice of the wind calls our wandering feet, \\
Through echoing forest and echoing street, \\
With lutes in our hands ever-singing we roam, \\
All men are our kindred, the world is our home. \\ \\
 ---\emph{Sarojini Naidu}, in Wandering Singers
\end{tabular}
\end{flushright}
\end{vplace}

\cleartorecto


\begingroup
\hypersetup{linkcolor=black}
\tableofcontents*
\endgroup




\mainmatter 

\chapterstyle{thesis} 

\pagestyle{Ruled} 

\chapter{Introduction}
\label{intro}

\begin{quotation}
\begin{center}  To doubt everything or to believe everything are two equally convenient solutions; both dispense with the necessity of reflection. ---\emph{Henri Poincar\'e} \end{center}
\end{quotation}


\section*{Thermal physics, gravity and quantum theory} \label{tgq}

There are three foundational pillars of physics, namely thermal physics\footnote{By thermal physics, we mean statistical physics, thermodynamics, and many-body theory in general. For some discussions on the significance of many-body physics in the context of quantum gravity, see for instance \cite{Oriti:2018tym}.}, gravity and quantum theory. Any fundamental theory of nature must be built atop them.\

A deep interplay between the three was unveiled by the discovery of black hole entropy \cite{Bekenstein:1972tm,Bekenstein:1973ur,Bekenstein:1974ax} and radiation \cite{Hawking:1974sw}. As a direct consequence, a multitude of new conceptual insights arose, along with many puzzling questions that continue to be investigated still after decades. In particular, a black hole is assigned physical entropy, 
\be S = \frac{A}{4 \ell_P^2} \ee
scaling linearly with the area $A$ of its horizon (to leading order), where $\ell_P = \sqrt{\hbar G/c^3}$ is the Planck length \cite{Bekenstein:1972tm,Bekenstein:1973ur,Wall:2018ydq,Harlow:2014yka,Carlip:2014pma}. This led to several distinct lines of thoughts, in turn leading to various lines of investigations, like holography \cite{tHooft:1993dmi,Susskind:1994vu} and thermodynamics of gravity \cite{Jacobson:1995ab,Bardeen:1973gs}. Further, early attempts at understanding the physical origin of this entropy made the relevance of quantum entanglement evident \cite{Bombelli:1986rw,Srednicki:1993im}, thus contributing significantly to the current prolific interest in connections between gravitational physics and quantum information theory \cite{Ryu:2006bv,VanRaamsdonk:2010pw,Bianchi:2012ev,Jacobson:2015hqa,Hoehn:2017gst,VanRaamsdonk:2016exw,Marolf:2017jkr,Jacobson:2018ahi}.

In fact, Bekenstein's original arguments \cite{Bekenstein:1973ur} were also information-theoretic in nature, utilising insights from Jaynes' realisation \cite{Jaynes:1957zza,Jaynes:1957zz} of equilibrium statistical mechanics based on maximising information entropy under a given set of macrostate constraints\footnote{As we will see in section \ref{GenGE}, these same insights are instrumental in our characterisation of statistical equilibrium in a background independent context \cite{Kotecha:2018gof,Kotecha:2019vvn}.}. Black hole entropy was understood as information entropy \cite{Bekenstein:1973ur}, quantifying our lack of knowledge about the specifics of the system, here of the detailed configuration of the black hole with respect to an exterior observer. 

For instance, a stationary (Kerr) black hole is classically characterised completely by its mass $M$, charge $Q$ and angular momentum $J$. An outside observer could in principle measure the macrostate of this peculiar thermodynamic system in terms of this set of observables, without detailed knowledge of its quantum gravitational microstate. They could write down its entropy as the Shannon or von Neumann entropy \cite{Bekenstein:1973ur}, 
\be S = -\braket{\ln \rho}_\rho \ee
in a statistical state $\rho$ such that $\{\braket{\mathcal{M}}_\rho=M,\braket{\mathcal{Q}}_\rho=Q,\braket{\mathcal{J}}_\rho=J\}$, where $\mathcal{M},\mathcal{Q}$ and $\mathcal{J}$ are observables defined on some suitable underlying state space of the system, and $\braket{.}$ denotes a statistical average. Then, entropy $S$ measures the uncertainty of the system being in a particular quantum microstate compatible with the given thermodynamic macrostate. We will return to a detailed discussion of these, and other related aspects of Jaynes' method in section \ref{GenGE}, particularly in the context of background independent systems. For now, let us mention two important features encountered here, which as we shall see later, are intrinsic also to the generalised equilibrium statistical mechanical framework developed as part of this thesis. Firstly, black hole entropy can be understood as a measure of the lack of knowledge of a given observer about the system, or a measure of inaccessibility of information \cite{Bekenstein:1973ur,Jacobson:2018nnf,Bombelli:1986rw,Bekenstein:1994bc}; and, this inaccessibility of (correlated\footnote{Correlations between the relevant degrees of freedom across an information barrier is critical for having thermality, as we will discuss more below. These correlations could arise due to interactions, say in classical statistical systems coupled to a heat bath, or even purely from entanglement between two quantum statistical systems without any interactions. In fact, in section \ref{threpcond} we will construct a family of thermal vacua, induced by a family of generalised Gibbs states, with entanglement between the underlying quantum gravitational degrees of freedom.}) information, here due to the presence of a horizon, is the reason for its thermality. Secondly, the notion of information and statistical states used here is subjective, in the sense that it refers to the state of knowledge of the given observer \cite{Bekenstein:1973ur,Jaynes:1957zza}. Thus, the associated notion of thermality is inherently observer-dependent. \

This brings us to a discussion of some fundamental and universal features related to thermality in gravitational and quantum settings in general, which we think are valuable also to bear in mind while investigating candidate quantum gravitational systems. The intriguing connections between thermality, gravity and quantum theory can be succinctly displayed in the following equation for the temperature associated with causal horizons,
\be \label{temphawking} T = \frac{a }{2\pi }\frac{ \hbar}{ k_B  c} \ee
where $a$ is the acceleration characterising the horizon, and $k_B$ is the Boltzmann's constant. In the context of black holes, this is the well-known Bekenstein-Hawking temperature with $a$ being the surface gravity, while in the setting of Rindler decomposition of Minkowski spacetime, it is the Unruh temperature. \

This is a remarkable formula, hinting at several important points. Notice that $T$ is independent of the precise details of any matter degrees of freedom. Even though standard derivations of this equation utilise some quantum matter field along with a chosen dynamical model, the final expression is evidently independent of them. This suggests that $T$ could be an inherent property of dynamical spacetime, and even more so because, besides the fundamental constants, it is completely characterised by the gravitational acceleration parameter $a$. In other words, this expression suggests that spacetime is hot, and any other matter field, if present, will then naturally equilibrate with it to acquire the same temperature \cite{Padmanabhan:2009vy}. This further suggests the existence of quantum microscopic degrees of freedom underlying a spacetime. Also, this temperature is intrinsically observer-dependent, due to the observer dependence of $a$. For instance, different Rindler observers with different accelerations will detect thermal radiation at different temperatures, according to the formula \eqref{temphawking}. In fact, spacetime thermodynamics is observer-dependent in general \cite{Padmanabhan:2014jta,Padmanabhan:2009vy,PhysRevD.69.064006}. \

A vital feature that is not totally explicit in equation \eqref{temphawking} is the fact that spacetime thermality is tightly linked to causal horizons, or null surfaces. In other words, it is tightly linked to the existence of information barriers, i.e. boundaries beyond which lies information that is inaccessible to a set of observers on the other side. In an algebraic setup for instance, this would be related to a pair of commuting algebras of observables associated with Kubo-Martin-Schwinger (KMS) states \cite{Haag:1992hx,Bratteli:1979tw,Bratteli:1996xq}. More generally, we notice that thermality originates from having inaccessible or hidden information that is \emph{correlated} with information in a region that is accessed by an observer. In the special case of spacetime thermality then, the region of this hidden information is naturally demarcated by a horizon, which in turn is determined fully by the causal structure. Specifically for the class of equilibrium KMS states, equivalently Gibbs states for finite number of degrees of freedom, this thermality is linked further to periodicity in the two-point correlation functions of the algebra of observables \cite{Haag:1992hx,Bratteli:1979tw,Bratteli:1996xq,Kubo:1957mj,Martin:1959jp,Haag:1967sg}. \

This connection between quantum correlations, information barriers and thermality can be illustrated with a simple, yet important and widely utilised example of a thermofield double state \cite{Takahasi:1974zn,Khanna:2009zz}. Consider the case of two bosonic, non-interacting oscillators, each described in the standard way by its own set of ladder operators, cyclic vacua, Fock Hilbert spaces and free (kinetic energy) Hamiltonians. The composite system is then given by ladder operators $\{\h{a}_1,\h{a}_1^\dag,\h{a}_2,\h{a}_2^\dag\}$ satisfying the following commutation algebra, $ [\h{a}_i,\h{a}_j^\dag] = \delta_{ij} \,,\, [\h{a}_i,\h{a}_j] = [\h{a}_i^\dag,\h{a}_j^\dag] = 0$, for $i,j=1,2$. Notice that the algebras of the two oscillators, generated by these ladder operators, commute with each other. The individual vacua, given by $\h{a}_i \ket{0_i} = 0$, specify a vacuum $\ket{\Omega} = \ket{0_1} \otimes \ket{0_2}$ of the composite system, which in turn generates the full Hilbert space of a tensor product form. Then, a thermofield double, is a vector state of the full system, defined by
\be \ket{\Omega_\beta} = \frac{1}{\sqrt{Z_\beta}} \sum_{n} e^{-\frac{\beta}{2}E_n} \ket{n}_1 \otimes \ket{n}_2  \ee
where $\ket{n}_i = (a_i^\dag)^n \ket{0_i} / \sqrt{n!}\,$ are the energy eigenstates of the individual oscillators with spectrum $E_n$ \cite{Takahasi:1974zn,Khanna:2009zz}. This is an entangled state, with (maximum\footnote{The entanglement entropy of a bipartite quantum system is maximised by its corresponding thermofield double state, for a given $\beta$ (see, for instance \cite{Khanna:2009zz}).}) quantum correlations between the two oscillators. Then, the connection between the three aspects that we noted above can be demonstrated most directly by the fact that neglecting the degrees of freedom of either one of the oscillators (via partial tracing of the thermofield double) results in a (maximally entropic) thermal Gibbs state at inverse temperature $\beta$. In other words, any observer restricted\footnote{This restricted access of a given observer to the observables of a subsystem, that commute with the observables of its complement, can be understood as having an information barrier between the two (in analogy with local observable algebras \cite{Haag:1992hx}).} to either one of these oscillators (in general, subsystems) will measure properties that are compatible with being in a thermal state, via observable averages of the associated restricted algebra; and, this is a direct consequence of the full system being in an entangled state $\ket{\Omega_\beta}$, unlike the separable vacuum $\ket{\Omega}$. \

Such entangled states naturally also occur in systems with many degrees of freedom, including quantum field theories on flat and curved spacetimes. Prime examples are the relativistic Minkowski vacuum for uniformly accelerated observers, and the Hartle-Hawking vacuum for stationary observers outside a black hole \cite{Israel:1976ur,Unruh:1976db,Bisognano:1976za,Hartle:1976tp,Sewell:1982zz}. Along the same lines, in the context of holographic theories, thermofield doubles of boundary conformal field theories are dual to bulk eternal AdS black holes \cite{Maldacena:2001kr}. In the context of discrete quantum gravity, we have constructed thermofield double vacua associated with a class of generalised Gibbs states \cite{Assanioussi:2019ouq}, as will be discussed later in section \ref{threpcond}. \

We thus see that the notions of thermality, observer-dependence, quantum correlations, and accessibility of information (in turn related to causality when working with spacetime manifolds), all of which we will also encounter in this thesis, become deeply intertwined at the interface of quantum theory, gravity and thermal physics. It is at this interface, we believe, that further key insights into the nature of gravity await our discovery. Particularly with regards to the topic of this thesis, the formulation and investigation of thermal aspects of candidate quantum gravity frameworks may be vital to gain a more fundamental and detailed understanding of physical systems, like quantum black holes and early cosmological universe that are used often as theoretical laboratories for foundational research. For example, what `hot' even means at Planck scales, where notions of energy and spacetime may be notably different or even absent, needs to be investigated rigorously.


\section*{Why search for quantum gravity}

Presently our best description of nature is dichotomous and incomplete. Quantum matter is described by the standard model of particle physics, while classical gravity by general relativity. We are yet to discover a falsifiable theory that consistently merges a quantum theory of matter with gravitational phenomena, despite many advances in the various candidate approaches over the past decades \cite{Kiefer:2004gr,Oriti:2009zz}.

Gravity, by its very nature, responds to mass and energy, including quantum fields, i.e. quantum matter and gravity cannot be screened from each other. Thus, physical phenomena cannot be inherently divided into purely gravitational and quantum sectors. However the two separate theories, as they stand presently, do not accommodate this fact. Moreover, general relativity treats matter, which we know is fundamentally quantum, as classical. Therefore, if physics is to give a fully consistent and accurate account of our universe, then such a dichotomous description is incomplete at the very least, and certainly cannot reflect any fundamental separability between the quantum and gravitational regimes, even in principle. Thus, it seems that the physical world cannot be completely described by this disunified set of frameworks. \cite{10.2307/3080960,10.1086/508946} \

Further, general relativity and quantum field theory are not valid at arbitrary energy scales. For instance in extreme environments, like the early universe or dynamical black holes, quantum fields will inevitably affect spacetime geometry to a significant extent and vice-versa. Also, even though we might not have data to which we can unambiguously assign a quantum gravitational origin, we certainly have phenomena that are still left unexplained by current theories, e.g. dark matter, dark energy, which could turn out to be low energy, non-perturbative effects of some underlying fundamental theory of quantum gravity. This further motivates the search for a unified framework \cite{10.2307/3080960,10.1086/508946,Kiefer2007,Nicolai:2013sz}, which is also expected to resolve divergences encountered in general relativity and quantum field theory \cite{Kiefer2007,Nicolai:2013sz}. \

The path to quantum gravity may also shed light on the open foundational problem of time \cite{Kiefer2007}. Time plays drastically different roles in quantum theory and general relativity \cite{doi:10.1002/andp.201200147,Hohn:2019cfk,bruknerQCaus,Kiefer:2009tq,Rovelli:2009ee,PhysRevD.43.442,wootters}. The two notions are intrinsically incompatible within our present, limited understanding of the concept. In the context of conventional quantum theory, it plays the role of a global, external parameter characterising fully the evolution of a system. On the other hand in general relativity, both time and space are dynamical. In particular, the dynamics is constrained and coordinate time is gauge. It no longer carries the same physical status as that of a non-gravitational system (including relativistic field theories on a flat background). Similarly, a theory of quantum gravity may also be fundamentally `timeless', in the sense that it may be devoid of an unambiguous notion of time, or may even display a complete absence of any time or clock variable. As we will see, the topic of this thesis, namely to develop a generalised framework for equilibrium statistical mechanics with subsequent applications in background independent discrete quantum gravity, is tightly linked with this issue. The notions of time, energy and temperature are inextricably intertwined. \

The search for a theory of quantum gravity is thus well-motivated and the need for it is largely acknowledged. But, what this theory is or could be, and how one should go about formulating it attracts numerous diverse methods and reasonings \cite{Kiefer:2004gr,Oriti:2009zz}, especially regarding which physical principles should be considered as foundational and indispensable to base the theory on. In this thesis, we are strictly concerned with fundamental discrete approaches to quantum gravity. Specifically, we work with the group field theory approach \cite{Reisenberger:2000zc,Freidel:2005qe,Oriti:2005mb,Oriti:2006se,Oriti:2011jm,Krajewski:2012aw,Oriti:2014yla}, which is a statistical field theory of candidate quanta of geometry of combinatorial and algebraic type \cite{Oriti:2013aqa,Chirco:2018fns,Kotecha:2019vvn}, the same type of quanta that are utilised in several other discrete formalisms as will be discussed later.


\section*{Outline of the thesis} \label{outline}

Background independence is a hallmark of general relativity that has revolutionised our conception of space and time. The picture of physical reality it paints is that of an impartial dynamical interplay between matter and gravitational fields. Spacetime is no longer a passive stage on which matter performs, but is an equally active performer in itself. Spacetime coordinates are gauge, thus losing their physical status of non-relativistic settings. In particular, the notion of time is modified drastically. It is no longer an absolute, global, external parameter uniquely encoding the full dynamics. It is instead a gauge parameter associated with a Hamiltonian constraint. \

On the other hand, the well-established fields of quantum statistical mechanics and thermodynamics have been of immense use in the physical sciences. From early applications to heat engines and study of gases, to modern day uses in condensed matter systems and quantum optics, these powerful frameworks have greatly expanded our knowledge of physical systems. However, a complete extension of them to a background independent setting, such as that for a gravitational field, remains an open issue \cite{Rovelli:1993ys,Connes:1994hv,Rovelli:2012nv,Kotecha:2019vvn}. The biggest challenge is the absence of an absolute notion of time, and thus of energy, which is essential to any standard statistical and thermodynamical consideration. This issue is particularly exacerbated in the context of defining statistical equilibrium, for the natural reason that the standard concepts of equilibrium and time are tightly linked. In other words, the constrained dynamics of a background independent system lacks a non-vanishing Hamiltonian in general, which makes formulating (equilibrium) statistical mechanics and thermodynamics, an especially thorny problem. This is a foundational issue, and tackling it is important and interesting in its own right. And even more so because it could provide useful insights into the very nature of fundamental quantum gravitational systems, and their connections with thermal physics. \

The importance of addressing these issues is further intensified in light of the open problem of emergence of spacetime in quantum gravity \cite{Oriti:2009zz,Carlip:2012wa,Oriti:2013jga,Oriti:2018dsg,Oriti:2018tym}. Having a quantum microstructure underlying a classical spacetime is a perspective that is shared, to varying degrees of details, by various approaches to quantum gravity such as loop quantum gravity (and related spin foams, and group field theories), simplicial gravity, and holographic theories, to name a few. \

Specifically within discrete non-perturbative approaches, spacetime is replaced by more fundamental entities that are discrete, quantum and pre-geometric, in the sense that no notion of smooth metric geometry and continuum manifold exists yet. The collective dynamics of such quanta of geometry, governed by some theory of quantum gravity is then thought to give rise to an emergent spacetime, corresponding to specific phases of the full theory. For instance, in analogy with condensed matter systems, our universe can be understood as a kind of a condensate that is brought into the existing smooth geometric form by a phase transition of a quantum gravitational system of pre-geometric `atoms' of space, with the cosmological evolution being encoded in effective dynamical equations for collective variables that are extracted from the underlying microscopic theory \cite{Hu:2005ub,Oriti:2016acw,Pithis:2019tvp,Gielen:2016dss,Gabbanelli:2020lme,Gielen:2014ila,Padmanabhan:2014jta}. Overall, this essentially entails identifying suitable procedures to extract a classical continuum from a quantum discretuum, and reconstructing effective general relativistic gravitational dynamics coupled with matter (likely with quantum corrections, potentially related to novel non-perturbative effects). This is the realm of statistical physics, which thus plays a crucial role even from the perspective of an emergent spacetime. \

The main technical strategy used in this thesis is to model discrete quantum spacetime as a many-body system \cite{Oriti:2017twl}, which in turn complements the view of a classical spacetime as a coarse-grained, macroscopic thermodynamic system. This formal suggestion is advantageous in multiple ways. It allows us to treat extended regions of quantum spacetime as built out of discrete building blocks, whose dynamics is determined by many-body mechanical models, here of generically non-local, combinatorial and algebraic type. It facilitates exploration of connections of discrete quantum geometries with quantum information theory and holography \cite{Bianchi:2012ev,Colafranceschi:2020ern,Chen:2021vrc,Colafranceschi:2021acz,Baytas:2018wjd,Livine:2017fgq,Chirco:2017xjb,Chirco:2019dlx,Chirco:2017wgl,Chirco:2017vhs}. Further, it makes possible implementing other many-body techniques, like algebraic Fock treatments and squeezing Bogoliubov transformations, for instance to find non-perturbative, possibly entangled vacua of the quantum gravitational system \cite{Assanioussi:2019ouq,Kegeles:2017ems,Bianchi:2016tmw}. It also allows for the development of a statistical mechanical framework for these candidate quanta of geometry (here, of combinatorial and algebraic type), formally based on their many-body mechanics \cite{Kotecha:2018gof,Chirco:2018fns,Assanioussi:2019ouq}. Such a framework naturally admits probabilistic superpositions of quantum geometries \cite{Kotecha:2018gof,Chirco:2018fns,Kotecha:2019vvn}; and facilitates studies of quantum gravitational states that incorporate fluctuations in relevant observables of the system, which can be interesting to study for example in the context of cosmology \cite{Marchetti:2020qsq,Gielen:2017eco,Assanioussi:2019ouq,Assanioussi:2020hwf}. \\

This thesis is devoted to investigations of aspects like these. In particular, we illustrate, the potential of and preliminary evidence for, a rewarding exchange between a suitable background independent generalisation of equilibrium statistical mechanics, and discrete quantum gravity based on a many-body framework. These are the two facets of interest to us, to which our original contributions belong, as reported in \cite{Kotecha:2018gof,Chirco:2018fns,Chirco:2019kez,Kotecha:2019vvn,Assanioussi:2019ouq,Assanioussi:2020hwf} and discussed in this thesis. Sections \ref{modstat}, \ref{FockExt} and \ref{genposgibbs}, and appendices \ref{appMod}, \ref{autapp}, \ref{appkmsgibb} and \ref{AppClosure}, in this thesis include details that are not reported in our previous works.\footnote{For \cite{Kotecha:2018gof}: Licensed under CC BY [creativecommons.org/licenses/by/3.0]. For \cite{Kotecha:2019vvn,Assanioussi:2019ouq}: Licensed under CC BY [creativecommons.org/licenses/by/4.0]. For \cite{Chirco:2018fns}: Reprinted excerpts and figures with permission from [Goffredo Chirco, Isha Kotecha, and Daniele Oriti, Phys. Rev. D, 99, 086011, 2019. DOI: 10.1103/PhysRevD.99.086011]. Copyright 2019 by the American Physical Society. For \cite{Assanioussi:2020hwf}: Reprinted excerpts and figures with permission from [Mehdi Assanioussi and Isha Kotecha, Phys. Rev. D, 102, 044024, 2020. DOI: 10.1103/PhysRevD.102.044024]. Copyright 2020 by the American Physical Society. Minor modifications are made for better integration into this thesis.}  \\

We begin in chapter \ref{GGS} with a discussion of a potential background independent extension of equilibrium statistical mechanics. In section \ref{sympmech}, we discuss the topic of presymplectic mechanics for many-body systems, in order to review the essentials to be utilised in later chapters, while also drawing attention to the role played by time in such systems. In section \ref{characG}, we clarify how to comprehensively characterise statistical states of the exponential Gibbs form, while placing the discussion within the broader context of the issue of background independent statistical equilibrium. After providing a succinct yet complete discussion of past proposals for generalised notions of statistical equilibrium in \ref{past}, we focus on the so-called thermodynamical characterisation for defining generalised Gibbs states in section \ref{maxent}, which is based on a constrained maximisation of information entropy. In sections \ref{modstat} and \ref{rem}, we detail further crucial and favourable properties of this particular characterisation, also in comparison with the previously recalled proposals. Subsequently in section \ref{GenTD}, we discuss aspects of a generalised thermodynamics based directly on the generalised equilibrium setup derived above, including statements of the zeroth and first laws. This chapter presents a (partial) general framework, which forms the basis of our subsequent applications in discrete quantum gravity in the following chapters. \

In chapter \ref{DQG}, we give an overview of the essentials of the many-body setup for the candidate quanta of geometry. The quanta considered here are combinatorial $d$-valent patches (elementary building blocks of graphs) dressed with algebraic data, which generate extended labelled graphs as generic boundary states. These types of states (dual to polyhedral complexes) are used in several discrete approaches, like loop quantum gravity, spin foams, group field theories and tensor models, dynamical triangulations and Regge calculus, which has motivated our choice of them also. Specifically in section \ref{effgft}, we show that group field theories, in their covariant formulations in terms of field theory partition functions, arise as effective statistical field theories under a coarse-graining of a class of generalised Gibbs density operators of the underlying system of an arbitrarily large number of these quanta. \

A group field theory (GFT) thus being a field theory of such quanta, then its quantum operator formulation naturally offers a suitable route to a quantum statistical mechanical framework; or its associated many-body classical phase space formulation, to look into its corresponding classical statistical mechanics. This is the setting for our investigations in the subsequent chapters, by utilising the formalism of GFT. \

Now, GFTs are background independent, in the radical sense of spacetime-free approaches to quantum gravity. However, they also present specific peculiarities, which are crucial in various analyses, particularly in our development of an equilibrium statistical mechanical framework for them. The base space for the dynamical fields of GFTs consists of Lie group manifolds, encoding discrete geometric as well as matter degrees of freedom. This is \emph{not} spacetime, and all the usual spatiotemporal features associated with the base manifold of a standard field theory are absent. As in other covariant systems, a physically sensible strategy for defining equilibrium can be to use internal dynamical variables, for example matter fields, as relational clocks with respect to which one defines dynamical evolution. Even in this case though, one does not expect the existence of a preferred material clock, nor, having chosen one, that this would provide a perfect clock, mimicking precisely an absolute Newtonian time coordinate. In the end, like standard constrained systems on spacetime, GFTs too are devoid of an external or even an internal variable that is clearly identified as a preferred physical evolution parameter. But the close-to-standard quantum field theoretic language used in GFTs, utilising many-body techniques in the presence of a base manifold (the Lie group, with associated metric and topology), imply the availability of some mathematical structures that are crucially shared with spacetime-based theories. This, then, allows us to move forward with the task of investigating their thermal aspects.

In chapter \ref{GFT}, we focus on the details of scalar group field theories, as required for the purposes of this thesis. In particular, we present the quantum operator formulation of bosonic GFTs associated with a degenerate vacuum i.e. a `no-space' state, with no geometric and matter degrees of freedom, and detail the construction of its corresponding Fock representation in sections \ref{fockspace} and \ref{usbas}. Subsequently in sections \ref{weylGFT} and \ref{AUT}, we provide an abstract Weyl algebraic formulation of the same system, and construct unitarily implementable translation automorphism groups. In sections \ref{classdep} and \ref{quantise}, we address the issue of extracting a suitable clock variable, i.e. the issue of deparametrization in group field theory.  \

In chapter \ref{TGFT}, we illustrate the applicability of the generalised statistical framework in discrete quantum gravity, based on the above many-body structure of GFTs. We present several concrete examples of classical and quantum generalised Gibbs states in \ref{gengibb}. Then for the class of states associated with positive (semi-bounded in general, see Remark in appendix \ref{posapp2}) and extensive operator generators, we construct their corresponding class of inequivalent, thermal representations in section \ref{threpcond}, along with their non-perturbative thermal vacua. These cyclic vacua are thermofield double states, which we came across in section \ref{tgq} above. Our construction is based on the use of Bogoliubov transformation techniques from the field of thermofield dynamics. We further identify and construct an interesting class of states to describe thermal quantum gravitational condensates in section \ref{CTS}. Equipped with these thermal condensates, we apply a specific kind of them, those which incorporate spatial volume fluctuations in quantum geometry, in the setting of GFT condensate cosmology in section \ref{GFTCC}. For a free GFT model, we derive the effective dynamical equations of motion in terms of relational clock functions, in sections \ref{eff1} - \ref{eff2}. Subsequently, we use these GFT equations of motion to derive relational generalised Friedmann equations, with quantum and statistical corrections, for homogeneous and isotropic cosmology in section \ref{eff3}. At late times, we recover the correct classical limit; while at early times, we observe a bounce between a contracting and an expanding phase, along with an early phase of accelerated expansion featuring an increased number of e-folds compared to past studies of the same model. \

We conclude in chapter \ref{CONC} with a summary and outlook.


\chapter{Generalised Statistical Equilibrium}
\label{GGS}

\begin{quotation}
Time is what keeps everything from happening at once. ---\emph{Ray Cummings}
\end{quotation}


\vspace{1.5mm}

\noindent What characterises statistical equilibrium? In a non-relativistic system, the answer is unambiguous. Equilibrium states are those which are stable under time evolution generated by the Hamiltonian ${H}$ of the system. In the algebraic description, these are the states that satisfy the KMS condition \cite{Kubo:1957mj,Martin:1959jp,Haag:1967sg,Haag:1992hx,Bratteli:1996xq}. For systems with finite number of degrees of freedom, KMS states take the explicit form of Gibbs states, whose density operators have the standard form, $e^{-\beta H}$ (see discussions in appendix \ref{appkmsgibb}). This characterisation is unambiguous because of the special role of the time variable and its conjugate energy in standard statistical mechanics, where time is absolute, and modelled as the unique, external parameter encoding the dynamics of the system.  \

Investigating this question in a background independent context, where the role of time is modified \cite{doi:10.1002/andp.201200147,Hohn:2019cfk,Kiefer:2009tq,Rovelli:2009ee,PhysRevD.43.442,Rovelli:1994rs}, is much more challenging and interesting. A complete framework for statistical mechanics in this setting is still missing. Classical gravity as described by general relativity (GR) is generally covariant. This means that space and time coordinates are gauge, and are not physical observables. Further, all geometric quantities, in particular temporal intervals, are dynamical, and generic solutions of the GR dynamics do not allow to single out any preferred time or space directions. This is the content of background independence in GR, and other modified gravity theories with the same symmetry content.  Specifically, coordinate time is no longer a universal, physical evolution parameter. The absence of an unambiguous notion of time evolution is even more conspicuous in some quantum gravity formalisms in which an even more radical setup is invoked, where the familiar spatiotemporal structures of GR like the differential manifold, continuum metric, standard matter fields etc. have disappeared.  How can one define an equilibrium thermal state then? \

Covariant statistical mechanics \cite{Rovelli:1993ys,Connes:1994hv,Rovelli:2012nv} broadly aims at addressing the issue of defining a statistical framework for constrained systems on spacetime. This issue, especially in the context of gravity, was brought to the fore in \cite{Rovelli:1993ys}, and developed subsequently in various studies of spacetime relativistic systems \cite{Rovelli:1993ys,Connes:1994hv,Rovelli:2012nv,Chirco:2013zwa,Rovelli:2010mv,Montesinos:2000zi,Chirco:2016wcs}, valuable insights from which have also formed the conceptual backbone of our first applications to discrete quantum gravity \cite{Kotecha:2018gof,Chirco:2018fns,Chirco:2019kez}. In this chapter, we present (tentative) extensions of equilibrium statistical mechanics to background independent\footnote{In the earlier works mentioned above \cite{Rovelli:1993ys,Connes:1994hv,Rovelli:2012nv,Chirco:2013zwa,Rovelli:2010mv,Montesinos:2000zi,Chirco:2016wcs}, such a framework is usually referred to as covariant or general relativistic statistical mechanics. But we will choose to call it background independent statistical mechanics as our applications to quantum gravity are evident of the fact that the main ideas and structures are general enough to be used in radically background independent systems devoid of any spacetime manifold or associated geometric structures and symmetries.} systems, laying out different proposals for a generalised statistical equilibrium, but emphasising on one in particular, based on which further aspects of a generalised thermodynamics are considered. \

In section \ref{sympmech}, we begin with a brief discussion surrounding the role of time in mechanics. The goal is not to give a thorough review of this vast subject of the nature and problem of time (see for example \cite{doi:10.1002/andp.201200147,Hohn:2019cfk,Rovelli:2009ee,Kiefer:2009tq}), but to simply introduce the basic structures of classical constrained systems in terms of their presymplectic and extended phase space descriptions (to be utilised later in sections \ref{depgft} and \ref{physeqm}), and importantly, in the process, to bring to attention some specific features of constrained systems, particularly in the context of defining a suitable notion of statistical equilibrium for them. Since the notions of equilibrium and time are strongly linked, reconsidering the role of time in mechanics may guide us to understand better its role in statistical mechanics and thermodynamics, and in fundamental (even, spacetime-free) theories of quantum gravity. \

In section \ref{GenGE}, we present generalised Gibbs states of the form $e^{-\sum_a \beta_a \mcO_a}$. We begin in section \ref{characG} with a detailed discussion of the defining characteristics of Gibbs states with the aim of generalising them to background independent systems \cite{Kotecha:2018gof}. In section \ref{past}, we touch upon the various proposals for statistical equilibrium put forward in past studies on spacetime covariant systems \cite{Rovelli:1993ys,Montesinos:2000zi,Rovelli:2012nv,Josset:2015uja,Haggard:2013fx}, in order to better contextualise our work. In section \ref{maxent}, we focus on the thermodynamical\footnote{Using the terminology of \cite{Kotecha:2018gof}, we call this a `thermodynamical' characterisation of equilibrium, to contrast with the customary KMS condition's `dynamical' characterisation. For a detailed discussion of these, we refer to section \ref{characG}. The main idea is that the various proposals for generalised Gibbs states can be divided in terms of these two characterisations from an operational standpoint. Which characterisation one chooses to use in a given situation depends on the information or description of the system that one has at hand. For instance, if the description includes a 1-parameter flow of physical interest, then using the dynamical characterisation, i.e. satisfying the KMS condition with respect to it, will define statistical equilibrium with respect to it. The procedures defining these two characterisations can thus be seen as `recipes' for constructing a Gibbs state, and which one is more suitable depends on our knowledge of the system.} characterisation \cite{Kotecha:2018gof}, based on Jaynes' information-theoretic characterisation of equilibrium \cite{Jaynes:1957zza,Jaynes:1957zz}. We devote sections \ref{modstat} - \ref{rem} to discuss various aspects of the thermodynamical characterisation, including highlighting many of its favourable features \cite{Chirco:2018fns,Kotecha:2019vvn}, also compared to the other proposals. In fact, we point out how this characterisation can comfortably accommodate the past proposals for Gibbs equilibrium \cite{Kotecha:2019vvn}. \

In section \ref{GenTD}, we define the basic thermodynamic quantities which can be derived immediately from a generalised Gibbs state, without requiring any additional physical and/or interpretational inputs. We clarify the issue of extracting a single common temperature for the full system from a set of several of them, and end with the zeroth and first laws of a generalised thermodynamics.


\section{Rethinking time in mechanics} \label{sympmech}

Presymplectic mechanics is a powerful framework that is manifestly covariant, in the sense that it does not require non-relativistic concepts like absolute time to describe the physical dynamics of a system. It is used widely to describe dynamical systems which are generally covariant, or more generally are constrained systems with a set of gauge symmetries and a vanishing canonical Hamiltonian. In this section, we review the required details of the structure of classical, finite presymplectic systems,  based primarily on discussions in \cite{Kiefer:2004gr,Kotecha:2018gof,PhysRevD.43.442,Rovelli:2004tv,e18100370,Rovelli:2001bq}, to be used subsequently in sections \ref{classdep} and \ref{physeqm}. Importantly, we draw attention to the role of time in mechanics, and re-emphasise the fact that it requires a rethinking in the context of background independence, especially in its connections to equilibrium statistical mechanics. In particular, we stress that time is what parametrizes a history, and this parametrization is neither absolute, nor global, nor non-dynamical in generic background independent settings. \

We discuss a reformulation of standard Hamiltonian mechanics in terms of a presymplectic system, which is then further described in terms of an extended symplectic phase space with the dynamics encoded in a Hamiltonian constraint. Using this simple class of standard Hamiltonian systems, which are equipped with a unique global notion of time, we illustrate the main features of a generic constrained particle system that is devoid of any such physical evolution parameter. We then move on to a brief discussion of deparametrization (i.e. a procedure to extract a clock variable, where there is none a priori), in order to anticipate the main ideas that will be encountered in the subsequent chapters in the context of quantum gravity. We will be succinct in our presentation, and focus mainly on the essential takeaways relevant for the topic of this thesis. 

\subsection{Preliminaries: Presymplectic mechanics}

\subsubsection*{Hamiltonian symplectic mechanics}
Consider a non-relativistic dynamical system defined by, a symplectic space of states $\Gamma$ equipped with a symplectic (closed, non-degenerate) 2-form $\omega$, and a Hamiltonian $H$ generating the time evolution $\alpha_t$, written together as $(\Gamma, \omega, H)$. 
By non-relativistic, we specifically mean that there exists a preferred, unique dynamical flow, denoted here by $\alpha_t$, encoding the complete physics of the system. In other words, there exists a physical evolution parameter, denoted here by $t$, which is unique, global and non-dynamical (thus, external to the system), i.e. a Newtonian time. \

In this thesis, we deal with systems with an arbitrarily large, but finite, number of quanta. Thus, here also we are concerned with multi-particle systems with a finite total number. In the present section, for simplicity, we further restrict discussions to a single particle defined on a 1-dimensional smooth manifold $\mcC$, which is its configuration space. This can be extended to a multi-particle system, and for a configuration space with a finite dimension greater than 1. Therefore, here $(\Gamma, \omega, H)$ describes a single classical particle, where $\Gamma = \mcT^*(\mcC)$ is a cotangent bundle over the configuration space.\

Phase space $\Gamma$ is a symplectic manifold with a symplectic 2-form $\omega$. Local coordinates on $\mcC$ are $(q)$, and correspondingly on $\Gamma$ are $(q,p)$. The symplectic form can be written using these coordinates as, $\omega = dp \w dq$. The dynamics is encoded in a smooth, Hamiltonian function $H: \Gamma \to \mathbb{R}$. The equations of motion are given by,
\be \label{symeom} \omega(\mcX_H) = -dH \ee
where the notation on the left hand side denotes the interior product of the symplectic 2-form with a vector field. This is a compact, geometric rewriting of Hamilton's equations of motion, which can be seen as follows. The Hamiltonian vector field $\mcX_H$, as defined by the above equation, is given by
\be \mcX_H = \partial_p H \partial_q - \partial_q H \p_p  \ee
where, $\p_q \equiv \p/\p q$ denotes a partial derivative. Now, let $t \in \mathbb{R}$ parametrize the integral curves of $\mcX_H$ on $\Gamma$. Then, we get
\be \p_p H = \frac{dq(t)}{dt} \,, \quad \p_q H = - \frac{dp(t)}{dt} \ee
which are the Hamilton's equations in a familiar form. Notice that the variable parametrizing the dynamical trajectories is what we \emph{call} time (here, Newtonian). 

\subsubsection*{Parametrized presymplectic mechanics}

Let us parametrize the dynamical trajectories $(q(t),p(t))$ generated by $H$, and consider instead, graphs of them $(t, q(t),p(t))$, in an arbitrary parametrization $(t(\tau),q(\tau),p(\tau))$. Then, these graphs are curves in the enlarged space $\Sigma :=  \mathbb{R} \times \Gamma$, with local coordinates $(t,q,p)$. We know that equation \eqref{symeom} encodes the dynamics on $\Gamma$. Then on $\Sigma$, this same dynamics takes the form,
\be \label{presymeom} \omega_\Sigma(\mcX) = 0 \ee
where,
\be \omega_\Sigma = \omega - dH \w dt \ee
is a closed, degenerate (presymplectic) 2-form on $\Sigma$, and the vector field is proportional to,
\be \mcX = \p_t + \mcX_H \,. \ee
Equation \eqref{presymeom} means that $\mcX$ is a null vector field on $\Sigma$. Its integral curves, called orbits\footnote{In general for $k \geq 1$ number of scalar first class constraints, we have $k$-dimensional gauge orbits (null surfaces) on the constraint surface $\Sigma$. See for instance \cite{Henneaux:1992ig}, for details.}, admit an arbitrary parametrization, denoted above by $\tau$. Therefore, $(\Sigma,\omega_\Sigma)$ is a presymplectic manifold, and $\Sigma$ is known as the constraint surface. Notice that equations \eqref{symeom} $\Leftrightarrow$ \eqref{presymeom} (denoting, if and only if). In this sense, of having the same physical dynamics, the descriptions $(\Gamma,\omega,H)$ and $(\Sigma,\omega_\Sigma)$ with quantities as defined above, are equivalent for a non-relativistic Hamiltonian system. But for our purposes, what is important is to realise that the latter description allows for more generality, being able to also describe generic constrained systems. In fact, constrained systems admit a description of the former kind only if it is deparametrized, thus admitting a time or a clock structure. 
We will return to aspects of deparametrization shortly below, and later in sections \ref{depgft} and \ref{clock}. For now, let us introduce further \emph{extended} structures, which will be of particular value to us in the context of classical group field theory in section \ref{depgft}. 

\subsubsection*{Extended symplectic mechanics} 

The main idea of the extended symplectic formulation is to take all relevant variables, including a time or a clock (if it is present), as being dynamical, thus subject to the dynamical constraint of the system. This is certainly a reasonable feature from an operational point of view, because any clock that one may use is made of matter, and is thus dynamical. \

For the case of a non-relativistic Hamiltonian system $(\Gamma,\omega,H)$, we have an extended configuration space $\mcC_\ex = \mathbb{R} \times \mcC$, along with the associated extended phase space $\Gamma_\ex = \mcT^*(\mcC_\ex) = \mathbb{R}^2 \times \Gamma$ with local coordinates $(t,p_t,q,p)$. Evidently, time variable $t$ is now internal to the system's description. The symplectic structure on $\Gamma_\ex$ is,
\be \omega_\ex = \omega + dp_t \w dt \ee 
and, the constraint function takes a specific form, given by
\be \label{ccdep} C = p_t + H(q,p) \,. \ee
It is important to remark that this particular form of $C$, namely being first order in clock momentum and zeroth order in clock variable, is characteristic of having a dynamically viable clock or time structure. In other words, a system described by a constraint function of the above form is deparametrized, with a clock variable $t$. Also, when the constraint takes the form in equation \eqref{ccdep}, the presymplectic structure takes a specific form given by, $\omega_\Sigma = \omega - dH \w dt$, that we saw before. Furthermore in this case, the null vector field of $\omega_\Sigma$ is always of the form,
\be \mcX_C = \p_t + \mcX_H \ee
which is a direct consequence of the following crucial properties of the constraint function (in turn, arising as a direct consequence of its specific form in equation \eqref{ccdep}),
\be \label{ordersc} \frac{\p C}{\p p_t} = 1\,, \quad \frac{\p C}{\p t} = 0 \,. \ee
We also see that the constraint surface in this case takes the form describing a foliation in $t \in \mathbb{R}$, of the canonical phase space $\Gamma$, that is
\be \label{sigfol} \Sigma = \mathbb{R} \times \Gamma . \ee
Both, equations \eqref{ordersc} and \eqref{sigfol}, are characteristic of a system equipped with a good clock structure. \

In general, a constrained system is described by an extended configuration space $\mcC_\ex$, extended phase space $\Gamma_\ex = \mcT^*(\mcC_\ex)$ with a given symplectic 2-form $\omega_\ex$, and a smooth constraint function
\be C: \Gamma_\ex \to \mathbb{R} \ee
which is not necessarily of the specific form in equation \eqref{ccdep} above. If a given system is equipped with a clock variable, then it is now one of the configuration variables in $\mcC_\ex$. However, existence of such a variable is not required for a well-defined description of the physical system. A constrained dynamical system in its extended description is thus given by $(\Gamma_\ex, \omega_\ex, C)$. Restricting to the surface where the constraint function vanishes gives the constraint surface,
\be \Sigma = \Gamma_\ex|_{C=0} \subset \Gamma_\ex  \ee
which is not necessarily of the specific form in equation \eqref{sigfol}. $\Sigma$ is a presymplectic manifold, with its degenerate 2-form defined via a pull-back, $ \omega_\Sigma = \imath^* \omega_\ex$, of an embedding map $\imath : \Sigma \to \Gamma_\ex$. The equations of motion are 
\be \omega_\Sigma(\mcX_C) = 0 \,. \ee
Lastly, the reduced physical space, which is the space of orbits, can be obtained by factoring out the kernel of the presymplectic form.

We thus have at hand three kinds of formulations for classical mechanical systems, namely: Hamiltonian symplectic $(\Gamma,\omega,H)$, presymplectic $(\Sigma,\omega_\Sigma)$, and extended symplectic $(\Gamma_\ex,\omega_\ex,C)$. For non-relativistic systems, or more generally for systems characterised by a Hamiltonian constraint of the form \eqref{ccdep}, these three are all equivalent. But importantly, only the last two are general enough for generic background independent systems, since their description of the dynamics does not rely on the choice, or even the existence, of an explicit time evolution variable. 

\subsection{Deparametrization} \label{depggs}

Given a constrained system $(\Gamma_\ex,\omega_\ex,C)$, we have stressed above that its description may not necessarily include a good clock structure. And that even without such a structure, the extended (or presymplectic) formulation describes fully the dynamical system. In such a priori `timeless' systems then, we may still want to extract a clock variable. This is called deparametrization. In the present context for instance, deparametrization is valuable to define physical equilibrium states with respect to clock Hamiltonians (see sections \ref{characG}, \ref{past} and \ref{physeqm} for more discussions and examples).

From the above discussions of constrained mechanical systems, especially in comparison with extended non-relativistic systems, we can identify the core, universal strategy behind deparametrization: to reformulate, either exactly or under certain approximations, the dynamical constraint of the system, to bring it to the form of equation \eqref{ccdep}, i.e.
\be \label{1deparamC1} C \longmapsto C_{\dep} = p_\lambda + H(q_{\text{can}},p_{\text{can}}) \ee
where $\lambda$ is the clock variable, $H$ is the clock Hamiltonian generating evolution in clock time $\lambda$, and $(q_{\text{can}},p_{\text{can}})$ denote the phase space variables of the canonical system to which the clock variable is external. 

A simple illustrative example is a classical relativistic particle on spacetime, with its covariant dynamics given by, \be C = p^2 - m^2 \,,  \ee
which is defined on an extended phase space $\Gamma_\ex$ coordinatised by $(x^\mu,p_\mu)$, with $\mu = 0,1,2,3$ and signature $(+,-,-,-)$. In this case, the complete dynamical, presymplectic description does not require deparametrization, i.e. given the extended phase space along with $C$, its constraint surface is well-defined. But the point is that this system is deparametrizable, which means that it is possible to bring $C$ to a manifestly non-covariant form, here without changing the physics that is captured completely by the constraint surface $(\Sigma,\omega_\Sigma)$. The full constraint can be rewritten as \be C_{\dep} = p_0 + \sqrt{\textbf{p}^2 + m^2} \ee which now describes the same relativistic particle system but in a fixed Lorentz frame where the configuration variable $x^0$ has been chosen as the clock variable and the corresponding clock dynamics is dictated by the Hamiltonian $H = \sqrt{\textbf{p}^2 + m^2}$.\

A different example of a system which by itself is non-deparametrizable, i.e. it doesn't naturally admit a good clock without further approximations, is the so-called timeless double pendulum \cite{PhysRevD.42.2638,PhysRevD.43.442,Rovelli:2004tv}. It is defined by an extended configuration space $\mcC_\ex = \mathbb{R}^2 \ni (q_1,q_2)$, phase space $\Gamma_\ex = \mcT^*(\mcC_\ex)$ with $\omega_\ex = \sum_{a=1,2}dp_a \w dq^a$, and a dynamical constraint\footnote{Notice that this system is fundamentally different from an analogous, non-relativistic one with Newtonian time $t$, which would instead be given by a constraint of the form, $C = p_t + \frac{1}{2}(q_1^2 + q_2^2 + p_1^2 + p_2^2) - k$.} 
\be C = \frac{1}{2}(q_1^2 + q_2^2 + p_1^2 + p_2^2) - k \ee 
for a real constant $k$. The constraint surface in this case is, $\Sigma = \Gamma_\ex|_{C=0} \cong S^3$, a 3-sphere with radius $\sqrt{2k}$. Clearly, $\Sigma$ is compact and does not admit a form \eqref{sigfol} with a foliation that is characteristic of having a time structure. Therefore, the system so described does not naturally have any internal variable which can play the role of time, \emph{before} any approximations. For instance, if one were to approximate the dynamics by the following constraint function,
\be C_{\dep} = p_2 + \frac{1}{2}(q_1^2 + p_1^2) - k \equiv p_2 + H(q_1,p_1)  \ee
then, in this dynamical regime (if it exists) described by $C_{\dep}$, variable $q_2$ acts as a good clock parameter with the physical evolution being generated by $H$. \

Finally, we note that in the case of a system composed of many particles, there may not exist a single global clock, even if each particle is deparametrized separately and thus equipped with its own individual clock. In this case then, one would expect to have to sync the different clocks in order to define a single common clock. This issue can be investigated in a general multi-particle setting, continuing from the discussions above \cite{Chirco:2013zwa,Chirco:2016wcs}. However, we will discuss it in detail directly in the context of classical GFTs \cite{Kotecha:2018gof} in section \ref{classdep}.


\section{Generalised Gibbs equilibrium} \label{GenGE}

Equilibrium states are a cornerstone of statistical mechanics and thermodynamics. They are vital in the description of macroscopic systems with a large number of microscopic constituents. In particular, Gibbs states have a vast applicability across a broad range of fields such as condensed matter physics, quantum information, tensor networks, and (quantum) gravity, to name a few. They are special, being the unique class of states in finite systems satisfying the KMS condition\footnote{The algebraic KMS condition \cite{Haag:1992hx,Bratteli:1996xq} is known to provide a comprehensive characterisation of statistical equilibrium in systems of arbitrary sizes, as long as there exists a well-defined 1-parameter dynamical group of automorphisms of the system. This latter point, of the required existence of a preferred time evolution of the system, is exactly the missing ingredient in generic background independent cases. Moreover, we observe that the automorphism group in the definition of KMS condition does not, technically, need to be identified with time evolution. Thus KMS equilibrium can be defined for more general symmetries, even though the standard definition would be physically more natural in systems with a unique definition of time. We refer to sections \ref{characG} and \ref{rem} for more details.}. Furthermore, usual coarse-graining techniques also rely on the definition of Gibbs measures. In treatments closer to hydrodynamics, one often considers the full (non-equilibrium) system as being composed of many interacting subsystems, each considered to be at local equilibrium. While in the context of renormalisation group flow treatments, each phase at a given scale, for a given set of coupling parameters is also naturally understood to be at equilibrium, each described by (an inequivalent) Gibbs measure. Given this physical interest in Gibbs states, the question then is how to define them for background independent systems. 
 

\subsection{Characterising Gibbs states} \label{characG}

Let us begin with a discussion on how to characterise completely statistical states of the Gibbs form, based on insights from both standard and covariant statistical mechanics \cite{Kotecha:2018gof}. This discussion is not restricted to any particular quantum gravity formalism, or even to classical or quantum sectors for systems on spacetime. Rather, it attempts to present the strategies employed in past studies for defining Gibbs states, along with a variety of examples, in a coherent way. The main goal here is to reconsider the standard characterisations that are well-known to us, and attempt to understand them from a broader perspective, in order to then generalise them to background independent systems, including discrete quantum gravity. \

Gibbs states, $\rho_\beta = \frac{1}{Z_\beta} e^{-\beta \mcO}$ (where $\mcO$ is not necessarily a Hamiltonian), can be categorised according to two main criteria, respectively primary and secondary: 

A. whether or not the state is a result of considering an associated pre-defined flow or transformation of the system; \

B. the nature of the functions or operators $\mcO$ in the exponent characterising the state, specifically whether these quantities encode the physical dynamics of the system, or refer to some structural properties. \

Categories A and B are mutually independent, in the sense that a single system could simultaneously be both of types A and B. Details of these two categories, their respective subclasses and related examples follow. \

Let us first look at the primary category A, which can be considered at a higher footing than the secondary B because the contents of classification under A are the actual construction procedures or `recipes' used to arrive at a resultant Gibbs state. Moreover, it is within A where we observe that the known Jaynes' entropy maximisation principle \cite{Jaynes:1957zza,Jaynes:1957zz} could prove to be especially useful in background independent contexts such as in discrete quantum gravity frameworks. Under A, we can identify two recipes to construct, or equivalently, ways to completely characterise, a Gibbs state depending on the information that we have at hand for a given system. 

\subsubsection*{A1. Dynamical characterisation: Use of Kubo-Martin-Schwinger condition}

The KMS condition \cite{Kubo:1957mj,Martin:1959jp,Haag:1967sg,Haag:1992hx,Bratteli:1996xq} is formulated in terms of a 1-parameter group of automorphisms of the system.  \

Let $\omega$ be an algebraic state\footnote{Algebraic states $\omega: \mathcal{A} \to \mathbb{C}$, are complex-valued, linear, positive, normalised functionals over an algebra. Linearity is: $\omega[z_1A + z_2B] = z_1\omega[A] + z_2\omega[B] \;\; \forall z_1,z_2 \in \mathbb{C}$. Positivity is: $\omega[A^\dag A] \geq 0 \;\; \forall A \in \mathcal{A}$. Normalisation is: $\omega[\mathcal{I}] = 1$, where $\mathcal{I}$ is the identity element of $\mathcal{A}$. \cite{Bratteli:1979tw} \label{FNstate}} over a C*-algebra\footnote{A C*-algebra is a norm complete *-algebra, equipped with the C*-norm property: $||A^* A|| = ||A||^2$. A normed *-algebra is a *-algebra equipped with a *-norm. A *-algebra is an algebra (i.e. vector space over $\mathbb{C}$, with an associative and distributive multiplication law) equipped with a *-operation (also called, involution or adjoint) such that: $A^{**}=A$, $(AB)^*=B^*A^*$, and $(z_1A + z_2B)^* = \bar{z}_1A^* + \bar{z}_2B^*$. A *-norm is a norm $||.||$ with one additional property: $||A^*|| = ||A||$. We note that in this thesis, we denote the *-operation by a dagger $\dag$. \cite{Bratteli:1979tw}} $\mathcal{A}$, and $\alpha_t$ be a 1-parameter group of *-automorphisms\footnote{A *-automorphism is a bijective *-homomorphism of the algebra into itself. We refer to appendix \ref{autapp} for details. \cite{Bratteli:1979tw}} of $\mathcal{A}$. Consider the strip $\mathfrak{I} = \{z \in \mathbb{C} \; | \; 0 < \text{Im}(z) < \beta, \;\beta \in \mathbb{R}_{\geq 0}\}$, and let $\bar{\mathfrak{I}}$ denote its closure, such that $\bar{\mathfrak{I}} = \mathbb{R}$ if $\beta = 0$. Then, $\omega$ is an $\alpha$-KMS state at value $\beta$, if for any pair $A,B \in \mathcal{A}$, there exists a complex function $F_{AB}(z)$ which is analytic on $\mathfrak{I}$, and continuous and bounded on $\bar{\mathfrak{I}}$, such that it satisfies the KMS boundary conditions: 
\be F_{AB}(t) = \omega[A \, \alpha_t(B)] \,, \quad F_{AB}(t + i\beta) = \omega[\alpha_t(B) \, A]  \ee
for all $t \in \mathbb{R}$ \cite{Bratteli:1996xq}. We note that an $\alpha$-KMS state satisfies stationarity, i.e. $\omega[\alpha_t A] = \omega[A]$ (for all $t$), which captures the simplest notion of equilibrium (see appendix \ref{appkmsgibb} for some more details). If in the given description of a system, one can identify a relevant set of transformations (not necessarily Hamiltonian) with respect to which one is interested in defining an equilibrium state, then one asks for the state to satisfy the KMS condition with respect to a continuous, unitarily implementable 1-parameter (sub-) group of the said transformations. This gives (uniquely, for a certain class of systems as detailed in appendix \ref{appkmsgibb}, e.g. finite systems on spacetime), a Gibbs state of the form $\rho \propto e^{-\beta \mathcal{G}}$, where  $\mathcal{G}$  is the self-adjoint generator of the flow, and $\beta \in \mathbb{R}$. Notice that the `inverse temperature' $\beta$ enters formally as the periodicity in the flow parameter $t$, regardless of its interpretation. \

Thus, this characterisation is strictly based on the existence of a suitable pre-defined flow of the configurations of the system and then imposing the KMS condition with respect to it. These transformations could correspond to physical or structural properties of the system (see category B below). Simple examples are, respectively, the physical time flow $e^{iHt}$ in a non-relativistic system where $H$ is the Hamiltonian, which gives rise to the standard Gibbs equilibrium state of the form $e^{-\beta H}$; and a $U(1)$ gauge flow $e^{i N \theta}$ where $N$ is a number operator, which would lead to an equilibrium state of the form $e^{-\beta N}$. 
 
\subsubsection*{A2. Thermodynamical characterisation: Use of maximum entropy principle under a given set of constraints} 

Consider a situation wherein the given description of a system does not include relevant symmetry transformations, or (even if such symmetries exist, which they usually do) that we are interested in those properties of the system which are not naturally associated to sensible flows, in the precise sense of being generators of these flows. Examples of the latter are observables such as area, volume or mass. Geometric operators like volume are of special interest in the context of quantum gravity since they may be instrumental for statistically extracting macroscopic geometric features of spacetime regions from quantum gravity microstates. In such cases then, what characterises a Gibbs state and what is the notion of equilibrium encoded in it? \

In order to construct a Gibbs state here, where operationally we may only have access to a set of constraints fixing the mean values of a set of functions or operators $\{\langle \mathcal{O}_a \rangle_\rho = \overline{\mathcal{O}_a}\}_{a=1,2,...,\ell}$ (in classical or quantum descriptions respectively), one must rely on Jaynes' principle \cite{Jaynes:1957zza,Jaynes:1957zz} of maximising the entropy $S[\rho] = -\langle \ln \rho \rangle_\rho$ while simultaneously satisfying the above constraints. The angular brackets denote statistical averages in a state $\rho$ defined on the state space of the system, be it a phase space in the classical description or a Hilbert space in the quantum one. Undertaking this procedure, one arrives at a Gibbs state $\rho = e^{-\sum_a \beta_a \mathcal{O}_a}$ (where one of the $\mathcal{O}$'s is the identity fixing the normalisation of the state as will be detailed below in section \ref{maxent}). We can see that here the inverse temperatures $\beta_a$ enter formally as Lagrange multipliers. Also, the averages $\overline{\mathcal{O}_a}$ with parameters $\beta_a$, and other quantities derived from them, can be understood as thermodynamic variables defining a macrostate of the system, and can take on the same formal roles as in usual statistical mechanics and thermodynamics. But their exact interpretation would depend on the context. The identification and interpretation of such \emph{relevant} quantities \emph{is} in fact the non-trivial aspect of the problem, particularly in quantum gravity. \

Notice that this characterisation is strictly independent of the existence of any pre-defined transformations or symmetries of the underlying microscopic system, as long as there is at least one function or operator (identified as relevant) whose statistical average is assumed (or known) to be fixed at a certain value. Consequently, this characterisation could be most useful in background independent settings, exactly since it is based purely on Jaynes' information-theoretic method. Finally, we note that in this characterisation a notion of equilibrium is implicit in the requirement that a certain set of observable averages remain constant, i.e. it is implicit in the existence of the constraints $\langle \mathcal{O}_a \rangle_\rho = \overline{\mathcal{O}_a}$ (see for example the discussion in section II in \cite{tribus}). We will undertake a detailed discussion of the thermodynamical characterisation, and formulate aspects of a generalised equilibrium statistical mechanical framework based on it, along with basic features of its generalised thermodynamics, in the remaining sections of this chapter. \

For now, let us summarise the above classifications and make some additional remarks. If the aim is to construct Gibbs states for a system of many quanta (whatever they may be), then the two characterisations, dynamical and thermodynamical, under category A offer us two formally independent strategies to do so. Based on our knowledge of a given system, we may prefer to use one over the other. If there is a known set of symmetries with respect to which one is looking to define equilibrium, then the technical route one takes is to construct a state satisfying the KMS condition with respect to (a 1-parameter subgroup of) the symmetry group. The result of using this recipe in a finite system (cf. remarks in appendix \ref{appkmsgibb}) is a Gibbs state $e^{-\beta \mathcal{G}}$, characterised by the generator $\mathcal{G}$ of the 1-parameter flow of these transformations and the periodicity parameter $\beta$. On the other hand, if one does not have interest in or access to any particular symmetry transformations of the system, but has a partial knowledge about the system in terms of a set of observable averages $ \overline{\mathcal{O}_a} $, then one employs the principle of maximising Shannon or von Neumann entropy under the given set of constraints. The resultant statistical state is again of a Gibbs exponential form $e^{-\sum_a \beta_a \mathcal{O}_a}$, now characterised by the set of observables $\mathcal{O}_a$ and Lagrange multipliers $\beta_a$. \

Notice that given a Gibbs state, constructed say from recipe A1, then once it is already defined, it also satisfies the thermodynamic condition of maximum entropy \cite{Haag:1992hx,Bratteli:1996xq}. Similarly, a Gibbs state defined on the basis of A2, after it is constructed also satisfies the KMS condition with respect to a flow that is derived from the state itself (due to Tomita-Takesaki theorem \cite{Haag:1992hx,Bratteli:1996xq}). For instance, consider the standard non-relativistic Gibbs state $ e^{-\beta H}$, which is constructed by satisfying the KMS condition with respect to unitary time translations $e^{iHt}$. That is, this state is classified as A1 since its construction relies on the KMS condition. But, once this state is defined, it is also the one that maximises the entropy under the constraint $\langle H \rangle = E$. Now consider an example of a state $ e^{-\beta V}$, where $V$ is say spatial volume. This state is derived as a result of maximising the information entropy under the constraint $\langle V \rangle = v$, hence it is classified as A2. Once this state is defined, one can extract a flow \emph{from} the state, with respect to which it will satisfy the KMS condition.\footnote{In the context of covariant statistical mechanics, the utility of this observation has been presented in \cite{Rovelli:1993ys,Connes:1994hv,Rovelli:2012nv}, and is the crux of the thermal time hypothesis.} This is the modular flow $e^{i\beta V \tau}$, where $\tau$ is the modular flow parameter.\footnote{In a classical phase space description of a system, the modular flow is the integral curve of the vector field $\mcX_V$ defined by $\omega(\mcX_V) = -dV$, where $\omega$ is the symplectic form and $V$ is a smooth function. In this case naturally the flow $e^{\tau \mcX_V}$ is in terms of the vector field $\mcX_V = \partial_\tau$. In the quantum C*-algebraic description (or in a specific Hilbert space representation of it), the modular flow is that of the Tomita-Takesaki theory.} Therefore, classifications A1 and A2 refer to the construction procedures employed as per the situation at hand. Once a Gibbs state is constructed using any one of the two procedures, then technically it will satisfy both the KMS condition with respect to a flow,\footnote{Whether the flow parameter has a reasonable physical interpretation is a separate issue, and would depend on the specific context.} and maximisation of entropy. \

Now, given that a Gibbs state can be constructed as a result of either the dynamical or the thermodynamical recipe, one can consider the nature of the functions or operators $\mathcal{O}$ that characterise it. This is the content of classification under category B. 

\subsubsection*{B1. Physical state} 

If $\mathcal{O}$ is associated with the physical dynamics of the system, i.e. it dictates, partially or completely, the dynamical model of the system, then the associated Gibbs state can be understood as encoding, partially or completely, a physical notion of equilibrium. In a standard non-covariant system, $\mathcal{O}$ would simply be the Hamiltonian. While in a covariant system, if it is deparametrizable with a suitable choice of a good clock, then $\mathcal{O}$ would be the associated clock Hamiltonian. Overall, this particular classification refers explicitly, and in the definition of the Gibbs state, to the physical dynamical evolution of the system. This is encoded in a relevant model-dependent function or operator, whether it is a conventional Hamiltonian or a deparametrizable constraint. 

\subsubsection*{B2. Structural state}

If $\mathcal{O}$ are quantities referring to kinematic or structural properties of the system, and not directly to the specific physical dynamical model of it, then the associated Gibbs states are understood to be structural. Examples of structural transformations are generic rotations or translations of the base manifold of the theory. Examples of quantities that are not necessarily associated to symmetry transformations would be observables like area or volume. 

\subsubsection*{Examples}

Four different types of Gibbs states can be constructed from combinations of classifications under categories A and B. Let us mention a variety of examples, including the ones constructed as part of this thesis.  \

\textbf{A1-B1:} Gibbs states with respect to physical time translations, as considered in standard statistical mechanics \cite{Landau:1980mil,Haag:1992hx,Bratteli:1996xq}; with respect to a clock time in deparametrized systems, with examples constructed in covariant statistical mechanics on spacetime \cite{Rovelli:2012nv,Josset:2015uja,Chirco:2016wcs} (section \ref{past}), and in group field theory quantum gravity \cite{Kotecha:2018gof} (section \ref{physeqm}). \

\textbf{A2-B1:} Gibbs states with respect to 4-momenta of relativistic particles in covariant statistical mechanics \cite{Montesinos:2000zi}; with respect to momentum map associated with diffeomorphism group for parametrized field theory \cite{Chirco:2019tig,melechirco}; with respect to a dynamical projector \cite{Oriti:2013aqa,Chirco:2018fns,Chirco:2019kez}, and kinetic and vertex operators \cite{Kotecha:2019vvn} in group field theory (section \ref{effgft}). \

 \textbf{A1-B2:} Gibbs states with respect to U(1) symmetry generated by the number observable in standard statistical mechanics \cite{Landau:1980mil,Haag:1992hx,Bratteli:1996xq}; with respect to internal translation automorphisms in algebraic group field theory \cite{Kotecha:2018gof} (section \ref{inttrans}). \
 
 \textbf{A2-B2:} Souriau's Gibbs states with respect to kinematical symmetries \cite{souriau1,e18100370}; Gibbs states with respect to gauge-invariant observables in covariant statistical mechanics on spacetime \cite{Montesinos:2000zi}; with respect to spatial volume for pressure ensemble in standard statistical mechanics \cite{Jaynes:1957zza}; with respect to geometric area and volume operators in loop quantum gravity \cite{Montesinos:2000zi,PhysRevD.55.3505}; with respect to positive, extensive operators in group field theory, e.g. spatial volume operator \cite{Kotecha:2018gof} (section \ref{posgibbs}); with respect to classical closure and half-link gluing constraints in group field theory \cite{Chirco:2018fns,Chirco:2019kez} (section \ref{congibb}). \\
 
In light of the fact that these classifications comprehensively span all possible kinds of Gibbs states, we call them generalised Gibbs states.


\subsection{Past proposals} \label{past}

We now give an overview \cite{Kotecha:2019vvn} of the different proposals for defining statistical equilibrium based on studies in covariant statistical mechanics, in order to better contextualise our proposal of the thermodynamical characterisation which is the focus of the subsequent sections. These proposals rely on various different principles originating in standard non-relativistic statistical mechanics, extended to a constrained setting. \

The first proposal \cite{Rovelli:1993ys,Chirco:2013zwa} was based on the idea of statistical independence of arbitrary (small, but macroscopic) subsystems of the full system, in a classical setting. The notion of equilibrium is taken to be characterised by the factorisation property of the state,  
\be \rho = \rho_1 \rho_2 \ee 
for any two subsystems 1 and 2; and where $\rho_1$ and $\rho_2$ are taken to be defined on mutually exclusive subregions of the full phase space, so that we additionally have  \be \{\rho_1,\rho_2\}=0 \,, \ee or equivalently, \be [\mcX_{\rho_1},\mcX_{\rho_2}]=0 \,, \ee 
where, $\omega(\mcX_{\rho_{1,2}}) = d\ln \rho_{1,2}$ , $\omega$ is the symplectic form on the phase space of the full system, and $\{.,.\}$ is the Poisson bracket. Then, the whole system is said to be at equilibrium if any one of its subsystems is statistically independent from all the rest. We note that this characterisation for equilibrium is related to an assumption of weak interactions \cite{Landau:1980mil}. \

This same dilute gas assumption is integral also to the Boltzmann method of statistical mechanics. It characterises equilibrium as the most probable distribution, that is one with maximum entropy. This method is used in \cite{Montesinos:2000zi} to study a gas of constrained particles\footnote{We remark that except for this one work, all other studies in spacetime covariant statistical mechanics are carried out from the Gibbs ensemble point of view.}. Even though this method relies on maximising the entropy like the thermodynamical characterisation, it is more restrictive than the latter, as will be made clear in section \ref{rem}. \

The work in \cite{Rovelli:2012nv} puts forward a characterisation for a physical equilibrium state. The suggestion is that, $\rho$ (itself a well-defined state on the physical, reduced state space) is said to be a physical Gibbs state if $(i)$ its modular Hamiltonian 
\be h = -\ln \rho \,, \ee 
is a smooth function on the physical state space, and $(ii)$ $h$ is such that there exists a (local) clock function $T(\textbf{x})$ on the extended state space, with its conjugate momentum $p_T(\textbf{x})$, such that $h$ is proportional to $p_T$. Importantly, when this is the case the modular flow (`thermal time' \cite{Rovelli:1993ys,Connes:1994hv}) is a geometric flow foliating spacetime, in which sense $\rho$ is said to be `physical'. Notice that the built-in strategy here is to define KMS equilibrium in a deparametrized system, since by construction it identifies a state's modular Hamiltonian with a (local) clock Hamiltonian on the base spacetime manifold. Thus it is an example of using the dynamical characterisation, associated with a clock Hamiltonian in a deparametrized system with constraint of the form \eqref{ccdep}.  \

Another strategy \cite{Josset:2015uja} is based on the use of the ergodic principle and introduction of clock subsystems to define clock time averages. This characterisation relies on the validity of a postulate, even if traditionally a fundamental one, like two of the previous ones that were based on the physical assumption of weak interactions. \

Finally, the proposal in \cite{Haggard:2013fx} interestingly characterises equilibrium by a vanishing information flow between interacting histories. The notion of information used is that of Shannon (entropy), 
\be I = \ln N ,\ee
where $N$ is the number of microstates traversed in a given history during interaction. Equilibrium between two histories 1 and 2 is encoded in a vanishing information flow, \be \delta I = I_2 - I_1 = 0 \,. \ee This characterisation of equilibrium is evidently information-theoretic, even if relying on an assumption of weak interactions. Moreover it is much closer to our thermodynamical characterisation, because the condition of vanishing $\delta I$ is nothing but an optimisation of information entropy. \

These different proposals, along with the thermal time hypothesis \cite{Rovelli:1993ys,Connes:1994hv}, have led to some remarkable results, like recovering the Tolman-Ehrenfest effect \cite{Rovelli:2010mv,Chirco:2016wcs}, relativistic J\"uttner distribution \cite{Chirco:2016wcs} and Unruh effect \cite{Martinetti:2002sz}. However, they all assume the validity of one or more principles, postulates or assumptions about the system. Moreover, none (at least presently) seems to be general enough like the proposal below, so as to be implemented in a full quantum gravity setup, while also accommodating within it the rest of the proposals.


\subsection{Thermodynamical characterisation} \label{maxent}

This brings us to the proposal of characterising a generalised Gibbs state based on a constrained maximisation of information (Shannon or von Neumann) entropy  \cite{Kotecha:2018gof,Chirco:2018fns,Chirco:2019kez,Kotecha:2019vvn}, along the lines advocated by Jaynes \cite{Jaynes:1957zza,Jaynes:1957zz} purely from the perspective of evidential statistical inference. Jaynes' approach is fundamentally different from other more traditional ones of statistical physics. So too is the thermodynamical characterisation, compared with the others outlined above, as will be exemplified in the following. It is thus a new proposal for background independent equilibrium \cite{Kotecha:2018gof,Haggard:2018thr}, which has the potential of incorporating also the others as special cases, from the point of view of constructing a Gibbs state \cite{Kotecha:2019vvn}. \

Consider a macroscopic system with a large number of constituent microscopic degrees of freedom. Our (partial) knowledge of its macrostate is given in terms of a finite set of averages $\{\langle \mcO_a \rangle = U_a\}$ of the observables that we have access to. Jaynes suggests that a fitting probability estimate (which, once known, will allow us to infer also the other observable properties of the system) is not only one that is compatible with the given observations, but also that which is least-biased in the sense of not assuming any more information about the system than what we actually have at hand, namely $\{U_a\}$. In other words, given a limited knowledge of the system (which is always the case in practice for any macroscopic system), the least-biased probability distribution compatible with the given data should be preferred. As shown below, this turns out to be a Gibbs distribution with the general form \eqref{genstate1}, which we refer to as generalised Gibbs states. \

Let $\Gamma$ be a symplectic phase space (be it extended or reduced), which is a cotangent bundle over a finite-dimensional, smooth configuration manifold. Consider a finite set of smooth real-valued functions $\mcO_a$ defined on $\Gamma$, with $a=1,2,...,\ell$. Denote by $\rho$ a smooth statistical density (real-valued, positive and normalised function) on $\Gamma$, to be determined. Then, the prior on the known macrostate gives a finite number of constraints,
\be \label{const} \langle \mcO_a \rangle_\rho = \int_{\Gamma} d\lambda \; \rho \, \mcO_a = U_a \,,  \ee
along with the normalisation constraint for the state,
\be \label{constnorm} \braket{1}_\rho = 1 \,,  \ee
where $d\lambda$ is a Liouville measure on $\Gamma$, and the integrals are taken to be well-defined. Further, $\rho$ has an associated Shannon entropy 
\be \label{en} S[\rho] = -\langle \ln \rho \rangle_\rho \;. \ee
By understanding $S$ to be a measure of uncertainty quantifying our ignorance about the details of the system, the corresponding bias is minimised (compatibly with the prior data) by maximising $S$ (under the set of constraints \eqref{const} and \eqref{constnorm}) \cite{Jaynes:1957zza}. Using the Lagrange multipliers technique, this amounts to finding a stationary solution for the following auxiliary functional
\be \label{auxfn}
L[\rho,\beta_a,\kappa] = S[\rho] - \sum_{a=1}^{\ell} \beta_a (\langle \mcO_a \rangle_{\rho} - U_a) - \kappa(\langle 1 \rangle_\rho - 1)
\ee 
where $\beta_a, \kappa \in \mathbb{R}$ are Lagrange multipliers. Then, requiring stationarity\footnote{Notice that requiring stationarity of $L$ with variations in the Lagrange multipliers implies fulfilment of the constraints \eqref{const} and \eqref{constnorm}. These two `equations of motion' of $L$ along with the one determining $\rho$ (coming from stationarity of $L$ with respect to $\rho$) provide a complete description of the system at hand.} of $L$ with respect to variations in $\rho$ gives a generalised Gibbs state
\be \label{genstate1}
 \rho_{\{\beta_a\}} = \frac{1}{Z_{\{\beta_a\}}} e^{-\sum \limits_{a=1}^{\ell} \beta_a \mcO_a} \ee 
 with partition function,
 \be Z_{\{\beta_a\}} \equiv \int_{\Gamma} d\lambda \; e^{-\sum_a \beta_a \mcO_a}  = e^{1+\kappa} \,,\ee 
where as is usual, normalisation multiplier $\kappa$ is a function of the remaining multipliers. The partition function $Z_{\{\beta_a\}}$ encodes all properties of the system in principle.

Analogous arguments hold for quantum mechanical systems and the above scheme can be implemented directly \cite{Jaynes:1957zz}, as long as the operators under consideration are such that the relevant traces are finite on a representation Hilbert space. Statistical states are density operators (self-adjoint, positive, and trace-class operators, with $\Tr(\rho)=1$) on the Hilbert space. Statistical averages for (self-adjoint) observables  $\h{\mcO}_a$ are now,
 \be \label{qconst2} \langle \h{\mcO_a} \rangle_\rho = \Tr(\h{\rho} \, \h{\mcO}_a) = U_a \,. \ee
Following the constrained optimisation method again gives a Gibbs density operator,
 \be \label{genstate2}  \h{\rho}_{\{\beta_a\}} = \frac{1}{Z_{\{\beta_a\}}} e^{-\sum \limits_{a=1}^{\ell} \beta_a \h{\mcO}_a} \,. \ee

A generalised Gibbs state can thus be defined, characterised fully by a finite set of observables of interest $\mcO_a$, and their conjugate generalised `inverse temperatures' $\beta_a$. Given this class of equilibrium states, it should be evident that some thermodynamic quantities (like generalised `energies' $U_a$) can be identified immediately. Aspects of a generalised thermodynamics will be discussed in section \ref{GenTD}.  \

We note that there are two key features of this characterisation. First is the use of evidential (epistemic) probabilities, thus taking into account the given evidence $\{U_a\}$. This interpretation, that statistical states are states of knowledge \cite{levine1979maximum,Uffink2011-UFFSPS,PhysRevA.75.032110}, is innate to Jaynes' method \cite{Jaynes:1957zza,Jaynes:1957zz}. Second is a preference for the least-biased (or most ``honest'') distribution out of all the different ones compatible with the given evidence. It is not enough to arbitrarily choose any that is compatible with the prior data. An aware observer must also take into account their own ignorance, or lack of knowledge honestly, by maximising the information entropy.


\subsection{Modular flows and stationarity} \label{modstat}

Given a generalised Gibbs state, a natural question that arises is, with respect to which flow or transformations of the system is this state stationary. We know that any density distribution or operator is stationary with respect to its own modular flow. In fact by the Tomita-Takesaki theorem \cite{Haag:1992hx,Bratteli:1996xq}, \emph{any} faithful algebraic state over a von Neumann algebra is KMS with respect to its own 1-parameter modular (Tomita) flow. \

In the classical case then, a generalised Gibbs state $\rho_{\{\beta_a\}}$ of the form \eqref{genstate1}, is stationary with respect to the local flow $e^{t\mcX_\rho}$ (parametrized by $t \in \mathbb{R}$) on a given symplectic phase space $\Gamma$, generated by vector field $\mcX_\rho$ that is defined by the equation
\be \label{modXh} \omega(\mcX_\rho) = -dh \,, \ee
where $h$ is the modular Hamiltonian of the state,
\be \label{moddh} h = \sum_a \beta_a \mcO_a \,. \ee
Then, stationarity with respect to the modular flow is evident from,
\be \mcX_\rho(\rho_{\{\beta_a\}}) = \{\rho_{\{\beta_a\}},h\} = 0  \ee
where, $\{.,.\}$ is a Poisson bracket on the algebra of smooth functions over $\Gamma$. Now, the individual flows generated by $\mcX_a$ are associated with the different observables via the defining equation
\be \label{modXO} \omega(\mcX_a) = -d\mcO_a \,. \ee
Then, we have that the modular vector field is in general a linear superposition of the separate ones, weighted by their respective inverse temperatures,
\be \label{supmod} \mcX_\rho = \sum_a \beta_a \mcX_a  \ee
using equations \eqref{modXh}, \eqref{modXO}, and, linearities of the symplectic form $\omega$ and the exterior derivative. In particular, $\rho_{\{\beta_a\}}$ is \emph{not} stationary in general with respect to the individual flows $\mcX_{a'}$ generated by $\mcO_{a'}$, that is
\be \mcX_{a'}(\rho_{\{\beta_a\}}) = \{\rho_{\{\beta_a\}},\mcO_{a'}\} \neq 0 \,. \ee

However, $\rho_{\{\beta_a\}}$ becomes stationary with respect to each of these individual flows $\mcX_{a'}$ (in addition to always being stationary with respect to the full modular flow), if the generators of the individual flows commute with each other, that is 
\be \label{indistation} \{\mcO_{a''}, \mcO_{a'}\}=0 \;\; (\text{for all} \; a',a'')  \quad \Rightarrow \quad \mcX_{a'}(\rho_{\{\beta_a\}}) = 0 \;\; (\text{for any} \; a') \,.  \ee 
See appendix \ref{appMod} for details. \

In the quantum case, a generalised Gibbs density operator $\h{\rho}_{\{\beta_a\}}$ of the form \eqref{genstate2}, satisfies the KMS condition with respect to its modular flow\footnote{Notice that $ \Delta \equiv e^{\h{h}} $ is the modular operator of Tomita-Takesaki theory \cite{Haag:1992hx,Bratteli:1996xq}.} $e^{i\h{h}t}$ (parametrized by $t\in \mathbb{R}$) on a given Hilbert space, generated by the modular Hamiltonian operator,
\be \label{moddhq} \h{h} = \sum_a \beta_a \h{\mcO}_a \,.\ee
Then, the state is stationary with respect to the modular flow,
\be [\h{\rho}_{\{\beta_a\}},\h{h}] = 0 \ee
while, it is not at equilibrium with respect to the separate flows in general,
\be [\h{\rho}_{\{\beta_a\}},\h{\mcO}_{a'}] \neq 0 \ee
where, $[.,.]$ denotes a commutator bracket. But as before, we have that
\be \label{indistation2} [\h{\mcO}_{a''},\h{\mcO}_{a'}] = 0 \;\; (\text{for all} \; a',a'')  \quad \Rightarrow \quad [\h{\rho}_{\{\beta_a\}},\h{\mcO}_{a'}] = 0 \;\; (\text{for any} \; a') \,. \ee
See appendix \ref{appMod} for details. \

Therefore, we see that generalised Gibbs states, both classical and quantum, are stationary with respect to the modular flow extracted from the state itself, as expected from statistical mechanics. But we also see that none of the individual observables are in any way preferred or special, thus giving a generalised notion of equilibrium, with all the characterising observables being on an equal footing. \

In the special case when the different observable generators $\mcO_a$ are independent of each other, then the state is additionally stationary with respect to each of the different flows generated by them. Also in this case, the individual flows commute with each other. Thus, the system retains its equilibrium properties even when it evolves along any of these directions $\mcX_a$, and not only along the specific direction defined by the superposition \eqref{supmod}. \

Finally, we remark on one aspect that seems to differentiate the quantum from the classical case discussed above. For an operational implementation of Jaynes' procedure in a quantum system, the constraints \eqref{qconst2} can exist only if the operators $\h{\mcO}_a$ all commute with each other, that is if the observables under study are compatible. One could thus argue, that since this requirement is operationally unavoidable for a quantum system, then a quantum generalised Gibbs state will always satisfy \eqref{indistation2}. A more detailed understanding, overall, of fundamental differences between the quantum and classical cases (for instance, \cite{Spekkens2016}) for generalised Gibbs states requires further investigation, and is left to future work.


\subsection{Remarks} \label{rem}

In this section we make additional remarks about the thermodynamical characterisation, highlighting many of its interesting features \cite{Kotecha:2019vvn}. \

We notice that this notion of equilibrium is inherently observer-dependent because of its use of the macrostate thermodynamic description of the system, which in itself is observer-dependent due to having to choose a coarse-graining, that is the set of macroscopic observables \cite{Jaynes1992,doi:10.1119/1.1971557}. Further, the role of information entropy is shown to be instrumental in defining (local\footnote{Local, in the sense of being observer-dependent.}) equilibrium states.\footnote{As noted by Jaynes: ``...thus entropy becomes the primitive concept with which we work, more fundamental even than energy...'' \cite{Jaynes:1957zza}.} It is also interesting to notice that Bekenstein's arguments \cite{Bekenstein:1973ur} can be observed to be influenced by Jaynes' information-theoretic insights surrounding entropy, and these same insights have now guided us in the issue of background independent statistical equilibrium. \

To be clear, the use of the most probable distributions as characterising statistical equilibrium is not new in itself. It was used by Boltzmann in the late 19th century, and utilised (also within a Boltzmann interpretation of statistical mechanics) in a constrained system in \cite{Montesinos:2000zi}. Nor is the fact that equilibrium configurations maximise the system's entropy, which was well known already since the time of Gibbs\footnote{But as Jaynes points out in \cite{Jaynes:1957zza}, these properties were relegated to side remarks in the past, not really considered to be fundamental to the theory, nor to the justifications for the methods of statistical mechanics. }. The novelty here is: in the revival of Jaynes' information-theoretic perspective, of deriving equilibrium statistical mechanics in terms of evidential probabilities, solely as a problem of statistical inference without depending on the validity of any further conjectures, physical assumptions or interpretations \cite{levine1979maximum,Jaynes:1957zza,Jaynes:1957zz}; and, in the suggestion that it is general enough to apply to genuinely background independent systems, including quantum gravity. Below we list some of these more valuable features.

\begin{itemize}

\item The procedure is versatile, being applicable to a wide array of cases (both classical and quantum), technically relying only on a sufficiently well-defined mathematical description in terms of a state space, along with a set of observables with dynamically constant averages $U_a$ defining a suitable macrostate of the system\footnote{In fact, in hindsight, we could already have anticipated a possible equilibrium description in terms of these constants, whose existence is assumed from the start.}.

\item Evidently, this manner of defining equilibrium statistical mechanics (and from it, thermodynamics) does not lend any fundamental status to energy, nor does it rely on selecting a single, special (energy) observable out of the full set $\{\mcO_a\}$. It can thus be crucial in settings where concepts of time and energy are dubious at the least, or not defined at all like in non-perturbative quantum gravity.

\item It has a technical advantage of not needing any (1-parameter) symmetry (sub-) groups of the system to be defined a priori, unlike the dynamical characterisation based on the KMS condition. 

\item It is independent of additional physical assumptions, hypotheses or principles that are common to standard statistical physics, and in the present context, to the other proposals of generalised equilibrium recalled in section \ref{past}. Some examples of these extra ingredients that we encountered above in section \ref{past} are ergodicity, weak interactions, statistical independence, and often a combination of them. These are not required in the thermodynamical characterisation.

\item It is independent of any physical interpretations attached (or not) to the observables involved. This further amplifies its appeal for use in quantum gravity where the geometrical and physical meanings of the quantities involved may not necessarily be clear from the start. 

\item One of the main features is the generality in the choice of observables $\mcO_a$ allowed naturally by this characterisation. This also helps accommodate the other proposals as special cases of this one. In principle, $\mcO_a$ and $U_a$ need only be mathematically well-defined in the given description of the system (regardless of whether it is kinematic i.e. working at the extended state space level, or dynamic, i.e. working with the physical state space), such that the resultant Gibbs state is normalisable. More standard choices include a Hamiltonian in a non-relativistic system, a clock Hamiltonian in a deparametrized system \cite{Kotecha:2018gof,Rovelli:2012nv,Josset:2015uja,Chirco:2016wcs}, and generators of kinematic symmetries like rotations, or more generally of 1-parameter subgroups of Lie group actions \cite{souriau1,e18100370}. Some of the more unconventional choices include observables like volume \cite{Kotecha:2018gof,Jaynes:1957zza,Montesinos:2000zi} (section \ref{posgibbs}), (component functions of the) momentum map associated with geometric closure of classical polyhedra \cite{Chirco:2018fns,Chirco:2019kez} (section \ref{entclosure}), half-link gluing (or face-sharing) constraints of discrete gravity \cite{Chirco:2018fns} (section \ref{glucondgibb}), a projector in group field theory \cite{Oriti:2013aqa,Chirco:2018fns}, kinetic and vertex operators in group field theory \cite{Kotecha:2019vvn} (section \ref{effgft}) and generic gauge-invariant observables (not necessarily symmetry generators) \cite{Montesinos:2000zi}.

For instance, \eqref{indistation} shows that the proposal of \cite{Rovelli:1993ys,Chirco:2013zwa} based on statistical independence, i.e. \be  [\mcX_{\rho_1}, \mcX_{\rho_2}] = 0  \ee can be understood as a special case of this one, when the state is characterised by a pair of observables that are defined from the start on mutually exclusive subspaces of the state space. In this case, their respective flows will automatically commute and the state will be said to satisfy statistical independence.

\end{itemize}

In section \ref{gengibb} we will present several examples of applying such notions of generalised equilibrium in background independent discrete quantum gravity, and take first steps towards thermal investigations of these candidate quantum gravitational degrees of freedom subsequently in sections \ref{threpcond} and \ref{GFTCC}.


\section{Generalised thermodynamics} \label{GenTD}

Traditional thermodynamics is the study of energy and entropy exchanges. But what is a suitable generalisation of it for background independent systems? This, like the question of a generalised equilibrium statistical mechanics which we have considered till now, is still open. In the following, we offer some insights gained from preceding discussions, including identifying certain thermodynamic potentials, and generalised zeroth and first laws \cite{Kotecha:2019vvn}.


\subsection{Thermodynamic potentials and multivariable temperature} \label{TDPot}

Thermodynamic potentials are vital, particularly in characterising the different phases of a system. The most important variable is the partition function $Z_{\{\beta_a\}}$, giving the free energy of the system
\be \Phi(\{\beta_a\}) := - \ln Z_{\{\beta_a \}} \,.  \ee 
 It encodes complete information about the system from which other thermodynamic quantities can be derived in principle. Notice that the standard definition of a free energy $F$ comes with an additional factor of a (single, global) temperature, that is, we normally have $\tilde{\Phi} = \beta F$. But for now, $\Phi$ is the more suitable quantity to define and not $F$ since we do not (yet) have a single common temperature for the full system. We will return to this point below in section \ref{commtemp}. Next is the entropy for generalised Gibbs states of the form \eqref{genstate1} or \eqref{genstate2}, which is given by
\be S(\{U_a\}) =  \sum_a  \beta_a U_a  -  \Phi \,.  \ee
Notice that Jaynes' method \cite{Jaynes:1957zza,Jaynes:1957zz}, which forms the basis of the thermodynamical characterisation, identifies information entropy (given in equation \eqref{en}) with the above thermodynamic entropy, since the former is the starting point of this procedure. Also, notice again the lack of a single $\beta$ scaling the whole equation at this more general level of equilibrium. Further, a set of generalised heats can be defined by varying $S$,
\be 
dS = \sum_a \beta_a (d U_a - \langle d \mcO_a \rangle) =: \sum_a \beta_a \,d Q_a   
\ee
where, $d U_a$ are assumed to be independent variations \cite{Jaynes:1957zza}. From this (at least part of the\footnote{By this we mean that the term $\langle d\mcO_a \rangle$, based on the \emph{same} observables defining the generalised energies $U_a$, can be seen as reflecting some work done on the system. But naturally we do not expect or claim that this is all the work that is/can be performed on the system by external agencies. In other words, there could be other work contributions, in addition to the terms $dW_a$. A better understanding of work terms in this background independent setup, will also contribute to a better understanding of the generalised first law presented below.}) work done on the system $dW_a$ \cite{Chirco:2018fns,Kotecha:2019vvn}, can be identified 
\be
dW_a := \langle d\mcO_a \rangle = \frac{1}{\beta_a} \int_{\Gamma} d\lambda \; \frac{\delta \Phi}{\delta \mcO_a} \,d\mcO_a 
\ee
where $\delta$ denotes functional derivatives.

From the setup of the thermodynamical characterisation presented in section \ref{maxent}, we can immediately identify $U_a$ as generalised `energies'. Jaynes' procedure allows these quantities to \emph{democratically} play the role of generalised energies. None had to be selected as being \emph{the} energy in order to define equilibrium. This a priori democratic status of the several conserved quantities can be broken most easily by preferring one over the others. In turn if its modular flow can be associated with a physical evolution parameter (relational or not), then this observable can play the role of a dynamical Hamiltonian. \ 

Thermodynamic conjugates to these energies are several generalised inverse temperatures $\beta_a$. By construction each $\beta_a$ is the periodicity in the flow of $\mcO_a$, in addition to being the Lagrange multiplier for the $a^{\text{th}}$ constraint in \eqref{const}. Moreover these same constraints can  determine $\beta_a$, by inverting the equations
\be \frac{\partial \Phi}{\partial \beta_a} = U_a \,, \ee
or equivalently from
\be \frac{\partial S}{\partial U_a} = \beta_a \,. \ee
In general, $\{\beta_a\}$ is a multi-variable inverse temperature. In the special case when $\mcO_a$ are component functions of a dual vector, then $\vec{\beta} \equiv (\beta_a)$ is a vector-valued temperature. For example, this is the case when $\vec{\mcO}\equiv \{\mcO_a\}$ are dual Lie algebra-valued momentum maps associated with Hamiltonian actions of Lie groups, as introduced by Souriau \cite{souriau1,e18100370}, and appearing in the context of classical polyhedra in \cite{Chirco:2018fns,Chirco:2019kez} (see section \ref{entclosure}).


\subsection{Single common temperature} \label{commtemp}

As we saw above, generalised equilibrium is characterised by several inverse temperatures, but an identification of a single common temperature for the full system is of obvious interest. This is the case \cite{Chirco:2018fns,Chirco:2013zwa,Kotecha:2019vvn} when the modular Hamiltonian
\be h = \sum_a \beta_a \mcO_a \ee
satisfies the thermodynamical characterisation via a single constraint 
\be \label{moddhfix} \langle h \rangle =  \text{constant}\ee 
instead of several of them \eqref{const}, resulting in a state associated with a rescaled modular Hamiltonian $\tilde{h} = \beta h$, now characterised by a single inverse temperature $\beta$,
\be \tilde{\rho}_\beta = \frac{1}{\tilde{Z}_\beta} e^{- \tilde{h}} = \frac{1}{\tilde{Z}_\beta} e^{-\beta h} \,. \ee
Clearly, this case is physically distinct from the previous, more general one. Here, we have a weaker, single constraint corresponding to the situation in which there would be an exchange of information between the different observables (so that they can thermalise to a single $\beta$). This can happen for instance when one observable is special (say, the Hamiltonian) and the rest are functionally related to it (like the volume or number of particles). Whether such a determination of a single temperature can be brought about by a more physically meaningful technique is left to future work. Having said that, it will not change the general layout of the two cases, with a single $\beta$ or a set of several $\beta_a$, as outlined above. \

One immediate consequence of extracting a single $\beta$ is regarding the free energy, which can now be written in the familiar form as
\be \Phi = \beta F \,. \ee
This is most directly seen from the expression for the entropy, 
\be \label{ds} \tilde{S} = -\langle \ln \tilde{\rho}_\beta\rangle_{\tilde{\rho}_\beta} = \beta \sum_a \beta_a \tilde{U}_a + \ln \tilde{Z} \;\;\;\Leftrightarrow\;\;\; \tilde{F} = \tilde{U} - \beta^{-1} \tilde{S} \ee
where $\tilde{U} = \sum_a \beta_a \tilde{U}_a$ is a total energy, and tildes mean that the quantities are associated with the state $\tilde{\rho}_\beta$. Notice that the above equation clearly identifies a single conjugate variable to entropy, the temperature $\beta^{-1}$.\

It is important to remark that in the above method to get a single $\beta$, we still didn't need to choose a special observable, say $\mcO'$, out of the given set of $\mcO_a$. If one were to do this, i.e. select $\mcO'$ as a dynamical energy (so that by extension it is a function of the other $\mcO_a$), then by standard arguments, the rest of the Lagrange multipliers will be proportional to $\beta'$, which in turn would then be the common inverse temperature for the full system. The point is that this latter instance is a special case of the former.


\subsection{Generalised zeroth and first laws} \label{zflaws}

We end with zeroth and first laws of generalised thermodynamics. The crux of the zeroth law is a definition of equilibrium. Standard statement refers to a thermalisation resulting in a single temperature being shared by any two systems in thermal contact. This can be extended by the statement that at equilibrium, all inverse temperatures $\beta_a$ are equalised. This is in exact analogy with all intensive thermodynamic parameters, such as the chemical potential, being equal at equilibrium.  \

The standard first law is basically a statement about conservation of energy. In the generalised equilibrium case corresponding to a set of individual constraints \eqref{const}, the first law is satisfied $a^{\text{th}}$-energy-wise, 
\be dU_a = dQ_a + dW_a \,. \ee
The fact that the law holds $a$-energy-wise is not surprising because the separate constraints \eqref{const} for each $a$ mean that observables $\mcO_a$ do not exchange any information amongst themselves. If they did, then their Lagrange multipliers would no longer be mutually independent and we would automatically reduce to the special case of having a single $\beta$ at equilibrium. \

On the other hand, for the case with a single $\beta$, variation of the entropy \eqref{ds} gives
\be  d\tilde{S} = \beta \sum_a \beta_a (dU_a - \langle d\mcO_a \rangle) =: \beta d\tilde{Q} \ee
giving a first law with a more familiar form, in terms of total energy, total heat and total work variations
\be d\tilde{U} = d\tilde{Q} + d\tilde{W} . \ee
As before, in the even more special case where $\beta$ is conjugate to a single preferred energy, then this reduces to the traditional first law. We leave the consideration of the second law for the generalised entropy to future work. We also leave a detailed study of the quantities introduced above and physical consequences of their corresponding thermodynamics to future work.


\begin{subappendices}

\section{Stationarity with respect to constituent generators} \label{appMod}

We show that a generalised Gibbs state is stationary with respect to each of the individual, constituent generators, $\mcO_a$ or $\h{\mcO}_a$, of the full modular hamiltonian, $h$ or $\h{h}$, if these generators of the individual flows all commute with each other. \\

\noindent \textbf{Lemma 1.}	 For a classical state $\rho_{\{\beta_a\}}$ given in \eqref{genstate1}, we have 
\be \{\mcO_{a''}, \mcO_{a'}\}=0 \;\; (\text{for all} \; a',a'')  \quad \Rightarrow \quad \mcX_{a'}(\rho_{\{\beta_a\}}) = 0 \;\; (\text{for any} \; a')   \ee 
i.e. implication in \eqref{indistation} is true.
 
 \medskip

\noindent \textbf{Proof 1.} The line of reasoning is summarised by the following set of relations: for any $a',a''$
\begin{align}
 \{\mcO_{a''}, \mcO_{a'}\}=0 &\Rightarrow \{h,\mcO_{a'}\}=0 \label{st2} \\
					  &\Leftrightarrow \{h^k,\mcO_{a'}\}=0 \label{st3} \\
					  &\Rightarrow \{e^{-h}, \mcO_{a'}\} = 0 \label{st4}
\end{align}
which are detailed in Lemmas $1.1-1.3$ below. $\hfill \square$ 

\medskip


\noindent \textbf{Lemma 1.1.} $\{\mcO_{a''}, \mcO_{a'}\}=0 \Rightarrow \{h,\mcO_{a'}\}=0$, as in relation \eqref{st2}. \medskip

\noindent \textbf{Proof 1.1.} Using equation \eqref{moddh} and linearity of the Poisson bracket, we have:
\begin{align}
 \{h,\mcO_{a'}\} &= \{ \sum_{a''} \beta_{a''} \mcO_{a''} , \mcO_{a'}  \} \\
 			&= \sum_{a''} \beta_{a''} \{ \mcO_{a''} , \mcO_{a'}  \} \,.
\end{align} $\hfill \square$  

\medskip


\noindent \textbf{Lemma 1.2.} $\{h,\mcO_{a'}\}=0 \Leftrightarrow \{h^k,\mcO_{a'}\}=0 $, for $k \in \mathbb{N}$, as in relation \eqref{st3}. \medskip

\noindent \textbf{Proof 1.2.} By induction, we have
\be \label{recclass} \{ h^k , \mcO_{a'}\} = kh^{k-1}\{h,\mcO_{a'}\}  \ee
where we have recursively used the product rule for the Poisson bracket, and commutativity of the associative product of the Poisson algebra of smooth functions over $\Gamma$. The implication ``$\Leftarrow$'' is true, since the equality in \eqref{recclass} holds for arbitrary non-zero $k$ and $h$.  $\hspace*{\fill} \square$  

\medskip


\noindent \textbf{Lemma 1.3.} $\{h^k,\mcO_{a'}\}=0 \Rightarrow \{e^{-h}, \mcO_{a'}\} = 0 $, as in relation \eqref{st4}. \medskip

\noindent \textbf{Proof 1.3.} Using the formal exponential expansion, and linearity of the Poisson bracket, we have
\begin{align}
\{e^{-h}, \mcO_{a'}\} &= \{ \sum_{k=0}^\infty \frac{(-1)^k}{k!} h^k , \mcO_{a'} \} \\
				&= \sum_{k=0}^\infty \frac{(-1)^k}{k!}  \{ h^k , \mcO_{a'}\} \,.
\end{align}  $\hfill \square$  

\medskip


\noindent \textbf{Lemma 2.} For a quantum state $\h{\rho}_{\{\beta_a\}}$ given in \eqref{genstate2}, we have
\be  [\h{\mcO}_{a''},\h{\mcO}_{a'}] = 0 \;\; (\text{for all} \; a',a'')  \quad \Rightarrow \quad [\h{\rho}_{\{\beta_a\}},\h{\mcO}_{a'}] = 0 \;\; (\text{for any} \; a')   \ee 
i.e. implication in \eqref{indistation2} is true.

\medskip

\noindent \textbf{Proof 2.} The proof follows the following line of reasoning: for any $a',a''$
\begin{align}
[\h{\mcO}_{a''}, \h{\mcO}_{a'}] = 0 &\Rightarrow [\h{h},\h{\mcO}_{a'}]=0 \label{2st1T} \\
						&\Rightarrow [\h{h}^k,\h{\mcO}_{a'}]=0 \label{2st2T} \\
						&\Rightarrow [e^{-\h{h}}, \h{\mcO}_{a'}] = 0 \label{2st3}
\end{align}
with details of each provided in the following Lemmas $2.1 - 2.3$, as done previously. $\hfill \square$  

\medskip

\noindent \textbf{Lemma 2.1.} $[\h{\mcO}_{a''}, \h{\mcO}_{a'}] = 0 \Rightarrow [\h{h},\h{\mcO}_{a'}]=0$, as in relation \eqref{2st1T}.   \medskip

\noindent \textbf{Proof 2.1.} Using equation \eqref{moddhq} and linearity of the commutator bracket, we have
\begin{align}
[\h{h},\h{\mcO}_{a'}] &= [ \sum_{a''} \beta_{a''} \h{\mcO}_{a''} , \h{\mcO_{a'}} ] \\
 			&= \sum_{a''} \beta_{a''} [ \h{\mcO}_{a''} , \h{\mcO}_{a'}  ] \,.
\end{align} $\hfill \square$  

\medskip


\noindent \textbf{Lemma 2.2.}  $ [\h{h},\h{\mcO}_{a'}]=0 \label{2st1} \Rightarrow [\h{h}^k,\h{\mcO}_{a'}]=0 $, as in relation \eqref{2st2T}. \medskip

\noindent \textbf{Proof 2.2.} By induction, we have
\be [\h{h}^k, \h{\mcO}_{a'}] = \sum_{j=0}^{k-1} \h{h}^{k-j-1}[\h{h},\h{\mcO}_{a'}]\h{h}^j  \ee 
where we have recursively used the commutator identity, $[ab,c] = a[b,c] + [a,c]b$. $\hfill \square$  

\medskip


\noindent \textbf{Lemma 2.3.} $[\h{h}^k,\h{\mcO}_{a'}]=0 \label{2st2} \Rightarrow [e^{-\h{h}}, \h{\mcO}_{a'}] = 0$, as in relation \eqref{2st3}.  \medskip

\noindent \textbf{Proof 2.3.} Using the formal exponential expansion, and linearity of the commutator bracket, we have
\begin{align}
[ e^{-\h{h}}, \mcO_{a'} ] &= [ \sum_{k=0}^\infty \frac{(-1)^k}{k!} \h{h}^k , \h{\mcO}_{a'} ] \\
				&= \sum_{k=0}^\infty \frac{(-1)^k}{k!}  [ \h{h}^k , \h{\mcO}_{a'} ] \,.
\end{align} $\hfill \square$

\end{subappendices}


\chapter{Many-Body Quantum Spacetime}
\label{DQG}

\begin{quotation}
\begin{center} The beginnings of all things are small. ---\emph{Marcus Tullius Cicero} \end{center}
\end{quotation}


\vspace{2mm}
\vspace{1pt}

\noindent Now that we have presented a potential generalisation of statistical equilibrium in the previous chapter, we move on to its application in discrete quantum gravity. As anticipated in section \ref{outline}, in this thesis a discrete quantum spacetime is modelled as a many-body system composed of a large number of elementary building blocks\footnote{We should emphasise that in any non-spacetime-based, background independent setting (like the present one), elementary quanta of geometry are not ``small'' in the conventional intuitive sense; and, ``large'' spacetime is not simply a sum of many ``small'' building blocks, in the additive sense. Rather, continuum spacetime must emerge from the collective dynamics, coarse graining and composite properties of the underlying constituents.}, or quanta of space \cite{Oriti:2017twl}. \

Specifically, the candidate atoms of space under consideration here are geometric convex $d$-polyhedra with $d$ bounding faces, dual to open $d$-valent nodes with its half-links dressed by suitable algebraic data \cite{Bianchi:2010gc}. This choice is motivated strongly by loop quantum gravity (LQG) \cite{Ashtekar:2004eh,Rovelli:2004tv,Bodendorfer:2016uat}, spin foams \cite{Perez:2012wv, rovelli_vidotto_2014}, group field theory \cite{Reisenberger:2000zc,Freidel:2005qe,Oriti:2005mb,Oriti:2006se,Oriti:2011jm,Krajewski:2012aw,Gielen:2021dlk}, dynamical triangulations \cite{Loll:2019rdj} and lattice quantum gravity \cite{Hamber:2009mt} approaches in the context of 4d models. Extended discrete quantum space can be built out of these fundamental `atoms' or `particles', via kinematical compositions (or boundary gluings), and spacetime via dynamical interactions (or bulk bondings). The perspective innate to such a many-body quantum spacetime is thus a constructive one, which is then naturally also extended to the statistical mechanics based on this many-body mechanics. \

As we will see, two types of data specify a mechanical model, combinatorial and algebraic. States and processes of a model are supported on combinatorial structures, here abstract\footnote{Thus not necessarily embedded into any continuum spatial manifold.} graphs and 2-complexes respectively; and algebraic dressings of these structures add discrete geometric and matter information. Thus, different choices of combinatorics and algebraic data give different mechanical models. For instance, the simplest 4d spin foam models (and their associated group field theories) are associated with: boundary combinatorics based on a 4-valent node (dual to a tetrahedron, under closure), bulk combinatorics based on a 4-simplex interaction vertex, and algebraic (or group representation) data of $SU(2)$ labelling the boundary 4-valent graphs and bulk 2-complexes. \

Clearly this is not the only choice, in fact far from it. The vast richness of possible combinatorics, compatible with our constructive point of view, is comprehensively illustrated in \cite{Oriti:2014yla}. 
And the various choices for variables to label the discrete structures with (so that they may encode some notion of discrete geometry, which notion depending exactly on the variables chosen and constraints imposed on them) have been an important subject of study, starting all the way from Regge \cite{Regge:1961px,Regge:2000wu,Dittrich:2008va,Freidel:2010aq,Rovelli:2010km,Dittrich:2008ar}. Accommodation of these various different choices is yet another appeal of our constructive many-body viewpoint and this framework. We clarify further some of these aspects in the following, but often choose to work with simplicial combinatorics and $SU(2)$ holonomy-flux data in the subsequent chapters. \

In this chapter, we use results from the previous sections to outline a framework of equilibrium statistical mechanics for these candidate quanta of geometry \cite{Kotecha:2018gof,Chirco:2018fns,Kotecha:2019vvn}. In particular, we show that a group field theory arises naturally as an effective statistical field theory from a coarse-graining of a class of generalised Gibbs configurations of the underlying quanta. In addition to providing an explicit quantum statistical basis for group field theories, this reinforces their status as being field theories for quanta of geometry \cite{Reisenberger:2000zc,Freidel:2005qe,Oriti:2005mb,Oriti:2006se,Oriti:2011jm}. Motivated by these discussions, we will present aspects of thermal states in group field theories in the following chapters.


\section{Atoms of quantum space and kinematics} \label{atomkin}

In the following we will refer to some of the combinatorial structures defined in \cite{Oriti:2014yla}. However we will be content with introducing them in a more intuitive manner, and not recalling the rigorous definitions as that is not required for the purposes of this thesis. We refer to \cite{Oriti:2014yla} for details.\footnote{For clarity, we note that the terminology used here is slightly different from that in \cite{Oriti:2014yla}. Specifically the dictionary between here $\leftrightarrow$ there is: combinatorial atom or particle $\leftrightarrow$ boundary patch; interaction/bulk vertex $\leftrightarrow$ spin foam atom; boundary node $\leftrightarrow$ boundary multivalent vertex $\bar{v}$; link or full link $\leftrightarrow$ boundary edge connecting two multivalent vertices $\bar{v}_1,\bar{v}_2$; half-link $\leftrightarrow$ boundary edge connecting a multivalent vertex $\bar{v}$ and a bivalent vertex $\h{v}$. This minor difference is mainly due to a minor difference in the purpose for the same combinatorial structures. Here we are in a setup where the accessible states are boundary states, for which a statistical mechanics is defined; and the case of interacting dynamics is considered as defining a suitable (amplitude) functional over the the boundary state space. On the other hand, the perspective in \cite{Oriti:2014yla} is more in a spin foam constructive setting, so that modelling the 2-complexes as built out of fundamental spin foam atoms is more natural there.}  \

\begin{figure}[t]
\centering
\includegraphics[width=3 in]{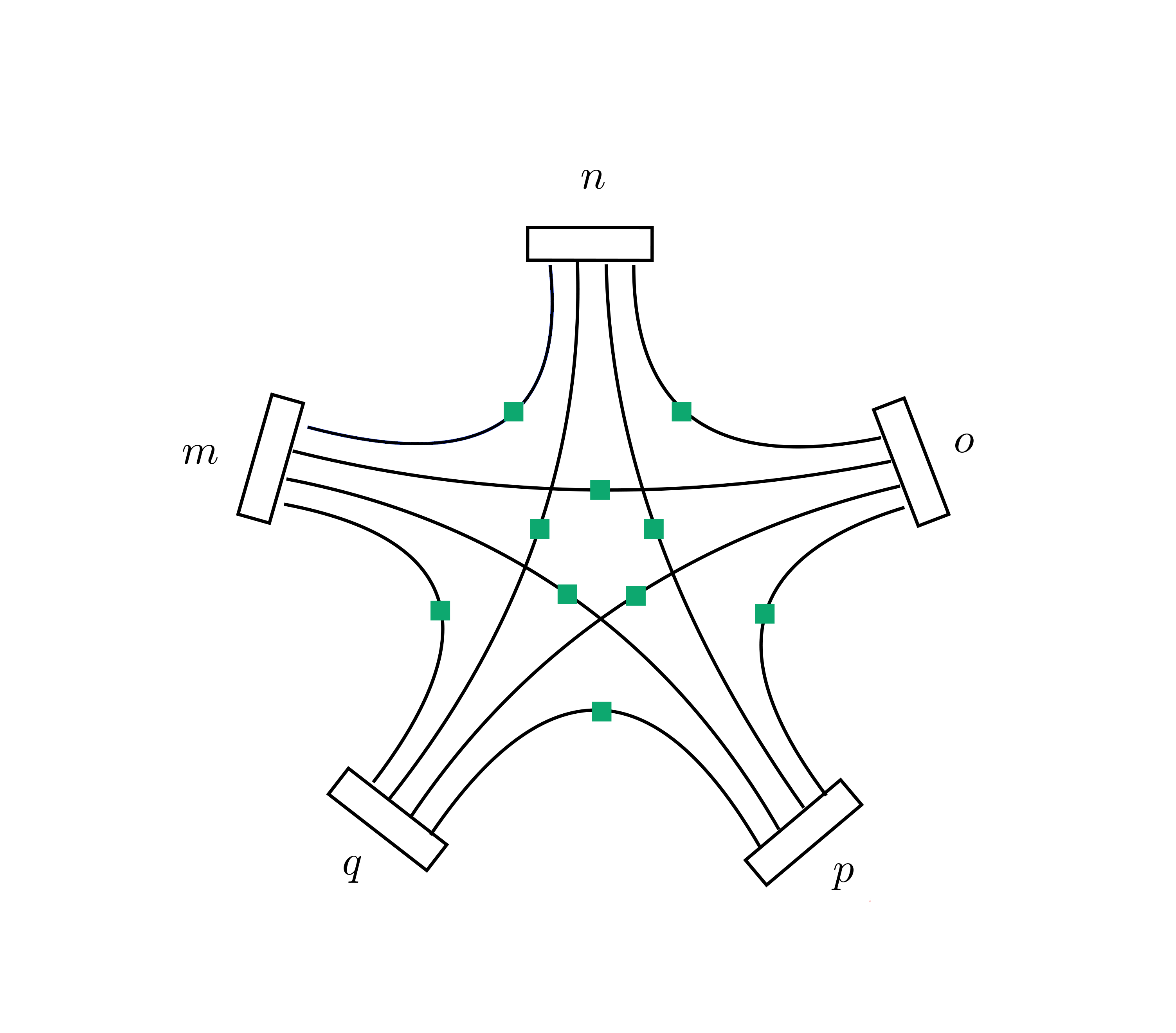}
\caption{Bisected boundary graph of a 4-simplex, as a result of non-local pair-wise gluing of half-links. Each full link is bounded by two 4-valent nodes (shown here as rectangles), and bisected by one bivalent node (shown here as squares). \cite{Kotecha:2019vvn}}\label{4sim}
\end{figure}

The primary objects of interest to us are boundary patches, which we take as the combinatorial atoms of space. To put simply, a boundary patch is the most basic unit of a boundary graph, in the sense that the set of all boundary patches generates the set of all connected bisected boundary graphs. A bisected boundary graph is simply a directed boundary graph with each of its full links bisected into a pair of half-links, glued at the bivalent nodes (for instance, see Figure \ref{4sim}). Different kinds of atoms of space are then the different, inequivalent boundary patches (dressed further with suitable data), and the choice of combinatorics basically boils down to a choice of the set of admissible boundary patches. Moreover, a model with multiple inequivalent boundary patches can be treated akin to a statistical system with multiple species of atoms.  \

The most general types of boundary graphs 
are those with nodes of arbitrary valence, and including loops. A common and natural restriction is to consider loopless structures, as they can be associated with combinatorial polyhedral complexes \cite{Oriti:2014yla}. As the name suggests, loopless boundary patches are those with no loops, i.e. each half-link is bounded on one end by a unique bivalent node (and on the other by the common, multivalent central node). A loopless patch is thus uniquely specified by the number of incident half-links (or equivalently, by the number of bivalent nodes bounding the central node). A $d$-patch, with $d$ number of incident half-links, is simply a $d$-valent node. Importantly for us, it is the combinatorial atom that supports geometric states of a $d$-polyhedron \cite{Bianchi:2010gc,Barbieri:1997ks,Baez:1999tk}. A further common restriction is to consider graphs with nodes of a single, fixed valence, that is to consider $d$-regular loopless structures. \

Let's take an example. Consider the boundary graph of a 4-simplex as shown in Figure \ref{4sim}. The fundamental atom or boundary patch is a 4-valent node. This graph can be constructed starting from five open 4-valent nodes (denoted $m,n,...,q$), and gluing the half-links, or equivalently the faces of the dual tetrahedra, pair-wise, with the non-local combinatorics of a complete graph on five 4-valent nodes. The result is ten bisected full links, bounded by five nodes. It is important to note here that a key ingredient of constructing extended boundary states from the atoms are precisely the half-link gluing, or face-sharing conditions on the algebraic data decorating the patches. For instance, in the case of standard LQG holonomy-flux variables of $\mcT^*(SU(2))$, the face-sharing gluing constraints are area matching \cite{Freidel:2010aq}, thus lending a notion of discrete classical twisted geometry to the graph \cite{Freidel:2010aq,Rovelli:2010km}. This is much weaker than a Regge geometry, which could have been obtained for the same variables if instead the so-called shape-matching conditions \cite{Dittrich:2008va} are imposed on the pair-wise gluing of faces, or equivalently half-links. Thus, kinematic composition (boundary gluings) that creates boundary states depends on two crucial ingredients, the combinatorial structure of the resultant boundary graph, and face-sharing gluing conditions on the algebraic data.
 
From here on we will restrict to a single boundary patch for simplicity, namely a gauge-invariant $d$-patch dressed with $SU(2)$ data, dual to a $d$-polyhedron \cite{Bianchi:2010gc,Barbieri:1997ks}. But it should be clear from the brief discussion above (and the extensive study in \cite{Oriti:2014yla}) that a direct generalisation of the present mechanical and statistical framework is formally possible also for these more enhanced combinatorial structures. \

\begin{figure}[t]
\centering
\includegraphics[width=3.5 in]{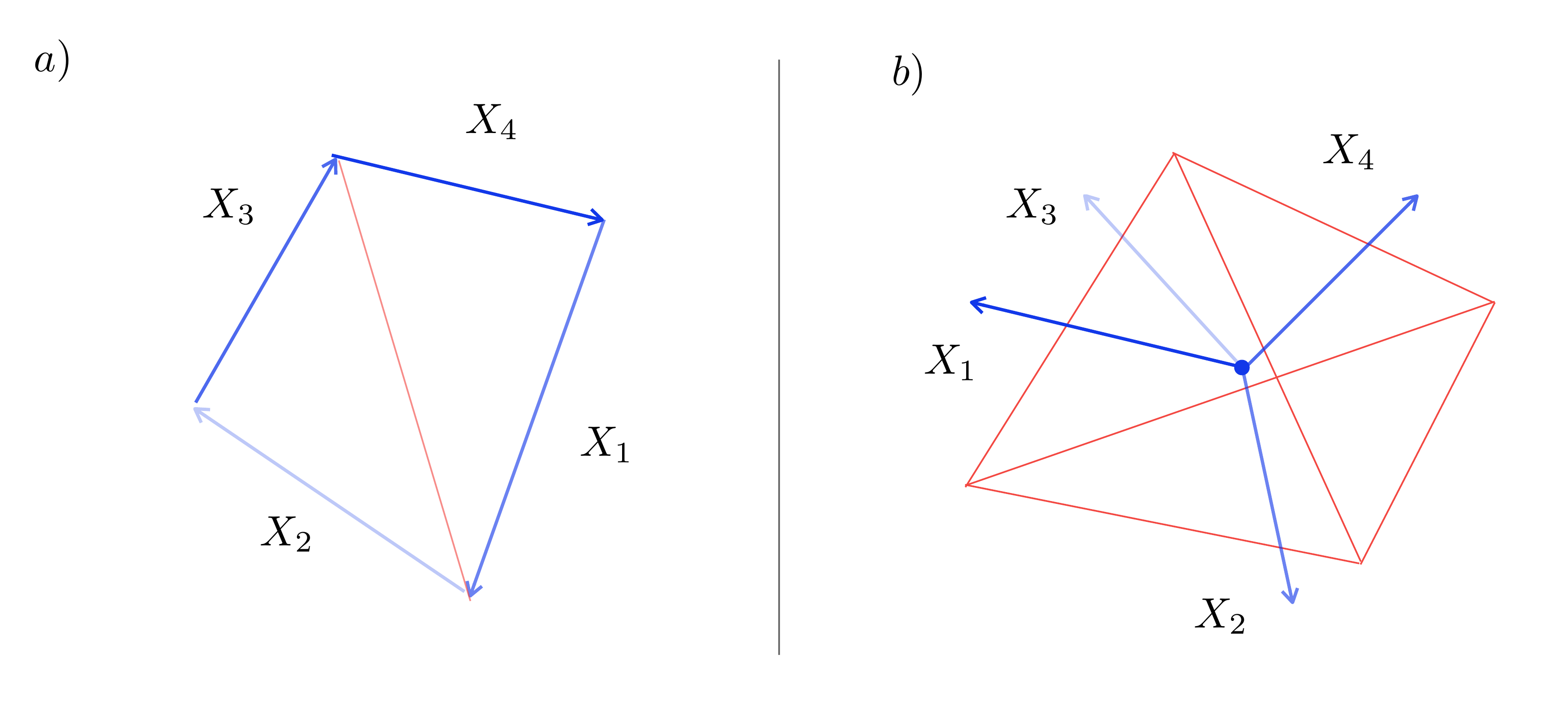}
\caption{$a)$ A convex polygon with a closed set of vectors $X_I$. The space of possible polygons in $\mathbb{R}^3$ up to rotations is a $(2d-6)$-dimensional phase space. For non-coplanar normals, the same data define also a unique polyhedron by Minkowski's theorem. $b)$ For $d=4$ we get a geometric tetrahedron. \cite{Chirco:2018fns} }\label{convex}
\end{figure}

The classical intrinsic geometry of a polyhedron with $d$ faces is given by the Kapovich-Millson phase space \cite{kapovich1996},
\be \mathcal{S}_d = \{(X_I) \in \su(2)^{*d} \;|\; \sum_{I=1}^d X_I=0, ||X_I|| = A_I\}/SU(2)  \ee
which is a $(2d-6)$-dimensional symplectic manifold. Now, for more generality, one could lift the restriction of fixed face areas, thereby adding $d$ degrees of freedom, to get the $(3d-6)$-dimensional space of closed polyhedra, modulo rotations in $\mathbb{R}^3$. For $d=4$, this is the 6-dim space of a tetrahedron \cite{Barbieri:1997ks, Baez:1999tk}, considered often in discrete quantum gravity contexts (Figure \ref{convex}). This space corresponds to the possible values of the 6 edge lengths of a tetrahedron, or to the 6 areas that include four areas of its faces and two independent areas of parallelograms identified by midpoints of pairs of opposite edges. This space is not symplectic in general, and to get a symplectic manifold from it, one can either remove the $d$ area degrees of freedom to get $\mathcal{S}_d$ again, or add another $d$ number of degrees of freedom, the $U(1)$  angle conjugates to the areas, to get the spinor description of the so-called framed polyhedra \cite{Livine:2013tsa}.\footnote{Along the lines shown below, one can thus also extend the statistical description to the case of the framed polyhedron system.} However, we are presently more interested in the statistical description of a collection of many connected, closed polyhedra. \

Let us then consider the space of closed polyhedra with a fixed orientation, and extend the phase space description so as to encompass the extrinsic geometric degrees of freedom, which we expect to play a role in the description of the coupling leading to a collective model. We know from above that the face normal vectors with arbitrary areas are elements of the dual algebra $\su(2)^* \cong \mathbb{R}^3$, which is a Poisson manifold with its Kirillov-Kostant Poisson structure \cite{Baez:1999tk}. We can supplement this data with conjugate degrees of freedom belonging to the group $SU(2)$ (thereby also doubling the dimension) and consider the phase space, 
\be \Gamma_{d\text{-poly}} = \mcT^*(SU(2)^d/SU(2)) \,. \ee 
Here, the quotient by $SU(2)$ imposes geometric closure of the polyhedron; equivalently, it constraints the area normals to close, $\sum_{I=1}^d X_I = 0$ (see also sections \ref{fockspace} and \ref{entclosure}). Specifically for $d=4$, we then have a single classical tetrahedron phase space given by,
\be \label{gammatet} \Gamma = \mcT^*(SU(2)^{ 4}/SU(2))  \ee
which encodes both intrinsic and extrinsic degrees of freedom (along with an arbitrary orientation in $\mathbb{R}^3$). Notice that the choice of domain manifold is essentially the choice of algebraic data. For instance, this could be $Spin(4)$ in Euclidean 4d settings, while $SL(2,\mathbb{C})$ in Lorentzian ones. Then, the states of a system of $N$ tetrahedra belong to the following direct product space,
\be \label{gammaN} \Gamma_N = \Gamma^{\times N} \ee
with an algebra composed of smooth real-valued functions defined on $\Gamma_N$. \cite{Kotecha:2018gof,Chirco:2018fns,Kotecha:2019vvn} 

On the other hand, in a quantum setting, each closed polyhedron face $I$ is assigned an $SU(2)$ representation label $j_I$ with its associated representation space $\mathcal{H}_{j_I}$, and the polyhedron itself with an intertwiner, i.e. an invariant tensor element of 
\be \mcH_{d\text{-poly}\{j_I\}} = \text{Inv} \bigotimes_{I = 1}^d \mathcal{H}_{j_I} \,.\ee 
This is the space of $d$-valent intertwiners with given fixed spins $\{j_I\}$ i.e. given fixed face areas, corresponding to a quantisation of $\mathcal{S}_d$. Then, quantisation of $ \Gamma_{d\text{-poly}}$ is the full space of $d$-valent intertwiners, given by 
\be \mcH_{d\text{-poly}} = \bigoplus_{\{j_I \in \mathbb{N}/2\}} \text{Inv} \bigotimes_{I =1}^d \mathcal{H}_{j_I} \,.\ee 
A collection of neighbouring quantum polyhedra has been associated to a spin network of arbitrary valence \cite{Bianchi:2010gc}, with the labelled nodes and links of the latter being dual to labelled polyhedra and their shared faces respectively.  \

Then for a quantum tetrahedron \cite{Barbieri:1997ks,Baez:1999tk}, the 1-particle Hilbert space is,
\be
\mcH_{\tet} = \bigoplus_{\{j_I \in \mathbb{N}/2\}} \text{Inv} \bigotimes_{I = 1}^4 \mcH_{j_I} 
\ee
with quantum states of an $N$-particle system belonging to the following tensor product space,
\be
\mathcal{H}_N = \mathcal{H}_{\tet}^{\otimes N} 
\ee
where, $\mcH_0 = \mathbb{C}$. We can equivalently work with the holonomy representation of the same quantum system in terms of $SU(2)$ group data,
\be \label{singleH}
\mathcal{H}_{\tet} = L^2(SU(2)^{4}/SU(2)) 
\ee
which is also the state space of a single gauge-invariant quantum of a group field theory defined on an $SU(2)^{4}$ base manifold (see section \ref{bosgft} for details). A further, equivalent representation could be given in terms of non-commutative Lie algebra (flux) variables $X_I \in \su(2)^*$ \cite{Baratin:2010wi,Baratin:2010nn}. \

On such multi-particle state spaces, mechanical models for a system of many tetrahedra can be defined via constraints among them. Typical examples would be non-local, combinatorial gluing constraints, possibly scaled by an amplitude weight. From the point of view of many-body physics, we expect these gluing constraints to be modelled as generic multi-particle interactions, defined in terms of tetrahedron geometric degrees of freedom. Different choices of these interactions would then identify different models of the system. The key ingredient of such interactions are the constraints which glue two faces of any two different tetrahedra, for instance by constraining the areas of adjacent faces to match and their face normals to align (see section \ref{glucondgibb}). More stringent conditions, imposing stronger matching of geometric data, as well as more relaxed ones, can also be considered, as will be discussed. What constitutes as gluing is thus a model-building choice, and so is the choice of which combinatorial pattern of gluings among a given number of tetrahedra is enforced. So, once the system knows how to glue two faces, then the remaining content of a model dictates how the tetrahedra interact non-locally to make simplicial complexes \cite{Chirco:2018fns,Kotecha:2019vvn,Chirco:2019kez}. More on this will be discussed in sections \ref{intdyn} and \ref{congibb}.\

Lastly, for a quantum multi-particle system, one can go a step further and define a Fock space to accommodate configurations with varying particle numbers. For bosonic\footnote{As for a standard multi-particle system, bosonic statistics corresponds to a symmetry under particle exchange. For the case when a system of quantum tetrahedra is glued appropriately to form a spin network, then this symmetry is understood as implementing the graph automorphism of node relabellings.} quanta, each $N$-particle sector is the symmetric projection of the full $N$-particle Hilbert space, so that a Fock space can be defined as, \be \mcH_F^{\tet} =  \bigoplus_{N \geq 0} \text{sym}\, \mathcal{H}_N . \ee
The associated Fock vacuum state $\ket{0_{\tet}}$ is degenerate with no discrete geometric degrees of freedom, the `no space' state. One choice for the algebra of operators on $\mcH_F^{\tet}$ is the von Neumann algebra of bounded linear operators $\mathcal{B}(\mcH_F^{\tet})$. A more common choice though is the larger *-algebra generated by ladder operators $\h{\varphi},\h{\varphi}^\dag$, which generate $\mcH_F^{\tet}$ by acting on the cyclic Fock vacuum, and satisfy a commutation relations algebra 
\be \label{ccrtemp} [\h{\varphi}(\vec{g}_1),\h{\varphi}^\dag(\vec{g}_2)] = \int_{SU(2)} dh \prod_{I=1}^4 \delta(g_{1I}h g_{2I}^{-1})  \ee
where $\vec{g} \equiv (g_I) \in SU(2)^4$ and the integral on the right ensures $SU(2)$ gauge invariance. In fact this is the Fock representation associated with a degenerate vacuum of an algebraic bosonic group field theory defined by a Weyl algebra \cite{Kotecha:2018gof,Oriti:2013aqa,Kegeles:2017ems}. We will return to a detailed discussion of these aspects in section \ref{bosgft}. \

Notice that this kind of formulation already hints at a second quantised language in terms of quantum fields of tetrahedra. This language can indeed be applied to the generalised statistical mechanics framework we have developed, to get a quantum statistical mechanics for tetrahedra (polyhedra in general), which will be the focus of the rest of the thesis.


\section{Interacting quantum spacetime and dynamics} \label{intdyn}

Coming now to dynamics, the key ingredients here are the specifications of propagators and admissible interaction vertices, including both their combinatorics, and functional dependences on the algebraic data i.e. their amplitudes. \

The combinatorics of propagators and interaction vertices can be packaged neatly within two maps, the bonding map and the bulk map respectively \cite{Oriti:2014yla}. A bonding map is defined between two bondable boundary patches. Two patches are bondable if they have the same number of nodes and links. Then, a bonding map between two bondable patches identifies each of their nodes and links, under the compatibility condition that if a bounding bivalent node in one patch is identified with a particular one in another, then their respective half-links (attaching them to their respective central nodes) are also identified with each other. So a bonding map basically bonds two bulk vertices via (parts of) their boundary graphs to form a process (with boundary). This is simply a bulk edge, or propagator. \

The set of interaction vertices can themselves be defined by a bulk map. This map augments the set of constituent elements (multivalent nodes, bivalent nodes, and half-links connecting the two) of \emph{any} bisected boundary graph, by one new vertex (the bulk vertex), a set of links joining each of the original boundary nodes to this vertex, and a set of two dimensional faces bounded by a triple of the bulk vertex, a multivalent boundary node and a bivalent boundary node. The resulting structure is an interaction vertex with the given boundary graph.\footnote{An interesting aspect is that the bulk map is one-one, so that for every distinct bisected boundary graph, there is a unique interaction vertex which can be defined from it \cite{Oriti:2014yla}.} The complete dynamics is then given by the chosen combinatorics, supplemented with amplitude functions that reflect the dependence on the algebraic data. \

The interaction vertices can in fact be described by vertex operators on the Fock space in terms of the ladder operators. An example vertex operator, corresponding to the 4-simplex boundary graph shown in Figure \ref{4sim}, is
\be \h{\mbV}_{\text{4sim}} = \int_{SU(2)^{20}} [dg]\;  \h{\varphi}^\dag(\vg_1) \h{\varphi}^\dag(\vg_2) {\msV}_{\text{4sim}}(\vg_1,...,\vg_5) \h{\varphi}(\vg_3) \h{\varphi}(\vg_4) \h{\varphi}(\vg_5)  \ee
where the interaction kernel ${\msV}_{\text{4sim}} = {\msV}_{\text{4sim}}(\{g_{ij}g_{ji}^{-1}\}_{i<j})$ (for $i,j=1,...,5$) encodes the combinatorics of the boundary graph. There are of course other vertex operators associated with the \emph{same} graph (that is with the same kernel), but including different combinations of creation and annihilation operators\footnote{This would generically be true for any second quantised operator \cite{fetter,Oriti:2013aqa}.}. \

It must be noted that generic configurations of this system do not admit an interpretation as quantised geometric fields. Therefore, geometric configurations cannot be presumed, and one would have to look for such phases to emerge within the full statistical description of the quanta of spacetime. This is an important and difficult open problem that we do not directly tackle in this thesis. \\

To summarise, a definition of kinematics entails: defining the state space, which includes specifying the combinatorics (choosing the set of allowed boundary patches, which generate the admissible boundary graphs), and the algebraic data (choosing variables to characterise the discrete geometric states supported on the boundary graphs); and, defining the algebra of observables acting on the state space. A definition of dynamics entails: specifying the propagator and bulk vertex combinatorics and amplitudes. Together they specify the many-body mechanics.


\section{Generalised equilibrium states} \label{GeneqmS}

Based on insights from chapter \ref{GGS} and the mechanical setup laid out above, we can formulate a statistical mechanical framework, and in particular generalised equilibrium statistical mechanics, for these discrete quantum geometric systems \cite{Kotecha:2018gof,Chirco:2018fns,Kotecha:2019vvn,Chirco:2019kez}. \

For a system of many classical tetrahedra (in general, polyhedra), a statistical state $\rho_N$ can be formally defined on the state space $\Gamma_N$. As we saw in section \ref{maxent}, if it satisfies the thermodynamical characterisation with respect to a set of functions on $\Gamma_N$ then it will be an equilibrium state. Further, a configuration with a varying number of tetrahedra can be described by a grand-canonical type state of the form
\be Z = \sum_{N \geq 0} e^{\mu N} Z_N \ee
where $Z_N = \int_{\Gamma_N} d\lambda \, \rho_N$, and $\mu$ is a chemical potential \cite{Chirco:2018fns}. Similarly for a system of many quantum tetrahedra, a generic statistical state $\h{\rho}$ is a density operator on $\mcH_F^{\tet}$ \cite{Kotecha:2018gof}. Generalised equilibrium states with a varying number of quanta are then given by,
\be Z = \Tr(e^{-\sum_a \beta_a \h{\mcO}_a + \mu \h{N}})\ee
where $\h{N} = \int d\vg \, \h{\varphi}^\dag(\vg) \h{\varphi}(\vg)$ is the number operator on $\mcH_F^{\tet}$. Such grand-canonical type boundary states are important because one would expect quantum gravitational dynamics to not be number conserving in general \cite{Chirco:2018fns,Oriti:2013aqa}. Also as pointed out in section \ref{rem}, what the precise content of equilibrium is depends crucially on which observables $\mcO_a$ are used to define the state. Operators of natural interest here are the ones encoding the dynamics, that is vertex and kinetic operators, as considered below in section \ref{effgft}. There are many other choices and types of observables one could consider in principle, as will be exemplified in chapter \ref{TGFT}. In fact, which ones are the relevant ones in a given situation is a crucial part of the whole problem. \

Given a density operator $\rho$, then the partition function $Z$, contains complete statistical information. From $Z$ can be defined an important thermodynamic potential, the free energy $\Phi = -\ln Z$. Entropy is $S = -\Tr(\rho \ln \rho)$. These thermodynamic variables are those whose construction does not really rely on the context in which the statistical mechanical framework is formulated, as we saw in section \ref{GenTD}. As emphasised in chapter \ref{GGS} and above, the remaining \emph{relevant} macrostate variables, like generalised energies $\braket{\mcO}$, need to be first identified depending on the specific system at hand. Then, a macrostate of the system is characterised by this set of thermodynamic variables (and others derived from them, say via Legendre transforms), whose compatible microstates are naturally the quantum states contributing to the statistical mixture. Having done so at a formal level, naturally the remaining challenging task is to identify a suitable physical interpretation for them. For example, if a thermodynamic volume potential is defined in analogy with usual quantum field theories, this would refer to the domain manifold of the group fields, i.e. the Lie group manifold. It would not be immediately related to spatial volumes, as deduced for example by the quantum operator considered later in sections \ref{posgibbs} and \ref{vgibbs}, which is motivated by the quantum geometric interpretation of the group field quanta.


\section{Effective statistical group field theory} \label{effgft} 

We end this chapter by making a direct link to the definition of group field theories using the above framework. Group field theories \cite{Freidel:2005qe,Oriti:2005mb,Oriti:2006se,Oriti:2011jm,Krajewski:2012aw,Gielen:2021dlk} are non-local field theories defined over Lie groups. Most widely studied models are for real or complex scalar fields, over copies of $SU(2)$ or $Spin(4)$. For instance, a complex scalar GFT over $SU(2)$ is defined by a partition function of the following general form, 
\be \label{ZGft} Z_{\text{GFT}} = \int [D\mu(\varphi,\bar{\varphi})] \;e^{-S_{\text{GFT}}(\varphi,\bar{\varphi})} \ee
where $\mu$ is a functional measure which in general is ill-defined, and $S_{\text{GFT}}$ is the GFT action of the form, \be S_{\text{GFT}} = \int_{G} dg_1 \int_G dg_2 \, {\msK}(g_1,g_2) \bar{\varphi}(g_1) \varphi (g_2) + \int_G dg_1 \int_G dg_2 \,...\; {\msV}(g_1,g_2,...) f(\varphi,\bar{\varphi}) \ee
where $g\in G$, and the kernel ${\msV}$ is generically non-local, which convolutes the arguments of several $\varphi$ and $\bar{\varphi}$ fields (written here in terms of a single function $f$). It defines the interaction vertex of the dynamics by enforcing the combinatorics of its corresponding (unique, via the inverse of the bulk map) boundary graph. \

$Z_{\text{GFT}}$ defines the covariant dynamics of a group field theory model encoded in $S_{\text{GFT}}$. In this section we show a way to derive such covariant dynamics from a suitable quantum statistical equilibrium description of a system of quanta of space defined in the previous sections \cite{Kotecha:2019vvn}. The following technique of using field coherent states is the same as used in \cite{Oriti:2013aqa,Chirco:2018fns}, but with the crucial difference that here we do not claim to define, or aim to achieve any correspondence (even if formal) between a canonical dynamics (in terms of a projector operator, related further to loop quantum gravity \cite{Oriti:2013aqa}) and a covariant dynamics (in terms of a functional integral). Instead we show a quantum statistical basis for the \emph{covariant} dynamics itself, and in the process, reinterpret the standard form of the GFT partition function \eqref{ZGft} as that of an effective statistical field theory arising from a coarse-graining, and further approximations, of the underlying statistical quantum gravity system. \
  
We discussed in \ref{intdyn} that the dynamics of the polyhedral atoms of space is encoded in the choices of propagators and interaction vertices, which can be written in terms of kinetic and vertex operators in the Fock description. In our present considerations with a single type of atom, namely $SU(2)$-labelled 4-valent node, let us consider the following generic kinetic and vertex operators,
\begin{align} \h{\mbK} &= \int_{SU(2)^8} [dg] \; \h{\varphi}^\dag (\vg_1) {\msK}(\vg_1,\vg_2) \h{\varphi}(\vg_2) \\
 \h{\mbV} &= \int_{SU(2)^{4N}} [dg] \; {\msV}_{\gamma}(\vg_1,...,\vg_N) \h{f}(\h{\varphi},\h{\varphi}^\dag) \end{align}
where $N > 2$ is the number of 4-valent nodes in the boundary graph $\gamma$, and $\h{f}$ is a function of the ladder operators with all terms of a single degree $N$. For example when $N=3$, this function could be $\h{f} = \lambda_1 \h{\varphi}\h{\varphi}\h{\varphi}^\dag + \lambda_2 \h{\varphi}^\dag \h{\varphi}\h{\varphi}^\dag$. As we saw before, in principle a generic model can include several distinct vertex operators. Even though what we consider here is the simple of case of having only one $\h{\mbV}$, the following treatment can be extended to the general case. \

Operators $\h{\mbK}$ and $\h{\mbV}$ have well-defined actions on the Fock space $\mcH_F^{\tet}$. Using the thermodynamical characterisation, we can consider the formal constraints\footnote{A proper interpretation of these constraints is left to future work.} $\langle \h{\mbK} \rangle = $ constant and $\langle \h{\mbV} \rangle =$ constant, to write down a generalised Gibbs state on $\mcH_F^{\tet}$,
\be \h{\rho}_{\{\beta_a\}} = \frac{1}{Z_{\{\beta_a\}}} e^{-\beta_1 \h{\mbK} - \beta_2\h{\mbV}} \ee
where $a=1,2$ and the partition function is,
\be \label{zsm} Z_{\{\beta_a\}} = \Tr (e^{-\beta_1 \h{\mbK} - \beta_2\h{\mbV}}) \,. \ee

An effective field theory can then be extracted from the above by using a basis of coherent states on $\mcH_F^{\tet}$ \cite{Oriti:2013aqa,Chirco:2018fns,klauderbook,Kotecha:2019vvn}. Field coherent states give a continuous representation on $\mcH_F^{\tet}$ where the parameter labelling each state is a wavefunction \cite{klauderbook}. For the Fock description outlined in section \ref{atomkin} (and detailed in the upcoming section \ref{bosgft}), the Fock vacuum is specified by 
\be \h{\varphi}(\vg) \ket{0_{\tet}} = 0 \,, \quad \forall \vg \,. \ee
The field coherent states are,
\be \ket{\psi} = e^{\h{\varphi}^\dag(\psi) - \h{\varphi}(\psi)}\ket{0_{\tet}} = e^{-\frac{||\psi||^2}{2}}e^{\h{\varphi}^\dag(\psi)}\ket{0_{\tet}}  \ee
where $\psi \in \mcH_{\tet}$, $||.||$ is the $L^2$ norm in $\mcH_{\tet}$, and $\h{\varphi}(\psi) = \int d\vg \, \bar{\psi}\h{\varphi}$ along with its adjoint are the smeared ladder operators. The set of all such states provides an over-complete basis for $\mcH_F^{\tet}$. A very useful property of these states is that they are eigenstates of the annihilation operator, 
\be \label{coheigstate} \h{\varphi}(\vg)\ket{\psi} = \psi(\vg)\ket{\psi} . \ee 
This property implies that, for an operator $\h{Q}(\h{\varphi},\h{\varphi}^\dag)$ as a function of the ladder operators, we have \cite{klauderbook},
\be \label{normord} \bra{\psi} : \h{Q}(\h{\varphi},\h{\varphi}^\dag) : \ket{\psi} = Q(\psi,\bar{\psi})  \ee
where, $: . :$ denotes normal ordering, i.e. ordering in which all $\h{\varphi}^\dag$'s are to the left of all annihilation operators $\h{\varphi}$'s, and $Q$ is the same function of $\psi$ and $\bar{\psi}$, as the operator $\h{Q}$ is of $\h{\varphi}$ and $\h{\varphi}^\dag$ respectively. \

The traces for the partition function and other observable averages can then be evaluated in this basis,
\begin{align} \label{cohZ}
\Tr(e^{-\beta_1 \h{\mbK} - \beta_2\h{\mbV}} \h{\mcO}) &= \int [D\mu(\psi,\bar{\psi})] \bra{\psi} e^{-\beta_1 \h{\mbK} - \beta_2\h{\mbV}} \h{\mcO}\ket{\psi} \\
 \text{with} \quad Z_{\{\beta_a\}} &= \Tr(e^{-\beta_1 \h{\mbK} - \beta_2\h{\mbV}} I)  \label{cohZ1}
\end{align}
where the resolution of identity and the coherent state functional measure are respectively given by \cite{klauderbook},
\begin{align} I &= \int [D\mu(\psi,\bar{\psi})] \ket{\psi}\bra{\psi} ,  \\
D\mu(\psi,\bar{\psi}) &= \lim_{K \to \infty}\prod_{k=1}^{K}\frac{d\,\text{Re}\psi_k \; d\,\text{Im}\psi_k}{\pi} \,. \end{align}
The set of all such observable averages formally defines the complete statistical system. In particular, the quantum statistical partition function can be reinterpreted as the partition function for a field theory (of complex-valued square-integrable fields) of the underlying quanta, which here are quantum tetrahedra, as follows \cite{Oriti:2013aqa,Chirco:2018fns,Kotecha:2019vvn}. \

For simplicity, let us first consider, $e^{-\beta \h{C}}$, associated with an operator $\h{C}(\h{\varphi},\h{\varphi}^\dag)$. Then, given $\h{C}$ and $\h{\mcO}$ as polynomial functions of the generators, with a given (but generic) choice of the operator ordering defining the exponential operator, the integrand of the statistical averages can be treated as follows:
\begin{align}
\bra{\psi}e^{-\beta\h{C}}\h{\mcO}\ket{\psi} &= \bra{\psi}\sum_{k=0}^{\infty}\frac{(-\beta)^k}{k!}  \h{C}^k \h{\mcO}\ket{\psi}  \\
&= \bra{\psi} :e^{-\beta \h{C}} \h{\mcO}: \ket{\psi} + \bra{\psi} :\text{po}_{{C},{\mcO}}(\h{\varphi},\h{\varphi}^\dag,\beta) : \ket{\psi}
\end{align}
where to get the second equality, we have used the commutation relations \eqref{ccrtemp} on each $\h{C}^k\h{\mcO}$, to collect all normal ordered terms $:\h{C}^k \h{\mcO}:$ 
giving the normal ordered $:e^{-\beta \h{C}} \h{\mcO}:$, and the second term is a collection of the remaining terms arising as a result of swapping $\h{\varphi}$'s and $\h{\varphi}^\dag$'s, which will then in general be a normal ordered series in powers of $\h{\varphi}$ and $\h{\varphi}^\dag$, with coefficient functions of $\beta$. The precise form of this series will depend on both $\h{C}$ and $\h{\mcO}$, hence the subscripts. 
Using equations \eqref{coheigstate} and \eqref{normord}, we have:
\be
\bra{\psi} :e^{-\beta \h{C}}\h{\mcO}: \ket{\psi} = e^{-\beta C[\psi,\bar{\psi}]} \mcO[\psi,\bar{\psi}] 
\ee
where $C[\psi,\bar{\psi}] = \bra{\psi} : \h{C} : \ket{\psi}$ and $\mcO[\psi,\bar{\psi}] = \bra{\psi} : \h{\mcO} : \ket{\psi}$; and, 
\be
 \bra{\psi} : \text{po}_{C,\mcO}(\h{\varphi},\h{\varphi}^\dag,\beta) : \ket{\psi} \equiv \bra{\psi} :\h{A}_{C,\mcO}(\h{\varphi},\h{\varphi}^\dag,\beta) : \ket{\psi} = A_{C,\mcO}[\psi,\bar{\psi},\beta] \,.
\ee
Thus, averages can be written as
\be \label{avgs} \Tr({e^{-\beta\h{C}} \h{\mcO}}) = \int [D\mu(\psi,\bar{\psi})] \; \left( e^{-\beta C[\psi,\bar{\psi}]} \mcO[\psi,\bar{\psi}] + A_{C,\mcO}[\psi,\bar{\psi},\beta] \right) . \ee
In particular, the quantum statistical partition function is 
\be \label{pf} Z = \int [D\mu(\psi,\bar{\psi})] \; \left(e^{-\beta C[\psi,\bar{\psi}]} + A_{C,\mbI}[\psi,\bar{\psi},\beta]\right) \;=:\; Z_{0} + Z_{\mcO(\hbar)} 
\ee
where, by notation $\mcO(\hbar)$ we mean only that this sector of the full theory encodes all higher orders in quantum terms relative to $Z_0$.\footnote{Further investigation into the interpretation, significance and consequences of this rewriting of $Z$ in discrete quantum gravity is left for future work.} 
This full set of observable averages (or correlation functions) \eqref{avgs}, including the above partition function, defines thus a statistical field theory of quantum tetrahedra (or in general, polyhedra with a fixed number of boundary faces), characterised by a combinatorially non-local statistical weight, that is, a group field theory. \
 
If we are further able to either reformulate exactly, or under suitable approximations, $A_{C,\mcO}$ in the following way,
 \be A_{C,\mcO} = A_{C,\mbI}[\psi,\bar{\psi},\beta] \,\mcO[\psi,\bar{\psi}]  \ee
 then, the partition function \eqref{pf} defines a statistical field theory for the algebra of observables $\mcO[\psi,\bar{\psi}]$, for a dynamical system of complex-valued square-integrable fields $\psi$, defined on the base manifold $SU(2)^4$.

The integrands in \eqref{cohZ} and \eqref{cohZ1} can then be treated and simplified along the lines above, in particular to get an effective partition function, 
\be Z_o = \int [D\mu(\psi,\bar{\psi})] \; e^{-\beta_1{\msK}[\psi,\bar{\psi}] - \beta_2 {\msV}[\psi,\bar{\psi}]} = Z_{\{\beta_a\}} - Z_{\mcO(\hbar)} \ee
where subscript $o$ indicates that we have neglected higher order terms, collected inside $Z_{\mcO(\hbar)}$, resulting from normal orderings of the exponent in $Z_{\{\beta_a\}}$. As before, the functions in the exponent are ${\msK} = \bra{\psi}:\h{\mbK}:\ket{\psi}$ and ${\msV} = \bra{\psi}:\h{\mbV}:\ket{\psi}$. It is then evident that $Z_o$ has the precise form of a generic GFT partition function of the form \eqref{ZGft}. It thus \emph{defines} a group field theory as an effective statistical field theory, that is \be Z_{\text{GFT}} := Z_o \,. \ee   

From this perspective, it is clear that the generalised inverse temperatures (which are basically the intensive parameters conjugate to the generalised energies in the generalised thermodynamical setting of section \ref{GenTD}) \emph{are} the coupling parameters defining the effective model, thus characterising the phases of the emergent statistical field theory, as would be expected. Moreover, from this purely statistical standpoint, we can understand the GFT action more appropriately as Landau-Ginzburg free energy (or effective `Hamiltonian', in the sense that it encodes the effective dynamics), instead of a Euclidean action which might imply having Wick rotated a Lorentzian measure, even in an absence of any such notions as is the case presently. Lastly, deriving like this the covariant definition of a group field theory, based entirely on the framework presented here, strengthens the statement that a group field theory is a field theory of combinatorial and algebraic quanta of space \cite{Oriti:2006se,Oriti:2011jm}.

\chapter{Group Field Theory}
\label{GFT}

\begin{quotation}
If the space void of all bodies, is not altogether empty; what is it then full of? Is it full of extended spirits perhaps, or immaterial substances, capable of extending and contracting themselves; which move therein, and penetrate each other without any inconveniency, as the shadows of two bodies penetrate one another upon the surface of the wall? ---\emph{Gottfried W. Leibniz}\footnote{As cited in \cite{hesse2005forces}, p161.}
\end{quotation}


\vspace{3.5mm}
\vspace{1pt}

\noindent We have discussed, in the previous chapter, how group field theories can arise effectively from a coarse graining of the underlying quantum statistical system. We now turn to GFTs proper, and introduce the structures that are relevant for our subsequent applications. \

Group field theories are field theories of combinatorial and algebraic quanta of geometry, formally defined by a statistical partition function 
\be Z_{\text{GFT}} = \int [D\varphi_\kappa] \; e^{-S_{\text{GFT}}(\{\varphi_\kappa\})}  \ee
for a set of group fields $\{\varphi_\kappa\}$. They are strictly related to various other approaches like loop quantum gravity \cite{Ashtekar:2004eh,Rovelli:2004tv,Bodendorfer:2016uat}, spin foams \cite{Perez:2012wv, rovelli_vidotto_2014}, causal dynamical triangulations \cite{Loll:2019rdj}, tensor models \cite{Gurau:2011xp} and lattice quantum gravity \cite{Hamber:2009mt}. Like in usual field theories, the kinematics is specified by a choice of the fields $\varphi_\kappa$, each defined in general over a domain space of direct products of Lie groups, and taking values in some target vector space. The dynamics is specified by propagators and interaction vertices encoded in a function $S_{\text{GFT}}$, which can be understood as a Landau-Ginzburg free energy function in the present statistical context \cite{Kotecha:2019vvn}; or as a Euclidean action from the point of view of standard quantum field theories \cite{Reisenberger:2000zc,Freidel:2005qe,Oriti:2006se,Oriti:2011jm,Oriti:2005mb}. However unlike in usual field theories, $S_{\text{GFT}}$ is non-local in general with respect to the base manifold. This non-locality is essential, and encodes the non-trivial combinatorial nature of the fundamental degrees of freedom and their dynamics. Moreover, the base manifold is not spacetime, but carries algebraic information associated with discrete geometric and matter degrees of freedom. Such a complete absence of any continuum spacetime structures a priori is a manifestation of background independence in group field theory, like in various other non-perturbative approaches to quantum gravity. \

The partition function $Z_{\text{GFT}}$ perturbatively generates Feynman diagrams that are labelled 2-complexes (dual to labelled stranded diagrams), with boundary states given by labelled graphs \cite{Reisenberger:2000zc,Freidel:2005qe,Oriti:2006se,Oriti:2011jm,Oriti:2005mb}. For the choice of models closer to loop quantum gravity and spin foam setups, the boundary states are abstract spin networks (but organised in a second quantised Hilbert space, that of a field theory \cite{Oriti:2013aqa}) and bulk processes are spin foams, both of which in turn are dual to polyhedral complexes when restricting to loopless combinatorics \cite{Bianchi:2010gc,Oriti:2014yla}. Thus, a group field theory generates discrete quantum spacetimes made of fundamental polyhedral quanta\footnote{As discussed previously in chapter \ref{DQG}, a quantised polyhedron with $d$ faces is dual to a gauge-invariant open $d$-valent spin network node \cite{Bianchi:2010gc}. The latter is in fact a special case (namely, a $d$-patch \cite{Oriti:2014yla}) of richer combinatorial boundary structures which can be treated analogously in our present setup, but in that case without related discrete geometric understanding of the same.}. \

We can describe the same structures from a many-body perspective \cite{Oriti:2017twl}, and treat as more fundamental an interacting system of many such quanta. This viewpoint enables us to import formal techniques from standard many-body physics for macroscopic systems, by treating a quantum polyhedron or an open spin network node as a single particle of interest. In fact, it has allowed for tangible explorations of connections of group field theory with quantum information theory and holography \cite{Colafranceschi:2020ern,Colafranceschi:2021acz,Chirco:2017xjb,Chirco:2019dlx,Chirco:2017wgl,Chirco:2017vhs}, and also with quantum statistical mechanics and thermal physics \cite{Assanioussi:2019ouq,Kotecha:2019vvn,Chirco:2018fns,Chirco:2019kez,Kotecha:2018gof,Kegeles:2017ems}. It has further allowed for importing ideas and tools from condensed matter theory, which has been crucial for instance in the development of GFT condensate cosmology \cite{Pithis:2019tvp,Oriti:2016acw,Gielen:2016dss,Gabbanelli:2020lme}. \

This same perspective further leads to a modelling of an extended region of discrete quantum space (a labelled graph) as a multi-particle state, and a region of dynamical quantum spacetime (a labelled 2-complex) as an interaction process. As we have emphasised before, this is the perspective that we employ throughout the thesis to use techniques from many-body physics and statistical mechanics, even when working with a radically different kind of system, one that is background independent, and devoid of any standard notion of space, time and other associated geometric structures and standard matter couplings. \

The simplest class of models are scalar theories, with fields $\varphi: G^{d} \rightarrow \mathbb{C}$, defined on a direct product of $d$ copies of the local gauge group of gravity. This is the Lorentz group $SL(2,\mathbb{C})$ in 4d or its Euclidean counterpart $Spin(4)$. $SU(2)$ is often used as the relevant subgroup in the context of quantum gravity, especially for models connected to LQG. In this thesis, $G$ is taken to be locally compact so that the Haar measure is defined (even though it would be finite only for compact groups); connected (for constructions in section \ref{inttrans}); and, unimodular so that the left- and right-invariant Haar measures coincide. These properties are satisfied by $SU(2), \mathbb{R}, Spin(4)$ and $SL(2,\mathbb{C})$, of which the first two are used in some specific examples later in the thesis. \

Further, we may be interested in group field theories which generate discrete spacetimes coupled to discretised real scalar matter fields. One way to couple a single real scalar degree of freedom is by extending the original configuration space $G^d$, encoding purely geometric data, by $\mathbb{R}$. By extension, $n$ number of scalar fields can be coupled by considering the base manifold $G^d \times \mathbb{R}^n$ \cite{Li:2017uao,Pithis:2019tvp,Gielen:2016dss,Oriti:2016qtz,Oriti:2016acw,Gabbanelli:2020lme,Gielen:2018fqv,Gielen:2020fgi}. Doing this we have assigned additional $n$ real numbers representing the values of the fields, to each quantum of the group field. Consequently, a GFT Feynman diagram (labelled 2-complex) is enriched by $n$ scalar fields that are discretised on the vertices of the boundary graphs (equivalently, edges of the bulk 2-complex). Thus we are concerned with a group field,
\be \varphi: G^{d} \times \mathbb{R}^{n} \rightarrow \mathbb{C}, \, (\vg,\vphi) \mapsto \varphi(\vg,\vphi) \ee
defined for arbitrary natural numbers $n\geq0$ and $d>0$ (before any physical restrictions coming from considerations in discrete quantum gravity). \

The main reasons for including matter in GFTs are natural. Any fundamental theory of gravity must include, or must be able to generate at an effective level, matter degrees of freedom if it is to eventually realistically describe the universe. Also, material reference frames can be used to define physical relational observables in background independent systems \cite{Rovelli:1990ph,PhysRevLett.72.446,Brown:1994py,tamborninoreview,Bojowald:2010qw,Hohn:2018toe,Hohn:2019cfk,Hoehn:2020epv}. For instance in GFT, relational reference frames defined by scalar fields have been used in the context of cosmology \cite{Li:2017uao,Pithis:2019tvp,Gielen:2016dss,Oriti:2016qtz,Oriti:2016acw,Assanioussi:2020hwf,Marchetti:2020umh,Gabbanelli:2020lme,Wilson-Ewing:2018mrp,Gielen:2020fgi}. In this thesis, we make use of relational frames in two different contexts. In section \ref{momgibbs}, we define Gibbs states first with respect to internal translations in $\vphi$, and then with respect to a clock Hamiltonian associated with one of these $\phi$'s, now being an external clock to the system after deparametrization as described in section \ref{depgft} below. Later in section \ref{GFTCC}, we first clarify the notion of relational frames in GFT by using smearing functions along the $\phi$ direction, and then apply them in the context of thermal condensate cosmology in GFT. \

In this chapter, we present a quantum operator formulation of bosonic group field theories, detailing those aspects which are relevant for our later constructions. In section \ref{fockspace}, we give a Fock space construction based on the degenerate vacuum (which we also encountered in the previous chapter). In section \ref{usbas} we present three orthonormal bases that will be useful for later investigations. Then in section \ref{weylGFT}, we give the Weyl algebraic formulation of the same system, along with the construction of their translation *-automorphisms in \ref{AUT}. Finally in section \ref{depgft}, we consider the issue of deparametrization in group field theory based on insights and tools from presymplectic mechanics of multi-particle systems.


\section{Bosonic group field theory} \label{bosgft}


\subsection{Degenerate vacuum and Fock representation} \label{fockspace}

Adopting a second quantisation scheme \cite{Bratteli:1979tw,Bratteli:1996xq,fetter}, states of a group field $\varphi(\vg,\vphi)$ can be organised in a Fock space $\mathcal{H}_F$ generated by a Fock vacuum $\ket{\Omega_F}$ and the ladder operators\footnote{Throughout the thesis, we may choose to neglect or reinstate the hat notation for operators without notice, in order to simplify or clarify the notation in the setup at hand.} $\{\varphi,\varphi^\dag\}$\cite{Oriti:2013aqa,Mikovic:2001yg}. The vacuum is specified by
\be \varphi(\vg,\vphi) \ket{\Omega_F} = 0 \,, \quad \forall \, \vg,\vphi \,.  \ee
This is a degenerate vacuum, with no quantum geometrical or matter degrees of freedom. It is a state like $\ket{0_{\tet}}$, the `no-space' state, which we encountered previously in section \ref{atomkin}. We also remark that later in section \ref{threpcond}, a degenerate vacuum of this kind is denoted by $\ket{0}$, in order to declutter the overall notation of the investigation there. \

A single quantum is created by acting on $\ket{\Omega_F}$ with the creation operator,
\begin{equation}
\varphi^\dag(\vg,\vphi) \ket{\Omega_F} = \ket{\vg,\vphi} 
\end{equation}
which is the state of a $d$-valent node whose links are labelled by group elements $\vg = (g_1,...,g_d)$ and the node itself by a set of real numbers $\vphi = (\phi_1,...,\phi_n)$. Then, a generic single-particle state with wavefunction $\psi$ is given by, 
\be
\ket{\psi} = \int_{G^d} d\vg \int_{\mathbb{R}^n} d\vphi \; \psi(\vg,\vphi) \, \ket{\vg,\vphi} 
\ee
where $\psi$ is an element of the single-particle Hilbert space. This Hilbert space is given by\footnote{Notice that even for compact $G$, the base manifold is non-compact along $\mathbb{R}$, which might then demand appropriate regularisation schemes depending on the case at hand e.g. operator \eqref{phimom} in section \ref{inttrans}.}
\be \mathcal{H} = L^2(G^d \times \mathbb{R}^n)  \ee 
or, when the geometric condition of closure is satisfied, by\footnote{We have chosen to denote both these Hilbert spaces by the same $\mcH$, because our technical results are independent of which one we choose. What changes is the interpretation of a single quantum of the group field, which can be understood as a convex polyhedron only with the latter choice.}
\be \mcH = L^2(G^d/G \times \mathbb{R}^n) \cong L^2(G^d/G) \otimes L^2(\mathbb{R}^n)  \,.  \ee
Closure corresponds to the field $\varphi$ being invariant under a diagonal right\footnote{The right gauge invariance is imposed in order to avail a quantum geometric interpretation of the quanta of the GFT field, as discussed before. An additional left gauge invariance can also be imposed, as has been done in studies in the context of homogeneous cosmologies in group field theory \cite{Gielen:2014ila,Gielen:2013kla,Gielen:2013naa,Pithis:2019tvp,Oriti:2016qtz,Assanioussi:2020hwf}. However, our entire setup, along with the technical results based on it, will mathematically follow through with or without (either or even both) these additional symmetries and their associated geometric interpretations.} action of $G$ on $G^d$, that is
\be \label{ginv} \varphi(g_I,\phi) = \varphi(g_I h, \phi) \,, \quad \forall \, h \in G  \ee
and allows for understanding the quanta of the field as convex polyhedra (in turn dual to gauge-invariant spin network nodes) \cite{Barbieri:1997ks,Baez:1999tk,Bianchi:2010gc}. This invariance effectively reduces the geometric part of the domain space to $G^d/G$, as evident in the expression for $\mcH$ above. In this thesis however, even when imposing closure, we will continue to use a redundant parametrization for convenience and consider the gauge-invariant functions to be defined on full $G^d$ while explicitly satisfying equation \eqref{ginv}.  \

Further, we choose to impose a symmetry under arbitrary particle exchanges on the states, that is bosonic statistics. In the spin network picture, this condition reflects the graph automorphism of vertex relabelling and is a natural feature to require. For the case at hand then, the Hilbert space for bosonic quanta is the Fock space
\be \label{fockhilbsp}
\mathcal{H}_F = \bigoplus_{N \geq 0} \text{sym}\, \mathcal{H}^{\otimes N} 
\ee
where $\mathcal{H}^{\otimes N}$ describes the $N$-particle sector, with $\mcH_0=\mathbb{C}$, and sym refers to the symmetric projection of it. This is the space that we encountered before in section \ref{atomkin}. We stress again that this Fock space contains arbitrary spin network excitations \cite{Oriti:2013aqa}, thus the quantum gravity structures are shared with loop quantum gravity even if organised in a different way. This means that defining proper statistical equilibrium states on this Fock space truly means defining non-perturbative statistical equilibrium states in a fully background independent context and within a fundamental theory of quantum gravity based on spin network states.  \

The ladder operators that take us between the different multi-particle sectors satisfy the commutation relations algebra,
\begin{align} \label{ladderCR}
[\varphi(\vg,\vphi), \varphi^\dag(\vg', \vphi')] &= \mathbb{I}(\vg,\vg')\delta(\vphi - \vphi') \\ 
[\varphi(\vg,\vphi), \varphi(\vg', \vphi')] &= [\varphi^\dag(\vg,\vphi), \varphi^\dag(\vg', \vphi')] = 0
\end{align}
where, $\mathbb{I}$ and $\delta$ are delta distributions for functions on $G^d$ and $\mathbb{R}^n$ respectively. The delta distribution on $G^d$ for right gauge-invariant functions takes the form, 
\be \mathbb{I}(\vg,\vg') = \int_{G}dh \; \prod_{I=1}^d \delta(g_I h g_I'^{-1})\,. \ee

The second quantised operators (see \cite{Oriti:2013aqa} for more details in the context of LQG) are elements of the unital *-algebra generated by $\{\varphi,\varphi^\dag,I\}$, where $I$ is the identity operator on $\mathcal{H}_F$, and the adjoint $\dag$ is the * operation. These elements are in general polynomials $\mathcal{O}(\varphi,\varphi^\dag,I)$ of the three. We denote this algebra by $\mathcal{A}_F$, with action on $\mathcal{H}_F$. 


\subsection{Useful bases} \label{usbas}

\subsubsection*{Spin-momentum} \label{spinmom}

For compact $G$, by Peter-Weyl theorem, the Hilbert space $L^2(G)$ can be decomposed into a sum of finite-dimensional irreducible representations of $G$ \cite{kowalski,Martin-Dussaud:2019ypf}. Then for $G=SU(2)$, a useful basis in $L^2(SU(2)^d)$ is given by the Wigner modes $\mbfD_{\vchi}$, labelled by a set of irreducible representation indices $\vchi$. Particularly, for right gauge-invariant functions on $SU(2)^d$ (satisfying \eqref{ginv}), this is given by the following set of Wigner functions 
\be \label{wigner} \mbfD_{\vchi}(\vg) = \sum_{\vec{m}'}C^{\vec{j}}_{\vec{m}' \iota} \prod_{I=1}^d \mbfD^{j_I}_{m_I m'_I}(g_I)  \ee 
where $\vchi \equiv (\vec{j},\vec{m},\iota)$, $j_I  \in \mathbb{N}/2$ are spin irreducible representation indices of $SU(2)$, $m_I,m'_I \in (-j_I,...,+j_I)$ are matrix indices in representation $j_I$, $\mbfD^{j_I}_{m_I m'_I}$ are complex-valued Wigner matrix coefficients (multiplied by a factor of $\sqrt{\text{dim}_{j_I}} = \sqrt{2j_I+1}$, for normalisation shown in \eqref{wignorm} below) in representation $j_I$, and $C^{\vec{j}}_{\vec{m}' \iota}$ are intertwiner basis elements indexed by $\iota$ arising due to the closure condition in \eqref{ginv} for $SU(2)$. Orthonormality and completeness are respectively given by,
\begin{align} \label{wignorm} \int d\vg \; \overline{\mbfD}_{\vchi}(\vg) \mbfD_{\vchi'}(\vg) &= \delta_{\vchi\vchi'} \;, \\
\sum_{\vchi} \overline{\mbfD}_{\vchi}(\vg) \mbfD_{\vchi}(\vg') &= \mathbb{I}(\vg,\vg') \;.  \end{align}
While for the coupled matter degrees of freedom in $L^2(\mathbb{R}^n)$, one can consider the standard Fourier basis, 
\be \mbfF(\vp,\vphi) = e^{-i\vp.\vphi} \,. \ee
The corresponding basis in $\mcH$ is then of the tensor product form, $(\mbfD_{\vchi} \otimes \mbfF(\vp))(\vg,\vphi)$, and the algebra generators are given by the appropriate smearing,
\begin{align}
\varphi_{\vchi}(\vp) &:= \varphi(\mbfD_{\vchi} \otimes \mbfF(\vp)) = \int d\vg d\vphi \; \overline{\mbfD}_{\vchi}(\vg) \overline{\mbfF}(\vp,\vphi) \varphi (\vg,\vphi) \,, \\
\varphi^\dag_{\vchi}(\vp) &= \varphi^\dag(\mbfD_{\vchi} \otimes \mbfF(\vp)) = \int d\vg d\vphi \; {\mbfD}_{\vchi}(\vg) {\mbfF}(\vp,\vphi) \varphi^\dag(\vg,\vphi) \,.
\end{align}
The algebra structure in \eqref{ladderCR} is preserved, and now takes the form
\begin{align} \label{spinCR}
[\varphi_{\vchi}(\vp), \varphi^\dag_{\vchi'}(\vp')] &= (2\pi)^n \delta_{\vchi \vchi'} \delta(\vp - \vp') \\
 [\varphi_{\vchi}(\vp), \varphi_{\vchi'}(\vp')] &= [\varphi^\dag_{\vchi}(\vp), \varphi^\dag_{\vchi'}(\vp')] =0
\end{align}
where $\delta_{\vchi \vchi'}$ is the Kronecker delta, and $\delta(\vp - \vp')$ is a Dirac delta distribution on $\mathbb{R}^n$. The Fock space $\mcH_F$ is then generated via actions of $\{\varphi_{\vchi}(\vp),\varphi^\dag_{\vchi}(\vp),I\}$ on the vacuum $\ket{\Omega_F}$.


\subsubsection*{Discrete index} \label{discind}

We retain the use of the Wigner basis mentioned above, labelled by discrete indices $\vchi$. For the matter part, let us consider a basis of complex-valued smooth functions $\mbfT_{\valpha}(\vphi)$ in $L^2(\mathbb{R}^n)$, labelled by a set of discrete indices $\valpha = (\alpha_1,...,\alpha_n) \in \mathbb{N}^n$, satisfying orthonormality and completeness,
\begin{align} \int d\vphi \; \overline{\mbfT}_{\valpha}(\vphi) \mbfT_{\valpha'}(\vphi) &= \delta_{\valpha \valpha'} \,,\\
 \sum_{\valpha} \overline{\mbfT}_{\valpha}(\vphi) \mbfT_{\valpha}(\vphi') &= \delta(\vphi-\vphi') \,.\end{align}
We thus have a complete orthonormal basis on $\mcH$ consisting of functions $(\mbfD_{\vchi} \otimes \mbfT_{\valpha})(\vg,\vphi)$ of the tensor product form. Then as before, the set of mode ladder operators can be defined by smearing the operators $\varphi , \varphi^\dag$ with this basis,
\begin{align} \label{a}  
 {a}_{\vchi\valpha} &:=   {\varphi}(\mbfD_{\vchi} \otimes \mbfT_{\valpha}) = \int d\vg d\vphi\; \overline{\mbfD}_{\vchi}(\vg)\overline{\mbfT}_{\valpha}(\vphi)   {\varphi}(\vg,\vphi) \,,\\
 \label{adag}   {a}^\dag_{\vchi\valpha} &=   {\varphi}^\dag(\mbfD_{\vchi} \otimes \mbfT_{\valpha}) = \int d\vg d\vphi\; {\mbfD}_{\vchi}(\vg) {\mbfT}_{\valpha}(\vphi)   {\varphi}^\dag(\vg,\phi) \,. 
 \end{align}
This essentially decomposes the operators $  {\varphi},  {\varphi}^\dag$ in terms of the modes $\mbfD_{\vchi} \otimes \mbfT_{\valpha}$, which can be seen directly by inverting the above two equations. The algebra relations are,
\begin{align} \label{RegCom0} [  {a}_{\vchi\valpha},  {a}^\dag_{\vchi' \valpha'}] &= \delta_{\vchi\vchi'} \delta_{\valpha \valpha'} \\
[  {a}_{\vchi\valpha},  {a}_{\vchi'\valpha'}] &= [a^\dag_{\vchi\valpha},a^\dag_{\vchi'\valpha'}] = 0 \,.
 \end{align}
We note that due to the choice of a discrete basis $\{\mbfT_{\valpha}(\vphi)\}_{\valpha}$ in $L^2(\mathbb{R}^n)$, the algebra commutator in \eqref{RegCom0} now produces Kronecker deltas, instead of the Dirac delta distribution of the original basis in equation \eqref{ladderCR}, or the momentum basis in equation \eqref{spinCR}. The consideration of a regular algebra instead of a distributional one, particularly for the $\phi$-modes, is an important feature in certain parts of this thesis where we deal with inequivalent representations obtained via thermal Bogoliubov transformations. Namely, the Kronecker delta $\delta_{\alpha\alpha'}$ is crucial in order to avoid divergences related to the coincidence limit $\phi \to \phi'$ of $\delta(\phi-\phi')$. As we will see in sections \ref{CTS} and \ref{vgibbs}, such terms with $\delta(\phi-\phi')$ arise naturally when calculating thermal expectation values of certain relevant observables, for example the average thermal number density.

The degenerate vacuum is again specified by, 
\be \label{degvac} {a}_{\vchi \valpha}\ket{\Omega_F} = 0 \,, \quad \forall J,\alpha \ee 
which generates the symmetric Fock space $\mcH_F$ by the action of the generators $\{  {a}_{\vchi\valpha},   {a}^{\dag}_{\vchi\valpha},  {1}\}$. For instance, a single particle ($N=1$), single mode state is,
\be \ket{\vchi,\valpha} \equiv \ket{\mbfD_{\vchi} \otimes \mbfT_{\valpha}} =   {a}^\dag_{\vchi\valpha} \ket{\Omega_F} \ee
while a single particle state with wavefunction $\psi(\vg,\phi) = \sum \limits_{\vchi,\valpha} \mbfD_{\vchi} \otimes \mbfT_{\valpha}(\vg,\phi)\psi_{\vchi\valpha} \in \mcH$ is,
\be \ket{\psi} =   {a}^\dag(\psi)\ket{\Omega_F} = \sum_{\vchi,\valpha}\psi_{\vchi\valpha}  {a}^\dag_{\vchi\valpha}\ket{\Omega_F}. \ee


\subsubsection*{Occupation number} \label{occunum}

A good basis to work with in the Fock space is the orthonormal occupation number basis. It is particularly useful because it is the eigenbasis of the number operator, and therefore of all extensive operators, in $\mathcal{H}_F$. This basis organises the states according to the number of particles occupying a given mode $(\vchi,\valpha)$. \

Utilising the commutation algebra relations \eqref{RegCom0}, a normalised multi-particle state with $n_{\vchi\valpha}$ number of particles in a single mode $(\vchi,\valpha)$ is given by,
\begin{equation}
\ket{n_{\vchi\valpha}} := \frac{1}{\sqrt{n_{\vchi\valpha}!}} (a^\dag_{\vchi\valpha})^{n_{\vchi\valpha}}\ket{\Omega_F} .
\end{equation}
Then, generic multi-particle states occupying several modes $\vchi_i,\valpha_i$ are given by,
\begin{equation}
\ket{\{n_{\vchi_i\valpha_i}\}} \equiv \ket{n_{\vchi_1\valpha_1}, n_{\vchi_2\valpha_2}, ..., n_{\vchi_i\valpha_i}, ...} := \frac{1}{\sqrt{\prod \limits_i \left(n_{\vchi_i\valpha_i}!\right)}} \prod_i (a^\dag_{\vchi_i\valpha_i})^{n_{\vchi_i\valpha_i}} \ket{\Omega_F} 
\end{equation}
where mode index $i \in \mathbb{N}$ is finite. Orthonormality relations are,
\be \langle \{n_{{\vchi}\valpha}\} | \{m_{{\vchi}'\valpha'}\} \rangle = \delta_{n_{{\vchi}\valpha}m_{{\vchi}'\valpha'}}\delta_{{\vchi}{\vchi}'}\delta_{\valpha\valpha'} \,. \ee
The number operator for a single mode $N_{\vchi\valpha} = a^\dag_{\vchi\valpha} a_{\vchi\valpha}$, counts its occupation number 
\be N_{\vchi_j\valpha_j}\ket{\{n_{\vchi_i\valpha_i}\}} = n_{\vchi_j\valpha_j} \ket{\{n_{\vchi_i\valpha_i}\}} \ee 
while the total number operator $N = \sum \limits_{\vchi,\valpha} N_{\vchi\valpha}$, counts the total number of particles,  \be N\ket{\{n_{\vchi_i\valpha_i}\}} = \left(\sum_{j} n_{\vchi_j\valpha_j} \right) \ket{\{n_{\vchi_i\valpha_i}\}} . \ee


\subsection{Weyl algebra} \label{weylGFT}

Operator norm is not defined on the *-algebra $\mathcal{A}_F$, because of the unboundedness of the bosonic ladder operators $\varphi$ and $\varphi^\dag$ \cite{Bratteli:1996xq}. As is standard practice in algebraic treatments of many-body quantum systems, we could work instead with exponentiated versions of these resulting in a unital C*-algebra, the Weyl algebra. \

The following Weyl reformulation of the GFT system is based on the well-known literature in algebraic quantum field theory \cite{Bratteli:1979tw,Bratteli:1996xq,sewellbook,strocchi2005book}. For the purposes of this thesis, this surrounds the Fock representation of many-body, non-relativistic systems, here suitably adapted to define an analogous setup for group field theory \cite{Kegeles:2017ems,Kotecha:2018gof}.\

The fields $\varphi$ and $\varphi^\dag$ are operator-valued distributions. The corresponding operators are defined by smearing them with functions\footnote{Naturally, the smearing is independent of basis, i.e. $\varphi(f) = a(f)$.}  $f \in \mcH$,
\begin{align}
\varphi(f) &:= \int_{G^d \times \mathbb{R}^n} d\vg \, d\vphi \; \bar{f}(\vg,\vphi) \varphi(\vg,\vphi) \\ \varphi^\dag(f) &\,= \int_{G^d \times \mathbb{R}^n} d\vg \, d\vphi \; f(\vg,\vphi) \varphi^\dag(\vg,\vphi) \,.
\end{align}
Depending on the investigation at hand, the smearing functions may also be required to satisfy additional decay properties (see for instance, equations \eqref{bcond} in section \ref{eff2}).

The commutation relations \eqref{ladderCR} take the following form,
\begin{align} \label{weylCR}
&[\varphi(f_1),\varphi^\dag(f_2)] = (f_1,f_2) \\
[\varphi(f_1) & ,\varphi(f_2)] = [\varphi^\dag(f_1),\varphi^\dag(f_2)] = 0 
\end{align}
where \be (f_1,f_2) = \int_{G^d \times \mathbb{R}^n} d\vg\, d\vphi \;\; \bar{f_1}(\vg,\vphi) f_2(\vg,\vphi) \ee is the $L^2$-inner product.\

Bosonic ladder operators are unbounded in the operator norm on $\mathcal{H}_F$, and they are defined on dense subsets of $\mcH_F$. Therefore, let us define the following self-adjoint operators\footnote{In terms of operators $\Phi$ and $\Pi$, the commutation relations take the form, $[\Phi(f_1),\Pi(f_2)] = i\,\text{Re}(f_1,f_2) \;,\; [\Phi(f_1),\Phi(f_2)] = [\Pi(f_1),\Pi(f_2)] = i\,\text{Im}(f_1,f_2)$.} in the common dense domain of $\varphi$ and $\varphi^\dag$, 
\begin{align} \Phi(f) &:= \frac{1}{\sqrt{2}}(\varphi(f) + \varphi^\dag(f)) \\
\Pi(f) &:= \frac{1}{i\sqrt{2}}(\varphi(f)-\varphi^\dag(f)) \end{align}
and consider their exponentiations, \be W_F(f) := e^{i\Phi(f)} \ee or equivalently\footnote{Since $\Pi(f) = \Phi(if)$, both $\varphi$ and $\varphi^\dag$ can be recovered from $\Phi$ or $\Pi$ alone. By convention then, $\Phi$ are usually chosen as the generators of this representation. \cite{Bratteli:1996xq}} $\widetilde{W}_F(f) = e^{i\Pi(f)}$. Then, these exponential operators:

\begin{itemize}
\item are unitary, $W_F(f)^\dag = W_F(f)^{-1} = W_F(-f)$ , and

\item satisfy Weyl relations, $W_F(f_1)W_F(f_2) = e^{-\frac{i}{2}\text{Im}(f_1,f_2)}W_F(f_1 + f_2)$.
\end{itemize}
In other words, $\{W_F(f) \,|\, f\in \mcH\}$ defines a Weyl system \cite{CHAIKEN196723,Bratteli:1979tw,Bratteli:1996xq} on the Hilbert space $\mathcal{H}_F$, over the space of test functions $f$. It defines a unitary representation of the GFT commutation  algebra on $\mcH_F$ with generators $\Phi$. \

Retaining this algebraic structure and forgetting (for now) the generators $\Phi$ which lend the concrete representation, an \emph{abstract} bosonic GFT system can be defined by the pair $(\mathcal{A},\mathcal{S})$, where $\mathcal{A}$ is the Weyl algebra\footnote{Fermionic statistics would correspond to a Clifford algebra.} generated by Weyl unitaries $W(f)$, and $\mathcal{S}$ is the space of algebraic states\footnote{Recall that algebraic states are linear, positive, normalised, complex-valued functionals over an algebra (see footnote \ref{FNstate}, on page \pageref{FNstate}).}. The defining relations of this algebra are, 
\begin{equation} \label{weylCR}
W(f_1)W(f_2) = W(f_2)W(f_1) \, e^{-i \, \text{Im}(f_1,f_2)} = W(f_1+f_2) \, e^{-\frac{i}{2} \text{Im}(f_1,f_2)}\end{equation}
where identity is $\mathcal{I}=W(0)$, and unitarity is $W(f)^{-1} = W(f)^\dag = W(-f)$. This is a unital C*-algebra, equipped with the C*-norm. The benefits of defining a quantum GFT system with an abstract Weyl algebra stem from the fact that some general results can be deduced, which are representation-independent (and would apply, for example, also to other inequivalent representations \cite{Kegeles:2017ems,Assanioussi:2019ouq}). This allows in particular for exploring structural symmetries at the level of the algebra formulated in terms of automorphisms. In the upcoming section \ref{AUT}, we will consider examples of automorphisms of $\mathcal{A}$ corresponding to structural symmetries (translations) of the underlying theory. The KMS condition with respect to these automorphisms will then lead to a definition of structural equilibrium states in section \ref{inttrans}, which encode stability with respect to the corresponding internal flows of the transformation under consideration.  \

The Fock system is now generated as the Gelfand-Naimark-Segal (GNS) representation \cite{strocchi2005book,Bratteli:1979tw,Bratteli:1996xq,Haag:1992hx} $(\pi_F,\mathcal{H}_F,\Omega_F)$ of the regular Gaussian algebraic state given by 
\be \label{algfst} \omega_F[W(f)]:= e^{-\frac{||f||^2}{4}} \,. \ee
Here $\mathcal{H}_F$ is the GNS representation space which is identical to the one that we constructed in the previous section directly using the ladder operators, via the following identities 
\be \label{pwfun} \pi_F(W(f)) = W_F(f) = e^{i\Phi(f)} \ee
for all $W(f) \in \mathcal{A}$. The vector state $\Omega_F$ is the cyclic GNS vacuum generating the representation space, $\pi_F(\mathcal{A})\ket{\Omega_F} \overset{\text{dense}}{\subset} \mathcal{H}_F$. It is the same degenerate Fock vacuum that was introduced earlier, the no-space state. \

The kinematic system can thus be defined by a pair consisting of, the algebra of bounded linear operators\footnote{The von Neumann algebra $\mathcal{B}(\mathcal{H}_F)$ is the closure of the C*-algebra $\pi_F(\mathcal{A})$ in the weak operator topology on $\mcH_F$. Weak operator topology is defined by continuity of the map $\mathcal{B}(\mathcal{H}_F) \ni A \mapsto (A\psi_1,\psi_2)$, for every $\psi_1,\psi_2 \in \mcH_F$. Further, the weak operator topology is weaker (or coarser) than the strong operator topology i.e. continuity of the map $\mathcal{B}(\mathcal{H}_F) \ni A \mapsto ||A\psi ||$, for every $\psi \in \mcH_F$. Thus, $\mathcal{B}(\mcH_F)$ is also strongly closed. \cite{strocchi2005book,sewellbook}} $\mathcal{B}(\mathcal{H}_F)$ on the Fock space or the algebra $\mathcal{A}_F$, both of which we encountered before in sections \ref{atomkin} and \ref{fockspace}; and, the space of normal states $\mathcal{S}_n$ over the algebra. 
Normal states are algebraic states $\omega_\rho$ induced by density operators $\rho$ on $\mathcal{H}_F$, i.e. $\omega_\rho[A] = \Tr(\rho A)$.


\subsection{Translation automorphisms} \label{AUT}

In section \ref{momgibbs} we will construct Gibbs states which are at equilibrium with respect to translations of the system along the base manifold, for which the relevant definitions and constructions are presented below \cite{Kotecha:2018gof}.

\subsubsection{$\mathbb{R}^n$-translations} \label{rntranS}

The natural translation map on $n$ copies of the real line,
\be
T_{\vphi} : G^d \times \mathbb{R}^n \rightarrow G^d \times \mathbb{R}^n , \, (\vg,\vphi') \mapsto (\vg,\vphi'+\vphi)
\ee
induces a linear map (over $\mathbb{C}$) on square-integrable functions as a shift to the right, 
\be
T^*_{\vphi} : f(\vg,\vphi') \mapsto (T^*_{\vphi} f)(\vg,\vphi') := (f\circ T_{-\vphi})(\vg,\vphi') \,.
\ee 
This is a regular, unitary representation of $\mathbb{R}^n$ on $\mcH$ \cite{fecko2006,kowalski}. Notice that $T^*_{\vphi}$ preserves the $L^2$ inner product due to translation invariance of the Lebesgue measure, i.e. for any $\vphi \in \mathbb{R}^n$, $(T^*_{\vphi} f_1, T^*_{\vphi} f_2) = (f_1,f_2)$. Notice also that $T^*_{\vphi}$ is linear on the space of functions $f$, i.e. for any $z_1,z_2\in \mathbb{C}$, we have $T_{\vphi}^*(z_1f_1 + z_2f_2) = z_1T_{\vphi}^*f_1+z_2T_{\vphi}^*f_2$. Then, let us define a linear map on the Weyl algebra, 
\begin{equation} \label{alphavphi}
\alpha_{\vphi} : \mathcal{A} \rightarrow \mathcal{A} , \, W(f) \mapsto W(T^*_{\vphi}f) \,.
\end{equation}
For each $\phi$, the map $\alpha_\phi$ defines a *-automorphism of $\mathcal{A}$, and the set of maps $\{\alpha_\phi\}_{\phi \in \mathbb{R}}$ forms a 1-parameter group. This defines a representation $\alpha$ of the group $\mathbb{R}$ in the group of automorphisms of the algebra Aut$(\mathcal{A})$, i.e. the map $\alpha : \mathbb{R} \rightarrow \text{Aut}(\mathcal{A}), \, \phi \mapsto \alpha_\phi$  preserves the algebraic structure of reals, $\alpha_{\phi_1 + \phi_2} = \alpha_{\phi_1}\alpha_{\phi_2}$. Extending this to $\mathbb{R}^n$, the maps $\{\alpha_{\vphi}\}_{\vphi\in\mathbb{R}^n}$ now form an $n$-parameter group, and $\alpha$ defines a representation of $\mathbb{R}^n$ in Aut$(\mathcal{A})$. See appendix \ref{autapp} for details.


\subsubsection{$G^d$-left translations} \label{gdtranS}

The natural left\footnote{Analogous statements hold for right translations.} translations on a group manifold are diffeomorphisms from $G$ to itself. On $G^d$, it is given by the smooth map,
\begin{equation}
L_{\vg} : (\vg', \vphi) \mapsto (\vg.\vg', \vphi) := (g_1g'_1,...,g_dg'_d, \vphi) 
\end{equation}
which induces a map on the space of functions,
\begin{equation}
L^*_{\vg}f(\vg',\vphi) := (f \circ L_{\vg^{-1}}) (\vg',\vphi) \,.
\end{equation}
This is a left regular, unitary representation of $G^d$ on $\mcH$ \cite{fecko2006,kowalski}. Notice that $L^*_{\vg}$ preserves the $L^2$-inner product, i.e. for any $\vg \in G^d$, $(L^*_{\vg}f_1,L^*_{\vg}f_2) = (f_1,f_2)$, using left-translation invariance of Haar measure on $G^d$. Also notice that $L^*_{\vg}$ is linear on the space of functions $f$, i.e. for any $z_1,z_2\in \mathbb{C}$, we have $L_{\vg}^*(z_1f_1 + z_2f_2) = z_1L_{\vg}^*f_1+z_2L_{\vg}^*f_2$. \

With this, let us define a linear transformation on the Weyl generators, 
\begin{equation} \label{alphavg}
\alpha_{\vg} (W(f)) := W(L^*_{\vg}f)\,.
\end{equation}
Then, map $\alpha_{\vg}$ defines a *-automorphism of $\mathcal{A}$. Also like for $\mathbb{R}^n$-translations, $\alpha: G^d \rightarrow \text{Aut}(\mathcal{A})$ is a representation of $G^d$ in the group of all automorphisms of the algebra as it preserves the algebraic structure, $\alpha_{\vg.\vg'} = \alpha_{\vg}\alpha_{\vg'}$. See appendix \ref{autapp} for details.


\subsubsection{Unitary representation} \label{unis}

Using the following known structural properties of GNS representation spaces \cite{Bratteli:1979tw,Haag:1992hx}, the automorphisms defined above can be implemented by unitary transformations in the Fock space $\mcH_F$ as follows.

\paragraph{$\alpha$-Invariant state.}

Let $\omega$ be an $\alpha$-invariant state, i.e. $\omega[\alpha A] = \omega[A]$ for all $A \in \mathcal{A}$, for some $\alpha \in \text{Aut}(\mathcal{A})$. Then it is known that \cite{Bratteli:1979tw,Haag:1992hx}, $\alpha$ is implemented by unitary operators $U_\omega$ in the GNS representation space $(\pi_\omega, \mathcal{H}_\omega, \Omega_\omega)$, defined by $U_\omega \, \pi_\omega(A) \, U_\omega^\dag = \pi_\omega(\alpha A)$ with invariance of the GNS vacuum $U_\omega\Omega_\omega = \Omega_\omega$. Similarly, for the general case when $\omega$ is invariant under a group of automorphisms, then $\alpha_g \, (g\in G)$ is implemented by a unitary representation $U_\omega$ of $G$ in $\mathcal{H}_\omega$, such that 
\begin{equation}\label{uni}
U_\omega(g) \, \pi_\omega(A) \, U_\omega^\dag(g) = \pi_\omega(\alpha_g A) \;, \quad U_\omega(g)\Omega_\omega = \Omega_\omega \,.
\end{equation}

\paragraph{Fock state.} 

We recall that the algebraic Fock state over $\mathcal{A}$ is given by equation \eqref{algfst}, with the associated GNS representation $(\pi_F, \mathcal{H}_F, \Omega_F)$. Then, any automorphism on $\mathcal{A}$ that is defined via a norm-preserving transformation on $\mathcal{H}$ will leave $\omega_F$ invariant. Thus $\omega_F$ is invariant under the class of norm-preserving transformations of the square-integrable test functions, including the translation automorphisms of the base manifold as defined above. Thus, automorphisms $\alpha_{\vphi}$ and $\alpha_{\vg}$ are implemented by groups of unitary operators in $\mathcal{H}_F$. \

From the unitary transformations \eqref{uni} as applied to Weyl generators in the Fock representation $\pi_F$, it is straightforward to see that the group field operators transform in a familiar way, 
\begin{align}
U_F(\vphi') \varphi(\vg,\vphi) U_F(\vphi')^{-1} &= \varphi(\vg,\vphi+\vphi') \\
U_F(\vg') \varphi(\vg,\vphi) U_F(\vg')^{-1} &= \varphi(\vg'\vg,\vphi) 
\end{align}
with analogous expressions for their adjoints. From here on the subscript $F$ on the unitary implementations of these translation automorphisms will be dropped with the understanding that in this setting $U$ refers only to the unitary representation of some group in target space $\mathcal{U}(\mathcal{H}_F)$ of unitary operators on Fock space. We remark that these transformations, defined here for $\pi_F(\mathcal{A})$, being bounded can also be extended to $\mathcal{B}(\mathcal{H}_F)$. \

Further, it is important to note (for subsequent use in section \ref{inttrans}) that the corresponding group homomorphism $U : G^d \to \mathcal{U}(\mathcal{H}_F)$, is strongly continuous in the Fock space. See appendix \ref{strcon} for details. \

Therefore, the system is now equipped with strongly continuous groups of unitary operators in the Fock representation that implement internal shifts of the underlying base manifold $G^d \times \mathbb{R}^n$.


\section{Deparametrization in group field theory} \label{depgft}

We have recalled the fundamental difficulties in defining equilibrium in generally covariant systems, due to the absence of preferred time variables in the previous chapters. A general strategy to solve those issues, in the description of the dynamics of such systems is to use matter degrees of freedom as relational clocks, under suitable approximations, and recast the general covariant dynamics in terms of a physical Hamiltonian associated with them. Here also we can consider the same general strategy as a way to tackle our (related) issue of defining statistical equilibrium states in quantum gravity. That is, we can consider the construction of states which are at equilibrium with respect to relational clocks, as will be done in section \ref{physeqm}. \

For this then, we first need to consider in detail aspects of deparametrization in group field theory to define a clock variable, which is the content of this section. Based on insights and tools of presymplectic mechanics as presented in section \ref{sympmech}, below we lay out the essentials for group field theory. As we will see, the resultant deparametrized, relational system will be `canonical' in clock time which now foliates the original GFT system discussed up until now. Algebra brackets \eqref{ladderCR} will be replaced with the corresponding equal-clock-time commutation relations, analogous to the equal-time commutation relations in a non-covariant system. Our interest in such a setup is natural because GFTs lack a preferred choice of an evolution parameter. The base space $G^d \times \mathbb{R}^n$ is chosen so as to facilitate a relational description of the system by coupling $n$ scalar fields \cite{Li:2017uao}. However, there are $n$ possible variables to choose from, and none is preferred over the others. By construction then, GFTs have a multi-fingered relational time structure in this sense.  \

The way we approach the task of deparametrizing is as follows \cite{Kotecha:2018gof}. We focus first on the classical description in section \ref{classdep}, starting with the case of a single group field particle in section \ref{one} and sketch how deparametrization works at this simple level, assuming that the GFT dynamics amounts to a specific choice of a scalar Hamiltonian constraint. Then in section \ref{two}, we consider the extension of the same deparametrization procedure for a system of many such particles, assumed as non-interacting. We then consider the quantisation of the resulting deparametrized system of many group field particles, arriving at the corresponding quantum multi-particle system in section \ref{quantise}, which is canonical with respect to a clock time (with respect to which one can then define relational equilibrium, in section \ref{physeqm}). Below we sketch the relevant steps of the overall construction \cite{Kotecha:2018gof}, and leave a more detailed analysis of the corresponding mathematical structures for GFTs to future work.


\subsection{Classical system} \label{classdep}
We begin with the investigation of deparametrization for a classical system \cite{Kotecha:2018gof}, utilising the framework of extended phase space and presymplectic mechanics as discussed in section \ref{sympmech}. Here, we face a similar issue of background independence with the associated absence of a preferred evolution parameter. The reason for undertaking classical considerations first is to utilise the existing knowledge already well-positioned to be imported to GFT due to the common structures encountered in any such multi-particle system, namely an extended symplectic phase space with a set of constraints, including a dynamical Hamiltonian one.


\subsubsection{Single-particle} \label{one}

The extended classical configuration and phase space for the single-particle sector in group field theory is 
\begin{align} \mathcal{C}_{\ex} &= G^d \times \mathbb{R}^n \ni (g^I,\phi^a) \\
\Gamma_{\ex} &= \mcT^*(\mathcal{C}_{\ex}) \cong G^d \times \mathbb{R}^{n} \times \g^{*d}  \times \mathbb{R}^n \ni (g^I,\phi^a,X_I,p_{\phi^a})  \end{align} 
where, $\g$ is the Lie algebra of $G$ and $\g^*$ is its dual vector space. Here, by a classical particle we mean a point particle living on the group base manifold, thus being described by a point on its phase space. States and observables are respectively points and smooth functions on $\Gamma_{\ex}$. Statistical states are smooth positive functions on the phase space, normalised with respect to the Liouville measure. The Poisson bracket on the space of observables defines its algebra structure. The symplectic 2-form on $\Gamma_{\ex}$ is,
\be \omega_{\ex} = \omega_G \,+\, \sum_a dp_{\phi^a} \wedge d\phi^a \ee 
where $\omega_G$ is a symplectic form on $\mcT^*(G^d)$. Let us assume that the covariant\footnote{Throughout sections \ref{classdep} and \ref{quantise}, by `covariant' we simply mean `not deparametrized', without any relation to diffeomorphisms on spacetime.} dynamics of this simple system is encoded in a smooth constraint function \be C_{\f} : \Gamma_{\text{ex}} \to \mathbb{R} \,, \ee and that there are no additional gauge symmetries. A single classical particle of the group field is thus described by $(\Gamma_{\ex}, \omega_{\ex}, C_{\f})$. The null vector field is defined by,
\be \omega_{\ex}(\mcY_{C_{\f}}) = -dC_{\f} \ee
while the constraint surface $\Sigma$ is characterised by $C_{\f} = 0$, with its presymplectic form given by, $\omega_\Sigma = \omega_{\ex}|_{\Sigma}$. The null orbits of $\omega_\Sigma$ are the graphs of physical motions encoding unparametrized correlations between the dynamical variables of the theory. These gauge orbits are integral curves of the vector field $\mcY_{C_{\f}}$, satisfying the equations of motion \be \omega_\Sigma(\mcY_{C_{\f}}) = 0 \,. \ee The set of all such orbits is the physical phase space $\Gamma_{\phy}$ that is projected down from $\Sigma$ and is equipped with a symplectic 2-form induced from $\Sigma$. Then $(\Gamma_{\phy}, \omega_{\phy})$ is the space of solutions of the system and a physical flow means a flow on this space. We notice again that here, a canonical time or clock structure is lacking, and is not required a priori. \ 

Deparametrizing this classical system, with respect to, say, the $c^{\text{th}}$ scalar field $\phi^c$, means reducing the full system to one wherein the field $\phi^c$ acts as a good clock. As anticipated in section \ref{sympmech}, this entails the following two separate approximations to $C_{\f}$,
\begin{align}
C_{\f} (g^I, \phi^a, X_I, p_{\phi^a}) &\approx p_{\phi^c} + \tilde{C}(g^I, \phi^a, X_I, p_{\phi^\alpha}) \label{depap1} \\
&\approx p_{\phi^c} + H(g^I, \phi^\alpha, X_I, p_{\phi^\alpha}) \label{depap2}
\end{align}
where the fixed index $c$ denotes `clock', and index $\alpha \in \{1,2,...,n-1\}$ runs over the remaining scalar field degrees of freedom that are not intended to be used as clocks and remain internal to the system. 
The first approximation retains terms up to the first order in clock momentum. At this level of approximation the part denoted by $\tilde{C}$ is a function of the clock time $\phi^c$. These two features mean that at this level of approximation $\phi^c$ behaves as a clock, but only locally since its momentum is not necessarily conserved in the clock time. Furthermore, by linearising in $p_{\phi^c}$, we have fixed a reference frame defined by the physical matter field $\phi^c$. At the second level, $\tilde{C}$ is approximated by a Hamiltonian $H$ that is independent of $\phi^c$, so that on-shell (i.e. $p_{\phi^c} = -H$) we have conservation of the clock momentum 
\be \partial_{\phi^c}p_{\phi^c} = 0 \,.\ee 
$H$ generates relational dynamics in $\phi^c$, which now acts as a global clock for this deparametrized system. We thus have a new system $(\Gamma_{\ex}, \omega_{\ex}, C_{\dep})$ after the above approximations, with 
\begin{equation} \label{deparamC}
C_\dep = p_{\phi^c} + H(g^I, \phi^\alpha, X_I, p_{\phi^\alpha})
\end{equation}
now deparametrized with respect to one of the extended configuration variables $\phi^c$ which takes on the role of a good clock variable. See also equations \eqref{ordersc} and the surrounding discussion. \

The presymplectic setup of this deparametrized system now takes on a structure mirroring that of a parametrized non-relativistic particle with a well-defined time. The constraint surface defined by vanishing of the relevant constraint, here $C_\dep=0$, now admits the topology of a foliation in clock time, \be \Sigma = \mathbb{R} \times \Gamma_{\can} \ni (\phi^c, g^I, \phi^\alpha, X_I, p_{\phi^\alpha}) \,. \ee 
As we have noticed before, this form of $\Sigma$ is a characteristic feature of a system with a clock structure. The reduced, canonical configuration and phase spaces are now given by, 
\begin{align} \mathcal{C}_{\can} &= G^d \times \mathbb{R}^{n-1} \,, \\
\Gamma_{\can} &= \mcT^*(\mathcal{C}_{\can}) \ni (g^I, \phi^\alpha, X_I, p_{\phi^\alpha}) \,.  \end{align} The function $H : \Gamma_{\can} \to \mathbb{R}$ is the clock Hamiltonian encoding relational dynamics in $\phi^c$, and one can subsequently define the standard Hamiltonian dynamics with respect to it.


\subsubsection{Multi-particle} \label{two}

We now want to extend the above deparametrization procedure beyond the one-particle sector. Let us consider the simplest case of two, non-interacting particles \cite{Chirco:2013zwa,Chirco:2016wcs,Kotecha:2018gof}. Let 
\be \Gamma^{(1,2)} = \mcT^*(\mathcal{C}^{(1,2)}_{\ex}) \ni (g^{(1,2)I},\phi^{(1,2)a},X^{(1,2)}_I,p^{(1,2)}_{\phi^a}) \ee 
be the extended phase spaces of particles 1 and 2 respectively. The extended phase space of the composite system is \be \Gamma = \Gamma^{(1)} \times \Gamma^{(2)} \ee with symplectic 2-form $\omega = \omega^{(1)} + \omega^{(2)}$. Notice that each particle is equipped with $n$ possible clocks. The aim is to select a single common clock for the composite system so as to then be able to define a common equilibrium for the total system.  \

Let the individual dynamics of each particle be given by constraints $C^{(1,2)}_{\f} : \Gamma^{(1,2)} \to \mathbb{R}$. Deparametrizing particle 1 with respect to say variable $\phi^{(1)c_1}$, and particle 2 with respect to say a different $\phi^{(2)c_2}$, gives the new constraints for each,
\begin{align}
C^{(1)} &= p^{(1)}_{\phi^{c_1}} + H^{(1)}(g^{(1)I},\phi^{(1)\alpha},X^{(1)}_I,p^{(1)}_{\phi^\alpha})    \\
C^{(2)} &= p^{(2)}_{\phi^{c_2}} + H^{(2)}(g^{(2)I},\phi^{(2)\alpha},X^{(2)}_I,p^{(2)}_{\phi^\alpha})   \label{particledep}
\end{align}
where $H^{(1,2)}$ are functions on the individual reduced phase spaces, 
\be \Gamma^{(1,2)}_{\can} = \mcT^*(\mathcal{C}^{(1,2)}_{\can}) \ni (g^{(1,2)I},\phi^{(1,2)\alpha},X^{(1,2)}_I,p^{(1,2)}_{\phi^\alpha}) \ee 
with symplectic 2-forms, 
\be \omega^{(1,2)}_{\can} = \omega_G^{(1,2)} + \sum_{\alpha} dp^{(1,2)}_{\phi^\alpha} \w d\phi^{(1,2)\alpha} \,. \ee 
This is a complete theoretical description of the deparametrized 2-particle system. However, it is inconveniently described in terms of two different clocks ascribed to each particle separately. We are seeking a single common clock. \

Such a clock can be defined by synchronizing the two individual clocks, via the imposition of an additional constraint as follows. Let us choose $\phi^{(1)c_1}$ to be the common clock, and write the second clock as a smooth function of the first, that is 
\be \label{synsim} \phi^{(1)c_1} = t \,, \quad \phi^{(2)c_2} = F(t) \,. \ee 
More generally, we can choose a common clock $t \in \mathbb{R}$, and sync the two separate clocks with it via two functions,
\be
F_i:  t \mapsto  \phi^{(i)c_i} \,, \quad i=1,2 \,.
\ee
Any such syncing function, which maps two clocks, is assumed to be a bijection. It is thus invertible. This is done to ensure that the procedure of syncing is consistent and well-defined. Say, clock 2 is synced with clock 1. What we mean by this is that, every reading from clock 2 should give a unique reading of clock 1. This is possible only if the map between the readings, that is $F$, is one-to-one and onto.    \

We work with the case associated with equations \eqref{synsim}, for simplicity. Then, syncing the two clocks can be imposed by a constraint of the form\footnote{In the more general case with two functions $F_1$ and $F_2$, the syncing constraint can be given by, $\mathfrak{s} = F_1^{-1}(\phi^{(1)c_1}) - F_2^{-1}(\phi^{(2)c_2})$. In this case, the common clock is $t = F_1^{-1}(\phi^{(1)c_1}) = F_2^{-1}(\phi^{(2)c_2})$, when this constraint is imposed.},
\be \mathfrak{s} := \phi^{(1)c_1} - F^{-1}(\phi^{(2)c_2}) \,. \ee

Imposing this constraint, i.e. $\mathfrak{s} = 0$, amounts to choosing a 1-parameter flow in the (here, 2-dim) space of the clock variables $\phi^{(1)c_1}$ and $\phi^{(2)c_2}$, which can themselves be understood as gauge parameters due to their association with dynamical constraints $C^{(1)}$ and $C^{(2)}$. Thus, this syncing can be seen as gauge fixing \cite{Chirco:2013zwa,Chirco:2016wcs}. Then the gauge-fixed 2-form of $\Gamma$ is, 
 \be \tilde{\omega} := \omega|_{\mathfrak{s}=0}  =   \omega_{\can} \;+\; dp_t \wedge dt \ee 
 where,
\be p_t = p^{(1)}_{\phi^{c_1}} + F'(t)p^{(2)}_{\phi^{c_2}} \ee
is the clock momentum of the single clock $t$, and 
\be \omega_{\can} = \omega^{(1)}_{\can} + \omega^{(2)}_{\can} \ee 
is the symplectic form on the canonical reduced phase space $\Gamma_{\can} = \Gamma^{(1)}_{\can} \times \Gamma^{(2)}_{\can}$ of the composite system. Here, prime $'$ denotes a total derivative with respect to $t$. \

Now, one can consider a reformulation\footnote{We have that $\{C^{(1)}=0,C^{(2)}=0\} \Leftrightarrow \{C=0,\Delta=0\}$.} of the constraints $C^{(1)}$ and $C^{(2)}$ \cite{Chirco:2013zwa,Chirco:2016wcs}, in terms of
\be C := C^{(1)} + C^{(2)}   \ee
along with, $\Delta :=  C^{(1)}-C^{(2)}$. Further, notice that on the constraint surface we have, $C^{(2)} = 0 \Leftrightarrow F'C^{(2)}=0$, for an arbitrary non-zero function $F'$. Then, the constraint ${C}$ can be rewritten as, $\tilde{C}:= C^{(1)} + F'C^{(2)}$. We thus have,
\be \tilde{C} = p_t + H^{(1)} + F'(t)H^{(2)}  \ee 
where the Hamiltonian $H^{(1)} + F'(t)H^{(2)}$ is independent of clock $t$ iff $F'(t)=k$, for $k$ an arbitrary non-zero real constant. In other words, $t$ is a good clock for the choice of affine gauge $F(t) = kt + \tilde{k}$. This gives,
\be \tilde{C} = p_t + H \ee
where, \be H = H^{(1)} + kH^{(2)} \,. \ee 

Now that the 2-particle system has been brought to the form of a standard Hamiltonian system with clock time $t$, the remaining elements for the complete extended symplectic description can be identified. The extended configuration and phase spaces are 
\begin{align} \mathcal{C}_{\ex} &= \mathbb{R} \times \mathcal{C}_{\can} \ni (t, g^{(1)I}, \phi^{(1)\alpha}, g^{(2)J}, \phi^{(2)\gamma}) \\
\Gamma_{\ex} &= \mcT^*(\mathcal{C}_{\ex}) \end{align} with $\omega_{\ex} = \tilde{\omega}$. The constraint function $\tilde{C}$ defines the presymplectic surface 
\be \Sigma = \mathbb{R} \times \Gamma_{\can} \ni (t, g^{(1)I},\phi^{(1)\alpha},X^{(1)}_I,p^{(1)}_{\phi^\alpha},g^{(2)J},\phi^{(2)\gamma},X^{(2)}_J,p^{(2)}_{\phi^\gamma}) \ee
with \be \omega_\Sigma = \omega_{\can} - dH \wedge dt \,. \ee Thus, $(\Gamma_\ex,\omega_\ex,\tilde{C})$ as given above provides a complete description of the 2-particle non-interacting system equipped with a single relational clock $t$. \

For an $N$-particle non-interacting system, each with $n$ possible clocks, the extension of the above procedure is direct. Select any one $\phi$ variable as a clock for each individual particle, i.e. bring the individual full constraints of each particle to deparametrized forms like in \eqref{particledep}. Then, given one clock per particle, identifying a global clock for all particles means choosing any one at random (call it $t$) and synchronizing the rest with this one via affine functions $F_2(t),...,F_{N}(t)$. This defines a relational system on $\mathcal{C}_{\ex} \ni (t, g^{(1)I}, \phi^{(1)\alpha},...,g^{(N)J}, \phi^{(N)\gamma})$, $\Gamma_{\ex} = \mcT^*(\mathcal{C}_{\ex})$, with constraint function $\tilde{C} = p_t + H$ on $\Gamma_{\ex}$, and Hamiltonian function $H = H^{(1)} + k_2H^{(2)} + ... + k_N H^{(N)}$ on $\Gamma_{\can}$. 


\subsubsection{Discussion}

Before moving on to quantisation, let us pause to make a few important remarks, and summarise sections \ref{one} and \ref{two} \cite{Kotecha:2018gof}. The following discussion is meant to: 
\begin{enumerate}
\item clarify that we are dealing with two different, `before' and `after' deparametrization, systems. The former is covariant (or constrained), while the latter is derived from the former via the deparametrization approximations in equations \eqref{depap1}-\eqref{depap2}.

In general the two systems are physically distinct. However, it is possible that deparametrization does not change the physical content of the theory. This corresponds to the case in which the deparametrization steps in \eqref{depap1}-\eqref{depap2} do not correspond to approximations of the dynamics, but to an exact re-writing of it, e.g. a relativistic particle (see section \ref{depggs} for simple illustrative examples); 

\item clarify that the latter, `after' deparametrization, system, includes within it a canonical system, which is eventually quantised. This system is `canonical' (or non-relativistic, see section \ref{sympmech}) with respect to the relational clock that is selected during the process of deparametrization;

\item clarify the conceptual (and notational) differences between these systems.

\end{enumerate}

For the 1-particle system the `before' picture is one wherein the system is fully covariant and the corresponding kinematics consist of the configuration space $\mathcal{C}_{\ex}^{\cov} = G^d \times \mathbb{R}^n \ni (g^I, \phi^a)$ and phase space $\Gamma_{\ex}^{\cov} = \mcT^*(\mathcal{C}_{\ex}^{\cov})$. Covariant (or constrained) dynamics is encoded in a dynamical constraint function $C_{\f}$ defined on $\Gamma_{\ex}^{\cov}$. Vanishing of the constraint function defines a constraint hypersurface in $\Gamma_{\ex}^{\cov}$. \

The `after' picture defines the second system, which includes the canonical one. The extended configuration space of the deparametrized system is $\mathcal{C}_{\ex}^{\dep} = \mathbb{R} \times (G^d \times \mathbb{R}^{n-1}) \ni (t, g^I, \phi^\alpha)$, where the 1-particle canonical configuration space is $\mathcal{C}_{\can} = G^d \times \mathbb{R}^{n-1}$. Here, we have denoted $\phi^c \equiv t$. Canonical variables are those dynamical variables of the original covariant system $\mathcal{C}_{\ex}^{\cov}$ which are \emph{not} used as clocks. The extended phase space is $\Gamma_{\ex}^{\dep} = \mcT^*(\mathcal{C}_{\ex}^{\dep})$. Deparametrized dynamics is encoded in a constraint function $C$ on $\Gamma_{\ex}^{\dep}$, of the form, $C = p_t + H$, where $H$ is a smooth function on the canonical phase space $\Gamma_{\can} = \mcT^*(\mathcal{C}_{\can})$.\footnote{See equation \eqref{deparamC}, or \eqref{1deparamC1}.} It is a genuine Hamiltonian defining dynamical evolution with respect to the relational clock $t$. The constraint surface (satisfying $C=0$) is $\Sigma = \mathbb{R} \times \Gamma_{\can}$, characterised by a foliation consisting of slices $\Gamma_{\can}$ along clock $t$. This form of $\Sigma$ and the existence of the canonical subsystem is a direct consequence of deparametrization. The canonical subsystem is thus absent for a generic non-deparametrized, constrained system $(\Gamma_{\ex}^{\cov}, \omega_{\ex}^{\cov}, C_{\f})$. \

Note that for the 1-particle system, the covariant and deparametrized kinematic descriptions, in the respective configuration spaces $\mcC_\ex^{\cov} = G^d \times \mathbb{R}^n$ and $\mcC_\ex^{\dep} =  \mathbb{R} \times (G^d \times \mathbb{R}^{n-1})$, are identical. As will be seen below, this does not hold for an $N$-particle system with $N>1$, when seeking a description with a single clock. In this case, the configuration spaces are $\mathcal{C}_{\ex, N}^{\cov} = (G^d \times \mathbb{R}^n)^{\times N} $ and $\mathcal{C}_{\ex,N}^{\dep} = \mathbb{R} \times (G^d \times \mathbb{R}^{n-1})^{\times N}$. \

For the non-interacting $N$-particle system, the `before' system consists of the covariant extended configuration space $\mathcal{C}_{\ex, N}^{\cov} = (G^d \times \mathbb{R}^n)^{\times N} \ni (g^{(1)I}, \phi^{(1)a}, ..., g^{(N)J}, \phi^{(N)b})$ and the associated phase space $\Gamma_{\ex, N}^{\cov} = \mcT^*(\mathcal{C}_{\ex, N}^{\cov}) = (\Gamma_{\ex}^{\cov})^{\times N}$. The covariant dynamics is encoded in a set of constraints $C_{\f}^{(1)},...,C_{\f}^{(N)}$, each defined on the respective copies of the 1-particle covariant extended phase space $\Gamma_{\ex}^{\cov}$. \

The `after' system is deparametrized with a single clock $t$. As before, existence of this clock structure means that the extended symplectic description takes on the form of a non-relativistic system. The extended configuration space is $\mathcal{C}_{\ex,N}^{\dep} = \mathbb{R} \times \mathcal{C}_{\can, N}$ where the $N$-particle reduced configuration space is $\mathcal{C}_{\can, N} = (G^d \times \mathbb{R}^{n-1})^{\times N}$, and the clock $t\in\mathbb{R}$ is an extended configuration variable. The extended phase space of the deparametrized system is $\Gamma_{\ex, N}^{\dep} = \mcT^*(\mathcal{C}_{\ex,N}^{\dep})$. The deparametrized dynamics is encoded in a constraint function, \begin{equation*}
C_N = p_t + H_N
\end{equation*} 
defined on $\Gamma_{\ex,N}^{\dep}$. Constraint surface $\Sigma = \mathbb{R} \times \Gamma_{\can,N}$ is characterised by $C_N=0$. The canonical phase space is $\Gamma_{\can, N} = \mcT^*(\mathcal{C}_{\can, N})$. Relational dynamics is encoded in the clock Hamiltonian defined on $\Gamma_{\can, N}$ given by, 
\begin{equation} \label{hamiltonianN}
H_N = \sum_{i=1}^N k_i H^{(i)} \,,
\end{equation}
for arbitrary real non-zero constants $k_i$ which encode the rates of synchronization between the $N$ different clocks, one per particle. Functions $H^{(i)}$ are single-particle clock Hamiltonians defined on the respective copies of single-particle reduced phase space $\Gamma_{\can}$. We can already anticipate that relational canonical Gibbs states are those that are KMS with respect to the $t$-flow of a clock Hamiltonian $H_N$. \

Finally, notice the similarity in the form of $H_N$ in equation \eqref{hamiltonianN} above, and the modular Hamiltonian with respect to a given set of observables in equation \eqref{moddh}. Based on the above discussion, one can understand the generalised temperatures $\beta_a$ in the latter as the rates at which the individual modular flow parameters, each associated with a different $\mcO_a$, sync with each other, in the case when there exists a single clock corresponding to the single constraint in equation \eqref{moddhfix}.


\subsection{Quantum system} \label{quantise}

We now move on to the quantisation of the above deparametrized non-interacting, many-body system. Our treatment is again limited to outlining the basic steps, which is sufficient at least for our present purposes \cite{Kotecha:2018gof}. In the context of deparametrization, since we are primarily interested in scalar degrees of freedom residing in copies of $\mathbb{R}$, we will continue to be content with omitting rigorous details about the symplectic structure on $\mcT^*(G)$, and its subsequent quantisation to a commutator algebra. We shall also not choose any specific quantisation map, and focus on the general ideas required to eventually define a $\phi$-relational Gibbs state. Further details can be found in \cite{Guedes:2013vi}, including examples of quantisation maps for $\mcT^*(G)$.\

Quantisation maps the phase space to a Hilbert space, and the classical algebra of observables as smooth (real) functions on the phase space to (self-adjoint) operators on the Hilbert space, with the Poisson bracket on the former being mapped to a commutator bracket on the latter. 

\subsubsection{Single-particle}

For the covariant 1-particle system, the phase space $\Gamma_{\ex}^{\cov}=\mcT^*(G^d \times \mathbb{R}^n)$ maps to, 
\be \mathcal{H} \equiv \mathcal{H}_{\cov} = L^2(G^d \times \mathbb{R}^n) \ee 
defined over the extended configuration manifold $\mcC_\ex^{\cov}=G^d \times \mathbb{R}^n$. This is the Hilbert space of a single quantum of geometry that we presented in earlier sections. Observables are the algebra of real-valued smooth functions on $\Gamma_{\ex}^{\cov}$, which map to operators on $\mathcal{H}$, with the Poisson structure on the former being mapped to the Heisenberg algebra on the latter. Specifically, for the matter degrees of freedom, this is 
\be \{\phi^a, p_{\phi^b}\} = \delta_{ab} \rightsquigarrow [\widehat{\phi^a}, \widehat{p_{\phi^b}}] = i\delta_{ab} \ee 
where the hat denotes some quantisation map. Notice here that all $n$ scalar fields are quantised. This is in contrast with the corresponding case of the canonical system, wherein the canonical phase space $\Gamma_{\can} = \mcT^*(G^d \times \mathbb{R}^{n-1})$ maps to a canonical Hilbert space,
\be \mathcal{H}_{\can} = L^2(G^d \times \mathbb{R}^{n-1}) \ee
with the algebra again mapping from functions on $\Gamma_{\can}$ to operators on $\mathcal{H}_{\can}$. But now, the brackets defining the algebra structure of the system are reduced by one in number, as a direct consequence of the reduction of the base space by one copy of $\mathbb{R}$, to which the clock variable belongs. Under quantisation we now have, 
\be \{\phi^\alpha, p_{\phi^\gamma}\} = \delta_{\alpha\gamma} \rightsquigarrow [\widehat{\phi^\alpha}, \widehat{p_{\phi^\gamma}}] = i\delta_{\alpha\gamma} \ee 
where $\alpha, \gamma = 1,...,n-1$. The commutator corresponding initially to the clock $\phi$-variable is now identically zero, that is \be [\widehat{\phi^c}, \widehat{p_{\phi^c}}] = 0 \,, \ee 
meaning that the corresponding degrees of freedom are treated as entirely classical; moreover, their intrinsic dynamics is trivialised. In other words, this quantum canonical system is one in which there exists a classical clock, which was quantum in the original quantum covariant system. Dynamics is defined via a Hamiltonian operator $\hat{H}$ on $\mathcal{H}_{\can}$ giving evolution with respect to the clock. 

\subsubsection{Multi-particle}

For the covariant $N$-particle system, $\Gamma_{\ex,N}^{\cov} = (\Gamma_{\ex}^{\cov})^{\times N}$ is mapped to \be \mathcal{H}_N = \mathcal{H}^{\otimes N} \,.\ee Algebra of smooth functions on $\Gamma_{\ex,N}^{\cov}$ is mapped to an operator algebra on $\mathcal{H}_N$, whose quantum matter fields now satisfy \be [\widehat{\phi^{(i)a}}, \widehat{p^{(j)}_{\phi^b}}] = i\delta_{ab}\delta_{ij} \ee
where $i,j=1,2,...,N$ denote the particle label. This is the multi-particle sector as considered in sections \ref{atomkin} and \ref{fockspace}. Again, we note that none of the $N$ particles have chosen a clock yet, that is all $n \times N$ number of scalar fields $\phi$ are quantum. In the corresponding canonical quantum system \be \mathcal{H}_{\can, N}= \mathcal{H}_{\can}^{\otimes N} \ee with the algebra of observables on it, the single clock variable $t$ (which is synced with all the separate clocks now carried by each particle) is classical. The Hamiltonian operator defining $t$-evolution, for fixed $N$, is given by the operator \be \hat{H}_N = \sum_{i=1}^N k_i \hat{H}^{(i)} \ee where $\hat{H}^{(i)}$ are the separate Hamiltonians of each particle scaled by the respective rates of syncing of the different clocks (or, generalised inverse temperatures \cite{Rovelli:2010mv}), and we have neglected interactions. \ 

In the multi-particle case, it is worthwhile to also look at the quantised extended deparametrized system. This consists of $\Gamma_{\ex,N}^{\dep}$ being quantised to $\mathcal{H}_{N}^{\dep} = L^2(\mathbb{R}\times \mathcal{C}_{\can,N})$ and the corresponding Heisenberg algebra has two additional generators (compared to the canonical system) satisfying $[\widehat{t}, \widehat{p_t}] = i$. Such a system is different from both the quantum extended covariant and the quantum canonical. In the former, there is no single clock variable. In the latter, there is one but it is no longer quantum. Quantising the extended deparametrized system is like quantising a non-relativistic particle at the level of its extended phase space, which includes Newtonian time and its conjugate momentum as phase space variables. Quantising Newtonian time to define the corresponding operator comes with its own set of conceptual and technical problems. However our case is fundamentally different because here $t$ is really a function of degrees of freedom that are understood as coupled scalar matter fields \cite{Li:2017uao}. \

Therefore for a multi-particle system, one ends up with three different quantum systems: quantum extended covariant, quantum extended deparametrized and quantum canonical. The last two each have a potential clock parameter, and going from the former to the latter is the step of making this variable classical and therefore treating it as a perfect, thus idealised, clock. This distinction between quantum extended covariant and extended deparametrized systems is absent in the simple 1-particle system because in this case deparametrization does not require the extra step of syncing the different clocks (as there is only one). It only requires choosing one out of $n$ so that the kinematics of both systems ends up being identical.  

\subsubsection{Fock extension} \label{FockExt}

We arrive now at the quantum Fock systems built out of the above $N$-particle systems. The covariant Fock system composed of the $N$-particle quantum extended covariant systems as described above is the GFT Fock representation as detailed in section \ref{bosgft}. \

On the other hand, a canonical Fock system (associated with a degenerate vacuum) is as follows. The canonical Hilbert space for bosonic quanta can be identified as 
\be \mathcal{H}_{\can, F} = \bigoplus_{N\geq0} \text{sym}\, \mathcal{H}_{\can}^{\otimes N} \;, \ee
which is generated by ladder operators acting on a cyclic vacuum, and satisfying the equal- (Fock-) clock-time commutation relations,
\be \label{canladderCR}
[\hat{\varphi}(t_F,\vg,\underline{\phi}),\hat{\varphi}^\dag(t_F,\vg',\underline{\phi}')] = \mathbb{I}(\vg,\vg')\delta(\underline{\phi} - \underline{\phi}')
\ee
with $[\hat{\varphi}, \hat{\varphi}] = [\hat{\varphi}^\dag, \hat{\varphi}^\dag] =0$, and $\mathbb{I}$ and $\delta$ being the respective delta distributions on $G^d$ and $\mathbb{R}^{n-1}$. Notation $\underline{\,\cdot\,}$ has been used to make explicit the difference between variables $\phi$ in canonical and covariant systems. Here $\underline{\phi} \equiv (\phi^1,...,\phi^{n-1})$ denotes canonical variables whereas earlier, $\vphi \equiv (\phi^1,...,\phi^{n})$ belonged to the covariant system in which all scalar fields were internal variables. $\vg$ continues to denote $(g_1,...,g_d)$. The associated canonical Weyl system is now based on test functions which are defined on the reduced configuration space $\mathcal{C}_{\can} = G^d \times \mathbb{R}^{n-1}$, and analogous constructions to those considered in section \ref{weylGFT} follow through. The *-algebra $\mathcal{A}_{\can,F}$ now consists of polynomial functions of the above canonical ladder operators, defined over the reduced base space $G^d \times \mathbb{R}^{n-1}$. For example, the number operator now takes the form, 
\be
\hat{N} = \int_{G^d\times \mathbb{R}^{n-1}} d\vg\,d\underline{\phi} \; \hat{\varphi}^\dag(\vg,\underline{\phi})\hat{\varphi}(\vg,\underline{\phi}) \,. \ee
Note that one can understand these quantities also as observables in the full theory, just computed at given values of the relational clock variable. The heuristic interpretation is valid, but the actual algebraic properties of these observables would be (potentially very) different. \

Particularly, from the point of view of the original non-deparametrized system described by the algebra \eqref{ladderCR}, one can already expect quantities in a $\phi$-frame to have singularities due to the presence of $\lim_{\phi \to \phi'} \delta(\phi-\phi')$. As we will see in section \ref{threpcond}, this feature is unavoidable in aspects of inequivalent thermal representations induced by generalised Gibbs states, and their subsequent applications in GFT condensate cosmology. In section \ref{clock} we will also suggest one possible way of introducing non-singular clock frames, through a class of smearing functions $t(\phi)$, and subsequently apply them to derive relational equations of motion for effective homogeneous and isotropic cosmology \cite{Assanioussi:2020hwf}. \

Let us make a further remark regarding deparametrization from the perspective of the full quantum non-deparametrized theory. The strategy employed above for a finite dimensional system is to start from a classical constrained system, deparametrize it to get a classical canonical system with respect to a relational clock, and then quantise the canonical system leaving the clock as classical. A more fundamental construction leading possibly to a more physical sort of (approximate) deparametrization is to begin from the complete quantum theory (in our case, a group field theory) in which all possible relational scalar fields are quantum. Then deparametrizing would mean to identify a relevant regime of the full theory in which one of the coupled scalar fields becomes semi-classical, and only then apply the deparametrization approximations outlined in the classical case to our full quantum system. For example, such a regime could be associated with a class of semi-classical coherent states peaking on a specific would-be clock variable \cite{Marchetti:2020umh}. We will briefly return to this point in sections \ref{clock} and \ref{eff2}, when we discuss in detail our setup of using smearing functions to define clock frames as reported in \cite{Assanioussi:2020hwf}. We leave further investigation of quantum matter reference frames and deparametrization in the present quantum gravitational system to future work \cite{GiacominiQMCO,Vanrietvelde20,delaHamette:2020dyi,Krumm:2020fws,Hohn:2018toe,Gambini:2004pe,RovelliQRSys}.  \

Lastly, comparing the algebra \eqref{canladderCR} to \eqref{ladderCR}, it is evident that the algebra in \eqref{canladderCR} describes a canonical setup. The nature of the time $t_F$ requires clarifications, which we now provide, and more work, which we leave for the future. A canonical Fock system requires a global time variable which is common to all the different multi-particle sectors, that is for a varying $N$. In other words, a clock variable, extracted somehow from the original covariant system, which in the reduced canonical system plays a role similar to the time in usual many-body quantum physics. In the case of GFTs, as we saw above, this is a relational variable (or a function of several such variables). To get a clock for the canonical Fock system then, one needs a Hamiltonian constraint operator defining some model, since the definition of a relational clock is always model-dependent due to the clock itself being one of the dynamical variables of the full extended system; and the Fock time variable must be accessible from all $N$-particle sectors, that is its construction and definition must be compatible with changing the total particle number. \

To see this, consider a system with two non-interacting particles, each equipped with its own clocks $\phi^{c_1}$ and $\phi^{c_2}$, along with their clock Hamiltonians $H^{(1)}$ and $H^{(2)}$ respectively. Let $t$ be the global clock time such that $\phi^{c_1} = F_1(t)=k_1t + \tilde{k}_1$ and $\phi^{c_2} = F_2(t)=k_2t + \tilde{k}_2$. Equivalently, $t = F_1^{-1}(\phi^{c_1}) = F_2^{-1}(\phi^{c_2})$. The $t$-clock Hamiltonian is $H_2 = k_1H^{(1)} + k_2H^{(2)}$. Now let's add a third particle to the mix, such that the resultant system remains non-interacting. Then in the new system, the global clock variable $t'$ will in general be different from $t$, corresponding to a changed relational dynamics given now by $H_3$ which has a non-zero contribution from the dynamics of the third particle also. However, notice that if this third particle, besides its own clock $\phi^{c_3}$, was also equipped with additional information, namely a syncing function $F_3$, then it would ``know'' how to sync with $t$. Therefore, for an arbitrarily large number $N$ of non-interacting particles, if each is equipped with an individual clock $\phi^{c_i}$ and a syncing function $F_i$, then a common clock time can be defined via, \be t_F = F_1^{-1}(\phi^{c_1}) = F_2^{-1}(\phi^{c_2}) = ... \ee where, in the present setting as detailed above, the set of functions $F_i$ are essentially put in by hand. In fact, we can expect that this syncing information may naturally be encoded within non-trivial interaction terms \cite{Chirco:2016wcs}, the investigation of which we leave to future work.


\begin{subappendices}

\section{*-Automorphisms for translations} \label{autapp}

We show that the map defined in \eqref{alphavg} is a *-automorphism of the GFT Weyl algebra with respect to $G^d$ translations. Notice that the case $G = \mathbb{R}$, $d=n$ is included within this more general case. Thus, the following proof is also applicable to the map \eqref{alphavphi}. \medskip

\noindent Let us first recall the required definitions \cite{Bratteli:1979tw}: \medskip

\noindent \textbf{Definition.} A *-homomorphism between two C*-algebras is a map $\pi: \mathcal{A}_1 \to \mathcal{A}_2$, which 
\begin{enumerate}
\item is complex linear $\pi(z_1 A + z_2 B) = z_1 \pi(A) + z_2 \pi(B)$ 

\item preserves algebra composition $\pi(AB) = \pi(A)\pi(B)$ 

\item preserves *-operation $\pi(A^*) = \pi(A)^*$ 
\end{enumerate}
\noindent for all $A,B \in \mathcal{A}_1$ and $z_1,z_2 \in \mathbb{C}$. The *-operation is what we have denoted as the $\dagger$-operation throughout this thesis.    \\

\noindent \textbf{Definition.} A *-isomorphism $\pi$ is a bijective *-homomorphism, i.e. ker$(\pi) = \{0\}$. \\ 

\noindent \textbf{Definition.} A *-automorphism $\alpha$ is a *-isomorphism of the algebra into itself, i.e. $\alpha: \mathcal{A} \to \mathcal{A}$, and ker$(\alpha)=\{0\}$. \\ 


\noindent \textbf{Lemma 1.} Map $\alpha_{\vg}$ as defined in equation \eqref{alphavg} is a *-homomorphism of the GFT Weyl algebra $\mathcal{A}$. \medskip

\noindent \textbf{Proof 1.} For the Weyl generators, map $\alpha_{\vg}$ (for every $\vg \in G^d$) is: \medskip

composition preserving,
\begin{align}
\alpha_{\vg}(W(f_1)W(f_2)) &= \alpha_{\vg}(W(f_1+f_2) \, e^{-\frac{i}{2} \text{Im}(f_1,f_2)}) \\
&= \alpha_{\vg}(W(f_1+f_2)) \, e^{-\frac{i}{2} \text{Im}(f_1,f_2)}  \\
&= W(L^*_{\vg}(f_1+f_2)) \, e^{-\frac{i}{2} \text{Im}(L^*_{\vg}f_1,\,L^*_{\vg}f_2)} \\
&= W(L^*_{\vg}f_1+L^*_{\vg}f_2) \, e^{-\frac{i}{2} \text{Im}(L^*_{\vg}f_1,\,L^*_{\vg}f_2)}  \\
&= W(L^*_{\vg}f_1)W(L^*_{\vg}f_2) = \alpha_{\vg}(W(f_1))\,\alpha_{\vg}(W(f_2)) 
\end{align}

*-operation preserving,
\begin{align}
\alpha_{\vg}(W(f)^\dag) = \alpha_{\vg}(W(-f)) = W(-L^*_{\vg}f) = W(L^*_{\vg} f)^\dag = (\alpha_{\vg}W(f))^\dag .
\end{align}

\noindent $\alpha_{\vg}$ satisfies linearity over $\mathbb{C}$ by definition. These properties are extended to the full algebra by linearity and composition.  $\hfill \square$  \\


\noindent \textbf{Lemma 2.} The *-homomorphism $\alpha_{\vg}:\mathcal{A} \to \mathcal{A}$ is a *-automorphism, i.e. ker$(\alpha_{\vg}) = \{0\}$. \medskip
 
\noindent \textbf{Proof 2.} Recall that, for any $\vg \in G^d$, ker$(\alpha_{\vg}) = \{ A \in \mathcal{A} \,|\, \alpha_{\vg}A = 0 \}$. Then, for a generic element $A \in \mathcal{A}$, given by a superposition of the basis elements $W(f)$, we have 
\be
\alpha_{\vg}A = \alpha_{\vg} \sum_i c_i W(f_i) = \sum_i c_i W(L^*_{\vg}f_i) 
\ee
where $c_i\in \mathbb{C}$, $i\in \mathbb{N}$. Recall that, $L^*:G^d \times \mcH \to \mcH$ is the regular representation of $G^d$ on the space of functions $\mcH$, which is closed under the action of $L^*_{\vg}$ (for all $\vg \in G^d$) \cite{fecko2006,kowalski}. Thus, $W(L^*_{\vg}f_i)$ are also elements of the same basis set.\footnote{Another simple way to see this is: for a given $\vg \in G^d$, $L_{\vg}^*f := f\circ L_{\vg^{-1}}$, i.e. map $L_{\vg}^*$ is a composition of $f$ with left translations on $G^d$, and these are diffeomorphisms on $G^d$ \cite{fecko2006,kowalski}.} Then, by using linear independence of the basis, and by definition of element $A$, we have
\be \sum_i c_i W(L^*_{\vg}f_i) = 0 \quad \Leftrightarrow \quad c_i = 0 \;\;\; \forall \, i \quad \Leftrightarrow \quad A = 0 \,.  \ee
Thus, $\alpha_{\vg}A = 0 \Leftrightarrow A = 0$. $\hfill \square$  \\


\noindent \textbf{Lemma 3.} Map $\alpha: G^d \rightarrow \text{Aut}(\mathcal{A})$ is a representation of $G^d$ into the group of all automorphisms of the algebra $\mathcal{A}$. \medskip

\noindent \textbf{Proof 3.} For any $\vg_1,\vg_2 \in G^d$, we have:
\be \label{comsec}
(L^*_{\vg_1 . \vg_2}f)(\vg) = f((\vg_1 . \vg_2)^{-1}. \vg) = f(\vg_2^{-1}.\vg_1^{-1}.\vg) = (L^*_{\vg_2}f)(\vg_1^{-1}.\vg) = (L^*_{\vg_1}L^*_{\vg_2}f)(\vg)  
\ee
and,
\be \label{comthi}
W(L^*_{\vg_1}L^*_{\vg_2}f) = \alpha_{\vg_1}(W(L^*_{\vg_2}f)) = \alpha_{\vg_1}\alpha_{\vg_2}(W(f))  \,.
\ee
Then, composition is preserved by $\alpha$: 
\be \alpha_{\vg_1 . \vg_2}(W(f)) = W(L^*_{\vg_1 . \vg_2}f) = W(L^*_{\vg_1}L^*_{\vg_2}f) = \alpha_{\vg_1}\alpha_{\vg_2}(W(f))  \ee
where, we have used results \eqref{comsec} and \eqref{comthi}. This is extended to the full algebra by linearity and composition. $\hfill \square$


\section{Strong continuity of unitary translation group} \label{strcon}

The existence of unitary groups in $\mcH_F$ has been established in section \ref{unis}, using the invariance of the Fock state $\omega_F$ under the translation automorphisms. Given $U(\vg) \in \mathcal{U}(\mcH_F)$, for $\vg \in G^d$, we show below that the map $\vg \mapsto U(\vg)$ is strongly continuous in $\mathcal{H}_F$. Notice that the case $G = \mathbb{R}$, $d=n$ with $T^*=L^*$, is included within this more general case. This proof is along the lines of that reported in \cite{Kotecha:2018gof}, but is detailed further for clarity. \\

\noindent \textbf{Lemma.} $\{U(\vg) \,|\, \vg \in G^d \}$ is a strongly continuous family of operators in $\mathcal{H}_F$, i.e.
\be ||(U(\vg_1) - U(\vg_2))\psi|| \to 0 \;, \quad \text{as} \;\; \vg_1 \to \vg_2 \ee 
for all $\psi \in \mathcal{H}_F$, and all $\vg_1, \vg_2 \in G^d$. \medskip

\noindent \textbf{Proof.} Let us begin with the set of basis vectors $\{W_F(f)\Omega_F \,|\, f\in \mcH \}$ in $\mcH_F$. Recall: from equation \eqref{pwfun}, 
\be W_F(f) = \pi_F(W(f)) \,, \ee 
and, from equations \eqref{uni} and \eqref{alphavg},
\be U(\vg)W_F(f)U(\vg)^{-1} = W_F(L^*_{\vg}f) \,,\quad U(\vg)\Omega_F = \Omega_F \,.\ee 

\noindent Then, for any $\vg_1,\vg_2 \in G^d$,
 \begin{align}
||(U(\vg_1)-U(\vg_2))W_F(f)\,\Omega_F||^2 &= ( U(\vg_1) W_F(f)\,\Omega_F  , U(\vg_1) W_F(f)\,\Omega_F ) \nonumber \\ 
& \;\; + (U(\vg_2) W_F(f)\,\Omega_F ,  U(\vg_2) W_F(f)\,\Omega_F ) \nonumber \\
& \;\; - (U(\vg_1) W_F(f)\,\Omega_F ,  U(\vg_2) W_F(f)\,\Omega_F ) \nonumber \\
& \;\; - (U(\vg_2) W_F(f)\,\Omega_F ,  U(\vg_1) W_F(f)\,\Omega_F )  \\
&= ||W_F(L^*_{\vg_1} f)\Omega_F||^2 + ||W_F(L^*_{\vg_2} f)\Omega_F||^2 \nonumber \\
& \;\; - 2 \, \text{Re}\,(W_F(L^*_{\vg_2} f) \Omega_F, W_F(L^*_{\vg_1} f)\Omega_F) \label{ResOne} \,.
\end{align}

\noindent Notice that, for any $\vg \in G^d$,
\be 
||W_F(L^*_{\vg} f))\Omega_F||^2 \leq ||W_F(L^*_{\vg} f)||^2\,||\Omega_F||^2 = ||W_F(L^*_{\vg} f)||^2 = 1 
\ee
where, the last equality is because $W_F(f)$ is a unitary operator, for any $f$ (see equation \eqref{pwfun}), and, $|| \pi_F(.) ||$ is the operator norm\footnote{For any bounded operator $A$ on $\mcH_F$, the operator norm is $||A|| := \sup \limits_{\Psi \in \mcH_F} \frac{||A\Psi ||}{||\Psi||} $.} on $\mcH_F$. This implies that,
\be \label{ResTwo}
||W_F(L^*_{\vg_1} f)\Omega_F||^2 + ||W_F(L^*_{\vg_2} f)\Omega_F||^2 \leq 2 \,.
\ee
Also, notice that, we have
\begin{align}
(W_F(L^*_{\vg_2} f) \Omega_F, W_F(L^*_{\vg_1} f)\Omega_F) &= ( \Omega_F, W_F(L^*_{\vg_2} f)^{\dag}W_F(L^*_{\vg_1} f)\Omega_F) \\
&= ( \Omega_F, \pi_F(W(- L^*_{\vg_2} f) W(L^*_{\vg_1} f))\Omega_F) \\
&= ( \Omega_F, \pi_F(W(- L^*_{\vg_2} f + L^*_{\vg_1} f) e^{-\frac{i}{2} \text{Im}(- L^*_{\vg_2} f, L^*_{\vg_1} f)} ) \, \Omega_F) \\
&= ( \Omega_F, W_F( L^*_{\vg_1} f  - L^*_{\vg_2} f) e^{\frac{i}{2} \text{Im}(L^*_{\vg_2} f, L^*_{\vg_1} f)} \Omega_F) \\
&= e^{\frac{i}{2} \text{Im}(L^*_{\vg_2} f, L^*_{\vg_1} f)} \omega_F[W(L^*_{\vg_1} f  - L^*_{\vg_2} f)] \\
&= e^{\frac{i}{2} \text{Im}(L^*_{\vg_2} f, L^*_{\vg_1} f)} e^{-|| L^*_{\vg_1} f  - L^*_{\vg_2} f ||^2/4} 
\end{align}
and its real component,
\begin{align}
\text{Re}\,(W_F(L^*_{\vg_2} f) \Omega_F, W_F(L^*_{\vg_1} f)\Omega_F) &= \text{Re} \left[ e^{\frac{i}{2} \text{Im}( L^*_{\vg_2} f, L^*_{\vg_1} f)} e^{-|| L^*_{\vg_1} f  - L^*_{\vg_2} f ||^2/4} \right] \\
&= e^{-|| L^*_{\vg_1} f  - L^*_{\vg_2} f ||^2/4} \cos \left( \frac{1}{2} \text{Im} (L^*_{\vg_2} f, L^*_{\vg_1} f) \right) .  \label{ResThree}
\end{align}

\noindent For the functions $f$, we have
\begin{align}
(L^*_{\vg_2} f, L^*_{\vg_1} f) &= \int_{G^d \times \mathbb{R}^n} d\vphi d\vg \;\; \overline{L^*_{\vg_2} f}(\vg) L^*_{\vg_1} f(\vg) \\
&= \int_{G^d \times \mathbb{R}^n} d\vphi d\vg \;\; \overline{f}(\vg_2^{-1}.\vg) f(\vg_1^{-1}.\vg) \\
&= \int_{G^d \times \mathbb{R}^n} d\vphi d\vec{h} \;\; \overline{f}(\vec{h}) f(\vg_1^{-1}.\vg_2.\vec{h}) \label{EQQ} \\
&\longrightarrow (f,f) = ||f||^2 \qquad \text{as} \; \vg_1 \to \vg_2 \label{ResFour}
\end{align}
where, translation invariance of Haar measure on $G^d$ gives equality \eqref{EQQ}, and smoothness of $f$ and of multiplication map on any Lie group gives \eqref{ResFour}. Further, we have,
\begin{align}
|| L^*_{\vg_1} f  - L^*_{\vg_2} f ||^2 &= || L^*_{\vg_1} f||^2 + || L^*_{\vg_2} f||^2 - 2\text{Re}\,(L^*_{\vg_2} f, L^*_{\vg_1} f) \\
&= 2||f||^2 - 2\text{Re}\,(L^*_{\vg_2} f, L^*_{\vg_1} f) \\
&\longrightarrow 0 \qquad \text{as} \; \vg_1 \to \vg_2 \label{ResFive}
\end{align}
using the result \eqref{ResFour}.\footnote{We remark that results \eqref{ResFive} and \eqref{ResFour} basically show strong continuity of representation $L^*$ of $G^d$ on $\mcH$. This is an expected feature of regular representations of locally compact Lie groups, on the space of square-integrable smooth functions defined on the same group \cite{kowalski}.} Therefore, for any $\vg_1,\vg_2 \in G^d$:
\begin{align}
||(U(\vg_1)-U(\vg_2))W_F(f)\,\Omega_F||^2  &\leq 2 \left[1 - e^{-|| L^*_{\vg_1} f  - L^*_{\vg_2} f ||^2/4} \cos \left( \frac{1}{2} \text{Im} (L^*_{\vg_2} f, L^*_{\vg_1} f) \right) \right] \nonumber \\
&\longrightarrow 0 \qquad \text{as} \; \vg_1 \to \vg_2   \label{strcongen}
\end{align}
using the results \eqref{ResOne}, \eqref{ResTwo},  \eqref{ResThree}, \eqref{ResFour} and \eqref{ResFive}. \medskip

\noindent Now, consider the dense domain $D = \text{Span}\{W_F(f)\Omega_F \,|\, f\in \mcH \} \subset \mcH_F$, 
and a vector $\psi \in D$ written generally as a superposition of the basis vectors,
\be \psi = \sum_i c_i W_F(f_i) \Omega_F \ee 
with coefficients $c_i \in \mathbb{C}$. Then, we have
\begin{align}
||(U(\vg_1)-U(\vg_2))\psi|| 
			&\leq \sum_i |c_i| \, ||(U(\vg_1)-U(\vg_2)) \, W_F(f_i) \Omega_F|| \\
			& \longrightarrow 0  \qquad \text{as} \; \vg_1 \to \vg_2
\end{align}
using the result \eqref{strcongen} for each $i$.  Thus, the family of unitary operators $\{U({\vg})\,|\, \vg \in G^d\}$ is strongly continuous in $D$. \medskip

\noindent Finally, notice that $U({\vg})$ (for all $\vg \in G^d)$ are bounded linear transformations, thus extended naturally to $\mathcal{H}_F$. Also, $D$ is dense in $\mcH_F$, i.e. for every $\Psi \in \mcH_F$ and arbitrarily small $\epsilon > 0$, $\exists \, \psi \in D$ such that $|| \Psi - \psi || < \epsilon$. Then, 
\begin{align}
||U(\vg_1)\Psi-U(\vg_2)\Psi|| 
&\leq ||U(\vg_1)\Psi -  U(\vg_1)\psi || + ||(U(\vg_1) -U(\vg_2))\psi || \nonumber \\
& \quad + ||U(\vg_2)\psi - U(\vg_2)\Psi  ||  \\
&= 2||\Psi - \psi || + ||(U(\vg_1) -U(\vg_2))\psi || \\
&< 2\epsilon + ||(U(\vg_1) -U(\vg_2))\psi || < 3\epsilon 
\end{align}   
using strong continuity in $D$ for the last inequality.  $\hfill \square$

\end{subappendices}

\chapter{Thermal Group Field Theory}
\label{TGFT}

\begin{quotation} All the previously enumerated theorems are strict consequences of a single proposition: that stable equilibrium corresponds to the maximum of entropy. That proposition, in turn, follows from the more general one that in every natural process the sum of the entropies of all participating bodies is increased. Applied to thermal phenomena this law is the most general expression of the second law of the mechanical theory of heat ... ---\emph{Max Planck}\footnote{As cited in \cite{kuhn1987}, p16.} \end{quotation}


\vspace{2.5mm}

\noindent Equipped with the many-body description of an arbitrarily large number of algebraic and combinatorial quanta of geometry from the previous sections, a statistical mechanics for them can formally be defined. Given a normal state $\omega_\rho$, then the quantities $\omega_\rho[A] = \Tr(\rho A)$ are the quantum statistical averages of operators $A$. If additionally, $A$ is self-adjoint and has a structure that admits some physical interpretation, then $\omega_\rho[A]$ are ensemble averages of observables $A$. $\omega_\rho$ is a statistical mixture of quantum states encoding discrete gravitational and scalar matter degrees of freedom associated with the group field.\footnote{Group field theory, being a second quantisation of loop quantum gravity \cite{Oriti:2013aqa}, then this reformulation of spin network degrees of freedom would allow us to define a statistical mechanics for them.} \

Equilibrium statistical states are known to play an important role in describing macroscopic systems. Particularly, Gibbs thermal states are ubiquitous in physics, being utilised across the spectrum of fields, ranging from phenomenological thermodynamics, condensed matter physics, optics, tensor networks and quantum information, to gravitational horizon thermodynamics, AdS/CFT and quantum gravity. Even generic systems not at equilibrium may be modelled via the notion of local equilibrium, wherein various subsystems of the whole are locally at equilibrium, while the whole system is not. Then, collective variables such as number and temperature densities, vary smoothly across these different patches, while having constant (equilibrium) values within a given subsystem. This basic idea underlies several techniques for coarse-graining in general, and as such displays again the usefulness of statistical equilibrium descriptions. Thus also in discrete quantum gravity, statistical equilibrium states would be of value, in ongoing efforts to get an emergent coarse-grained spacetime, based on techniques from finite-temperature quantum and statistical field theory. Also, equilibrium statistical mechanics provides a fundamental microscopic foundation for thermodynamics. Therefore, a statistical framework for some quantum gravitational degrees of freedom would also facilitate identification of thermodynamic variables with geometric interpretations originating in the underlying fundamental theory, to then make contact with studies in spacetime thermodynamics. Construction of several different examples of generalised Gibbs states in the present quantum gravity system is the focus of the first part of this chapter. \

Further, an increasing number of studies are hinting toward intimate links between geometry and entanglement. Spacetime is thought to be highly entangled, with quantum correlations in the underlying quantum gravitational system being crucial. Particularly in discrete approaches, a geometric phase of the universe is expected to emerge from a pre-geometric one via a phase transition, which must then also be highly entangled. The emergent phase must also encode a suitable notion of semi-classicality, for example by introducing coherent states. Moreover, fluctuations in relevant observables are expected to be important in the physics of a quantum spacetime, and thus must be included in the description of the system. In the second part of this chapter, we show how to construct quantum gravitational phases in group field theory (for now, kinematically) using techniques from thermofield dynamics, which display all these features, namely entanglement, coherence and statistical fluctuations in given observables. \

Finally, the ultimate goal of any theory of quantum gravity is to describe the known physics, while also providing novel falsifiable predictions on a measurable scale. An important arena in this respect is cosmology, with features such as singularity resolution and inflation representing crucial checkpoints for any viable model based on an underlying theory of quantum gravity. It is thus important for any candidate theory to find a suitable continuum and semi-classical regime within the full theory, in which standard cosmology can be approximated, up to effective corrections of quantum gravitational origin. In group field theory, such a regime has been suggested via a class of condensate phases of the system \cite{Pithis:2019tvp,Oriti:2016acw,Gielen:2016dss,Gabbanelli:2020lme}. In the last part of this chapter, we consider a class of thermal condensates to derive an effective model of homogeneous and isotropic cosmology from non-interacting GFT dynamics, with corrections of quantum and statistical origins to the classical relational Friedmann equations. \

We begin in section \ref{gengibb}, with the construction of different examples of generalised Gibbs states associated with a variety of generators, utilising the many-body formalism introduced previously. In section \ref{posgibbs}, we consider a class of positive, extensive operators in the GFT Fock representation induced by a degenerate vacuum. This class includes the special case of a volume operator. The corresponding states are defined using the maximum entropy principle. We show that these states naturally admit Bose-Einstein condensation to the single-particle ground state of the generator (which corresponds to a low-spin phase in the spin representation). In section \ref{momgibbs}, we consider the KMS condition for Gibbs states. We define equilibrium states with respect to momentum generators for translations on the group base manifold, both for internal translations corresponding to the C*-automorphisms in section \ref{inttrans}; and external translations corresponding to clock Hamiltonians after deparametrization in section \ref{physeqm}. In section \ref{congibb}, we use the maximum entropy principle to consider a classical system of tetrahedra, fluctuating in terms of the geometric condition of closure in section \ref{entclosure}; and in terms of half-link gluing conditions in section \ref{glucondgibb}, leading to statistically fluctuating twisted geometries. \

In the next section \ref{threpcond}, we begin with an overview of the relevant essentials of the formalism of thermofield dynamics in \ref{tfd}. We use this formalism in the subsequent sections to present a systematic extension of group field theories, for constructing finite temperature equilibrium phases associated with generalised Gibbs states. Using the setup of bosonic group field theory coupled to a scalar matter field as presented in section \ref{bosgft} above, we give a description for the zero temperature phase based on the degenerate vacuum in \ref{zero}. Then focusing on positive extensive operators as generators for generalised Gibbs states, we construct their corresponding thermal vacua (thermofield double states), and the inequivalent phases generated by them, in section \ref{finite}. In section \ref{CTS} we introduce the class of coherent thermal states and give an overview of their useful properties.  \

Finally in section \ref{GFTCC}, we analyse a free GFT model for effective cosmology, based on the above construction of equilibrium thermal representations, and using coherent thermal states as candidates for thermal quantum gravitational condensates. In section \ref{vgibbs}, we explicate the choice of the state, based on which we derive the GFT effective equations of motion in \ref{eff1}. In sections \ref{clock} and \ref{eff2}, we reformulate the effective dynamics in terms of relational clock functions, implemented as smearing functions along the $\phi$ direction of the base manifold. We show that this provides a suitable non-singular generalisation of the relational frame used in previous works in terms of the coordinate $\phi$. We further derive the effective generalised Friedmann equations for flat homogeneous and isotropic cosmology in section \ref{EFhic}, recover the correct classical general relativistic limit in a late time regime in \ref{lateev}, and characterise the early time evolution through an assessment of singularity resolution and accelerated expansion in \ref{earlyev}. We close with a discussion of some aspects surrounding the inclusion of interactions and its implications at the level of the effective thermal GFT models in section \ref{disc}.


\section{Generalised Gibbs states} \label{gengibb}

In the study of bulk properties of a system of many discrete constituents, Gibbs states provide the simplest description of the system, that of equilibrium. Generalised Gibbs states can be written as $e^{-\sum_a \beta_a \mathcal{O}_a}$, where $\mathcal{O}_a$ are operators that are of interest in the situation at hand whose state averages $\langle \mathcal{O}_a \rangle$ are fixed, and $\beta_a$ are the corresponding intensive parameters that characterise the equilibrium configuration (see chapter \ref{GGS} for details). Naturally, this leaves open the possibilities for the precise choice of the different observables. In fact, which ones are relevant in any given situation is an important part of the broader problem of investigating the statistical mechanics of quantum gravity for an emergent, thermodynamical spacetime with features compatible with semi-classical studies. In this section we present examples of generalised Gibbs states in discrete quantum gravity, by applying insights from discussions on generalised statistical equilibrium (as presented in chapter \ref{GGS}) in group field theory (as presented in chapters \ref{DQG} and \ref{GFT}).


\subsection{Positive extensive operators} \label{posgibbs}

\subsubsection{General} \label{genposgibbs}

A particularly interesting class of operators for which the corresponding Gibbs states are well-defined, are positive extensive operators on $\mcH_F$ of the form
\be \label{posopp}  {\mcP} = \sum_{{\vchi},\valpha} \lambda_{{\vchi}\valpha}  {a}^\dag_{{\vchi}\valpha} {a}_{{\vchi}\valpha} \,, \quad \lambda_{{\vchi},\valpha} \in \mathbb{R}_{\geq 0} \ee
where labels $\vchi,\valpha$ denote the discrete basis introduced in section \ref{discind}. Notice that $\mcP$ is self-adjoint, by definition. Further $\mcP$ is an extensive operator, which means that it is proportional to the size of the system \cite{fetter}, i.e. the total number of quanta, or spin network nodes. Also, $\mcP$ is a one-body operator \cite{fetter} on the Fock space, which means that its total action on any multi-particle state is additive, with irreducible contributions coming from individual actions on a single particle $\ket{\vchi,\valpha}$. Both these features are manifest in the fact that modes of $\mcP$ scale as the number density operator. A typical example of an extensive, one-body operator in a standard many-body quantum system is the total kinetic energy. In group field theory, extensive one-body operators are a second quantisation \cite{fetter} of those loop quantum gravity operators which are diagonal in some intertwiner basis, such as the spatial volume operator \cite{Ashtekar:1997fb,Oriti:2013aqa,Kotecha:2018gof}. \

Using the thermodynamical characterisation (section \ref{maxent}), the corresponding Gibbs state is
\be \label{posextrho} \rho_{\beta} = \frac{1}{Z_\beta} e^{-\beta \mcP} \,.  \ee
On a Fock space however, it is more natural to define states of the grand-canonical type, with fluctuating particle number, that is
\be \label{gcposrho} \rho_{\beta,\mu} = \frac{1}{Z_{\beta,\mu}} e^{-\beta (\mcP - \mu N)} \,. \ee
These are self-adjoint, positive and trace-class (density) operators on $\mcH_F$, for $0 < \beta < \infty$ and $\mu < \min (\lambda_{{\vchi}\valpha})$ for all ${\vchi}$ and $\valpha$. See appendix \ref{posapp2} for details. Further, using the occupation number basis introduced in section \ref{occunum}, the partition function can be evaluated to give,
\be \label{zgc} Z_{\beta,\mu} = \prod_{{\vchi},\valpha} \frac{1}{1-e^{-\beta (\lambda_{{\vchi}\valpha} - \mu)}}  \ee
as shown in appendix \ref{posapp2}. This is a grand-canonical state of quanta of the $a$-operators. Like in standard statistical mechanics, this state essentially describes a gas of these quanta of space with a changing total number in a given system. Notice that, for a constant $\mu$, the number operator in \eqref{gcposrho} simply implements a shift in the spectrum of $\mcP$ by $\mu$, thus allowing for an overall replacement $\lambda_{{\vchi}\valpha} - \mu \mapsto \lambda_{{\vchi}\valpha}$, as will be done later in sections \ref{threpcond} and \ref{GFTCC}. \

It is evident that by construction, the parameter $\beta$ controls the strength of statistical fluctuations in $\mcP$, regardless of any other interpretations. Note however, that one can reasonably inquire about its geometric meaning, especially if the operator $\mcP$, which is its thermodynamic conjugate, has a clear geometric interpretation. It would thus be interesting to investigate this aspect in a concrete example where the choice of the observable is adapted to a physical context, like cosmology (see for instance \cite{Assanioussi:2020hwf}, or section \ref{GFTCC} in this thesis).


\subsubsection{Spatial volume} \label{volposgibbs}

The volume observable plays a crucial role in quantum gravity. We recall that in loop quantum gravity there exist different proposals for a volume operator (see \cite{Ashtekar:1997fb,Bianchi:2010gc,Haggard:2011qvx,Rovelli:1994ge} and references therein), but in each its spectral values are attached to the nodes  of the labelled graphs. In other words, an elementary quantum of volume of space is assigned to a single node of a boundary graph. Since the GFT Fock space $\mcH_F$ is the second quantisation of the spin network degrees of freedom \cite{Oriti:2013aqa}, here a volume eigenvalue is associated to a quantum of the group field. Let us then consider a Gibbs state with respect to a volume operator $V$. As will be clear below, this is a special case of the states \eqref{gcposrho} defined above, with operator $\mcP$ now chosen to be the volume operator. \

We note that classical geometric observables, such as spatial volumes or areas of hypersurfaces, are not necessarily completely well-defined in a diffeomorphism invariant (background independent) context. This is because such quantities are at most diffeomorphism covariant, thus failing to represent a physical gauge-invariant observable. This open issue of understanding physical observables in general relativity, arises also within any background independent quantum gravity approach. Having said that, one could still construct mathematically well-defined quantum operators, formally associated with certain classical quantities in a given quantum framework, such as the volume of space. In this sense, one could still define operators with a geometric interpretation in a background independent context, like in group field theory or loop quantum gravity. \cite{Assanioussi:2019ouq} \

In line with the original work reported in \cite{Kotecha:2018gof}, here we choose to neglect the scalar matter degrees of freedom taking values in $\mathbb{R}^n$, but it should be evident from section \ref{genposgibbs} above that an extension to this case is directly possible. The main reasons for this choice are the following. First, in this example we are not particularly interested in matter degrees of freedom or any relational reference frames that they may define. For instance, in section \ref{GFTCC} in the context of cosmology, we will reinstate this dependence in order to define relational evolution. Second, the volume of a quantum of space is a geometric quantity expected to depend primarily on the group representation data $\vchi$. This is not to say that the volume of the corresponding emergent spacetime manifold would not depend on matter, which it of course does according to GR. This is also the standard choice made in LQG. With this in mind, the choice of independence of the volume observable from matter degrees of freedom should be viewed as a first step that is simple enough to investigate geometric properties of a theory of fundamental discrete constituents of spacetime.  \\

\noindent \textbf{Volume Gibbs state} \\

\noindent Given a multi-particle boundary (graph) state, the total volume operator should basically count the number of particles (nodes) $n_{\vchi}$, in each mode $\vchi$, and multiply this number by the volume eigenvalue $v_{\vchi}$ associated to a single particle (node) in that mode. It is thus an extensive, one-body operator, given by
\begin{equation} \label{vol}
V = \sum_{\vchi} v_{\vchi} \, a^\dag_{\vchi} a_{\vchi} \,.
\end{equation}

As long as the perspective of attaching a quantum of space to a graph node holds, a volume operator in a Fock space formulation of states of labelled graphs will always be of this form. This operator is diagonal in the occupation number basis (section \ref{occunum}), with action
\be V\ket{\{n_{\vchi_i}\}} = \left(\sum_{i'} v_{\vchi_{i'}} n_{\vchi_{i'}} \right) \ket{\{n_{\vchi_i}\}} . \ee

Keeping in mind our understanding of $v_{\vchi}$ as the volume of a quantum polyhedron with faces coloured by $SU(2)$ representation data $\vchi$, the following reasonable assumptions are made. First, the single-mode spectrum is chosen to be real and positive, i.e. \be v_{\vchi} \in \mathbb{R}_{> 0} \ee
for all $\vchi$. This implies that the spectrum of the total volume operator $V$ is real and non-negative because it is simply a result of scaling $v_{\vchi}$ with occupation numbers $n_{\vchi} \in \mathbb{N}_{\geq 0}$. Therefore by construction, $V$ is a positive, self-adjoint element of $\mathcal{A}_F$. Positivity ensures boundedness from below and therefore the existence of at least one ground state. Second, we assume uniqueness (or, non-degeneracy) of the single-particle ground state, i.e. for $v_0 := \min(v_{\vchi})$, and $V\ket{\vchi_0} = v_0 \ket{\vchi_0}$, we have
\be v_{\vchi} = v_0 \Leftrightarrow \vchi = \vchi_0 \,. \ee 

Naturally, the precise value of $v_0$ depends on the specifics of the spectrum $v_{\vchi}$, which in turn depends on the specific quantisation scheme used to define the operator. Our results are independent of these specifics. Notice that the uniqueness assumption would fail if the degenerate zero eigenvalue for $v_{\vchi}$ is included in the spectrum, because this could correspond to several different spin configurations. We stress however that non-degeneracy of the single-particle ground state $\ket{\vchi_0}$ is assumed only for conceptual consistency in the upcoming scenario of Bose-Einstein condensation. In fact our technical results, particularly the definition of the state, will hold even without both these assumptions. This is clear from the discussions and proofs presented in section \ref{genposgibbs} that are valid for spectrum $\lambda_{\vchi,\valpha} \in \mathbb{R}_{\geq 0}$, thus including possibly degenerate zero eigenvalues.  \

Then, the corresponding equilibrium state on $\mcH_F$ is given by, 
\begin{equation}\label{volumerho}
\rho = \frac{1}{Z} \, e^{-\beta (V-\mu N)}
\end{equation}
with real parameters $\mu$ and $0 < \beta < \infty$, and $Z$ as given in \eqref{partFUN}. The details in appendix \ref{posapp2} show that, for $\mu < v_0$, the above $\rho$ is a legitimate density operator. \

Let us pause to consider what it means to define such a Gibbs state, as generated by the volume operator. Referring back to the discussion in section \ref{GenGE}, specifically to the thermodynamical characterisation in a background independent setting, a state like \eqref{volumerho} can be best understood as arising from the principle of maximisation of entropy, $S = - \langle \ln \rho \rangle$ of the system, under the constraints $\langle I \rangle = 1$, $\langle V \rangle = \mathbf{V}$ and $\langle N \rangle = \mathbf{N}$, without any need for a pre-defined flow. Parameters $\beta$ and $\mu$ enter formally as Lagrange multipliers. From a purely statistical point of view, the corresponding physical picture, intuitively, would be that of a system in contact with a bath, which exchanges quantities corresponding to the operators $V$ and $N$.\footnote{In the case at hand, exchange of particles inevitably leads to exchange of volume (and vice-versa), because the particles themselves carry the quanta of volume. In fact, $\mu$ is just a constant shift in the volume spectrum in this example.} The macroscopic description of the system is then given by the averages $\mathbf{V}$ and $\mathbf{N}$, along with the intensive parameters $\beta$ and $\mu$\footnote{Formally $\beta$ and $\mu$ parametrise the class of Gibbs states \eqref{volumerho}. Presently no attempt is made to assign/require any additional interpretations to/of them.} which characterise the equilibrium phase. \

From a more general information-theoretic point of view, the system can be understood as being bi-partite, with the quanta (here, of the $a_{\vchi}$ field) in the state $\rho$ describing the (sub) system of interest. In this case then, the full system can be understood as being in an entangled state, with a fixed large number of nodes. 
Then, partial tracing over the complement would give a (reduced) mixed state for the (sub) system of interest. We will discuss some of these aspects further in section \ref{threpcond}. A precise characterisation of the complementary subsystem (often interpreted as a thermal bath in statistical mechanics), boundary effects due to spin network links puncturing the boundary surface, consequences of entanglement across the boundary surface, and, the exact conditions and process for thermalisation of the subsystem to a Gibbs state are left to future work.

Given our framework, defining a state like \eqref{volumerho} is justified from both a quantum statistical and information-theoretic perspectives, as noted above. Now in the context of quantum gravity, such a state could be interesting to consider for the following reason. Instead of an arbitrary pure (spin network) state, a mixed state associated with geometric operators like volume or area would be expected to better describe the physical state of a region of space, wherein the corresponding macroscopic volumes or areas of regions are given by statistical averages, $\langle V \rangle_\rho$ and $\langle A \rangle_\rho$; and, such averages characterise, at least partially, the geometric macrostate (see for instance discussions in \cite{Montesinos:2000zi}). In other studies for example, a similar perspective is held with the aim of defining `geometric' entropies, with respect to area measurements of boundary spin network links in \cite{PhysRevD.55.3505}, volume measurements of bulk links in \cite{Astuti:2016dmk}, and in various LQG-inspired analyses of quantum black holes microstates \cite{Perez:2017cmj,DiazPolo:2011np}. Within the framework described here, such geometrical entropies arise naturally as the information (von Neumann) entropy of a statistical state $\rho$ associated with geometric observables. \\

\noindent \textbf{Condensation to low-spin phase} \\

\noindent In addition to the link to macroscopic geometry that such states could provide, we show that they can lead to interesting phases purely as a result of the collective behaviour of the underlying quanta. Specifically, we show below that the volume Gibbs state as defined in \eqref{volumerho} admits a condensed phase that is populated majorly by quanta in the lowest possible spin configuration $\vchi_0$. We also comment on a special sub-class of such condensates, the commonly encountered spin-1/2 phase, which is characterised by isotropic\footnote{All links incident on an isotropic node are labelled by the same spin.} SU(2) spin network nodes and almost all links labelled by the same $j=1/2$. \

As before, the occupation number basis of $\mathcal{H}_F$, being the eigenbasis of \eqref{vol}, can be used for computations. From equation \eqref{zgc}, it is clear that the partition function in \eqref{volumerho} is given by,
\begin{equation} \label{partFUN}
Z = \sum_{\{n_{\vchi_i}\}} \bra{{\{n_{\vchi_i}\}}} \prod_i e^{-\beta ( v_{\vchi_i} - \mu ) n_{\vchi_i} } \ket{{\{n_{\vchi_i}\}}} = \prod_{\vchi} \frac{1}{1 - e^{-\beta ( v_{\vchi} - \mu ) } } \,.
\end{equation}

This partition function can be observed to have the same form as that of a gas of free non-relativistic bosons in a Gibbs thermal state, which is defined in terms of the standard free Hamiltonian (total kinetic energy) operator \cite{stringaribook}. In our case however, the simplicity of the state is not a statement about its underlying dynamics, or the result of some controlled approximation of some dynamics\footnote{In fact, the present example is without the use of any specific dynamical ingredients.}. For the simple case of the volume operator, unless our geometrical perspective of assigning a quantum of volume to a graph node changes, the corresponding operator will always be a one-body extensive operator (of the form \eqref{vol}) in a Fock space formulation, and whose corresponding Gibbs partition function will be reminiscent of an ideal Bose gas. As is clear from section \ref{genposgibbs}, such a partition function will arise for any operator of the general form \eqref{posopp}. Another example is the kinetic part of a GFT action with a Laplacian term, $S_K = \int d\vg \; \varphi^\dag(\vg) (-\sum_{I=1}^d \Delta_{g_I} + m^2) \varphi(\vg)$, which is used often in the literature. This operator would naturally contribute to the system's Hamiltonian, in cases when a suitable Legendre transform with respect to a chosen clock variable is possible \cite{Wilson-Ewing:2018mrp}. Since the $SU(2)$ Wigner expansion modes $\mbfD_{\vchi}(\vg)$ are eigenstates of the Laplacian, we have, $S_K = \sum_{\vchi} (A_{\vchi} + m^2)a^\dag_{\vchi}a_{\vchi}$, where $A_{\vchi} = \sum_{I=1}^d j_I(j_I + 1)$.   \

\noindent Now, the average total number of particles in state \eqref{volumerho} is\footnote{In terms of the thermodynamic free energy $F = \langle V - \mu N \rangle_\rho - \beta^{-1}S  = -\frac{1}{\beta}\ln Z$, the average particle number is, $\langle N \rangle_\rho = - \frac{\partial F}{\partial \mu}$, like in standard thermodynamics.},
\begin{equation}
\mathbf{N} = \langle N \rangle_\rho = \sum_{\vchi}  \frac{1}{e^{+\beta ( v_{\vchi} - \mu )}-1 } \,.
\end{equation} 
It is clear that the dominant term in the above series corresponds to the ground state with the smallest eigenvalue $v_0$. Therefore, as $\mu \rightarrow v_0$, the average occupation number of the ground state,
\be \mathbf{N}_0 =  \frac{1}{e^{+\beta ( v_0 - \mu )}-1 }  \ee 
diverges, and the system undergoes condensation. This results in a macroscopic occupation of the single-particle state $\ket{\vchi_0}$ with volume $v_0$. A low-spin condensate phase thus arises naturally as a quantum statistical process in a system with a large number of quanta of geometry. Notice that this is analogous to standard Bose-Einstein condensation in a non-relativistic gas of free bosons \cite{stringaribook}. The order parameter can now be directly seen as the non-zero expectation value of the group field operator, i.e.
\begin{align}
\langle \varphi(\vg) \rangle_{\rho} &= \langle \, \sum_{\vchi} \psi_{\vchi}(\vg) a_{\vchi} \, \rangle_\rho = \langle \, \psi_{\vchi_0}(\vg)a_{\vchi_0} + \sum_{\vchi \neq \vchi_0} \psi_{\vchi}(\vg) a_{\vchi} \,\rangle_\rho \\
&\overset{\mu \rightarrow v_0}{\longrightarrow} \langle \, \psi_{\vchi_0}(\vg) \sqrt{\mathbf{N}_0} \, \rangle_{\text{cond}} = \sqrt{\mathbf{N}_0} \,\psi_{\vchi_0}(\vg) 
\end{align}
where $\ket{\text{cond}} \approx \ket{\mathbf{N}_0,0,...} = \ket{\psi_{\vchi_0}}^{\otimes \mathbf{N}_0}$ is the condensate state, and we have used the Bogoliubov approximation\footnote{An intuitive reasoning behind the Bogoliubov approximation is as follows. The action of say the annihilation operator on the condensate state is, $a_{\vchi_0}\ket{\mathbf{N}_0,...} = \sqrt{\mathbf{N}_0}\ket{\mathbf{N}_0-1,...}$. For a large enough system we have that $\mathbf{N}_0 \gg 1$, thus giving $a_{\vchi_0}\ket{\mathbf{N}_0,...} \approx \sqrt{\mathbf{N}_0}\ket{\mathbf{N}_0,...}$. Therefore the action of the ladder operators is simply to multiply the state by $\sqrt{\mathbf{N}_0}$. Hence, in this case, the operators can themselves be approximated by $\sqrt{\mathbf{N}_0}I$.} \cite{stringaribook} wherein the ladder operators of the ground state mode $a_{{\vchi}_0} \sim \sqrt{\mathbf{N}_0}$ are of the order of the size of the condensed part. $\psi_{\vchi_0}$ is the so-called condensate wavefunction. \
 
The single-particle state $\ket{\vchi_0}$ characterising the condensate corresponds to a set of $SU(2)$ spin labels, encoding the ground state data of the chosen quantum volume operator. A special class of such condensates is then for the choice of isotropic nodes of and a ground state corresponding to a minimum spin $j_0 = 1/2$. Then, the above is a mechanism, which is rooted purely in the quantum statistical mechanics of group field quanta, for the emergence of a spin-1/2 phase. This configuration has been identified and used often as the relevant sector in LQG for loop quantum cosmology, and also in GFT condensate cosmology.  \

Finally, we note that given the nice properties of the operator $V$ that mimics the Hamiltonian of a system of non-interacting bosons in a box \cite{stringaribook}, the result that this system condenses to the single-particle ground state is not surprising in retrospect. Still, this simple example illustrates the potential of considering collective, statistical features that are inherent in the perspective that spacetime has a fundamental microstructure consisting of discrete quantum gravity degrees of freedom. It also illustrates the usefulness of the GFT perspective of a many-body discrete quantum spacetime, and the consequent reformulation of spin network degrees of freedom in a Fock space. \

For completeness, below we make a few specific remarks, in relation to similar results of obtaining spin-1/2 phases in the setting of GFT cosmology \cite{Gielen:2016uft,Pithis:2016wzf}. These studies \cite{Gielen:2016uft,Pithis:2016wzf} are carried out at a mean field level, in terms of an effective collective variable (condensate wavefunction $\sigma$) of the underlying theory. While, our analysis as shown above \cite{Kotecha:2018gof} is directly at the level of the microscopic statistical theory, and not at an effective or approximate level. Further, the condensate state here is derived to be $\ket{\psi_{\vchi_0}}^{\otimes \mathbf{N}_0}$, and not assumed to be a coherent state from the start, which is a standard choice made in condensate cosmology (thus, also in \cite{Gielen:2016uft,Pithis:2016wzf}). In fact, the starting point here is a maximally mixed state $\rho$. Moreover, the result here is more robust, with a low-spin phase arising as a universal model-independent feature of a class of thermal states characterised by extensive operators. Also, there is no restriction to isotropic nodes, unlike in the aforementioned studies in condensate cosmology. Finally, the low-spin phase is shown here to emerge already for only geometric degrees of freedom, without any coupling to scalar matter field. This is true also for the study in \cite{Pithis:2016wzf}, which however, as pointed out above, is still at the mean field level and restricted to isotropic configurations, along with the choice of a specific class of models. We contrast this with the study in \cite{Gielen:2016uft} wherein a spin-1/2 phase is shown to emerge from the particular wavefunction solution $\sigma(\phi)$, in an asymptotic regime of relational evolution $\phi \rightarrow \pm \infty$. Due to this crucial reliance on the inclusion of scalar matter field $\phi \in \mathbb{R}$ in the GFT base manifold, it would seem that the consequent analysis is also restricted to models coupled to matter.


\subsection{Momentum operators}  \label{momgibbs}

In this section, we turn our attention to the KMS condition and Gibbs states. As first examples in group field theory, we consider the relatively simple case of translation automorphisms along the base manifold, here $G^d \times \mathbb{R}^n$. The automorphisms, and their unitary representations $U(\vphi)$ and $U(\vg)$ on $\mathcal{H}_F$, have been defined earlier in section \ref{AUT}. \

Let us begin with an important remark, which is independent of GFT, about the relation between Gibbs states and the KMS condition. In algebraic quantum statistical mechanics for finite-sized matter systems on spacetime, it is known that the unique normal KMS states, with respect to a given well-defined automorphism group, are Gibbs states (see Remark 1 in appendix \ref{appkmsgibb}) \cite{Haag:1992hx,Emch1980,Strocchi:2008gsa}. Along similar lines in appendix \ref{appkmsgibb}, we have shown that: given an algebraic system, described by a concrete C*-algebra $\mfA$ which is irreducible on a Hilbert space $\mcH_{\mfA}$, equipped with a strongly continuous 1-parameter group of unitary transformations $U(t) = e^{i\mcG t}$, then the unique normal KMS state over $\mfA$, with respect to $U(t)$ at value $\beta$, is of the Gibbs form $e^{-\beta \mcG}$. It further follows (see the Corollary in appendix \ref{appkmsgibb}) that this conclusion, of uniqueness of Gibbs states for a given automorphism group, also holds for an irreducible representation of an abstract C*-algebra. \

Since the above result relies solely on certain algebraic structures, we observe that it can also be applied to the GFT system at hand  \cite{Kotecha:2018gof}. Since the Fock representation $\pi_F$ is irreducible on the GFT Weyl algebra $\mathcal{A}$, we have that: given an automorphism group $\alpha_t$, with its strongly continuous unitary representation $U(t)$ on $\mathcal{H}_F$, then the unique normal KMS state on $\pi_F(\mathcal{A})$, with respect to $\alpha_t$, is a Gibbs state characterised by the generator of the transformation. This extends naturally to the algebra $\mathcal{B}(\mcH_F)$ of bounded linear operators on the GFT Fock space. Then, such states provide an explicit realisation of KMS states in the present quantum gravitational system.\footnote{Such states also amount to a concrete realisation of the thermal time hypothesis \cite{Rovelli:1993ys,Connes:1994hv}, since they may be understood as defining implicitly a notion of time, when the corresponding automorphism implements physical evolution. However, much remains to be done in order to elucidate and analyse in detail their physical meaning and potential applications in discrete quantum gravity.}


\subsubsection{Internal translations} \label{inttrans}

Let $\mathbb{G}$ be a connected Lie group. Recall that any connected Lie group is path-connected because as a smooth manifold it is locally-path-connected \cite{fecko2006}. Thus any two points on $\mathbb{G}$ can be connected by a continuous curve.\footnote{Notice that the groups relevant in GFT, namely $SL(2,\mathbb{C}), Spin(4), SU(2)$ and $\mathbb{R}$, are all connected and simply connected, so that their direct product groups are also connected spaces.} The natural curves to consider on any Lie group are the 1-parameter groups generated via the exponential map. Then, let 
\be g_X(t) = \exp(tX) \,,  \quad X \in \mathfrak{G}, \forall t \in \mathbb{R} \ee
be a 1-parameter subgroup in $\mathbb{G}$, such that 
\be g_X(0) = e \,, \quad \left. \frac{dg_X}{dt}\right|_{t=0} = X \ee 
where $e \in \mathbb{G}$ is the identity, and $\mathfrak{G} = \mathcal{T}_e \mathbb{G}$ is the Lie algebra of $\mathbb{G}$. The generators of generic left translation flows, $g_X(t,g_0)=e^{tX}g_0 = L_{e^{tX}}g_0$, are the right-invariant vector fields, $\mathfrak{X}$. The set of all such vector fields is isomorphic to the Lie algebra by right translations $R_g$ on $\mathbb{G}$, that is 
\be \{R_{g*}X \;|\; X \in \mathfrak{G}, g\in \mathbb{G}\} = \{\mathfrak{X}(g)\} \,.\ee  

\noindent The map \be g_X : \mathbb{R} \to \mathbb{G},\, t \mapsto g_X(t) \ee is a continuous group homomorphism, preserving additivity of the reals, i.e. \be g_X(t_1)g_X(t_2) = g_X(t_1 + t_2)\,.\ee 
Now, let \be U: \mathbb{G} \to \mathcal{U}(\mathfrak{H}) \ee be a strongly continuous unitary representation of $\mathbb{G}$ in a Hilbert space $\mathfrak{H}$, where $\mathcal{U}(\mathfrak{H})$ is the group of unitary operators on $\mathfrak{H}$. Then, given the standard setup above \cite{fecko2006}, the map 
\be U_X := U\circ g_X : t \mapsto U(g_X(t)) \ee is a strongly continuous 1-parameter group of unitary operators in $\mathcal{U}(\mathfrak{H})$. The group property is straightforward to see from, 
\be U(g_X(t_1)) U(g_X(t_2)) = U(g_X(t_1)g_X(t_2)) = U(g_X(t_1+t_2)) \ee 
so that in terms of $U_X$, we have the expected form, 
\be U_X(t_1)U_X(t_2) = U_X(t_1 + t_2) \,.\ee 
See appendix \ref{appcontgrp} for proof of continuity \cite{Kotecha:2018gof}. Applying Stone's theorem to this strongly continuous group of unitary operators leads to the existence of a self-adjoint (not necessarily bounded) generator $\mathcal{G}_X$ defined on $\mathfrak{H}$, such that
\be \label{uxtaut}
U_X(t) = e^{-i\mathcal{G}_X t} \,.
\ee
By construction, operator $\mathcal{G}_X$ implements infinitesimal translations of quantum states in $\mathfrak{H}$, along the direction of the integral flow of $\mathfrak{X}$. It is thus understood as a momentum operator. \

Then, given the 1-parameter group of unitary transformations $U_X(t)$ in \eqref{uxtaut}, equilibrium states on $\mathfrak{H}$ are states that satisfy the KMS condition with respect to $U_X(t)$. Further, if $\mathcal{G}_X$ has a discrete spectrum $x_i$ (e.g. on compact $\mathbb{G}$ or on a compact subspace of locally-compact $\mathbb{G}$) such that $\sum_{i} e^{-\beta x_i}$ converges, and if the algebra under consideration is irreducible on $\mathfrak{H}$, then as detailed in appendix \ref{appkmsgibb}, this KMS state must be of the following Gibbs form, 
\be \label{GibbsGFT}
\rho_{X} = \frac{1}{Z} e^{-\beta \mathcal{G}_X}  
\ee
where $\beta$ is the periodicity in the flow parameter $t$. Notice that $\rho_{X}$ is characterised by both the periodicity $\beta$ and the algebra vector $X$. Therefore, the corresponding notion of equilibrium has an intrinsic dependence on the curve used to define it. \

Further, the construction above holds independently of whether $\mathbb{G}$ is abelian, or not. The detailed Lie algebra structure determines whether the system retains its equilibrium properties on the entire $\mathbb{G}$, or not. In other words, it determines whether the system is stable under arbitrary translation perturbations. The state $\rho_X$, as defined by the curve $g_X(t)$, remains invariant under translations to anywhere on $\mathbb{G}$ if and only if $\mathbb{G}$ is abelian. Otherwise, the system is at equilibrium \emph{only} along the curve which defines it. To see this, let us perturb a system at identity $e$ in state $\rho_X$, so that it leaves its defining trajectory $g_X(t)$, and reaches another point $h \in \mathbb{G}$ which is not on $g_X(t)$, i.e. $h \notin \{g_X(t)\;|\;t \in \mathbb{R}\}$. Since $\mathbb{G}$ is connected, any element of it can in general be written as a product of exponentials, i.e. 
\be h = \exp{Y_1}...\exp{Y_\kappa} \ee for some finite $\kappa \in \mathbb{N}$, and $Y_1,..,Y_\kappa \in \mathfrak{G}$. Given the unitary representation $U$, left translation by $h$ is implemented in the Hilbert space as,
\be U(h) = U(\exp{Y_1})...U(\exp{Y_\kappa}) = \exp(U_*(Y_1))...\exp(U_*(Y_\kappa)) \ee
where, $U_*$ is an anti-hermitian representation\footnote{For a finite-dimensional $\mathfrak{H}$, we recall that \cite{fecko2006}: a representation $\mathcal{D}$ of a Lie group $G$ on $\mathfrak{H}$, induces a unique representation $\mathcal{D}_*$ of the Lie algebra $\mathcal{T}_e G$ on $\mathfrak{H}$, such that $ \mathcal{D}(\exp X) = \exp (\mathcal{D}_* X)$ (for any $X \in \mathcal{T}_e G$). In fact, $\mathcal{D}_*$ is a differential map (or, push-forward) corresponding to the Lie group homomorphism $\mathcal{D} : G\to GL(\mathfrak{H})$, where $GL(\mathfrak{H})$ is the general linear group of all invertible linear operators on $\mathfrak{H}$. Then, for a unitary representation $U$, the derived representation $U_*$ is anti-hermitian, i.e. $U_*(X)^\dag = -U_*(X)$, as expected. While for an infinite-dimensional complex Hilbert space $\mathfrak{H}$, there are more subtleties involved (as expected) but similar structures can be defined as long as $U$ is a strongly continuous representation of $G$. Specifically, an anti-hermitian algebra representation $U_*$ is now defined on a dense subspace of $\mathfrak{H}$, the so-called G{\aa}rding domain \cite{moretti}. We do not delve into these details further, mainly because we do not expect them to qualitatively change our result in equation \eqref{QUAL}, and also because these details are outside our scope presently.} of $\mathfrak{G}$. This acts on the density operator as,
\be
U(h)^{-1} \rho_X U(h) = e^{-U_*(Y_\kappa)}...e^{-U_*(Y_1)} \; e^{- i \beta U_*(X)} \; e^{U_*(Y_1)}...e^{U_*(Y_\kappa)} 
\ee
where, $\mathcal{G}_X = iU_*(X)$. For non-abelian $\mathbb{G}$, clearly $[X,Y] \neq 0$ for arbitrary $X,Y \in \mathfrak{G}$. Thus, we have 
\be \label{QUAL} U(h)^{-1} \rho_X U(h) \neq \rho_X \ee 
in the general, non-abelian case. While for abelian $\mathbb{G}$, all Lie brackets are zero and equality will hold for arbitrary $h$, thus leaving the state invariant. So overall, the notion of equilibrium that we have defined here, with respect to translations on $\mathbb{G}$, is curve-wise or direction-wise, with the ``direction'' being defined on each point of the manifold by the vector field $\mathfrak{X}$ (uniquely associated with a given $X \in \mathfrak{G}$). For non-abelian $\mathbb{G}$, an equilibrium state can be defined only \emph{along} a particular direction\footnote{This is also the case, for example, for the Unruh effect treated via the Bisognano-Wichmann construction \cite{Bisognano:1976za}. In that case, the symmetry group $\mathbb{G}$ is the Lorentz group, acting on the base manifold which is a Rindler wedge of the Minkowski spacetime, and the KMS state of an accelerated observer depends on the specific trajectory generated by a 1-parameter flow of boosts (in say $x^1$ direction) taking the form $\Lambda_{k_1}(t) = e^{t a k_1}$, where $k_1$ is the boost generator, and acceleration $a$ parametrises the strength of this boost. Another example from this thesis, is the classical Gibbs distribution with respect to the closure condition, in section \ref{entclosure}.}. \

Applying this to the GFT system, we take: Hilbert space $\mathfrak{H}$ to be the Fock space $\mathcal{H}_F$; Lie group $\mathbb{G}$ to be, either $\mathbb{R}^n$ for internal scalar field translations, or $G^d$ for group left translations, both acting on the base manifold $G^d \times \mathbb{R}^n$ as detailed in sections \ref{rntranS} and \ref{gdtranS} respectively; and, the strongly continuous unitary groups $U(g)$ are those constructed in section \ref{unis}, which implement the translation automorphisms of the GFT Weyl algebra in $\mathcal{H}_F$. The form of the generators on the Fock space is,
\begin{equation} \label{fockgen}
\mathcal{G}_X := i\int_{G^d \times \mathbb{R}^n} d\vg\,d\vphi \; \varphi^\dag(\vg,\vphi) \mathcal{L}_{\mathfrak{X}}\varphi (\vg,\vphi) 
\end{equation}
which is motivated from momentum operators for spatial translations in Fock representations of scalar field theories on spacetime. Here, $\mathfrak{X}$ is the right-invariant vector field on $\mathbb{G}$ corresponding to the vector $X \in \mathfrak{G}$, and related to it by right translations as, $\mathfrak{X}(g) = R_{g*}X$ (for $g \in \mathbb{G}$); and, $\mathcal{L}_{\mathfrak{X}}$ denotes the Lie derivative with respect to the vector field $\mathfrak{X}$. Then, equilibrium states for the GFT system are KMS states with respect to the automorphism group \eqref{uxtaut}, where the generator is given by \eqref{fockgen}. Further, this state will be of the Gibbs form in \eqref{GibbsGFT}, with generator \eqref{fockgen} (which is suitably regularised when required, see below for an example).


\subsubsection*{Equilibrium in internal $\phi$-translations}

As a specific example of the momentum Gibbs states considered above, we now present those states that are in equilibrium with respect to flows on $\mathbb{R}^n$ part of the base space \cite{Kotecha:2018gof}. These are of particular interest from a physical perspective. First, we anticipate that in light of the interpretation of $\vphi \equiv (\phi_1,...,\phi_a,...,\phi_n) \in \mathbb{R}^n$ as $n$ number of minimally coupled scalar fields \cite{Li:2017uao,Oriti:2016qtz}, the corresponding momenta that generate these internal translations are the scalar field momenta, so there is an immediate meaning to the variables. Second, and more important, the scalar values $\phi$ can be used as relational clocks, as in GFT cosmology \cite{Pithis:2019tvp,Oriti:2016acw,Gielen:2016dss,Oriti:2016qtz}, thus their translations can be related directly to dynamical evolution after deparametrization. This is also the reason why we call the $\vphi$-translations here as being \textit{internal}, to contrast with the case where the system is deparametrized with respect to one of these scalar fields, so that the resulting translations along this clock variable then become external to the system (thus defining relational evolution, see section \ref{depgft}), and with respect to which then physical clock equilibrium can be defined (as done below in section \ref{physeqm}). The momentum of the clock scalar field defined within the reduced, deparametrized system then also gives the clock Hamiltonian. \

The basis of invariant vector fields on $\mathbb{G} = \mathbb{R}^n$ is $\{\frac{\partial \,\;\;}{\partial \phi^a}\}$ in cartesian coordinates $(\phi^a)$. These are generated by the set of basis vectors of the Lie algebra $\{E_a\}$. The full set of invariant vector fields is then generated by linearity. For a generic tangent vector, $X = \sum \limits_{a=1}^n \lambda^a E_a$, the corresponding invariant vector field is $\mathfrak{X} = \sum \limits_{a=1}^n \lambda^a \partial_a$. Then directly for the basis elements, generators \eqref{fockgen} take the simple, familiar form,
\begin{align}
 \mathcal{G}_{E_a} &= i \int d\vg\,d\vphi \; \varphi^\dag(\vg,\vphi) \frac{\partial}{\partial \phi^a} \varphi (\vg,\vphi) \\
 &= \sum_{\vchi} \int \frac{d\vp}{(2\pi)^n}  \; p_a \, \varphi^\dag(\vchi,\vp) \, \varphi(\vchi,\vp) 
\end{align}
where $p_a \in \mathbb{R}$, and we have used the spin-momentum basis introduced in section \ref{spinmom} for the second equality.\footnote{We remark that the operators $\mathcal{G}_{E_a}$ as constructed above are the same as those used in the GFT cosmology framework, and introduced in \cite{Oriti:2016qtz} for the case $a=1$.} Then, infinitesimal translations of the operators are generated in the expected way, e.g.
\be \partial_{\phi^a} \varphi (\vg,\vphi) = i[ \mathcal{G}_{E_a}, \varphi (\vg,\vphi)] \,. \ee

\noindent Notice that operators $\mathcal{G}_{E_a}$ have a continuous and unbounded spectrum, as expected for any quantum mechanical momentum operator \cite{Woit:2017vqo,Gieres_2000,moretti}. Then in order to consider well-defined quantities, we naturally need to consider suitable IR and UV regularisations. As per standard practice \cite{Woit:2017vqo,Gieres_2000,dimock_2011}, let us: put our system in a finite box of size $L^n$ in $\mathbb{R}^n$, with periodic boundary conditions, which results in a discrete momentum spectrum,\footnote{Recall that, from periodicity in momentum eigenfunctions, we have \cite{Woit:2017vqo,Gieres_2000,dimock_2011}: $e^{i p \phi} = e^{i p (\phi + L)} \Rightarrow e^{ipL} = 1 \Rightarrow p = \frac{2\pi}{L}M$ (where $M \in \mathbb{Z}$).} i.e. for each $a$, we have $p_a = \frac{2\pi}{L}M_a$, where $M_a \in \mathbb{Z}$; and, put a high momentum cut-off, which renders the spectrum bounded from above, i.e. for each $a$, $\exists$ a finite $M_{a0}$, such that $p_{a0} = \frac{2\pi}{L}M_{a0} = \max (p_a)$. Denoting the resultant regularised operator by $P_a$, we have
\be \label{phimom}
P_a = \sum_{\vchi,\vp}  \; p_a \, \varphi^\dag(\vchi,\vp) \, \varphi(\vchi,\vp)
\ee
which is diagonal in the occupation number basis,
\be
P_a\ket{\{n_{\vchi_i,\,\vp_i}\}} = \left( \sum_{i'} p_{a,{i'}} \,n_{\vchi_{i'},\,\vp_{i'}} \right) \ket{\{n_{\vchi_i,\,\vp_i}\}}
\ee
where $p_a \in \frac{2\pi}{L}\mathbb{Z}$, and $p_{a,{i}}$ is the $a^{\text{th}}$ component of the ${i}^{\text{ th}}$ mode, i.e. $\vp_{i} = (p_1,...,p_a,...,p_n)_{i}$. Then, operators
\be \label{RHOMOM} \rho_a = \frac{1}{Z} e^{-\beta (P_a - \mu N)} \ee
are generalised Gibbs states for $-\infty < \beta < 0$ and $\mu > p_{a0}$. Here, $\beta$ is negative because $P_a$ is bounded from above (see the Remark in appendix \ref{posapp2}). Given these $P_a$ along with the above ranges for $\beta$ and $\mu$, the proofs for boundedness, positivity and normalisation of $\rho_a$ proceed in analogy with those in appendix \ref{posapp2}, to which we refer for details. Finally, we note that $\rho_a$ as defined above, would naturally depend on the IR and UV cut-off parameters. Then, like in standard quantum statistical mechanics, one would consider the infinite volume and momentum limits to check the independence (or, specific dependences) of quantities of interest, e.g. correlation functions, on boundary conditions and regularisation. We leave further investigation of these states to future work.


\subsubsection{Clock evolution} \label{physeqm}

The states defined above, generated by momenta \eqref{phimom}, encode equilibrium with respect to \emph{internal} $\phi_a$-translations. These flows are structural, and so are the resultant equilibrium states, being devoid of any physical model-dependent information, in particular of a specific choice of dynamics. Below we present states, based on discussions in section \ref{depgft}, that are at equilibrium with respect to a clock Hamiltonian encoding relational dynamics, wherein the $\phi$-translations take on the role of an \emph{external} clock evolution \cite{Kotecha:2018gof}.

Recall from sections \ref{depggs} and \ref{depgft} that obtaining a good canonical structure in terms of a relational clock, i.e. deparametrization, amounts to the approximation
\begin{equation}
C_{\f} \overset{\text{deparam.}}{\longrightarrow} C = p_t + H_N 
\end{equation}
for a constrained classical system (see section \ref{classdep}). Then, the constraint surface post-deparametrization is $\Sigma = \Gamma_{\ex}|_{C=0} = \mathbb{R} \times \Gamma_{\can,N}$, which has the characteristic structure of foliation in clock time $t \in \mathbb{R}$. Clock Hamiltonian is $H_N$, which is a smooth function on $\Gamma_{\can,N}$. Then, relational Gibbs distributions can in principle be defined on the canonical phase space $\Gamma_{\can,N}$, taking the standard form
\be 
\rho_{\can} = \frac{1}{Z_{\can}} e^{-\beta H_N} \ee
where $\beta \in \mathbb{R}$ and $H_N$ are assumed to be such that $Z_{\can}$ converges. This state is at equilibrium with respect to the flow $\mcY_{H_N}$ that is parametrized by the clock time $t$. Equivalently, the state 
\be \rho = \frac{1}{Z} e^{-\beta H_{\phy}} \,, \quad \pi^* H_{\phy} = H_N \ee 
is a Gibbs state on the reduced, physical phase space $\Gamma_{\phy,N}$, where $\pi$ is the projection from constraint surface $\Sigma$ to $\Gamma_{\phy,N}$. This state is naturally at equilibrium with respect to the flow generated by $\mcY_{H_{\phy}}$. Notice that this flow on $\Gamma_{\phy,N}$ is determined by the Hamiltonian flow on $\Sigma$, that is $ \mcY_{H_{\phy}} = -\pi_*(\partial_t) $,
using $\pi_*(\mcY_C) = 0$ and $\mcY_C = \partial_t + \mcY_{H_N}$. \

In the corresponding quantum system outlined in section \ref{quantise}, formal constructions of the corresponding Gibbs density operators follow directly. A relational Gibbs state is a density operator on $\mathcal{H}_{\can,N}$ of the form $\widehat{\rho}_{\can} \propto e^{-\beta \hat{H}_N}$, which can further be extended to the Fock space. Finally we remark that, heuristically we expect these relational states to be obtained from some reduction of the states defined in section \ref{inttrans} above with respect to internal $\phi$-translations, through the impositions of the dynamical constraint of the theory (i.e. being on-shell with respect to the given dynamics), and of the deparametrizing approximations on the same constraint. We leave detailed investigations of a rigorous link between these two classes of states to future work.


\subsection{Constraint functions} \label{congibb}

In this section, we turn briefly to some explorations in a classical setting. From traditional statistical mechanics, we know that an important difference between canonical and microcanonical distributions over a phase space, is that the former includes also statistical fluctuations around a constant energy surface. In other words, a microcanonical distribution is characterised completely by a single energy shell, $H=E$, while a canonical distribution is characterised by a constant \emph{average} energy, $\braket{H} = E$. Therefore, in a canonical (or a grand-canonical) state, the same amount of internal energy may be distributed over many more microstates. Further, canonical (and grand-canonical) states are often technically easier to handle, compared to the microcanonical one, to study physical properties of a system. \

Motivated by this, one might inquire if statistical distributions may be utilised to study the physics of a constrained system more conveniently, in an approximate or effective way, by considering a weaker constraint equation $\braket{C}=0$, instead of the exact one $C=0$ that specifies the presymplectic constraint surface. Then, a generalised Gibbs distribution with respect to a finite set of scalar constraint functions $\{C_a\}_{ a=1,2,...,k}$ on an extended phase space $\Gamma_\ex$, would encode statistical fluctuations around the constraint surface. In this sense, the constraint information might be encoded partially or weakly in a statistical distribution that is defined on the full unconstrained extended state space. \

This might be interesting to consider also because in a system with many degrees of freedom, observable averages correspond to statistical averages in generic mixed quantum states. However, in a constrained system it may be difficult to precisely find the physical state space, i.e. the space of all gauge-invariant states. In this case, we could consider instead states that satisfy the dynamical constraint only effectively, by fulfilling the weaker condition $\braket{C} = 0$, instead of the exact one; and subsequently evaluate observable averages in this approximately physical state. Then, from the maximum entropy principle, one is naturally led to a state of the form \eqref{gibbsconst} below. \

We say that a system satisfies a set of constraints weakly if $\langle C_a \rangle_\rho = U_a$, for some statistical state $\rho$ on an extended phase space $\Gamma_\ex$, and real constants $U_a$. Then by the thermodynamical characterisation (section \ref{maxent}), the state $\rho$ is given by
\be \label{gibbsconst} \rho_{\{\beta_a\}} = \frac{1}{Z_{\{\beta_a\}}} e^{-\sum \limits_{a=1}^k \beta_a C_a} \,. \ee

On the other hand, we say that a system satisfies a set of constraints strongly if $C_a = U_a$. This identifies the constraint submanifold $\Sigma =  \Gamma_\ex|_{\{C_a = U_a\}}  \subset \Gamma_\ex$. From a statistical viewpoint, this same condition can be written in terms of a microcanonical density on $\Gamma_\ex$, 
\be \label{micro}
\rho_{\{U_a\}} = \frac{1}{Z_{\{U_a\}}} \prod \limits_{a=1}^k \delta(C_a-U_a)  \ee
where $Z_{\{U_a\}} = \text{vol}(\Sigma)$.\footnote{Notice that the factorisation property of the state (equivalently, of the full partition function) is a consequence of the fact that we are considering a system of constraints each defined independently on $\Gamma_\ex$.}
Then for a fixed $b$, we have
\be
C_b = U_b \;\;\Leftrightarrow \;\; \langle C_b \rangle_{\rho_{\{U_a\}}} = \int_{\Gamma_\ex} d\lambda \; C_b \, \rho_{\{U_a\}} = U_b \;.
\ee

Therefore, the effective and exact impositions of constraints can be seen from a statistical perspective, as respectively either defining a Gibbs state of the form \eqref{gibbsconst}, or a microcanonical state of the form \eqref{micro}.\footnote{Like in standard statistical mechanics, these two states, $\rho_{\{\beta_a\}}$ and $\rho_{\{U_a\}}$, are related by a Laplace transform between their respective partition functions, with their scalar component coefficients, $\beta_a$ and $U_a$, being the conjugate variables under the transform, i.e. $(\mathcal{L}Z_U)(\beta) = \tilde{Z}(\beta) \equiv Z_\beta$.} In the following, we consider examples of the general form \eqref{gibbsconst}, in a system of classical tetrahedra.


\subsubsection{Closure condition} \label{entclosure}

As a first example, we consider the relatively simple case of the closure constraint for a single classical tetrahedron\footnote{The case of a classical $d$-polyhedron (section \ref{atomkin}) and its associated closure condition can be treated in a completely analogous manner to this one.} \cite{Chirco:2018fns,Chirco:2019kez}. Recall that the symplectic phase space of intrinsic geometries of a convex tetrahedron is given by the 2-dim Kapovich-Millson phase space \cite{kapovich1996,Barbieri:1997ks,Baez:1999tk},
\be \mathcal{S}_4 = \{(X_I) \in \su(2)^{*4} \cong \mathbb{R}^{3\times 4} \;|\; ||X_I|| = A_I \,, \sum_{I=1}^4 X_I=0 \}/SU(2)  \ee
where $\su(2)^*$ is the dual space of the Lie algebra $\su(2)$; and, $X_I$ are the face normals of the four triangles in $\mathbb{R}^3$, with fixed areas $A_I$ (for $I=1,2,3,4$). In particular, notice that the four co-vectors $X_I$ (and the surfaces associated to them, as orthogonal to each of them) close, that is
\be \label{STRCLO} \sum_{I=1}^4 X_I=0 \ee
thus giving a closed convex tetrahedron in $\mathbb{R}^3$, modulo rotations. This is the closure constraint, which allows us to understand geometrically a set of 3d vectors as the normal vectors to the faces of a tetrahedron, and thus to fully capture its intrinsic geometry in terms of them (see figure \ref{convex}) \cite{kapovich1996,Barbieri:1997ks,Baez:1999tk,Conrady:2009px}. Now along the lines described above, we are interested in imposing this closure constraint effectively (on average) via a Gibbs distribution defined on an extended phase space. Then from a statistical perspective, we can interpret the strong fulfilment of closure (given above in \eqref{STRCLO}) as defining a microcanonical state with respect to this constraint, and therefore a generalised Gibbs state as encoding a weak fulfilment of the same constraint. Below we show that such a state can technically indeed be defined, but further investigation of any possible physical consequences of such a definition, in discrete gravity, is left to future work. \

Thus, let us remove the closure condition \eqref{STRCLO} from $\mathcal{S}_4$, to define the phase space for an \emph{open} tetrahedron,
\begin{align} \Gamma_{\{A_I\}} &= \{(X_I) \in \su(2)^{*4} \cong \mathbb{R}^{3\times 4} \;|\;  ||X_I|| = A_I\}   \\ &\cong S_{A_1}^2 \times ... \times S^2_{A_4}  \end{align}
where each $S^2_{A_I}$ is a 2-sphere
with radius $A_I \in \mathbb{R}_{>0}$. This is the extended phase space of interest, with respect to the closure constraint. It is the space of sets of four oriented triangles in $\mathbb{R}^3$, with face normals $X_I$ (and areas $A_I$) that are not constrained to close; and, $SU(2)$ acts on it diagonally by rotation\footnote{This is the adjoint action of $SU(2)$ on $\Gamma_{\{A_I\}}$: $SU(2) \times \Gamma_{\{A_I\}} \ni g, (X_I) \mapsto (gX_I g^{-1}) \in \Gamma_{\{A_I\}}$.} \cite{Conrady:2009px}. \

On the extended phase space $\Gamma_{\{A_I\}}$, let us write the closure condition as a smooth map\footnote{In more mathematically inclined literature on classical mechanics, this is known as a moment or momentum map; it is a generalisation of generators, of Hamiltonian actions on a phase space, to the case of generic non-abelian Lie groups \cite{souriau1,e18100370,fecko2006}. In the present case, $J$ generates a diagonal action of $SU(2)$ on $\Gamma_{\{A_I\}}$, i.e. simultaneous rotations of the spheres $S^2_{A_I}$ \cite{Conrady:2009px,e18100370}.} $J : \Gamma_{\{A_I\}}\to \su(2)^*$, defined by
\be \label{mom}
J(m) := \sum_{I=1}^4 X_I \,, \quad ||X_I|| = A_I 
\ee
where $m = (X_1,...,X_4) \in \Gamma_{\{A_I\}}$ denotes a point on the extended phase space manifold. We note that the symplectic reduction of $\Gamma_{\{A_I\}}$ with respect to the zero level set, $J=0$, gives back the Kapovich-Millson phase space $\mathcal{S}_4 = \Sigma/SU(2)$, where $\Sigma = J^{-1}(0) = \{m \in \Gamma_{\{A_I\}} \,| \, J(m) = 0\}$ is the constraint surface \cite{Conrady:2009px,kapovich1996}.  \

Then, a generalised Gibbs state with respect to closure for an open tetrahedron can be defined by: maximising the entropy functional \eqref{en} under normalisation \eqref{constnorm}, and the following $\su(2)^*$-valued constraint,
\be \label{jconst}
\langle J \rangle_\rho = \int_{\Gamma_{\{A_I\}}} d\lambda \, \rho J = U
\ee
where $\rho(m)$ is a statistical density on $\Gamma_{\{A_I\}}$, 
and $U \in \su(2)^*$ is a constant. Specifically, it can be defined by optimising the function \eqref{auxfn}, which in the present case takes the form,
\be 
L[\rho,\beta,\kappa] = - \braket{\ln \rho}_\rho - \beta . (\braket{J}_{\rho} - U) - \kappa(\langle 1 \rangle_\rho - 1)
\ee
where now the Lagrange multiplier $\beta \in \su(2)$ for the constraint on $J$ is algebra-valued \cite{e18100370}, and, $b.x$ denotes an inner product between elements $b \in \su(2)$ of the algebra and $x \in \su(2)^*$ of its dual. 
Then, 
\be \frac{\delta L}{\delta \rho} = 0 \quad \Rightarrow \quad \rho = e^{-(\beta . J + 1+\kappa)} \ee
which gives,
\be \label{closuregibbs}
\rho_{\beta} = \frac{1}{Z_{\beta}} e^{-\beta \cdot J} \,.
\ee
The equilibrium partition function is given by, 
\begin{align} Z_{\beta} = e^{1+\kappa} &= \int_{\Gamma_{\{A_I\}}} d\lambda \; e^{-{\beta \cdot J}}  \\
&=  \prod_{I=1}^4 \frac{4\pi A_I}{||\beta||} \sinh (A_I ||\beta||)  \label{closZb} 
\end{align}
for which, the details are provided in appendix \ref{AppClosure}. \

Notice that, $\beta.J : \Gamma_{\{A_I\}} \to \mathbb{R}$ is a smooth real-valued function on the phase space, since the components $\beta_a$ and $J_a$ (with $a=1,2,3$) in any basis of the algebra or its dual, are real-valued, and scalar component functions $J_a(m)$ are smooth. We can further write it as, 
\be J_\beta(m) = \beta.J(m) \ee 
for $m \in \Gamma_{\{A_I\}}$.\footnote{$J_\beta$ is the so-called comomentum map, associated with the momentum map $J$, for any $\beta \in \su(2)$ \cite{e18100370}.} Then, it is evident that $J_\beta$ is a modular Hamiltonian, i.e. $J_\beta = - d \ln \rho_\beta$, with respect to which the state $\rho_\beta$ is at equilibrium (cf. discussions in section \ref{modstat}). As expected, $J_\beta$ defines a vector field $\mathfrak{X}_\beta$ on $\Gamma_{\{A_I\}}$ via the equation,
\be \omega(\mathfrak{X}_\beta) = - d J_\beta \ee
where $\omega$ is the symplectic 2-form on $\Gamma_{\{A_I\}}$. $\mathfrak{X}_\beta$ is the fundamental vector field corresponding to the vector $\beta \in \su(2)$. The state $\rho_\beta$ is at equilibrium with respect to translations along the integral curves of $\mathfrak{X}_\beta$ on the base manifold $\Gamma_{\{A_I\}}$. In other words, $\rho_\beta$ encodes equilibrium with respect to the one-parameter flow characterised by $\beta$, which is a generalised vector-valued temperature \cite{souriau1,e18100370}. As encountered earlier in this thesis too, this is an expected feature of any equilibrium state associated with an action of a non-abelian Lie group (for instance, see section \ref{momgibbs}).\footnote{Another example is the well-known case of accelerated trajectories on Minkowski spacetime, where thermal equilibrium is established along Rindler orbits defined by the boost isometry, where $\beta$ encodes the strength of acceleration and defines the Unruh temperature. } In fact, our construction here \cite{Chirco:2018fns,Chirco:2019kez} is an example of Souriau's generalised Gibbs ensembles \cite{souriau1,e18100370} in a simplicial geometric context, associated with a Lie group (diagonal $SU(2)$) action for a first class (closure) constraint. Applications of the many results and insights from Souriau's generalisation, and the corresponding Lie group extension of thermodynamics \cite{souriau1,e18100370,Chirco:2019tig,melechirco}, in the present simplicial geometric setting are left to future work.


\subsubsection{Gluing conditions} \label{glucondgibb}

We now turn to a system of many classical closed tetrahedra, and consider gluing conditions which constraint a set of disconnected tetrahedra to form an extended simplicial complex \cite{Chirco:2018fns,Kotecha:2019vvn}. The main motivation behind this consideration is the group field theory approach, which advocates for a many-body treatment of building blocks of spacetime, as we have seen repeatedly in this thesis. From this perspective, statistical distributions may naturally give rise to discrete quantum gravity partition functions, with the dynamical information being encoded in the statistical weight (cf. section \ref{effgft}). Here, we consider statistical weights characterised by gluing constraints on the $SU(2)$ data. The resulting distribution would then be a superposition of simplicial complexes, admitting a notion of a certain type of discrete geometry, on average. We note that our treatment below is formal. More work remains to be done for a complete understanding of the resulting distributions, including normalisability and further consequences in discrete quantum gravity \

In the following, the key ingredient is a set of gluing conditions on the state space of many disconnected tetrahedra. As we will see, the same gluing process can be encoded in terms of dual graphs, which is the 1-skeleton of the cellular complex (dual to the simplicial complex) of interest. The geometry of the initial set of tetrahedra, as well as of the resulting simplicial complex, is captured by the $\mcT^*SU(2)$ data introduced earlier in section \ref{atomkin}. We will perform our construction in terms of these data first. A more refined characterisation of the same notion of discrete geometry can be obtained in terms of the so-called twisted geometry decomposition \cite{Freidel:2010aq, Rovelli:2010km}, which we will connect with at a second stage, to suggest further research directions based on our construction.   \\

\noindent \textbf{Setup} \\

\noindent Let $\gamma$ denote an oriented, 4-valent closed graph with $L$ number of oriented links and $N$ number of nodes. Each link $\ell$ is dressed with $\mcT^*SU(2) \cong SU(2) \times \su(2)^* \ni (g_\ell,X_\ell)$ data, with variables satisfying invariance under diagonal $SU(2)$ action at each node $n$, thus satisfying closure. $\gamma$ is dual to a simplicial complex $\gamma^*$, with links dual to triangular faces $\ell$ and nodes dual to tetrahedra $n$. The source and target nodes (tetrahedra) sharing a directed link (face) $\ell$ are denoted by $s(\ell)$ and $t(\ell)$ respectively. A state $\{(g_\ell,X_\ell)\}_{\gamma}$ on the graph $\gamma$ is then an element  of $\Gamma_\gamma = \mcT^*SU(2)^L//SU(2)^N$, where the double quotient denotes symplectic reduction of $\mcT^*SU(2)^L$ with respect to closure at all nodes. Such configurations admit a notion of discrete geometry called twisted geometries \cite{Freidel:2010aq, Rovelli:2010km}, the details of which we will return to below. The geometry so-defined is potentially pathological, in the sense that the resulting simplicial complex may not be fully specified in terms of metric data, that is its associated edge lengths, as a Regge geometry \cite{Regge:2000wu,Regge:1961px} would be. For our purposes, though, this characterisation suffices to show how a statistical state can formally be constructed based on encoding gluing, and possibly other constraints, on the initially disconnected tetrahedra. 

\begin{figure}[t]
\centering
\includegraphics[width=3.5 in]{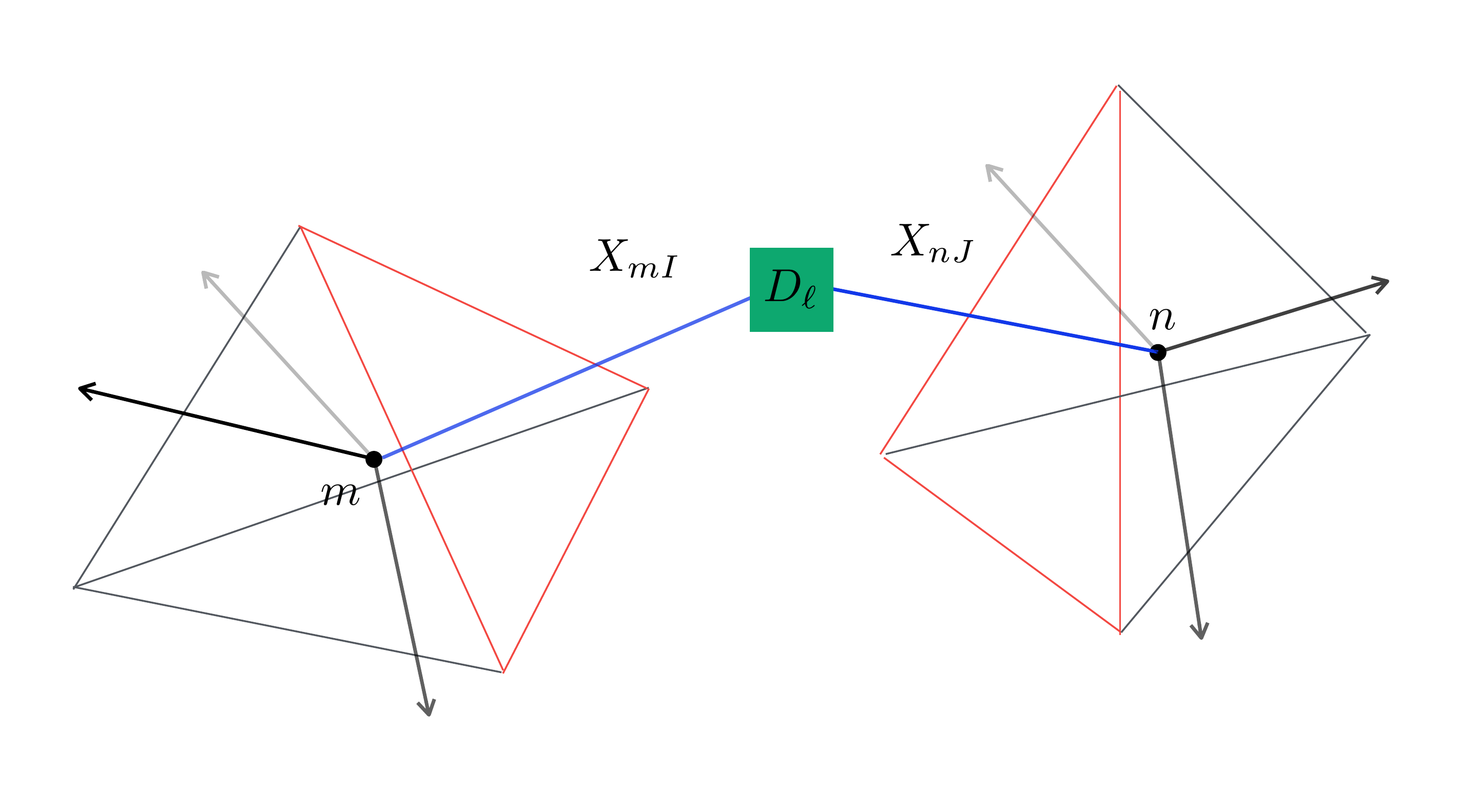}
\caption{Gluing of adjacent faces in neighbouring tetrahedra via constraints on holonomy-flux variables. \cite{Chirco:2018fns}}\label{gluings}
\end{figure}

\begin{figure}
\centering
  \includegraphics[width=3 in]{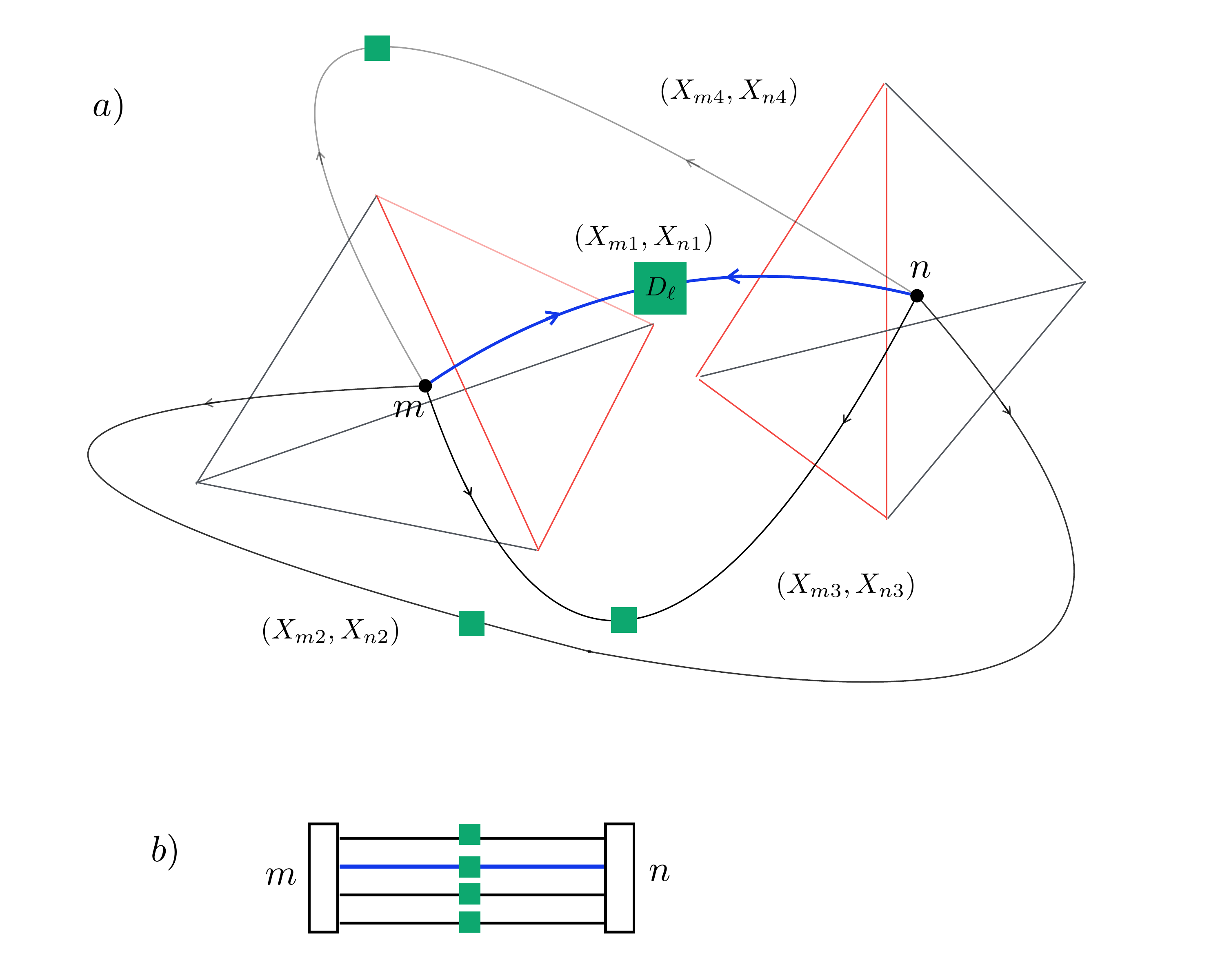}
\caption{$a)$ Dipole gluing in a system of two tetrahedra. $b)$ Combinatorics of the dipole gluing. \cite{Chirco:2018fns}}\label{dip}
\end{figure}

\begin{figure}
\centering
  \includegraphics[width=2.8 in]{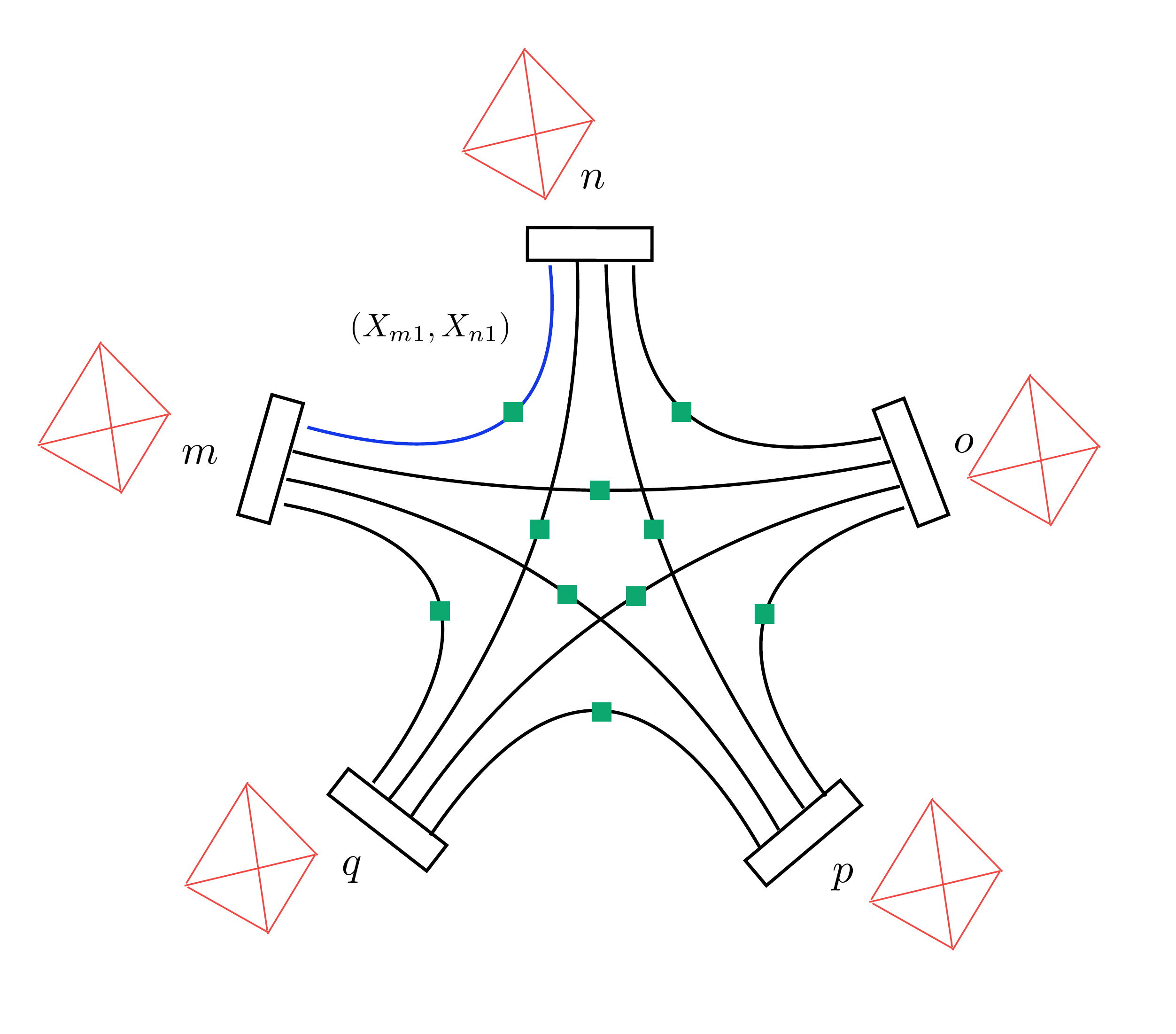}
  \caption{Resultant 4-simplex from combinatorial gluing between faces of five tetrahedra. Equivalently, the complete graph on five 4-valent nodes. \cite{Chirco:2018fns}}
  \label{4sym}
\end{figure}

To understand better the gluing process, and the corresponding constraints, let us begin with a single closed classical tetrahedron $n$. Recall that its state space is given by,
\be \Gamma = \mcT^*(SU(2)^4/SU(2)) \ee
as in equation \eqref{gammatet}. As discussed before in section \ref{atomkin}, $\Gamma$ is the space where 3d rotations have not been factorised out. This essentially means that each such tetrahedron is equipped with an arbitrary (orthonormal) reference frame determining its overall orientation in its $\mathbb{R}^3$ embedding. In the holonomy-flux representation, the four triangular faces $\ell |_n$ of a given tetrahedron $n$, are labelled by the four pairs $(g_\ell, X_\ell)$. In the dual picture, we have a single open graph node $n$, with four half-links $\ell |_n$ incident on it and each labelled by $(g_\ell, X_\ell)$. Each half-link is bounded by two nodes, one of which is the central node $n$, common to all four $\ell$, and the other is a bounding bivalent node. This is a 4-patch labelled with $\mcT^*(SU(2))$ data (see section \ref{atomkin}). For example, below: in Figure \ref{dip} $(a)$, the dipole graph is composed of two such 4-valent nodes, with the bivalent nodes shown as green squares; and in Figure \ref{4sym}, the 4-simplex graph is composed of five 4-valent nodes, with the various bivalent nodes shown as green squares. Each $\ell$ is oriented outward (by choice of convention) from the common 4-valent node $n$, which then is the source node for all four half-links. Then in the holonomy-flux parametrisation, each half-link $\ell$ is labelled by $(g_\ell,X_\ell)$. \

Let us denote the $I^{\text{th}}$ half-link belonging to an open node $n$ by $(nI)$, where $I=1,2,3,4$. Equivalently, $(nI)$ denotes the $I^{\text{th}}$ face of tetrahedron $n$. 
Two tetrahedra $n$ and $m$ are said to be neighbours (see Figure \ref{gluings}) if at least one pair of faces, $(nI)$ and $(mJ)$, are adjacent, i.e the variables assigned to the two faces satisfy the following constraints,
\be \label{gluing} g_{(nI)}g_{(mJ)} = e\,, \quad X_{(nI)}+X_{(mJ)}= 0 \,. \ee 

A given classical state associated to the connected graph $\gamma$ can then be understood as a result of imposing the constraints \eqref{gluing} on pairs of half-links (or, faces) in a system of $N$ open nodes (or, disconnected tetrahedra). That is, $\gamma$ is a result of imposing $L$ number each of $SU(2)$-valued and $\su(2)^*$-valued constraints, which we denote by $C$ and $D$ respectively. This is a total $6L$ number of $\mathbb{R}$-valued constraint component functions $\{C_{\ell,a} \,, D_{\ell,a}\}_\gamma$, for $\ell=1,2,...,L$ and $a=1,2,3$. For instance, creation of a full link $\ell = (nI,mJ)$ involves matching the fluxes, component-wise, by imposing the three constraints $D_{\ell,a} = X^a_{(nI)}+X^a_{(mJ)} = 0$, as well as restricting the conjugate parallel transports to satisfy $C_{\ell,a} = (g_{(nI)}g_{(mJ)})^a - e^a = 0$. Naturally the final combinatorics of $\gamma$ is determined by which half-links are glued pairwise, which is encoded in which specific pairs of such constraints are imposed on the initial data. \

As an example, consider the dipole graph in Figure \ref{dip}. This can be understood as imposing constraints on pairs of half-links of two open 4-valent nodes. Here $L=4$, thus we have at hand four constraints $D_\ell$ on flux variables,
\begin{align} \label{dipcon}
X_{(11)} + X_{(21)} &= 0 \,,	\quad	X_{(12)} + X_{(22)} = 0\,,	\nonumber \\ 
X_{(13)} + X_{(23)} &= 0\,,		\quad	X_{(14)} + X_{(24)} = 0 \,.
\end{align}
This corresponds to a set of $3\times 4$ component constraint equations $D_{\ell,a}=0$. Similarly for holonomy variables. \

As another example, consider a 4-simplex graph made of five 4-valent nodes as shown in Figure \ref{4sym}. The combinatorics is encoded in the choice of pairs of half links that are glued. Here $L=10$, corresponding to ten constraints $D_\ell$ on the flux variables, 
\begin{align} \label{4simp}
X_{(12)}+X_{(21)} &= 0 \,, \quad  X_{(13)}+X_{(31)} = 0 \,, \nonumber \\
X_{(14)}+X_{(41)} &= 0 \,, \quad  X_{(15)}+X_{(51)} = 0 \,, \nonumber \\
X_{(23)}+X_{(32)} &= 0 \,, \quad  X_{(24)}+X_{(42)} = 0 \,, \nonumber \\
X_{(25)}+X_{(52)} &= 0 \,, \quad  X_{(34)}+X_{(43)} = 0 \,, \nonumber \\
X_{(35)}+X_{(53)} &= 0 \,, \quad X_{(45)}+X_{(54)} = 0 \,.
\end{align}
As before, this corresponds to 30 component equations for the flux variables, and another 30 for holonomies.  \\

\noindent \textbf{Statistical mixtures} \\

\noindent When the above constraints are satisfied exactly, that is $\{C_{\ell,a} = 0 \,, D_{\ell,a} = 0\}_\gamma$ for all $\ell, a$, then this system of $N$ tetrahedra admits a twisted geometric interpretation based on the resultant simplicial complex. But as discussed previously, there is a way of imposing these constraints only on average, that is $\{\langle C_{\ell,a} \rangle_\rho = 0 \,, \langle D_{\ell,a} \rangle_\rho = 0\}_\gamma$.

This manner of imposing effective constraints and maximising the Shannon entropy results in a generalised Gibbs state, parametrised by $6L$ number of generalised temperatures, 
\be\label{rhocomp} \rho_{ \{\gamma, \alpha,\beta\}} \propto e^{-\sum \limits_{\ell=1}^{L} \sum \limits_{a=1}^3 (\alpha_{\ell,a} C_{\ell,a} + \beta_{\ell,a} D_{\ell,a})} \equiv e^{-{G}_\gamma ({\alpha},{\beta})}
 \ee
 where $\alpha,\beta \in \mathbb{R}^{3L}$ are multivariable inverse temperatures. The creation of a full link $\ell$ is thus associated with two $\mathbb{R}^3$-valued temperatures, $\alpha_\ell \equiv \{\alpha_{\ell,a}\}_{a=1,2,3}$ and $\beta_\ell \equiv \{\beta_{\ell,a} \}_{a=1,2,3}$. Notice that the constraints $\{C_{\ell,a}, D_{\ell,a}\}_\gamma$ are smooth functions on $\Gamma_N = \Gamma^{\times N}$, and thus $\rho_{ \{\gamma, \alpha,\beta\}}$ is a state of the $N$ particle system. For instance, for the 4-valent dipole graph of Figure \ref{dip} with flux constraints \eqref{dipcon} and the corresponding holonomy ones, we have 
 \be \label{GdiP} G_{\text{dip}}(\alpha,\beta) = \sum_{\ell=1}^4 \sum_{a=1}^3 \alpha_{\ell,a} C_{\ell,a} (g_{(1\ell)},g_{(2\ell)}) + \beta_{\ell,a} D_{\ell,a} (X_{(1\ell)},X_{(2\ell)})  \,. \ee \

We could further consider the case of assigning a single temperature for all three components $a$. Then, one pair of $\mathbb{R}$-valued parameters, $\alpha_{\ell}$ and $\beta_{\ell}$, controls each link $\ell$, instead of three pairs. We would thus get,
\be\label{rogam} \rho_{\{\gamma,\alpha,\beta\}} \propto e^{-\sum \limits_{\ell=1}^L \alpha_\ell C_\ell + \beta_\ell D_\ell } \ee 
where $\alpha,\beta \in \mathbb{R}^L$, $C_\ell = \sum_{a} C_{\ell,a}$ and $D_\ell = \sum_{a} D_{\ell,a}$ \cite{Chirco:2018fns}. Making such different choices is a non-trivial move (cf. discussions in section \ref{commtemp}); notice that $\{C_{\ell,a}=0,D_{\ell,a}=0\}_{\gamma} \Rightarrow \{C_\ell=0,D_\ell=0\}_{\gamma}$ but the converse is not true. Latter is thus a weaker condition than the former. States \eqref{rhocomp} and \eqref{rogam} correspond to these two sets of conditions respectively, associated with the constraints, $s_1 = \{\langle C_{\ell,a} \rangle = 0,\langle D_{\ell,a} \rangle = 0 \}_\gamma$ and $s_2 = \{\langle C_\ell \rangle = 0,\langle D_\ell \rangle = 0 \}_\gamma$, in the entropy maximisation procedure. \

If we were further to extract a single common temperature, say $\beta_\gamma$, then this would correspond to a set with a single constraint, $s_3 = \{\langle G_\gamma \rangle = 0\}_\gamma$ (cf. discussions in section \ref{commtemp}). Being associated with a sum of all constraints, $s_3$ is in turn weaker than both $s_1$ and $s_2$. Then, the state with respect to $s_3$ is naturally of the form $\rho_{\beta_\gamma} \propto e^{-\beta_\gamma G_\gamma}$. Notice that, in the weaker cases $s_2$ and $s_3$, the corresponding microcanonical states cannot be understood as giving back a state in $\Gamma_\gamma$, i.e. as giving a twisted geometry (unlike $s_1$). This is because making $\gamma$ requires imposing $\{C_{\ell,a}=0,D_{\ell,a}=0\}_{\gamma}$ (which is a microcanonical state of $s_1$), and not either of the two weaker conditions.  \

A state such as \eqref{rhocomp} is a statistical mixture of configurations where the ones which are glued with the combinatorics of $\gamma$, and thus admit a discrete geometric interpretation, are weighted exponentially more than those which are not. In this sense, such a state illustrates a statistically approximate notion of discrete (here, twisted) geometry. Notice that states like \eqref{rogam} and $\rho_{\beta_\gamma}$ would encode an even weaker notion of statistically fluctuating (approximate) geometries, even if that, because in general even those configurations satisfying $\{C_\ell=0,D_\ell=0\}$ or $G_\gamma = 0$ exactly would not necessarily correspond to a graph in $\Gamma_\gamma$. \

We can further generalise the above to include different `interaction' terms, each corresponding to a given combinatorial pattern of gluings associated with a different graph $\gamma$. Let us first take into account all possible graphs $\{\gamma\}_N$ with a fixed number of nodes $N$. To each $\gamma$ in this set corresponds a gluing $G_\gamma(\alpha,\beta)$ as a function of several temperatures, which control the fluctuations in the internal structure of the graph $\gamma$. We can then think of a statistical mixture of the different graphs, each being represented by its corresponding gluing function $G_\gamma$ and weighted by coupling parameters $\lambda_\gamma$. Such a Gibbs distribution can be formally written as,
\be \label{rhon}
\rho_N = \frac{1}{Z_N(\lambda_\gamma,\alpha,\beta)} e^{-\sum \limits_{\{\gamma\}_N} \frac{1}{k_\text{sym}} \lambda_\gamma G_\gamma(\alpha,\beta)}
\ee
where $k_\text{sym}$ factors out any repetitions due to underlying symmetries, say due to graph automorphisms \cite{Chirco:2018fns}. The choice of the set $\{\gamma\}_N$ is a model-building choice, analogous to choosing different interaction terms in the Hamiltonian of a standard many-body system. \

This can heuristically be generalised further to a superposition of such internally fluctuating twisted geometries for an $N$ particle system, but now including contributions from distinct graphs, each composed of in general a \emph{variable} number of nodes $M$. In line with analogous expressions in standard many-body theory, a state of this kind can formally be written as, 
\be \label{trho} \rho_N = \frac{1}{Z_N(M_{\max},\lambda_\gamma,\alpha,\beta)} e^{-\sum \limits_{M=2}^{M_{\max}} \sum \limits_{\{\gamma\}_M} \frac{1}{k_\text{sym}} \lambda_\gamma \sum \limits_{i_1 \neq ... \neq i_M =1}^{N} G_\gamma (\vec{g}_{i_1},\vec{X}_{i_1},...,\vec{g}_{i_M},\vec{X}_{i_M}; \alpha,\beta)} \ee
where $i$ is the particle index, and $M_{\max} \leq N$  \cite{Kotecha:2019vvn}. The value of $M_{\max}$ and the set $\{\gamma\}_M$ for a fixed $M$ are model-building choices. The first sum over $M$ includes contributions from all admissible (depending on the model, determined by $M_{\max}$) different $M$-particle subgroups of the full $N$ particle system, with the gluing combinatorics of the different boundary graphs with $M$ nodes. The second sum is a sum over all admissible boundary graphs $\gamma$, with a given fixed number of nodes $M$. And, the third sum takes into account all $M$-particle subgroup gluings (according to a given fixed $\gamma$) of the full $N$ particle system. We note that the state \eqref{trho} is a further generalisation of the state \eqref{rhon} above, specifically the latter is a special case of the former for the case of a single term $M=M_{\max}=N$ in the first sum. Let us consider a simple example, to understand better the definition of \eqref{trho}. Let the total number of particles be $N=3$; and, the model be defined by a single, two-particle interaction, i.e. $M_{\max} = 2$. Let this interaction be characterised by a single 2-particle graph, say a dipole graph, i.e. the set of admissible graphs is $ \{\gamma_{\text{dip}}\}$. Then, in the exponent: the first two sums contribute with a single term each; and, the last sum is
\be \sum \limits_{i_1=1}^3 \sum \limits_{i_2 \neq i_1 = 1}^3 G_{\text{dip}}(\vec{g}_{i_1},\vec{X}_{i_1},\vec{g}_{i_2},\vec{X}_{i_2}; \alpha,\beta) \ee
where, $G_{\text{dip}}(\vg_1,\vec{X}_1,\vg_2,\vec{X}_2; \alpha,\beta) \equiv G_{\text{dip}}(\alpha,\beta)$ as given in equation \eqref{GdiP} above. This sum consists of six terms in total, which reduce to three for indistinguishable particles. This corresponds to the three possible pairings of the total $N=3$ particles, i.e. the three number of 2-particle subgroups in a system with total 3 particles; and, each 2-particle subgroup interacts by $G_{\text{dip}}$.\

Finally, allowing for the system size $N$ to vary, we get a grand-canonical state,
\be Z(\mu,M_{\max},\lambda_\gamma,\alpha,\beta) = \sum_{N \geq 0} e^{\mu N} Z_N(M_{\max},\lambda_\gamma,\alpha,\beta) \ee 
where $\mu$ is the Lagrange parameter for $N$, and $Z _N$ is the canonical partition function for a finite number of tetrahedra, but including different graph contributions given by the state \eqref{trho}. The resultant set of parameters $\{\mu,M_{\max},\lambda_\gamma,\alpha,\beta\}$ are thus directly derived from the underlying microscopic model, as should be the case in any statistical consideration. \\

\noindent \textbf{Discussion} \\

\noindent Now that we have discussed states for many classical tetrahedra, involving gluing constraints that impose a discrete geometric interpretation, we conclude with a brief discussion of some model-building strategies in the context of simplicial gravity. 

Any such strategy would have to be based on a clear understanding of how simplicial geometry is encoded in the data that is used. For instance, in the case of holonomy-flux geometries discussed above, the simplicial geometric content is precisely that of classical twisted geometries \cite{Freidel:2010aq,Rovelli:2010km}. In other words, the holonomy-flux variables admit an equivalent parametrisation in terms of twisted geometries variables. This relies on the fact that the link space $\mcT^*SU(2)$ can be decomposed as $S^2 \times S^2 \times \mcT^*S^1 \ni (\N_{s(\ell)},\N_{t(\ell)},A_\ell,\xi_\ell)$, modulo null orbits of the latter, and up to a $\mathbb{Z}_2$ symmetry. The variables are related by the following canonical transformations,  
\be
g = \n_s e^{\xi \tau_3} \n_t^{-1} \,,\quad X = A \n_s \tau_3 \n_t^{-1}
\ee
where $\n_{s,t} \in SU(2)$ are those elements which in the adjoint representation $R$ rotate the vector $z \equiv (0,0,1)$ to give $\mathbb{R}^3$ vectors $\N_{s,t}$ respectively. That is $\N_{s,t} = R(\n_{s,t}).z$, or equivalently $\n_{s,t} \tau_3 \n_{s,t}^{-1} = \sum_{a=1}^3 \N^a_{s,t}\tau_a$ respectively for $s$ and $t$, where $\tau_a = -\frac{i}{2}\sigma_a$ are generators of $\su(2)$, with Pauli matrices $\sigma_a$. Vectors $\N_{s(\ell)}$ and $\N_{t(\ell)}$ are unit normals to the face $\ell$ as seen from two arbitrary, different orthonormal reference frames attached to $s(\ell)$ and $t(\ell)$ respectively. $A_\ell$ is the area of $\ell$, and $\xi_\ell$ is an angle which encodes (partial\footnote{The remaining two degrees of freedom of extrinsic curvature are encoded in the normals $\N_{s(\ell)}$ and $\N_{t(\ell)}$ \cite{Freidel:2010aq}. For instance, in the subclass of Regge geometries, $\xi_\ell$ is proportional to the modulus of the extrinsic curvature \cite{Rovelli:2010km}.}) extrinsic curvature information. So, a closed twisted geometry configuration supported on graph $\gamma$ is: an element of $\; \bigtimes_{\ell} \mcT^*S^1 \bigtimes_{n} \mathcal{S}_4$, where $\mathcal{S}_4$ is the Kapovich-Millson phase space of a tetrahedron given a set of face areas; each link of $\gamma$ is labelled with $(A_\ell, \xi_\ell)$; and each node of $\gamma$ is labelled with four area normals (in a given reference frame) that satisfy closure. \cite{Freidel:2010aq,Rovelli:2010km} \

A twisted geometry is in general discontinuous across the faces, and so is the one described in terms of holonomy-flux variables, because both contain the same information. Face area $A_\ell$ of a shared triangle is the same as seen from tetrahedron $s(\ell)$ or $t(\ell)$ on either side, but the edge lengths when approaching from either side may differ in general \cite{Freidel:2010aq}. That is, the shape of the triangle $\ell$, as seen from the two tetrahedra sharing it, is not constrained to match. If additional shape-matching conditions \cite{Dittrich:2008va} were satisfied, then we would instead have a proper Regge (metric) geometry on $\gamma^*$, which is then a subclass of twisted geometries. These shape-matching conditions can be related to the so-called simplicity constraints, which are central in all model building strategies in the context of spin foam models, and whose effect is exactly to enforce geometricity (in the sense of metric and tetrad geometry) on discrete data of the holonomy-flux type, characterising (continuum and discrete) topological BF theories. \

The gluing constraints in equation \eqref{gluing}, take the form of the following constraints in twisted geometry variables,
\begin{align} A_{(nI)} - A_{(mJ)} = 0 \,&,\quad \xi_{(nI)} + \xi_{(mJ)} = 0 \,, \nonumber \\
  \N_{s(nI)} - \N_{t(mJ)} = 0 \,&,\quad  \N_{t(nI)} - \N_{s(mJ)} = 0 \,. \label{gluingtwisted} \end{align}
The result is naturally the same as in the holonomy-flux case: half-links $(nI)$ and $(mJ)$ which satisfy the above set of six component constraint functions (in either of the parametrisations) are glued\footnote{Gluing the two half-links is essentially superposing one over the other in terms of aligning their respective reference frames. This is evident from the constraints for the normal vectors $\N$ which superposes the target node of one half-link on the source node of the other, and vice-versa.} to form a single link $\ell \equiv (nI,mJ)$. Equivalently, the two faces of the initially disconnected tetrahedra are now adjacent. \

The more refined parametrisation used in the twisted geometry language would allow for a model-building strategy leading, for example, to statistical states in which only some of the gluing conditions are imposed strongly, while others are imposed on average. In the same spirit of achieving greater geometrical significance of the statistical state that one ends up with, this construction scheme can be applied with additional constraints, beyond the gluing ones that we illustrated above. For instance, starting with the space of twisted geometries on a given simplicial complex dual to $\gamma$, one could consider imposing (on average) also shape-matching constraints, or simplicity constraints, to encode an approximate notion of a Regge geometry using a Gibbs statistical state. This would be the statistical counterpart of the construction of spin foam models, i.e. discrete gravity path integrals in representation theoretic variables \cite{Perez:2012db,Baratin:2011hp,Finocchiaro:2018hks}, based on the formulation of gravity as a constrained BF theory. These lines of investigations are left to future work. \

Lastly, we remark that the quantum counterparts of the above states can formally be defined on $\mcH_F$, and the general discussion above is applicable \cite{Chirco:2018fns,Oriti:2013aqa}. The basic ingredient of gluing is again to define face sharing conditions. For instance, the classical constraints of equation \eqref{gluing} can be implemented by group averaging of wavefunctions \cite{Oriti:2013aqa},  
\be \label{gluing2}
 \Psi_\gamma (\{g_{(nI)}g_{(mJ)}^{-1}\}) = \prod_{(nI,mJ)|_\gamma} \int_{SU(2)} dh_{(nI,mJ)} \; \psi(\{ g_{(nI)}h_{(nI,mJ)} , g_{(mJ)}h_{(nI,mJ)} \}) \ee
where $\psi \in \mathcal{H}_N$ is a wavefunction for a system of generically disconnected $N$ tetrahedra. So, a wavefunction defined over full links $(nI,mJ)$ of a graph $\gamma$ is a result of averaging over half-links $(nI)$ and $(mJ)$ by $SU(2)$ elements $h_{(nI,mJ)}$. The same can also be implemented in terms of fluxes $X$, using a non-commutative Fourier transform between the holonomy and flux variables \cite{Oriti:2013aqa,Baratin:2010wi,Baratin:2010nn}.


\section{Thermofield doubles, thermal representations and condensates} \label{threpcond}

Quantum gravitational phases that display features like entanglement, coherence and observable statistical fluctuations are important in the study of semi-classical and effective continuum descriptions of quantum spacetime, especially in physical settings like cosmology and black holes. In this section, we construct entangled, thermofield double states $\ket{\Omega_\rho}$, associated with a generalised Gibbs state $\rho$. These states naturally encode statistical fluctuations with respect to the given observables of interest, characterising $\rho$. We further construct their corresponding inequivalent representations. Subsequently, we introduce a family of thermal condensates built upon the thermal vacua $\ket{\Omega_\rho}$, as coherent configurations of the underlying quanta. \cite{Assanioussi:2019ouq}


\subsection{Preliminaries: Thermofield dynamics} \label{tfd}

Thermofield dynamics (TFD) is an operator framework for finite temperature quantum field theory \cite{Takahasi:1974zn,Matsumoto:1985mx,Umezawa:1982nv,Umezawa:1993yq,Khanna:2009zz}. One of its main advantages lies in the fact that its formulation parallels that of zero temperature quantum field theory. Thus powerful tools of the temperature-independent setup can be translated to the thermal case, including perturbative Feynman diagrammatic techniques, symmetry breaking analyses, and for what concerns us here, Fock space techniques. TFD has been applied in various fields such as superconductivity, quantum optics, and string theory. 

The core idea of the TFD formalism is to represent statistical ensemble averages as temperature-dependent vacuum expectation values. That is, given a system described by a Hilbert space $\mathfrak{h}$ and an algebra of observables $\mathfrak{A}_\mathfrak{h}$ on $\mathfrak{h}$, one looks for a vector state $\ket{\Omega_\rho}$ (thermal vacuum) in a Hilbert space $\mathfrak{H}$, corresponding to a density operator ${\rho}$ on $\mathfrak{h}$, such that the following condition holds for all observables $A$ of the system,
\be \label{req} \Tr_{\mathfrak{h}}( {\rho}  {A}_{\mathfrak{h}}) = \bra{\Omega_\rho} {A}_{\mathfrak{H}}\ket{\Omega_\rho}_{\mathfrak{H}}  \ee
where the subscript $\mathfrak{h} \; \text{or} \; \mathfrak{H}$ denotes a suitable representation of the operator in the respective Hilbert spaces. Here, we are mainly concerned with statistical equilibrium, thus with density operators of the Gibbs form $e^{-\beta  {\mcO_\mathfrak{h}}}$ as discussed in the previous sections. Further, notice that the vector state $\ket{\Omega_\rho}$ encodes the same information as the density operator $\rho$, as reflected in equations \eqref{req} holding for the full algebra. Thus, like in standard treatments of systems in a state $\rho$, statistical fluctuations in the corresponding state $\ket{\Omega_\rho}$ can be investigated for instance in terms of variances of relevant observables in this state. \ 

A vector state satisfying equation \eqref{req} can only be defined in an extended Hilbert space, by supplementing the original degrees of freedom with the so-called tilde conjugate degrees of freedom \cite{Takahasi:1974zn}. Importantly, this doubling, or in general an enlargement of the space of the relevant degrees of freedom, is a characteristic feature of finite temperature description of physical systems. This was discovered also in algebraic quantum field theory for equilibrium statistical mechanics \cite{Haag:1967sg}. In fact, equation \eqref{req} is strongly reminiscent of the construction of a GNS representation induced by an algebraic statistical state \cite{Bratteli:1979tw,Bratteli:1996xq,Haag:1992hx,Strocchi:2008gsa}, with the vector state $\ket{\Omega_\rho}$ being the cyclic vacuum of a thermal representation. These intuitions have indeed led to tangible relations between the two formalisms, with the tilde degrees of freedom of TFD being understood as those of the conjugate representation of KMS theory in the algebraic framework \cite{Ojima:1981ma,Matsumoto:1985mx,Landsman:1986uw,Celeghini:1998sy,Khanna:2009zz}. 

As we have noticed previously, these structures are also encountered commonly in quantum information theory \cite{nielsen_chuang_2010}, which in turn is utilised heavily in various areas of modern theoretical physics, like holography. Specifically, constructing a state $\ket{\Omega_\rho}$ is simply a purification of $ {\rho}$. A prime example of a vector state is the thermofield double state, which is the purification of a Gibbs density operator. These states are used extensively in studies probing connections between geometry, entanglement, and more recently, complexity \cite{VanRaamsdonk:2016exw,Chapman:2018hou}. For instance in AdS/CFT, this state is important because an eternal AdS black hole bulk is dual to a thermofield double in the boundary quantum theory \cite{Maldacena:2001kr}. \\

We now review the basics of the TFD formalism \cite{Takahasi:1974zn,Matsumoto:1985mx,Umezawa:1982nv,Umezawa:1993yq,Khanna:2009zz}, based on which we will later carry out our construction of thermofield doubles and their associated inequivalent representations in GFT. Here we consider a simple example of an oscillator, which will be extended to a field theory setup directly for GFTs in the subsequent sections. \

Consider a single bosonic oscillator, described by ladder operators $ {a}$ and ${a}^\dag$, satisfying the commutation algebra,
\begin{gather}
 [ {a}, {a}^\dag] = 1 \, ,\quad [ {a}, {a}]  = [ {a}^\dag, {a}^\dag]= 0
\end{gather}
with the $a$-particle vacuum being specified by \be {a}\ket{0} = 0 \,.\ee 
A Fock space $\mfh$ is generated by actions of polynomial functions of the ladder operators on $\ket{0}$. Thermal effects are then encoded in density operators defined on $\mfh$. In particular, an equilibrium state at inverse temperature $\beta$ is a Gibbs state,
\be \label{qmrho}  {\rho}_\beta = \frac{1}{Z} e^{-\beta  {H}} \ee
where $ {H}$ is a Hamiltonian operator on $\mfh$, possibly of grand-canonical type. But as discussed in chapter \ref{GGS} and section \ref{gengibb} above, in the context of background independent systems one may work (at least formally) with other observables in place of $H$. \

This system is extended by including tilde degrees of freedom, spanning a Hilbert space $\tilde{\mfh}$ generated by the ladder operators $ {\tilde{a}}$ and ${\tilde{a}}^\dag$ satisfying the same bosonic algebra,
\be [ {\tilde{a}}, {\tilde{a}}^\dag] = 1 \,, \quad  [ {\tilde{a}}, {\tilde{a}}] = [ {\tilde{a}}^\dag, {\tilde{a}}^\dag] = 0 \ee
acting on the tilde-vacuum given by,
\be {\tilde{a}}\ket{\tilde{0}} = 0 \,. \ee
All tilde and non-tilde degrees of freedom commute with each other, that is 
\be [ a,  {\tilde{a}}]=[  {a},  {\tilde{a}}^\dag]= [  {a}^\dag,  {\tilde{a}}]=[  {a}^\dag,  {\tilde{a}}^\dag]=0 \;. \ee
The Hilbert space $\tilde{\mfh}$ is conjugate to the original one via the following tilde conjugation rules of thermofield dynamics (or equivalently, via the action of the modular conjugation operator of KMS theory \cite{Ojima:1981ma}): $(AB)\tilde{\,} = \tilde{A}\tilde{B}$, $\;\; (A^\dag)\tilde{\,} = \tilde{A}^\dag $, $\;\; (\tilde{A})\tilde{\,} = A $, $\;\;  (z_1 A + z_2 B)\tilde{\,} = \bar{z}_1 \tilde{A} + \bar{z}_2\tilde{B} $, $\;\;\text{and}\; \ket{0}\tilde{\,} = \ket{\tilde{0}}$, $\;$
for all non-tilde and tilde operators defined on $\mfh$ and $\tilde{\mfh}$ respectively, and $z_1,z_2 \in \mathbb{C}$. \

The zero temperature (or the limiting case of its inverse, $\beta=\infty$) phase of the system is described by the enlarged Hilbert space,
\be \mfH_\infty = \mfh \otimes \tilde{\mfh} \ee
built from the Fock vacuum,
\be \ket{0_\infty} = \ket{0} \otimes \ket{\tilde{0}} \ee
by actions of the ladder operators $a,a^\dag,\tilde{a},\tilde{a}^\dag$, such that
\be a\ket{0_\infty} = \tilde{a}\ket{0_\infty} = 0 \,.\ee

\noindent Temperature is introduced via thermal Bogoliubov transformations of the algebra generators, \be\{a,a^\dag,\tilde{a},\tilde{a}^\dag\}_{\beta = \infty} \mapsto \{b,b^\dag,\tilde{b},\tilde{b}^\dag\}_{0 < \beta < \infty}\ee given by
\begin{align} \label{qmbog1}
b &= \cosh [\theta(\beta)] \, a - \sinh [\theta(\beta)] \, \tilde{a}^\dag  \\ \label{qmbog2}
\tilde{b} &= \cosh [\theta(\beta)] \, \tilde{a} - \sinh [\theta(\beta)] \, {a}^\dag 
\end{align}
along with analogous expressions for their adjoints $b^\dag$ and $\tilde{b}^\dag$. The Bogoliubov transformations are canonical, thus leaving the algebra unchanged. Then, the $\beta$-ladder operators also satisfy bosonic commutations relations, given by
\begin{align} \label{qmbcr1}
[ b, b^\dag] = [ \tilde{b}, \tilde{b}^\dag] = 1 \\ \label{qmbcr2}
[ b, \tilde{b} ]=[ b,  \tilde{b}^\dag]=0
\end{align}
along with their adjoints. The temperature-dependent annihilation operators now specify a thermal vacuum,
\be \label{qmbvac} b\ket{0_\beta} = \tilde{b}\ket{0_\beta} = 0  \ee
which is cyclic for a thermal Hilbert space $\mfH_\beta$.\footnote{We note that $\ket{0_\beta}$ is an example of a two-mode squeezed state, which can be restored to its full generality most directly by adding a net phase difference between the $\cosh$ and $\sinh$ terms in the transformations \eqref{qmbog1} and \eqref{qmbog2}.}  \

The Hilbert space $\mfH_\beta$ can naturally be organised as a Fock space with respect to the $\beta$-dependent ladder operators that create and annihilate $b$-quanta over the thermal background $\ket{0_\beta}$. Like in any other Fock space construction, one can define useful classes of states in $\mfH_\beta$. We will return to this point in section \ref{CTS} where we define one such interesting class of states in the quantum gravity system, namely the coherent thermal states. \

The thermal Bogoliubov transformations \eqref{qmbog1} are parametrized by $\theta(\beta)$, which must thus encode complete information about the corresponding statistical state. In the present case, it must be uniquely associated with the Gibbs state $\rho_\beta$, which has a well known characteristic Bose number distribution,
\be \Tr_{\mfh}(\rho_\beta a^\dag a) = \frac{1}{e^{\beta \omega}-1} \ee
using a Hamiltonian of the form $H = \omega a^\dag a$, and the number operator $N=a^\dag a$ of the original non-tilde system (which is the physical system of interest, see below). This is how $\theta$ is usually determined in TFD, by using equation \eqref{req} for the number operator. Then, the right hand side of equation \eqref{req} for the number operator gives
\be \bra{0_\beta}a^\dag a \ket{0_\beta}_{\mfH_\beta} = \sinh^2[\theta(\beta)]   \ee
using inverse Bogoliubov transformations. This specifies $\theta$ via the equation,
\be \frac{1}{e^{\beta \omega}-1} = \sinh^2[\theta(\beta)] \,. \ee

The non-tilde degrees of freedom can be understood as being physically relevant in the sense that they describe the subsystem of interest, which is accessible to an observer. In other words, it is the subsystem under study in a given situation. Then, the physically relevant observables are naturally those that belong to the algebra restricted to the non-tilde degrees of freedom, which can be retrieved from the full description by partially tracing away the complement (here, the tilde degrees of freedom). Thus, one is interested in observable averages of the form $\bra{0_\beta} \mcO(a,a^\dag) \ket{0_\beta}$, for operators $\mcO$ that are in general polynomial functions of the generators of the physical (in the above sense) non-tilde algebra. Notable geometric examples include relativistic quantum field vacua in Minkowski and Schwarzschild spacetimes, where the non-tilde algebra has support on spacetime regions exterior to the respective horizons. Here, the tilde subsystems are CPT conjugates of the non-tilde ones, and belong to the interior of the horizons \cite{Israel:1976ur,Kay:1985zs,Sewell:1982zz}. Another physical interpretation of the tilde subsystem that is more common in condensed matter theory, is that of a thermal reservoir \cite{Umezawa:1982nv,Strocchi:2008gsa}. \

In the present discrete quantum gravity context, for now we retain the elementary, quantum information-theoretic interpretation of the non-tilde and tilde degrees of freedom, simply as describing a given subsystem and its complement respectively, without assigning any further geometric meanings. \

The two sets of ladder operators are related to each other explicitly via equations \eqref{qmbog1} and \eqref{qmbog2}. This suggests that their respective vacua are also related by a corresponding transformation. Indeed they are, via the following unitary transformation\footnote{The form of this unitary operator, along with equation \eqref{qmuvac}, shows that the thermal vacuum $\ket{0_\beta}$ is a two-mode squeezed state in which $a\tilde{a}$-pairs have condensed \cite{Umezawa:1993yq}.}
\be \label{qmu} U(\theta) = e^{\theta(\beta)(a^\dag \tilde{a}^\dag - a \tilde{a})} \,. \ee
In terms of this thermal squeezing operator, we then have
\be \label{qmuvac} \ket{0_\beta} = U(\theta)\ket{0_\infty} \ee
and
\begin{align}  b &= U(\theta) \, a\, U(\theta)^{-1} \,,\\ \label{qmub}
 \tilde{b} &= U(\theta) \, \tilde{a}\, U(\theta)^{-1} \,. \end{align}

It is clear then that such unitary operators map the different $\beta$-representations into each other. In finite quantum systems this is simply a manifestation of von Neumann's uniqueness theorem. However, when extending the above setup to quantum field theory, one would expect that representations at different temperatures are inequivalent. This is certainly the case for standard physical systems in general. It is also true in the present quantum gravity case, as we show in section \ref{finite}. Lastly, we note that in the field theory extension, equations \eqref{qmbog1}-\eqref{qmbvac} hold mode-wise and still remain well-defined. Together, they describe the $\beta$-phase of the system. However, the operator $U(\theta)$ is no longer well-defined in general (before any cut-offs). Thus, without any suitable regularisation, equations \eqref{qmu}-\eqref{qmub} technically do not hold in full field theory.


\subsection{Degenerate vacuum and zero temperature phase} \label{zero}

The zero temperature phase of the system is based on an enlargement of the Fock representation of the bosonic algebra associated with the degenerate vacuum (as described in sections \ref{atomkin} and \ref{bosgft}), along the lines presented in \ref{tfd} above, but generalised here to a field theory. \

We consider here (mainly for convenience of the subsequent application in GFT cosmology in section \ref{GFTCC}) the Hilbert space for a single quantum to be given by the state space of geometries of a quantum polyhedron with a single real scalar coupling, that is 
\be \mcH =  L^2(SU(2)^d/SU(2)) \otimes L^2(\mathbb{R}) \ee
where the quotient by $SU(2)$ ensures closure. Then in the discrete index basis introduced in section \ref{discind}, we have the following mode ladder operators
\begin{align} 
 {a}_{\vchi\alpha} &=  \varphi(\mbfD_{\vchi} \otimes \mbfT_{\alpha}) = \int_{SU(2)^d \times \mathbb{R}} d\vg d\phi\; \overline{\mbfD}_{\vchi}(\vg)\overline{\mbfT}_\alpha(\phi)   {\varphi}(\vg,\phi)   \label{a}  \\
  {a}^\dag_{\vchi \alpha} &=  \varphi^\dag(\mbfD_{\vchi} \otimes \mbfT_{\alpha}) = \int_{SU(2)^d \times \mathbb{R}} d\vg d\phi\; {\mbfD}_{\vchi}(\vg) {\mbfT}_\alpha(\phi)   {\varphi}^\dag(\vg,\phi)  \label{adag}  
 \end{align}
which satisfy,
\be \label{RegCom} [  {a}_{\vchi \alpha},  {a}^\dag_{\vchi'\alpha'}] = \delta_{\vchi\vchi'} \delta_{\alpha \alpha'} \ee
and $[  {a},  {a}]=[a^\dag,a^\dag]=0$. As mentioned previously in \ref{discind}, we find that the use of discrete indices can help in avoiding $\delta$ distributional divergences in relevant quantities. We will return to this point in the upcoming sections. Recall that the vacuum is given by, 
\be \label{degvac} {a}_{\vchi\alpha}\ket{0} = 0 \,,\quad \forall \, \vchi,\alpha \ee 
which is the degenerate vacuum $\Omega_F$ that we have been considering till now, but here with a notational change $\ket{0} \equiv \ket{\Omega_F}$. It generates the Fock space $\mcH_F$ of equation \eqref{fockhilbsp}.

The zero temperature phase is then given by extending the above with the conjugate representation space $\tilde{\mcH}_F$, as discussed in section \ref{tfd}. This gives the zero temperature ($\beta = \infty$) description in terms of a Hilbert space,
\be \mcH_\infty = \mcH_F \otimes \tilde{\mcH}_F \ee
which is a Fock space on the cyclic vacuum 
\be \ket{0_\infty} = \ket{0} \otimes \ket{\tilde{0}} \ee
with ladder operators $ \{ a, {a}^\dag, \tilde{a},  {\tilde{a}}^\dag \}_{{\vchi},\alpha}$ that satisfy,
\begin{align}
[  {a}_{{\vchi}\alpha},  {a}^\dag_{{\vchi}'\alpha'}] &= \delta_{{\vchi}{\vchi}'} \delta_{\alpha \alpha'} \\
[  {\tilde{a}}_{{\vchi}\alpha},  {\tilde{a}}^\dag_{{\vchi}'\alpha'}] &= \delta_{{\vchi}{\vchi}'} \delta_{\alpha \alpha'}
\end{align}
and $[  {a},  {a}]=[  {\tilde{a}},  {\tilde{a}}]=[  {a},  {\tilde{a}}]=[  {a},  {\tilde{a}}^\dag]=0$. The non-tilde operators describing the system of interest are those defined in \eqref{a} and \eqref{adag}, while the tilde operators of the complement are given by,
\begin{align}   {\tilde{a}}_{{\vchi}\alpha} &= \int_{SU(2)^d \times \mathbb{R}} d\vg d\phi\; {\mbfD}_{\vchi}(\vg){\mbfT}_\alpha(\phi)  {\tilde{\varphi}}(\vg,\phi) \,, \\
   {\tilde{a}}^\dag_{{\vchi}\alpha} &= \int_{SU(2)^d \times \mathbb{R}} d\vg d\phi\; \overline{\mbfD}_{\vchi}(\vg) \overline{\mbfT}_\alpha(\phi)  {\tilde{\varphi}}^\dag(\vg,\phi) \,. \end{align}
The vacuum satisfies
\be    {a}_{{\vchi}\alpha}\ket{0_\infty} =    {\tilde{a}}_{{\vchi}\alpha}\ket{0_\infty} = 0 \,, \quad \forall \, {\vchi},\alpha \,. \ee
The action of all polynomial functions of non-tilde and tilde ladder operators on $\ket{0_\infty}$ generates $\mcH_\infty$, all in complete analogy with standard $\mcH_F$, including the construction of multi-particle states, coherent states, squeezed states, and so on.


\subsection{Thermal squeezed vacuum and finite temperature phase} \label{finite}

The familiar way to include thermal effects is with statistical states, as density operators in a given representation. As discussed in section \ref{tfd}, an equivalent way is with their corresponding vector states (thermal vacua) in an enlarged representation of the system, in which the additional degrees of freedom are integral for encoding finite temperature effects. \

Here we are interested in generalised Gibbs density operators for describing equilibrium phases of the quantum gravity system, as discussed in the previous sections. Specifically, we consider the well-behaved class of states defined by extensive, positive operators $\mcP$ for group field theories coupled with scalar matter presented in section \ref{posgibbs}. For these states the above machinery of TFD can be applied directly, to construct various different phases characterised by the corresponding thermofield double vacua. In the context of cosmology for instance, we will be interested in choosing $\mcP$ to be a spatial volume operator (see section \ref{GFTCC}) \cite{Assanioussi:2020hwf}. For now, we leave it as this more general class of states, as reported in \cite{Assanioussi:2019ouq}. \

From discussions in \ref{tfd}, we know that the ensemble average for number density is useful for the construction of the associated thermal representation. For the class of states \eqref{gcposrho} generated by operators \eqref{posopp}, it is given by the characteristic Bose distribution,
\be \label{num} \Tr_{\mcH_F}(\rho_{\beta, \mu} a^\dag_{{\vchi}\alpha}a_{{\vchi}\alpha}) = \frac{1}{e^{\beta (\lambda_{{\vchi}\alpha}-\mu)}-1}  \ee
from which the average total number $\braket{N}$ of polyhedral quanta can be obtained by summing over all ${\vchi}$ and $\alpha$. Partial sums over either ${\vchi}$ or $\alpha$ would give average number densities $\braket{N_\alpha}$ or $\braket{N_{\vchi}}$ respectively. In the context of relational dynamics, for instance in GFT cosmology \cite{Oriti:2016qtz,Oriti:2016acw,Gielen:2016dss,Pithis:2019tvp}, quantities like $\braket{N_\alpha}$ are strictly related to relational observables as functions of the matter variable $\phi$ \cite{Assanioussi:2020hwf}, the details of which are included in section \ref{clock}. \

Now that we have chosen a suitable class of Gibbs states of equation \eqref{gcposrho}, we can proceed to define its associated thermal phase generated by a thermal vacuum $\ket{0_\beta}$ and $\beta$-dependent ladder operators $\{b_{{\vchi}\alpha},b^\dag_{{\vchi}\alpha},\tilde{b}_{{\vchi}\alpha},\tilde{b}^\dag_{{\vchi}\alpha}\}_\beta$, along the lines detailed in section \ref{tfd}. \

Thermal Bogoliubov transformations\footnote{More general two-mode squeezing transformations can also be considered by taking a net phase difference between the two contributions, say $e^{i\delta}$ scaling the $\sinh$ terms.} give the new ladder operators, mode-wise,
\begin{align} \label{bog1}
b_{{\vchi}\alpha} &= \cosh [\theta_{{\vchi}\alpha}(\beta)] \, a_{{\vchi}\alpha} - \sinh [\theta_{{\vchi}\alpha}(\beta)] \, \tilde{a}_{{\vchi}\alpha}^\dag  \\ \label{bog2}
\tilde{b}_{{\vchi}\alpha} &= \cosh [\theta_{{\vchi}\alpha}(\beta)] \, \tilde{a}_{{\vchi}\alpha} - \sinh [\theta_{{\vchi}\alpha}(\beta)] \, {a}_{{\vchi}\alpha}^\dag 
\end{align}
along with their adjoints $b_{{\vchi}\alpha}^\dag$ and $\tilde{b}_{{\vchi}\alpha}^\dag$. Inverse transformations are,
\begin{align} \label{invbog1}
a_{{\vchi}\alpha} &= \cosh [\theta_{{\vchi}\alpha}(\beta)] \, b_{{\vchi}\alpha} + \sinh [\theta_{{\vchi}\alpha}(\beta)] \, \tilde{b}_{{\vchi}\alpha}^\dag  \\ \label{invbog2}
\tilde{a}_{{\vchi}\alpha} &= \cosh [\theta_{{\vchi}\alpha}(\beta)] \, \tilde{b}_{{\vchi}\alpha} + \sinh [\theta_{{\vchi}\alpha}(\beta)] \, {b}_{{\vchi}\alpha}^\dag 
\end{align}
and their respective adjoints. The $\beta$-dependent annihilation operators specify the thermal vacuum via
\be b_{{\vchi}\alpha}\ket{0_\beta} = \tilde{b}_{{\vchi}\alpha}\ket{0_\beta}  = 0 \,, \ee
thus giving the finite temperature Hilbert space $\mcH_\beta$. Notice that the state $\ket{0_\beta}$ is a concrete example of a (class of) thermofield double state(s) in discrete quantum gravity. It is an entangled state, with quantum correlations between pairs of $a_{{\vchi}\alpha}$ and $\tilde{a}_{{\vchi}\alpha}$ polyhedral quanta.\

Further, using equations \eqref{req}, \eqref{num}, and
\be \label{thnum} \bra{0_\beta}a_{{\vchi}\alpha}^\dag a_{{\vchi}\alpha}\ket{0_\beta}_{\mcH_\beta} = \sinh^2 [\theta_{{\vchi}\alpha}(\beta)] \,,\ee
the parameters $\theta_{{\vchi}\alpha}$ can be determined from 
\be \label{num2} \sinh^2[\theta_{{\vchi}\alpha}(\beta)] = \frac{1}{e^{\beta (\lambda_{{\vchi}\alpha} - \mu)}-1} \,.  \ee
Note that the singular case in equation \eqref{num2} (or in \eqref{num}) can be understood as Bose-Einstein condensation to the ground state of $\mathcal{P}$ (equation \eqref{posopp}) in the present thermal gas of quantum gravitational atoms. Such a phenomenon was reported first in the context of a volume Gibbs state, and was used to show a model-independent, purely statistical mechanism for the emergence of a low-spin phase \cite{Kotecha:2018gof} (see section \ref{volposgibbs}).

Lastly, the $\beta$-phase that we have constructed here, being described kinematically by $ \{\ket{0_\beta}, b_{{\vchi}\alpha},b^\dag_{{\vchi}\alpha},\tilde{b}_{{\vchi}\alpha},\tilde{b}^\dag_{{\vchi}\alpha} \}$, is inequivalent to the zero temperature phase that is described by $\{  \ket{0_\infty}, a_{{\vchi}\alpha},a^\dag_{{\vchi}\alpha},\tilde{a}_{{\vchi}\alpha},\tilde{a}^\dag_{{\vchi}\alpha} \}$. This can be seen directly from the transformation equations between the two vacua as follows:
\begin{align} \ket{0_\beta} &= U(\theta)\ket{0_\infty} \\ &= e^{\sum_{{\vchi},\alpha}\theta_{{\vchi}\alpha}(a_{{\vchi}\alpha}^\dag\tilde{a}_{{\vchi}\alpha}^\dag - a_{{\vchi}\alpha}\tilde{a}_{{\vchi}\alpha})} \ket{0_\infty} \\
&= e^{-\sum_{{\vchi},\alpha} \ln \cosh\theta_{{\vchi}\alpha}}e^{\sum_{{\vchi},\alpha}a^{\dag}_{{\vchi}\alpha}\tilde{a}^\dag_{{\vchi}\alpha}\tanh \theta_{{\vchi}\alpha}} \ket{0_\infty} \\
&= \prod_{{\vchi},\alpha} \frac{1}{ \cosh\theta_{{\vchi}\alpha} } \times e^{\sum_{{\vchi},\alpha}a^{\dag}_{{\vchi}\alpha}\tilde{a}^\dag_{{\vchi}\alpha}\tanh \theta_{{\vchi}\alpha}} \ket{0_\infty} \,. \end{align}
The pre-factor of the product of inverse $\cosh$ functions vanishes in general, without any cut-offs in the modes. This means that the overlap between the two vacua is zero, and the two representations built upon them are inequivalent. In other words, this transformation in field theory is ill-defined in general due to an infinite number of degrees of freedom, giving rise to the inequivalent representations describing distinct phases of the system.


\subsection{Coherent thermal condensates} \label{CTS}

We now define a family of coherent states in these thermal representations of GFT \cite{Assanioussi:2019ouq}. We understand them as defining thermal quantum gravity condensates, expected to be relevant in the studies of semi-classical and continuum approximations in discrete quantum gravity models based on polyhedral quanta of geometry. 
Indeed, unlike the purely thermal state $\ket{0_\beta}$, coherent states can encode a notion of semi-classicality, with which one can attempt to extract effective dynamics from an underlying quantum gravity model. For instance in group field theory, it has been shown that a coherent condensate phase of the $a$-quanta can support FLRW cosmological dynamics, and thus represents a viable choice of a quantum gravitational phase relevant in the cosmological sector \cite{Oriti:2016qtz,Oriti:2016acw,Gielen:2016dss,Pithis:2019tvp,Kegeles:2017ems}. But a coherent state over the degenerate vacuum \eqref{degvac} is unentangled, and we expect a geometric phase of the universe to be highly entangled. Moreover these states cannot in themselves encode statistical fluctuations in different observable quantities. Therefore, from a physical point of view, the construction of coherent thermal states is amply justified.

Coherent thermal states \cite{1985JOSAB...2..467B,MANN1989273,mann,Khanna:2009zz} are a coherent configuration of quanta over the thermal vacuum, implemented by displacing $\ket{0_\beta}$ with displacement operators of the form,
\be \label{dis} D_a(\sigma) = e^{a^\dag(\sigma) - a(\sigma)}  \ee
for $\sigma \in \mcH$. To recall, the usual coherent states $\ket{\sigma} \in \mcH_F$ of $a$-particles are,
\be \label{zerocs} \ket{\sigma} := D_a(\sigma) \ket{0} , \ee
while $\ket{\tilde{\sigma}} \in \tilde{\mcH}_F$ for $\tilde{a}$-particles are,
\be \ket{\tilde{\sigma}} := D_{\tilde{a}}(\sigma) \ket{\tilde{0}} . \ee
The tilde in the ket notation $\ket{\tilde{\sigma}}$ simply means that the state is an element of the conjugate Hilbert space, and $D_{\tilde{a}}$ is a displacement operator of the same form as \eqref{dis} but for tilde ladder operators.    \

The most useful property of these states is that they are eigenstates of their respective annihilation operators,
\begin{align}
a_{{\vchi}\alpha}\ket{\sigma} = \sigma_{{\vchi}\alpha}\ket{\sigma} \\
\tilde{a}_{{\vchi}\alpha}\ket{\tilde{\sigma}} = \sigma_{{\vchi}\alpha}\ket{\tilde{\sigma}}
\end{align}
which is at the heart of their extensive use as robust, most classical-like, quantum states.

Notice that under the tilde conjugation rules stated in section \ref{tfd}, we have
\be (\ket{\sigma} \otimes \ket{\tilde{0}})\tilde{\,} = \ket{0} \otimes \ket{\tilde{\bar{\sigma}}} \ee
in $\mcH_\infty$. That is, coherent states $\ket{\sigma} \in \mcH_F$ and $\ket{\bar{\sigma}} \in \tilde{\mcH}_F$, are conjugates of each other. Then, the following state
\be \ket{\sigma,\bar{\sigma};\infty} := \ket{\sigma} \otimes \ket{\tilde{\bar{\sigma}}} = D_a(\sigma)D_{\tilde{a}}(\bar{\sigma})\ket{0_\infty} \in \mcH_\infty \ee
of the full system at zero temperature is self-conjugate (or conjugate-invariant), i.e.
\be \ket{\sigma,\bar{\sigma};\infty}\tilde{\,} = \ket{\sigma,\bar{\sigma};\infty}. \ee

In the finite temperature phase then, coherent thermal states \cite{1985JOSAB...2..467B,MANN1989273,mann,Khanna:2009zz} are defined as the following self-conjugate\footnote{This is a natural feature to require in coherent thermal states in light of the fact that such states when understood as quantum gravitational vacua along with their associated GNS representations, are expected to be invariant under the tilde conjugation rule, i.e. $\ket{\Omega_{\sigma,\bar{\sigma};\beta}}\tilde{\,} = \ket{\Omega_{\sigma,\bar{\sigma};\beta}}$, as is well-known in algebraic KMS theory.} states encoding coherence in the original $a$ degrees of freedom over the thermal vacuum,
\be \label{cts} \ket{\sigma,\bar{\sigma};\beta} := D_a(\sigma)D_{\tilde{a}}(\bar{\sigma})\ket{0_\beta} \in \mcH_\beta \,. \ee
Being elements of $\mcH_\beta$, as expected they are eigenstates of the $\beta$-annihilation operators $b_{{\vchi}\alpha}$ with temperature-dependent eigenfunctions,
\begin{align}
b_{{\vchi}\alpha}\ket{\sigma,\bar{\sigma};\beta} = (\cosh\theta_{{\vchi}\alpha} - \sinh\theta_{{\vchi}\alpha})\sigma_{{\vchi}\alpha}\ket{\sigma,\bar{\sigma};\beta} ,\\
\tilde{b}_{{\vchi}\alpha}\ket{\sigma,\bar{\sigma};\beta} = (\cosh\theta_{{\vchi}\alpha} - \sinh\theta_{{\vchi}\alpha})\bar{\sigma}_{{\vchi}\alpha}\ket{\sigma,\bar{\sigma};\beta} .
\end{align}

It is clear from the above eigenstate equations, along with inverse transformations \eqref{invbog1} and \eqref{invbog2}, that states \eqref{cts} are not eigenstates of the annihilation operator $a$ of the original system. This is precisely how the expectation values of physical non-tilde operators $\mcO(a,a^\dag)$ display non-trivial thermal and coherence properties simultaneously. For instance, the average number density is,
\be \label{expNaa} \bra{\sigma,\bar{\sigma};\beta}a^\dag_{{\vchi}\alpha}a_{{\vchi}\alpha}\ket{\sigma,\bar{\sigma};\beta} = |\sigma_{{\vchi}\alpha}|^2 + \sinh^2[\theta_{{\vchi}\alpha}(\beta)] \ee
which contains both, the usual coherent condensate number density and an additional thermal contribution associated with the statistical state. \

It is important to note that our use of the basis ${\mbfD}_{\vchi}\otimes {\mbfT}_\alpha$, in particular of the discrete basis $\{\mbfT_\alpha\}_{\alpha \in \mathbb N}$ for $L^2(\mathbb{R})$, in order to develop the finite-temperature GFT formalism in terms of the ladder operators (equations \eqref{a}-\eqref{adag} and \eqref{qmbog1}-\eqref{qmbog2}), was crucial. The observables that one might consider in a chosen model must admit domains of definition which contain the sector of Hilbert space that one is interested in, here coherent thermal states. Otherwise, we quickly run into divergences and ill-defined expressions. This in particular applies to $\phi$-dependent operators, not smeared with $\mbfT_\alpha$. For instance, if one considers the number density operator as a function of $\phi$,
\begin{align}
 N_{\vchi}(\phi) &= \int d\vg  d\vg' \; \overline{\mbfD}_{\vchi}(\vg') {\mbfD}_{\vchi}(\vg) {\varphi}^\dag(\vg,\phi) {\varphi}(\vg',\phi) \\
 &= a_{\vchi}^\dag(\phi) a_{\vchi}(\phi) 
\end{align}
then, the calculation of the expectation value in a coherent thermal state would give,
\begin{align} \label{deltan}
 \langle N_{\vchi}(\phi) \rangle_{\sigma,\bar{\sigma};\beta} = |\sigma_{\vchi}(\phi)|^2 + \sinh^2[\theta_{\vchi}(\phi)]\ \delta(\phi - \phi) 
\end{align}
which is clearly ill-defined, due to the presence of the Dirac delta distribution $\delta(0)$ in the thermal part evaluated at the singular point. We thus need to consider smeared observables such as the operator $a^\dag_{{\vchi}\alpha}a_{{\vchi}\alpha}$ in \eqref{expNaa}, where now the thermal contribution contains a well-defined Kronecker delta $\delta_{\alpha \alpha}$ coefficient instead. \

Lastly, we note that the use of such condensates, say with respect to a spatial volume observable in cosmology (see section \ref{GFTCC}), can be understood as implementing both semi-classical and continuum approximations \cite{Assanioussi:2020hwf,Oriti:2016qtz,Oriti:2016acw,Gielen:2016dss,Pithis:2019tvp}. Specifically, a semi-classical approximation resides in considering a coherent state, with its characteristic single-particle wavefunction being the relevant dynamical collective variable, while a continuum approximation resides in considering a non-perturbative (inequivalent) condensate phase of the quantum gravity system, with (formally) an infinite number of underlying quanta.\footnote{In general, identifying suitable semi-classical and continuum limits in discrete quantum gravity is a known open issue, and we do not address it directly in this thesis.} The novel feature of the coherent thermal phase though is that of having statistical fluctuations associated with the thermal vacuum, in addition to quantum fluctuations inherent in any quantum state. Statistical fluctuations in general are inevitable in macroscopic systems and could be non-trivial in both the semi-classical and continuum limits. This is unlike purely quantum fluctuations which are expected to be negligible in a semi-classical limit, but of significance especially at early times \cite{Gielen:2017eco}.


\section{Thermal condensate cosmology} \label{GFTCC}

In this section, we utilise the results presented above, to discuss an effective cosmological model incorporating statistical fluctuations of quantum geometry \cite{Assanioussi:2020hwf}. This investigation builds directly on the works in \cite{Assanioussi:2019ouq,Oriti:2016qtz,deCesare:2016rsf}, with the aim of evaluating some preliminary consequences of the presence of such fluctuations in cosmological evolution that is extracted from an underlying GFT dynamical model. \

This is brought about by the use of thermal condensates \cite{Assanioussi:2019ouq}, which we understand as describing a phase of the universe in which not all quanta have condensed. Such a phase seems likely in any reasonable geometrogenesis scenario, in which the universe transitions from a primordial pre-geometric hot thermal phase, to a phase with an approximate notion of continuum and macroscopic geometry (here, encoded in the notion of a condensate \cite{Oriti:2016acw}), but naturally with a leftover thermal cloud in general, of quanta that have not condensed. In other words, here, we understand a pure, zero temperature GFT condensate, that has been used extensively in previous works in GFT cosmology \cite{Oriti:2016qtz,deCesare:2016rsf,Pithis:2019tvp,Gielen:2016dss}, as describing a suitable macroscopic phase only at very late times of the system's evolution, and not throughout. What we work with instead is an intermediary phase that would be expected to arise in a transition between a hot pre-geometric phase and a pure condensate. Thus in this study, we present a scenario wherein the universe is modelled as a quantum gravitational condensate of elementary quanta of geometry, along with a thermal cloud of the same quanta over it; and in which, an early time phase that is dominated by the thermal cloud, and a late time phase that is dominated by the condensate, are generated dynamically.\

We start by presenting the effective free GFT dynamics in a condensate phase with fluctuating geometric volume. We then introduce the notion of a reference clock function, and reformulate the setup, including the effective dynamical equations of motion, in terms of functional quantities with respect to a generic class of these clock functions. Based on this, we present an effective, relational homogeneous and isotropic cosmological model, and discuss its late and early times properties. \

Specifically, we work with a free GFT model coupled with a real-valued scalar field, and thermal condensates with a static (non-dynamical) thermal cloud. We derive effective generalised evolution equations for homogeneous and isotropic cosmology, which include correction terms originating in the underlying quantum gravitational and statistical properties of the system. At late times, we recover the correct general relativistic limit of relational Friedmann equations. At early times, we get a bounce between contracting and expanding phases, and a phase of accelerated expansion that is characterised by an increased number of e-folds compared with previously reported numbers for the same class of free models. 


\subsection{Condensates with volume fluctuations} \label{vgibbs}

Since we are interested in the homogeneous and isotropic cosmological sector, the main observable of interest is the volume operator, in particular the volume associated to a (spatial) hypersurface given by a foliation parametrized by a clock function (see section \ref{clock}). Recall that (see section \ref{volposgibbs}) the GFT volume operator on $\mcH_F$ is,   
\be
 V:= \sum_{{\vchi},\alpha} v_{\vchi}\ a^\dagger_{{\vchi}\alpha} a_{{\vchi}\alpha}
\ee
where $v_{\vchi} \in \mathbb{R}_{> 0}$ is the volume assigned to a single quantum with a configuration ${\vchi} = (\vec{j},\vec{m},\iota)$, which is the representation data associated with $SU(2)^4$.\footnote{In our present application of the results of previous chapters to GFT cosmology, we take the base manifold to be $SU(2)^4 \times \mathbb{R}$, in order to facilitate a direct comparison of results with past works which make the same choice of the base manifold.} This is an extensive positive operator on $\mcH_F$, and its action on any multi-particle state gives the total volume by summing up the volume contribution $v_{\vchi}$ from each quantum, as discussed previously. \

Let us choose a statistical state of the form \eqref{gcposrho} such that the generator is the volume operator above, i.e. a volume Gibbs state of the form
\begin{align} \label{crho}  {\rho}_\beta = \frac{1}{Z_\beta} e^{-\beta  V} \end{align} 
which encodes a statistically fluctuating volume of quantum spacetime \cite{Kotecha:2018gof,Kotecha:2019vvn}. Notice that a constant shift in the spectrum, by the chemical potential $\mu$, is implicit in the above expression, since it will not be important in the following investigations. We refer to section \ref{volposgibbs} for more discussions surrounding this class of states. \

The quantum gravity condensate that we are interested in is a coherent thermal state of the form \eqref{cts}, but associated specifically with the volume Gibbs state in equation \eqref{crho}. In other words, we are interested in a state $\ket{\sigma,\bar{\sigma};\beta}$, which is specified by two functions: the condensate wavefunction $\sigma \in \mcH$, and the Bogoliubov parameter $\theta_{{\vchi}\alpha}(\beta)$ that is identified by the Bose number distribution of the state \eqref{crho},
\be \label{bose}
\sinh^2\left[ \theta_{{\vchi}\alpha} (\beta) \right] = \frac{1}{e^{\beta v_{\vchi}} - 1 } \,.
\ee
Notice that since the spectrum of $V$ is independent of the modes ${\mbfT}_\alpha$, the functions $\theta_{{\vchi}\alpha}$ are also independent of them. Thus $\theta_{{\vchi}\alpha} = \theta_{\vchi}$, and we will drop the labels $\alpha$ in quantities associated with $\theta$ from here on. Also, notice the following important property of our chosen state,
\be \label{zerolim}
\lim_{\beta \to \infty} \ket{\sigma,\bar{\sigma};\beta} = \ket{\sigma,\bar{\sigma}} = D_a(\sigma) \tilde{D}_a(\bar{\sigma}) \ket{0,\tilde{0}} 
\ee
due to which all results of the previous works in GFT cosmology are reproduced here, when the fluctuations are turned off completely. \

In the present context of extracting effective cosmological models from a candidate background independent theory of quantum gravity, the use of relational observables is of utmost importance. As mentioned previously, past works in GFT cosmology have interpreted and used the base manifold coordinate $\phi$ as a relational matter clock, and considered quantities such as $N(\phi)$ as relational observables. However we have also noticed that, in the present setting, such quantities (e.g. see equation \eqref{deltan}) contain divergences related to occurrences of the ill-defined $\delta(\phi=0)$ distributions. This had in turn prompted us to use a discrete basis ${\mbfT}_\alpha$, as a first step for the inclusion of thermal fluctuations in the context of GFT condensates.\footnote{Further details about this aspect and the definition of a basis-independent clock are discussed below in section \ref{clock}.} It follows that in this basis, we are interested in $\alpha$-dependent quantities defined by a partial sum over ${\vchi}$, such as 
\be V_\alpha = \sum_{{\vchi}} v_{\vchi}\ a^\dagger_{{\vchi}\alpha} a_{{\vchi}\alpha} \ee
and its statistical average in the thermal condensate,
\be
\langle V_\alpha \rangle_{\sigma,\bar{\sigma};\beta} = \sum_{\vchi} v_{\vchi} (|\sigma_{{\vchi}\alpha}|^2 + \sinh^2\left[ \theta_{{\vchi}} (\beta) \right]) 
\ee
which includes statistical fluctuations in volume, in addition to the condensate volume.\

Finally we note that, since the thermal cloud is modelled in terms of the volume Gibbs state whose Bose distribution is independent of $\alpha$ (which as we will see in section \ref{clock}, is strictly related to the evolution parameters for relational dynamics), we are essentially working in a first approximation wherein the thermal cloud is non-dynamical, and only the condensate part of the full system is dynamical. We will return to this point later in sections \ref{earlyev} and \ref{disc}.


\subsection{Effective group field theory dynamics} \label{eff1}

A generic GFT action with a local kinetic term and a non-local interaction term (higher than quadratic order in the fields) takes the form,
\begin{align} \label{action}
 S = \int d\vec{g}d\phi \, \bar{\varphi}(\vec{g},\phi) \msK(\vec{g},\phi) \varphi(\vec{g},\phi) \;+\; S_{\text{int}}[\varphi,\bar{\varphi}] \end{align} 
and gives the following classical equation of motion,
\begin{align} \label{EoMCl20} \msK(\vec{g},\phi) \varphi(\vec{g},\phi) \;+\; \frac{\delta S_{\text{int}}[\varphi,\bar{\varphi}]}{\delta \bar{\varphi}(\vec{g},\phi)} = 0 \,. \end{align} 
In the corresponding quantum theory on $\mcH_F$, the operator equation of motion is
\begin{align} \msK(\vec{g},\phi) \h{\varphi}(\vec{g},\phi) + \widehat{\frac{\delta S_{\text{int}}[\varphi,\bar{\varphi}]}{\delta \bar{\varphi}(\vec{g},\phi)}} = 0 \end{align} 
with some choice of operator ordering (and the hat notation reinstated temporarily). An effective equation of motion can then be derived from the above operator equation by taking its expectation value in a class of quantum states. Here, we take the coherent thermal states introduced above as this class of states, implementing a notion of semi-classical and continuum approximations\footnote{In line with previous works, we understand the implementation of semi-classical and continuum approximations in the specific sense of using the class of coherent states (which are well-known to be the most classical quantum states, with a peak on a pair of classical conjugate variables), and a condensate phase (described by a collective condensate variable, and with a non-zero order parameter), respectively.}. We thus consider,
\begin{align} \label{EoMEff20} \bra{\sigma,\bar{\sigma};\beta} \msK(\vec{g},\phi) \h{\varphi}(\vec{g},\phi) + \widehat{\frac{\delta S_{\text{int}}[\varphi,\bar{\varphi}]}{\delta \bar{\varphi}(\vec{g},\phi)}} \ket{\sigma,\bar{\sigma};\beta} = 0 \,.\end{align}

As a first step to investigate the role of statistical fluctuations of quantum geometry in condensate cosmology, we focus here only on the free part. This would allow us to display clearly the impact of non-zero thermal fluctuations. In other words, any difference in results that we find, as compared to previous zero temperature free theory studies, could then be attributed directly to the presence of these statistical fluctuations. Therefore, restricting to the kinetic term, we obtain
\begin{align}
 \bra{\sigma,\bar{\sigma};\beta} \msK(\vec{g},\phi) {\varphi}(\vec{g},\phi) \ket{\sigma,\bar{\sigma};\beta}
= \msK(\vec{g},\phi) \sigma(\vg,\phi)  \,.
\end{align}
Further using the Peter-Weyl decomposition for $\sigma$,
\begin{align}
\sigma(\vg,\phi) &= \sum_{{\vchi}} {\mbfD}_{{\vchi}}(\vg) \sigma_{\vchi}(\phi) \,,
\end{align}
and considering the following kinetic term (which is a standard choice in studies in GFT cosmology, see for instance \cite{Gielen:2018xph} and references therein),
\begin{align} \label{kin}
\msK = \msK_0(\vec{g}) + \msK_1(\vec{g})\partial_\phi^2
\end{align} 
such that
\begin{align}
\msK_0(\vec{g}) ({\mbfD}_{{\vchi}}(\vg) \sigma_{\vchi}(\phi) ) &=  B_{\vchi}{\mbfD}_{{\vchi}}(\vg) \sigma_{\vchi}(\phi) \\ 
\msK_1(\vec{g})\partial_\phi^2 ({\mbfD}_{{\vchi}}(\vg) \sigma_{\vchi}(\phi) ) &= A_{\vchi} {\mbfD}_{{\vchi}}(\vg)\partial_\phi^2\sigma_{\vchi}(\phi) 
\end{align}
we obtain the following equations of motion,
\begin{align} \label{EqM}
\partial_\phi^2\sigma_{\vchi}(\phi) - M_{\vchi} \sigma_{\vchi}(\phi) = 0 \,,\quad \forall {\vchi} 
\end{align} 
where $M_{\vchi} := - B_{\vchi}/A_{\vchi}$ . Notice that here, the dynamical variable is the $\phi$-dependent condensate wavefunction, $\sigma(\phi)$.

We see that the free GFT dynamical equation of motion in a coherent thermal state i.e. equation \eqref{EqM}, is identical to the case in \cite{Oriti:2016qtz}, where one considers a simple coherent state \eqref{zerocs} in $\mcH_F$ with no thermal cloud. But as we can already anticipate, observable averages (like volume) will have thermal contributions in general, consequently modifying their evolution equations.

This concludes the derivation of the effective GFT equation of motion using a thermal coherent state in a $\phi$-basis. However as we saw above, calculations with observables in this basis leads to singularities in the $\phi$-dependent quantities. This brings us to the question of defining and applying a suitable time reference frame (a clock), and offer a preliminary interpretation of the resultant quantities.


\subsection{Smearing functions and reference clocks} \label{clock}

As we have emphasised before, the use of $\phi$ as a reference clock is not possible here since the quantities of interest, like $\braket{V(\phi)}$, are mathematically ill-defined. This prompted us to define quantities like $\braket{V_\alpha}$ instead. Below, we generalise this even further and introduce generic\footnote{While satisfying certain boundary conditions, see equations \eqref{bcond}.} square-integrable, complex-valued smooth functions, 
\be t(\phi) = \sum_{\alpha} t_\alpha {\mbfT}_\alpha(\phi) \ee
in order to define observables and their dynamics as functionals of $t(\phi)$ (which will later be interpreted as relational). This brings us back to the aspect of smearing. \

In the quantum operator setup described in section \ref{threpcond}, instead of smearing the algebra generators with a set of basis functions $(\mbfD_{{\vchi}}\otimes \mbfT_{\alpha})(\vg,\phi)$, we could instead smear with a complete set of more general smearing functions $F(\vg,\phi)$ (usually also satisfying additional analyticity and sufficient decay properties). In particular for the $\phi$ variable, this would amount to smearing with smooth functions, say $t(\phi)$. This would result in an equivalent, but basis-independent algebraic setup, as commonly encountered in Weyl C*-algebraic theory associated with bosonic quanta\footnote{See section \ref{weylGFT} for more details on a Weyl algebraic formulation in group field theory \cite{Kotecha:2018gof,Kegeles:2017ems}.}. For our actual purposes, we retain the use of the Wigner basis ${\mbfD}_{\vchi}(\vg)$, in order to retain also the associated geometric interpretation of (functions of) the spin labels ${\vchi}$, which is standard in both GFT and LQG. But, in the $\phi$ direction we smear with a function $t(\phi)$. We are thus interested in smeared operators of the form, 
\begin{align} 
 a_{\vchi}(t) &= \int_{SU(2)^4 \times \mathbb{R}} d\vg d\phi \, \overline{{\mbfD}}_{\vchi}(\vg) \overline{t}(\phi) {\varphi}(\vg,\phi) \\
 a^\dag_{\vchi}(t) &= \int_{SU(2)^4 \times \mathbb{R}} d\vg d\phi \, {{\mbfD}}_{\vchi}(\vg) {t}(\phi) {\varphi}^\dagger(\vg,\phi) 
\end{align}
which are now understood as \emph{functional} (relational) ladder operators, with respect to the function $t(\phi)$. By extension, the volume operator now takes the form,
\be
 V_t := \sum_{{\vchi}} \ v_{\vchi}\ a^\dagger_{{\vchi}}(t) a_{{\vchi}}(t)
\ee
which is interpreted as the operator associated to a spatial slice labeled by the function $t$. \ 

Notice that in general, $t$-relational operators, and their expectation values, are non-local functions of their $\phi$-relational counterparts. For instance, the average volume in a thermal condensate state is
\be
 \braket{V_t}_{\sigma,\bar{\sigma};\beta} = \sum_{\vchi} v_{\vchi} \left( |\sigma_{\vchi}(t)|^2 + \sinh^2{[\theta_{\vchi}]}||t||^2 \right)
 \ee
where $||t||^2 = (t,t)_{L^2(\mathbb{R})}$ and,
\be \label{sigt}  \sigma_{{\vchi}}(t) := \int_{\mathbb{R}} d\phi \; \overline{t}(\phi)\sigma_{\vchi}(\phi) = (t , \sigma_{\vchi})_{L^2(\mathbb R)} \,. \ee
Then, the quantity $\braket{V_t}_{\sigma,\bar{\sigma};\beta}$ can be expressed in terms of a non-local function of $\phi$, i.e.
\be \label{vnonloc}
 \braket{V_t}_{\sigma,\bar{\sigma};\beta} = \sum_{\vchi} v_{\vchi} \int_{\mathbb{R}^2} d\phi d\phi' \, t(\phi)\overline{t}(\phi') \braket{N_{\vchi}(\phi,\phi')}_{\sigma,\bar{\sigma};\beta} 
\ee
where
\be N_{\vchi}(\phi,\phi') := a_{\vchi}^\dag(\phi)a_{\vchi}(\phi') \ee 
is the off-diagonal number density operator with expectation value,
\be 
 \braket{N_{\vchi}(\phi,\phi')}_{\sigma,\bar{\sigma};\beta} =  \bar{\sigma}_{\vchi}(\phi)\sigma_{\vchi}(\phi') + \sinh^2[\theta_{\vchi}]\delta(\phi-\phi') \,.
\ee
This generic non-locality of $t$-relational quantities with respect to $\phi$, for instance in \eqref{vnonloc}, is reasonable to expect simply as a technical feature that is characteristic of changing reference frames in general. \

Lastly, the smearing functions $t(\phi)$ are understood as defining reference clock frames, the reasons for which are made more clear in the next section. For now, we note that such a treatment is compatible with the fact that at the level of a field theory, which a GFT is, we would expect a relational clock variable to be more reasonably defined as a genuine function, like $t(\phi)$, rather than simply a parameter $\phi$ (which is here a coordinate of the base manifold). Having said that, we strictly refrain from assigning any further \emph{physical} interpretation to the function $t$, especially from the spacetime point of view, unlike the coordinate $\phi$ which has been motivated as a minimally coupled scalar matter field in previous works (see for instance \cite{Li:2017uao}). 


\subsection{Relational functional dynamics} \label{eff2}

We now come to the important task of expressing the effective GFT equations of motion in terms of the smearing functions. The goal is to arrive at a consistent dynamical description of the present system in a $t$-relational reference frame. Let us first reiterate our main line of reasoning. Smearing functions $t(\phi)$ are used in order to avoid divergences in the $\phi$-frame, e.g. in relational quantities like $\braket{V(\phi)}_{\sigma,\bar{\sigma};\beta}$. This leads to observables like $\braket{V_t}_{\sigma,\bar{\sigma};\beta}$. The condensate functional $\sigma_{\vchi}(t)$ defined in \eqref{sigt}, then naturally takes on the role of the dynamical collective variable, instead of $\sigma_{\vchi}(\phi)$. Therefore, the equations of motion \eqref{EqM} in terms of the variable $\phi$, must be rewritten suitably in terms of functions $t$, as follows.

We are seeking a differential equation of motion for $\sigma_{\vchi}(t)$, encoding the same dynamics as \eqref{EqM}. We begin by noticing that the mass term in \eqref{EqM} can be written in terms of $\sigma_{\vchi}(t)$ simply as
\be  M_{\vchi} \sigma_{\vchi}(t) = \int_\mathbb{R} d\phi \,  M_{\vchi} \overline{t}(\phi) \sigma_{\vchi}(\phi)  \ee
using the smearing. Therefore, as before, we see that smearing might offer us a way forward. We then smear the equations \eqref{EqM}, with an arbitrary square-integrable complex-valued smooth function $t(\phi)$, obtaining 
\be \label{smearEqM} \int_\mathbb{R} d\phi \, \overline{t}(\phi) \partial^2_\phi \sigma_{\vchi}(\phi) - M_{\vchi} \sigma_{\vchi}(t) = 0 \,, \quad \forall {\vchi} \,. \ee 
Now, in order to get a description completely in the $t$-frame, we require a suitable derivative operator with a well-defined action on functionals of $t$. For this, notice that,
\begin{align} \label{nt2new}
\int_\mathbb{R} d\phi \, \overline{t}(\phi) \partial^2_\phi \sigma_{\vchi}(\phi) &= \left(- \int_\mathbb{R} d\phi \, \partial_\phi \overline{t} \; \frac{\delta}{\delta \overline{t}(\phi)} \right)^2 \sigma_{{\vchi}}(t)  \nonumber \\
&=: \mathbf{d}_t^2 \sigma_{{\vchi}}(t)
\end{align}
where we have used integration by parts, and the following boundary conditions for the smearing functions,
\be \label{bcond} 
\lim_{\phi \to \pm \infty}t(\phi) = 0 \; , \quad \lim_{\phi \to \pm \infty}\partial_\phi t(\phi) =0 \,. \ee
The operator $\mathbf{d}_t$ might seem to be a good choice for the functional derivative \cite{engel2013density,parr1994density}
that we are looking for. However, recall that we are working with complex-valued smearing functions. Thus, in our context, generic functionals of them depend on both $t$ and $\bar t$, which are considered to be independent variables, e.g. the norm $|\sigma_{\vchi}(t)|^2 = \overline{\sigma_{\vchi}(t)}\sigma_{\vchi}(t)$ depends on two variables, $t$ and $\bar{t}$. The operator $\mathbf{d}_t$ must thus be extended by the conjugate term to obtain the hermitian differential operator
\be \nabla_t :=
- \int_\mathbb{R} d\phi \left( \partial_\phi \overline{t} \; \frac{\delta}{\delta \overline{t}(\phi)} +  \partial_\phi t \; \frac{\delta}{\delta t(\phi)} \right) . \ee 
Notice that, as required, we get an equation in terms of $\nabla_t$ that is analogous to \eqref{nt2new} above, i.e.
\be \label{nt2}
 \nabla_t^2 \sigma_{{\vchi}}(t) = \int_\mathbb{R} d\phi \, \overline{t}(\phi) \partial^2_\phi \sigma_{\vchi}(\phi) \,.
 \ee 
Therefore, the equations of motion \eqref{EqM} can be equivalently expressed as
\begin{align} \label{SEoM}
 \nabla_t^2 \sigma_{{\vchi}}(t) - M_{\vchi} \sigma_{{\vchi}}(t) &= 0 \,, \quad \forall {\vchi}
\end{align}  
for all square-integrable smooth functions $t(\phi)$ satisfying the boundary conditions \eqref{bcond}.

Note that if one was working with a dynamical model based on higher (than 2) order derivatives in $\phi$, or in general is interested in extending this setup to include arbitrary higher order generalisations of equation \eqref{nt2} above, then the boundary conditions \eqref{bcond} must be supplemented by vanishing of all higher order derivatives of $t$ in the limit $\phi \to \pm \infty$. In such a case then one could work with the space of Schwartz functions for instance, as the relevant set of smearing functions. However in the present analysis, we do not need to restrict to this special subspace of smooth functions, and the conditions \eqref{bcond} are sufficient. \

Few remarks are in order concerning the operator $\nabla_t$ and the associated $t$-relational setup. The operator $\nabla_t$ is a functional differential operator, consisting of functional derivatives with respect to $\bar{t}$ and $t$ \cite{engel2013density,parr1994density}. The flow induced by it is not on the GFT base manifold (in contrast with $\phi$), nor on a given spacetime, but rather on the space of smearing functions. Recalling that functional derivatives can be understood as generalisation of directional derivatives, then $\nabla_t$ essentially defines a flow with components along the directions of $(-\partial_\phi \overline{t})$ and $(-\partial_\phi t)$. Further, by construction this operator satisfies,
\be \label{opSAT} \nabla_t \ell(t) = \int_\mathbb{R} d\phi \, \overline{t}(\phi) \partial_\phi \ell(\phi)  \ee
for any function $\ell(\phi) \in L^2(\mathbb{R})$, where $\ell(\phi)$ satisfies \eqref{bcond}, and $\ell(t) := (t, \ell)_{L^2(\mathbb{R})}$. Notice that equation \eqref{opSAT} straightforwardly gives equation \eqref{nt2} used above.  The property in \eqref{opSAT} is important because it motivates the use of smearing functions as relational clock fields. As we have shown above, the $t$-relational dynamical quantities and equations are derived from an appropriate smearing of their (possibly non-local) $\phi$-relational counterparts. In particular, the $t$-functional equations of motion \eqref{SEoM} are simply the smearing of the $\phi$-dependent equations \eqref{EqM}. The interpretation of the smearing can then be clarified, as a first step, by considering a limiting case where the $t$-relational setup reduces to the usual $\phi$-relational one. Namely, if one takes a delta distribution\footnote{Note that a distribution would not satisfy the boundary conditions \eqref{bcond}, and also the operator $\nabla_t$ would not be well defined. However, this peculiar case is to be understood only as a limit, for instance by considering the limit of vanishing width for a family of Gaussian functions.} peaked on $\phi$, that is $t(\phi') = \delta(\phi' - \phi)$, then the full $t$-relational setup introduced above naturally reduces to the $\phi$-relational one that is used in all previous works in GFT cosmology. For instance, all the smeared quantities take their usual forms as functions of $\phi$, e.g. $\sigma_{\vchi}(t) = \sigma(\phi)$, $a_{\vchi}(t) = a_{\vchi}(\phi)$, $\braket{N_{\vchi}(t)}_{\sigma,\bar{\sigma};\beta} = \braket{N_{\vchi}(\phi)}_{\sigma,\bar{\sigma};\beta}$. 

Along these lines, one can motivate specific choices of smooth clock functions peaked around points of the base manifold, namely values of $\phi$, for instance Gaussian functions. Such choices could then be interpreted as the implementation of a {\emph deparametrization} procedure at the level of the background independent quantum theory. One could further understand the selection of a relational clock as a restriction to a special sector of physical states in the full (non-deparametrized) quantum theory, as was suggested in \cite{Kotecha:2018gof}. However, in general, one would expect to be able to realise such mechanisms in possibly different ways. For instance, in the present setting, this would correspond to a special choice of smearing functions $t$; while a different possibility using coherent states is explored in \cite{Marchetti:2020umh}, in the context of zero temperature ($\beta=\infty$) GFT condensate cosmology. The complete details of mechanisms for deparametrization, how they relate to each other, and if there could be preferred choices, are interesting queries that are left for future investigations. In this article however, we proceed without any further restriction to a specific class of $t$ functions, and work with the general case. We note that the added generality may also allow for potential switching between relational reference frames in GFT, which is an expected feature of any background independent system devoid of an absolute notion of time or space (see for instance \cite{Hohn:2019cfk,Hohn:2018toe,Hohn:2018iwn}, and references therein).

Furthermore, we notice that the $t$-relational setup presented here is constructed from the full non-deparametrized operator formulation of GFT, with the algebra satisfying \eqref{ladderCR}, and the deparametrization with respect to a relational clock field is implemented via introduction of smearing functions $t(\phi)$, as discussed above. Specifically, the kinematic description of the system is fully covariant, i.e.\ no preferred clock parameter $\phi$ (from possibly several ones \cite{Kotecha:2018gof,Gielen:2020fgi}) or function $t(\phi)$ is chosen as \emph{the} clock. The dynamical description (equations \eqref{EoMCl20}-\eqref{EoMEff20}) is derived using the principle of least action, without the use of any relational Hamiltonian. This setup is then technically different from the one used in some recent works like \cite{Adjei:2017bfm,Wilson-Ewing:2018mrp}. The relational frame used in these other studies, as in all previous works in GFT cosmology \cite{Pithis:2019tvp,Oriti:2016acw,Gielen:2016dss}, is defined with respect to the parameter $\phi$, which as discussed above (see also \cite{Assanioussi:2019ouq}) may lead to divergences. Also, the studies in \cite{Adjei:2017bfm,Wilson-Ewing:2018mrp} are based on a canonical quantization of already deparametrized classical GFT models. Specifically, the kinematic description is canonical with respect to a chosen clock variable $\phi$, with the algebra based on equal $\phi$-time commutation relations. Subsequently, the dynamical description is derived from a clock Hamiltonian. Having said that, the descriptions based on these two, a priori technically different, setups could eventually be related, since they encode the physics of a given system before and after deparametrization. This question is however tightly connected to the open issue of time in quantum gravity. The investigation of this possible relation may help in addressing the question of how physical time emerges in the present background independent theory of quantum gravity.

Returning to the equations of motion \eqref{SEoM}, let us use the standard polar decomposition
\begin{align}
 \sigma_{{\vchi} }(t) = \zeta_{\vchi}^t e^{i \eta_{\vchi}^t}
\end{align}
where $\zeta_{\vchi}^t = \sqrt{ \sigma_{{\vchi} }(t)  \overline{\sigma_{{\vchi} }(t)}}$ is the modulus and $\eta_{\vchi}^t = \tan^{-1} \left( \frac{\text{Im} \, \sigma_{{\vchi} }(t)}{ \text{Re} \, \sigma_{{\vchi} }(t)} \right)$ is the phase of the condensate functional $\sigma_{{\vchi} }(t) \in \mathbb{C}$. Note that the quantities $\zeta_{\vchi}^t$ and $\eta_{\vchi}^t$ do not correspond to the smearing of the modulus $\zeta_{\vchi}(\phi)$ and the phase $\eta_{\vchi}(\phi)$ of the condensate function $\sigma_{{\vchi} }(\phi)$, which were used in the context of GFT cosmology in previous works \cite{Oriti:2016qtz}. Separating the real and imaginary parts of equations \eqref{SEoM}, we obtain
\begin{align}
\nabla_t^2 \zeta_{\vchi}^t - \zeta_{\vchi}^t (\nabla_t \eta_{\vchi}^t)^2 - M_{\vchi} \zeta_{\vchi}^t = 0 \\
 2\nabla_t \zeta_{\vchi}^t \nabla_t \eta_{\vchi}^t + \zeta_{\vchi}^t \nabla_t^2 \eta_{\vchi}^t = 0
\end{align}
for all ${\vchi}$. These two equations imply the existence of two constants of motion, as in the case of $\beta=\infty$ free theory \cite{Oriti:2016qtz}, given by
\begin{align} \label{ej}
 E_{\vchi} & = (\nabla_t \zeta_{\vchi}^t)^2 + (\zeta_{\vchi}^t)^2 (\nabla_t \eta_{\vchi}^t)^2 - M_{\vchi} (\zeta_{\vchi}^t)^2\\
 Q_{\vchi} & = (\zeta_{\vchi}^t)^2 \nabla_t \eta_{\vchi}^t \label{qj}
\end{align}
satisfying $\nabla_t E_{\vchi} = 0$ and $\nabla_t Q_{\vchi} = 0$.



\subsection{Effective cosmology with volume fluctuations} \label{eff3}




\subsubsection{Effective homogeneous and isotropic cosmology} \label{EFhic}

Our investigation is based on four ingredients: the choice of dynamics (here, free GFT); the choice of quantum states in the full theory as an approximate solution (here, coherent thermal states based on the chosen Gibbs state); the choice of relational observables (here, as functionals of $t$); and, the choice of a subclass of condensate wave functions. We have addressed the first three points in sections \ref{vgibbs}-\ref{eff2} above. This brings us to the last one, which we address precisely in line with past studies, as follows. A notion of homogeneity in the present non-spatiotemporal background independent setting resides in: $(i)$ using a coherent state, as the relevant condensate phase for studying the effective cosmology extracted from a GFT model, and $(ii)$ imposing an additional left diagonal symmetry on the condensate wavefunction, i.e. $\sigma(hg_i,\phi) = \sigma(g_i,\phi) , \; \forall h \in SU(2)$. In other words, it resides in the facts that: the collective dynamics is encoded in a left- and right-invariant single-particle wavefunction $\sigma$, which is also the order parameter of the condensate $\braket{a_{\vchi}(t)}_{\sigma,\bar{\sigma};\beta} = \sigma_{\vchi}(t)$, where now ${\vchi} \equiv (\vec{j},\iota_L,\iota_R)$; and, each $a$-quantum in the condensate is being described by the same wavefunction $\sigma$. Further, a notion of isotropy is implemented by: fixing the spins at each 4-valent node to be equal; fixing the two intertwiners $\iota_L,\iota_R$ to be equal (the geometric interpretation of which remains to be understood); and, choosing a special class of intertwiners, namely the eigenvectors of the volume operator with the highest eigenvalue. We refer to past works for detailed discussions on these aspects, for instance \cite{Pithis:2019tvp,Oriti:2016qtz,Oriti:2016acw,Gielen:2014ila,Gielen:2013kla,Gielen:2013naa}. \

These restrictions imply that the condensate function is entirely determined by the value of a single $SU(2)$ spin $j$. It follows that the equations of motion \eqref{SEoM} reduce to one equation for each value of the spin label $j$,
\begin{align}  \label{EqMj}
 \nabla_{t}^2\sigma_{j}(t) - M_j \sigma_{j}(t) = 0 \,, \quad \forall j\in \mathbb{N}/2 \,.
\end{align} 
Consequently we have, for all $j\in \mathbb{N}/2$
\begin{align}
\quad \nabla_t^2 \zeta_j^t - \zeta_j^t (\nabla_t \eta_j^t)^2 - M_j \zeta_j^t = 0 \label{CofM1}\\
 2\nabla_t \zeta_j^t \nabla_t \eta_j^t + \zeta_j^t \nabla_t^2 \eta_j^t = 0 \label{CofM2}
\end{align}
with the same conserved charges \eqref{ej} and \eqref{qj}, now labelled by the spin $j$.  \

Having set all the ingredients for a dynamical analysis, we can now proceed with the derivation of the effective dynamical equations for the average volume $\langle V_t \rangle$ in a coherent thermal state of the form \eqref{cts}, which include geometric volume fluctuations as discussed in \ref{vgibbs}. For simplicity of notation, we will drop the label $t$ on relational quantities (like $\zeta$, $\eta$ and volume averages) in the following. Relational volume average is given by
\be \label{Vt}
 {\mbfV}:=\langle V_t \rangle = \sum_j v_j (\zeta_j^2 + s_j^2||t||^2)
\ee
where $s_j := \sinh\left[ \theta_{j}(\beta)\right]$. Using the effective equations of motion \eqref{CofM1} and \eqref{CofM2}, and the expressions for the constants of motion \eqref{ej} and \eqref{qj}, we obtain
\begin{align}
 {\mbfV}' &:=  \nabla_t {\mbfV} = \, 2 \sum \limits_j v_j \zeta_j \nabla_t \zeta_j  \\  
 &=  \, 2 \sum \limits_j v_j \zeta_j\ \text{sgn}(\zeta'_j)\sqrt{E_j - \frac{Q_j^2}{\zeta_j^2} + M_j \zeta_j^2} \label{dVt} \\ \nonumber \\
\label{d2Vt}
 {\mbfV}'' &:= \nabla_t^2 {\mbfV} = 2 \sum \limits_j v_j (E_j + 2 M_j \zeta_j^2)
\end{align}
where we have used $\nabla_t ||t||^2 = 0$. From here on we shall assume $||t||^2=1$ for convenience. \

Then, the effective generalised Friedmann\footnote{We recall that the relational Friedmann equations of motion in general relativity, for spatially flat FLRW spacetime, with a minimally coupled massless scalar field $\phi$, are: $$ \left(\frac{1}{3V} \frac{dV}{d\phi} \right)^2 = \frac{4\pi G}{3} \,, \quad \frac{1}{V}\frac{d^2V}{d\phi^2} = 12\pi G \,.$$ For a short review in the context of GFT cosmology, we refer to the appendix in \cite{Oriti:2016qtz}.} equations, including both quantum gravitational and statistical volume corrections are
\begin{align} \label{dVt1}
 \left(\frac{{\mbfV}'}{3{\mbfV}}\right)^2 &=  \frac{4}{9}\left( \frac{ \sum_j v_j \zeta_j\ \text{sgn}(\zeta'_j)\sqrt{E_j - \frac{Q_j^2}{\zeta_j^2} + M_j \zeta_j^2} }{  \sum_j v_j \zeta_j^2 + \sum_j v_j s_j^2 } \right)^2 \\
 \frac{{\mbfV}''}{{\mbfV}} &= \frac{2 \sum_j v_j (E_j + 2 M_j \zeta_j^2)}{\sum_j v_j \zeta_j^2 + \sum_j v_j s_j^2} \label{d2Vt1}
\end{align}
which represent the relational evolution for the volume associated to a foliation labeled by the function $t$. Compared to the analogous equations obtained in \cite{Oriti:2016qtz}, the main difference arises due to the expression \eqref{Vt} for the average volume where there appears an additional statistical contribution $s_j^2$, which as we have described above, originates directly from the quantum statistical mechanics of the underlying theory.


\subsubsection{Late times evolution} \label{lateev}

In the following, we will make use of the quantities below that are formally defined as number densities corresponding to the different parameters characterising the different phases of the system,
\begin{align}
n_\co(j) = \zeta_j^2 \, &, \quad n_E(j) = \frac{E_j}{M_j} \,, \\
n_\tth(j) = s_j^2  \, &, \quad n_Q(j) = \frac{Q_j}{\sqrt{M_j}} \,. 
\end{align}
Different physical regimes can then be described in terms of relative strengths of these parameters. We note that $n_\tth$ (equal to \eqref{bose}) and $n_\co$ are the actual number densities of the thermal and condensate parts of the full system, and are therefore non-negative.  \

The domain, $n_\co(j) \gg n_E(j), n_Q(j) $, can be understood as a classical limit where the volume is large but curvature is small \cite{Oriti:2016qtz}. In this regime, we have
\begin{align} \label{sclimit}
 {\mbfV}' &\approx 2 \sum \limits_j \text{sgn}(\zeta'_j) \, v_j \sqrt{M_j} \, \zeta_j^2 \\
 {\mbfV}'' &\approx 4 \sum \limits_j v_j M_j \zeta_j^2
\end{align}
giving the corresponding generalised evolution equations,
\begin{align} \label{GENEVEQ}
 \left(\frac{{\mbfV}'}{3{\mbfV}}\right)^2 &= \frac{4}{9} \left( \frac{  \sum_j  \text{sgn}(\zeta'_j) \, v_j \sqrt{M_j} \, \zeta_j^2 }{ \sum_j v_j \zeta_j^2 + \sum_j v_j s_j^2} \right)^2\\
 \frac{{\mbfV}''}{{\mbfV}} &= \frac{4 \sum_j v_j M_j  \zeta_j^2}{\sum_j v_j \zeta_j^2 + \sum_j v_j s_j^2} \;. \label{GENEVEQtwo}
\end{align}

Further, notice that the thermal contribution ${\mbfV}_\tth := \sum_j v_j n_\tth(j)$ is invariant under variations in the time function $t$, that is $\nabla_t {\mbfV}_\tth = 0$. Hence, as the full system evolves such that the condensate number density $n_\co(j)$ increases monotonously in time, then eventually we will reach the domain where the condensate part, ${\mbfV}_\co := \sum_j v_j n_\co(j)$, dominates the thermal cloud, that is ${\mbfV}_\co \gg {\mbfV}_\tth $. This is the non-thermal limit, $n_\tth \ll n_\co$. Together, we thus have a classical and non-thermal, late times regime
\be \label{flrwlim} n_\co \gg n_\tth, n_Q, n_E \,.  \ee
Further, let us assume
\begin{align}\label{ClassCond}
 \forall j,\quad \text{sgn}(\zeta'_j)=\pm 1 \ ,\ M_j \equiv M=3 \pi G 
\end{align}
where $G$ is Newton's gravitational constant. Then from equations \eqref{GENEVEQ} and \eqref{GENEVEQtwo}, we obtain
\begin{align}
 \left(\frac{{\mbfV}'}{3{\mbfV}}\right)^2 &= \frac{4 \pi G}{3}\\
 \frac{{\mbfV}''}{{\mbfV}} &= 12 \pi G 
\end{align}
which are the relational Friedmann equations for spatially flat FLRW spacetime in general relativity. We have thus recovered the correct classical limit at late times, when the quantum gravity system is in a thermal condensate phase such that both \eqref{flrwlim} and \eqref{ClassCond} hold.
Physically, this regime where the condition \eqref{flrwlim} is satisfied, i.e.~ the contribution coming from the condensate is dominant while the statistical fluctuations are subdominant, corresponds to a phase that effectively mimics a system in a zero temperature condensate\footnote{Note that there could also be a classical regime where the statistical fluctuations are not subdominant. In other words, statistical fluctuations may be important even in regimes where quantum fluctuations are negligible. In the present setting, this would correspond to the case when $n_\co \gg n_E, n_Q$ holds true, but the interplay between $n_\co$ and $n_\tth$ is still relevant.}. Consequently, as shown above, in this regime we simply get the zero temperature condensate cosmology obtained in previous works \cite{Oriti:2016qtz,Oriti:2016ueo,Pithis:2019tvp}. In this sense, the use of zero temperature condensates like $\ket{\sigma}$ can be understood more generally as a thermal quantum gravity condensate being in a dynamical regime where the condensate dominates, $n_\co \gg n_\tth$.


\subsubsection{Early times evolution} \label{earlyev}

We now look at the evolution equations \eqref{dVt1} and \eqref{d2Vt1} in a different phase, in which the thermal contributions and quantum corrections become relevant. For consistency, the choice \eqref{ClassCond}, which recovers the classical limit giving the correct late time behaviour, is assumed in the following. \

Notice that the expression \eqref{dVt} for ${\mbfV}'$ admits roots. Namely, there exist solutions $\{\zeta_j^o\}_j$ such that
\be
 {\mbfV}'=0 \,.
\ee
The solution is given explicitly in terms of $M$ and the constants of motion $E_j,Q_j$ as
\be
n_\co^{o}(j) = -\frac{1}{2}n_E(j) + \sqrt{\frac{1}{4}n_E(j)^2 + n_Q(j)^2} \;, \quad  \forall j 
\ee
where we have ignored the negative solutions, since $n_\co$ is the number density of the condensate and thus must be non-negative. At this stationary point, the total volume is 
\be {\mbfV}^o = \sum_j v_j (n_\co^{o}(j) + n_\tth(j)) \ee 
which is clearly non-zero due to the non-vanishing thermal contribution in the present finite $\beta$ case, even if the condensate contribution were to vanish. However, as it happens, even $n_\co^{o}$ does not vanish as long as $E \neq 0$. In particular, $n_\co^{o} \neq 0$ even if $Q = 0$. To see this, notice that if $Q = 0$, then $E < 0$, which is evident from their expressions in equations \eqref{ej}-\eqref{qj}, assuming a positive $M$ as required by the correct classical limit \eqref{ClassCond}. In this case then, $n_\co^{o} = |n_E|$. In the general case with $Q \neq 0$, both positive and negative $E$ are allowed in principle, but in each case we again have $n_\co^{o} > 0$. Thus, the expectation value ${\mbfV}$ does not vanish when ${\mbfV}' = 0$, implying the existence of a non-vanishing minimum\footnote{This is indeed a minimum since $0 < {\mbfV}''|_{n_\co^{o}}$.} ${\mbfV}^o$ of ${\mbfV}$ throughout the evolution. Physically, in the context of homogeneous and isotropic cosmology, this means that the singularity has been resolved, and that the effective evolution displays a bounce, with a non-zero minimum of the spatial volume. \

Further, this ensures a transition between two phases of the universe characterised by the sign $\text{sgn}(\zeta'_j)$, describing a contracting universe ($\text{sgn}(\zeta'_j)<0$) and an expanding one ($\text{sgn}(\zeta'_j)>0$). Each of these phases behaves according to the general relativistic FLRW evolution in the classical and non-thermal limits \eqref{flrwlim}, that is, when $\zeta_j$ (equivalently the condensate contribution $n_\co(j)$ to the volume) becomes very large with respect to all the constants of motion and the thermal contributions. However, as expected, these two phases display a non-standard evolution in general, especially when close to the bounce. This is the regime where $\zeta_j$ is comparable in magnitude to the other quantities present in the model. This leads to the particularly important question about the presence of accelerated expansion and its magnitude. \

To address this question, we proceed with a simplified analysis, where we make the approximation  of selecting a single spin mode \cite{Oriti:2016qtz, deCesare:2016rsf}, thus dropping the sum over all spins in the various expressions. In this case, the generalised equations of motion \eqref{dVt1} and \eqref{d2Vt1} reduce to,
\begin{align} 
 \left(\frac{{\mbfV}'}{{\mbfV}}\right)^2 &= 4M + 4\left(\frac{E_j \zeta_j^2 - Q_j^2 - 2 M \zeta_j^2 s_j^2 - M s_j^4}{(\zeta_j^2 + s_j^2)^2}\right) \label{v'} \\
 \frac{{\mbfV}''}{{\mbfV}} &= 4M + 2\left(\frac{E_j - 2 M s_j^2}{\zeta_j^2 + s_j^2} \right) \label{v''} \;.
\end{align}
Now, the magnitude of a phase of accelerated expansion can be estimated in terms of: the number of e-folds\footnote{This is analogous to the standard expression for the number of e-folds in classical general relativity, $$ \mcN = \ln \frac{a_{\text{end}}}{a_{\text{beg}}} = \ln \left(\frac{V_{\text{end}}}{V_{\text{beg}}}\right)^{1/3} $$ where $a=V^{1/3}$ is the scale factor in terms of spatial volume.} \cite{deCesare:2016rsf} given by,
\begin{align}
 {\mcN} := \frac{1}{3} \ln \left( \frac{{\mbfV}_{\text{end}}}{{\mbfV}_{\text{beg}}}\right) &= \frac{1}{3} \ln \left( \frac{n_\co^{\text{end}} + n_\tth}{n_\co^{\text{beg}} + n_\tth}\right)
\end{align}
where ${\mbfV}_{\text{beg}}$ and ${\mbfV}_{\text{end}}$ are the average total volumes at the beginning and end of the phase of accelerated expansion respectively; and, in terms of an acceleration\footnote{This is motivated by the expression for acceleration in classical GR. We recall that in spatially flat Friedmann cosmology with a massless scalar field $\phi$, the relative acceleration with respect to proper time $\tau$, written in terms of $\phi$-derivatives is given by $$ \frac{\ddot{a}}{a} = \frac{1}{3}\left[ \frac{\ddot V}{V} - \frac{2}{3}\left( \frac{\dot V}{V}\right)^2 \right] = \frac{1}{3}\left(\frac{\pi_\phi}{V}\right)^2 \left[ \frac{\partial^2_\phi V}{V} - \frac{5}{3}\left( \frac{\partial_\phi V}{V}\right)^2 \right]  $$ where, dots denote $\tau$-derivatives, $V = a^3$, $\pi_\phi$ is conjugate momentum of scalar field $\phi$, and $\dot{\phi} = \pi_\phi/V$ has been used to change variables from $\tau$ to $\phi$. We refer to \cite{deCesare:2016rsf,deCesare:2016axk} for details.} parameter \cite{deCesare:2016rsf} given by,
\be
\mfa := \frac{{\mbfV}''}{{\mbfV}} - \frac{5}{3}\left( \frac{{\mbfV}'}{{\mbfV}} \right)^2 .
\ee
Using equations \eqref{v'} and \eqref{v''} above, we get
\be
\mfa =  -\frac{8}{3}M + 2M \left(\frac{n_E - 2n_\tth }{n_\co + n_\tth}\right)\left(1 - \frac{10}{3}\frac{n_\co}{n_\co + n_\tth}\right)  + \frac{20}{3}M \left( \frac{n_Q^2 + n_\tth^2}{(n_\co + n_\tth)^2} \right)  .
\ee
Assuming that the bounce is the starting point of the expansion phase, we have
\begin{align}
 n_\co^{\text{beg}} = n_\co^o 
\end{align}
for any $j$. Then, it is straightforward to check that acceleration is positive at the beginning, i.e. $\mfa|_{\text{beg}} > 0$, as required. \

Now, the end of accelerated expansion is characterised by $\mfa|_{\text{end}} = 0 \,$, 
which gives,
\be \label{zend}
n_\co^{\text{end}} = \frac{3}{4} n_\tth - \frac{7}{8} n_E + \sqrt{ \frac{49}{64} n_E^2 +\frac{5}{2}n_Q^2 + \frac{9}{16} \left(n_\tth^2 -  n_E n_\tth \right)} 
\ee
assuming an expanding phase of the universe, i.e. ${\mbfV}_{\text{end}} > {\mbfV}_{\text{beg}}$, and non-negativity of $n_\co^{\text{end}}$ even when $n_\tth$ is negligible. The number of e-folds can thus be estimated by,

\begin{align}
e^{3\mcN} = \frac{ \frac{7}{4}\left(n_\tth - \frac{1}{2}n_E\right) + \sqrt{\frac{49}{64}n_E^2 + \frac{5}{2}n_Q^2 + \frac{9}{16}\left( n_\tth^2 - n_En_\tth \right) } }{ \left( n_\tth - \frac{1}{2}n_E \right) + \sqrt{\frac{1}{4}n_E^2 + n_Q^2 } } \,.
\end{align}

Since the quantities $n_E$, $n_Q$ and $n_\tth$ are all independent, we can observe three interesting regimes, giving approximate numerical values for $\mcN$:
\begin{align}
n_\tth, n_Q \ll n_E \ &:\qquad {\mcN} \approx \frac{1}{3} \ln\left( \frac{7}{4} \right) \approx 0.186\\
n_\tth , n_E \ll n_Q \ &:\qquad {\mcN} \approx \frac{1}{6} \ln\left(\frac{5}{2} \right) \approx 0.152\\
n_E , n_Q \ll n_\tth \ &:\qquad {\mcN} \approx \frac{1}{3} \ln\left( \frac{5}{2} \right) \approx 0.305
\end{align}
The upper bound on $\mcN$ in the previous zero temperature free theory analysis \cite{deCesare:2016rsf} is $0.186$, while here for finite $\beta$ free theory it is $0.305$, achieved in an early time limit. This difference is attributed to the only new aspect that we have introduced in the model, the thermal cloud of quanta of geometry. This shows that the number of e-folds can be increased, even without a non-linear dynamics. This fact is in contrast with the previous conclusions \cite{deCesare:2016rsf}, that non-linear interaction terms in the GFT action are necessary to increase $\mathcal{N}$. The interaction terms are naturally accompanied by their corresponding coupling constants. These are free parameters which can then be fine-tuned to essentially give the desired value for $\mcN$, as in \cite{deCesare:2016rsf}. \

However, it remains true that the increase in $\mcN$ achieved in our present case is very minimal, and still not sufficient to match the physical estimates of $\mcN \sim 60$. Nevertheless, we expect this to be overcome by going beyond a static thermal cloud. In other words, a dynamical thermal cloud, which would be expected to be left over from a geometrogenesis phase transition (of an originally unbroken, pre-geometric phase), could have the potential to provide a viable mechanism for an extended phase of geometric inflation. We note that the implementation of dynamical statistical fluctuations would require special care in order to avoid pathological behaviours with regards to the use of relational clock functions $t(\phi)$. This would be in addition to the standard requirement of having a stable macroscopic phase, wherein fluctuations in the relevant observables are sufficiently subdominant or even decaying, throughout the evolution of the universe. We briefly discuss the issue of extending to a dynamical thermal cloud in section \ref{disc} below, but its complete investigation is left to future work. \

Finally, we expect that removing the restriction to a single spin mode in the calculation above, would not alter the qualitative conclusion that a static thermal cloud is insufficient to generate a satisfactory number of e-folds to match the observational estimate. However, a relaxation of the condition \eqref{ClassCond} for $\text{sgn}(\zeta'_j)$, by considering a non-homogeneous distribution of the sign with respect to the modes $j$ while preserving the classical limit manifest in the emergence of Friedmann equations at large volumes, might give rise to a larger ratio between the volumes at the end and at the beginning of the phase of accelerated expansion, and consequently a larger number of e-folds.


\subsection{Remarks} \label{disc}

We have presented above an effective, relational homogeneous and isotropic cosmological model based on the use of a condensate phase with fluctuating geometric volume, and an introduction of reference clock functions. In particular, we have considered a non-interacting class of group field theory models, and condensates that are characterised by non-dynamical thermal clouds. Below, we conclude with some discussions surrounding these two aspects, namely having an overall free dynamics and considering a thermal condensate with a static cloud, which are in fact mutually related.

There are three main features of our specific choice of state, namely a coherent thermal state of the form \eqref{cts} at inverse temperature $\beta$. Firstly, $\beta$ is assumed to be constant. Secondly, the average number of quanta in this state splits neatly into a condensate and a non-condensate part, such that the zero temperature limit gives a pure $\beta$-independent condensate (see also equation \eqref{zerolim}), i.e.
\begin{align}
\langle a^\dag  a \rangle_{\beta} &= n_\co + n_{\text{non-co}} \\
 \lim_{\beta \to \infty} \langle a^\dag a \rangle_{\beta} &= n_\co 
 \end{align}
where in the present work the non-condensate part is taken to be thermal and at equilibrium at inverse temperature $\beta$. Having such a split is not only convenient in doing computations but also adds clarity to the expressions in subsequent analyses when considering the interplays between the two. Moreover, having this $\beta \to \infty$ limit is crucial for recovering the results of past studies in GFT cosmology. Thirdly, the expectation value of the field operator in a coherent thermal state is temperature independent, i.e.
\be \label{orderpar} \braket{a_{{\vchi}\alpha}}_{\sigma,\bar{\sigma};\beta}   = \sigma_{{\vchi}\alpha}  = \braket{a_{{\vchi}\alpha}}_{\sigma} \ee
thus being identical to the zero temperature case. At first sight this seems contrary to our expectation that the condensate would be affected by the presence of a thermal cloud, which is indeed true in general. But what is also true is that, the independence of $\sigma_{{\vchi}\alpha}$ from $\beta$ in the present case, is entirely compatible with our current approximation of neglecting interactions and taking $\beta$ constant.
We know that when temperature is switched on, quanta from the condensate are depleted into the thermal cloud. Now in a generally interacting case, both the thermal cloud and the condensate are interacting and dynamical by themselves, while also interacting with each other. However, in the case of free dynamics, the thermal cloud will not interact with the condensate part, in addition to the quanta also being free within each part separately. Thus even though our state includes a thermal cloud, the coherent condensate part (described fully by its order parameter $\sigma$) will be unaffected by it, indeed as depicted by equation \eqref{orderpar} above. This is further reasonable in light of having a constant $\beta$, because if temperature were to change, say to increase, then we would expect more quanta to be depleted into the thermal cloud, and thus expect the state of the condensate, i.e. its order parameter, to also change. \

So overall, considering this class of states, in which the order parameter is temperature independent, is a reasonable approximation when the temperature is constant and interactions are neglected. Not including interactions ensures that the thermal cloud doesn't affect the condensate, while constant $\beta$ ensures that the amount of depletion is also constant, so together the condensate can indeed be approximated by a $\beta$-independent order parameter. In such a case we may be missing out on some interesting physics, however we take this case as a first step towards further investigations in the future. \

Finally, we note that the interesting case of a dynamically changing $\beta$ is also left for future studies. In such a case, the expected dominance of the condensate part over the thermal cloud at late times would not only be determined by a dynamically increasing condensate (as is the case in the present work), but also by what would be a dynamically decreasing temperature as the universe expands.


\begin{subappendices}

\section{Gibbs states for positive extensive generators} \label{posapp2}

We show that the operator $e^{-\beta(\mcP - \mu N)}$ (see equation \eqref{gcposrho}) is bounded, positive and trace-class on $\mcH_F$, and subsequently calculate its trace. Notice that it is  self-adjoint by definition, due to the self-adjointness of the exponent. In the following, we use the orthonormal occupation number basis $\{\ket{\{n_{\vchi_i\valpha_i}\}}\}$ as introduced in section \ref{occunum}, and denote the modes by $i \equiv \vchi_i,\valpha_i$ for convenience. \cite{Kotecha:2018gof} \\


\noindent \textbf{Lemma 1.} Operator $e^{-\beta(\mcP-\mu N)}$ is bounded in the operator norm on $\mathcal{H}_F$, for $0 < \beta < \infty$ and $\mu \leq \lambda_0$, where $\lambda_0 = \min(\lambda_{\vchi,\valpha})$ and $\mcP$ as defined in \eqref{posopp}.  \medskip

\noindent \textbf{Proof 1.} An operator $A$ is called bounded if there exists a real $k \geq 0 $ such that $||A\psi|| \leq k||\psi|| $ for all $\psi$ in the relevant Hilbert space. Recall that,
\be
e^{-\beta (\mcP-\mu N)} \ket{\{n_i\}} = e^{-\beta \sum_{i'} (\lambda_{i'} - \mu) n_{i'}} \ket{\{n_i\}} 
\ee
where, $\ket{\{n_{\vchi_i\valpha_i}\}}$ are the basis vectors. Then, for a generic state 
\be \ket{\psi} = \sum \limits_{\{n_i\}} c_{\{n_i\}} \ket{\{n_i\}} \,\in \mcH_F \ee 
with coefficients $c_i \in \mathbb{C}$, we have
\begin{align} 
||e^{-\beta (\mcP-\mu N)} \psi||^2 &= || \sum_{\{n_i\}} c_{\{n_i\}} e^{-\beta \sum_i (\lambda_i - \mu) n_i} \ket{\{n_i\}} ||^2 \\ 
&= \sum_{\{n_i\}, \{n_{i'}\}}  \bar{c}_{\{n_{i'}\}} c_{\{n_i\}} e^{-\beta \sum_{i'} (\lambda_{i'} - \mu) n_{i'}} e^{-\beta \sum_i (\lambda_i - \mu) n_i} \langle \{n_{i'}\} | \{n_i\} \rangle \\ 
&= \sum_{\{n_i\}} | c_{\{n_i\}} |^2  e^{- 2 \beta \sum_i (\lambda_i - \mu) n_i} \leq \sum_{\{n_i\}} | c_{\{n_i\}} |^2 = || \psi ||^2 
\end{align}
using orthonormality of basis; and, $0 < e^{- 2 \beta \sum_i (\lambda_i - \mu) n_i} \leq 1$ (since $\beta \sum_i (\lambda_i - \mu) n_i \geq 0$, where $\mu \leq \lambda_i$ for all $i$). $\hfill \square$ \\


\noindent \textbf{Lemma 2.} Operator $e^{-\beta(\mcP-\mu N)}$ is positive on $\mathcal{H}_F$, for $0 < \beta < \infty$ and $\mu \leq \lambda_0$. \medskip

\noindent \textbf{Proof 2.} A bounded self-adjoint operator $A$ is positive if $\langle \psi | A \psi \rangle \geq 0$ for all $\psi$ in the relevant Hilbert space. Then, for any $\psi \in \mcH_F$, we have
\begin{align} 
\bra{\psi} e^{-\beta (\mcP - \mu N)} \ket{\psi} &= \sum_{\{n_i\}, \{n_{i'}\}}  \bar{c}_{\{n_{i'}\}} c_{\{n_i\}} \bra{\{n_i\}} e^{-\beta \sum_{i'} (\lambda_{i'} - \mu) n_{i'}} \ket{\{n_{i'}\}} \\ 
&= \sum_{\{n_i\}}  |c_{\{n_i\}}|^2 e^{-\beta \sum_i (\lambda_i - \mu) n_i} \geq 0 \,.
\end{align} $\hfill \square$ \\


\newpage 

\noindent \textbf{Lemma 3.} Operator $e^{-\beta (\mcP-\mu N)}$ is trace-class on $\mathcal{H}_F$, for $0 < \beta < \infty$ and $\mu < \lambda_0$. Further, its trace is given by, 
\be Z_{\beta,\mu} = \prod_{\vchi,\valpha} \frac{1}{1-e^{-\beta(\lambda_{\vchi\valpha}-\mu)}} \ee 
as in equation \eqref{zgc}.  \medskip

\noindent \textbf{Proof 3.}\footnote{This proof is along the lines of that reported in \cite{Kotecha:2018gof}, but is detailed here further, for grand-canonical-type states, for clarity and consistency.} 
A bounded operator $A$ on a Hilbert space is trace-class if $\Tr(|A|) < \infty$, where $|A| := \sqrt{A^\dag A}$. Further, for a self-adjoint positive $A$, we have\footnote{We remark that there is, in fact, an iff equivalence between the following: for any bounded operator $B$, and some orthonormal basis $\{e_n\}$ in a complex Hilbert space, we have \cite{beltrametti,moretti} $$\sum_n || B e_n || < \infty \Leftrightarrow \Tr(|B|) < \infty \,.$$ However for the purposes of the present proof, we only need the forward implication and the result \eqref{APOS}, i.e. the implication shown in \eqref{IMPTC}.} 
\be \label{IMPTC} \Tr(A)< \infty \Rightarrow \Tr(|A|) < \infty \,. \ee  One way to see the validity of \eqref{IMPTC} is as follows \cite{beltrametti,moretti,reedsimon,Bratteli:1979tw}. Let $\{e_i\}$ be an orthonormal basis in a complex Hilbert space, such that $A\ket{e_i} = a_i\ket{e_i}$. Notice that, for self-adjoint positive $A$, we have
\begin{align}
\sum_i || Ae_i || = \sum_i \sqrt{(e_i,A^\dag Ae_i)} &= \sum_i \sqrt{(e_i,A^2 e_i)} \label{ASE} \\
&= \sum_i \sqrt{a_i^2} \label{AEB} \\
&= \sum_i a_i = \Tr(A) \label{APOS}
\end{align}
where we have used the self-adjointness of $A$ in \eqref{ASE}, and positivity in \eqref{APOS}. Then,
\begin{align}
\Tr(|A|) = \sum_i (e_i,|A|e_i) &\leq \sum_i |(e_i,|A|e_i)| \\
&\leq \sum_i || \, |A|e_i \,||  \label{CSIE} \\
&= \sum_i || A e_i || = \Tr(A) \label{RESPOS}
\end{align}
where we have used the Cauchy-Schwarz inequality in \eqref{CSIE}, result \eqref{APOS} in \eqref{RESPOS}, and the following identity \cite{reedsimon},
\be ||\,|A|\psi\,||^2 = (\psi,|A|^2\psi) = (\psi,A^\dag A \psi) = ||A\psi||^2 \ee
for any vector $\psi$ in the given Hilbert space. Thus $\Tr(|A|)  \leq \Tr(A)$, which implies \eqref{IMPTC}. In our case with $A=e^{-\beta (\mcP-\mu N)}$, it is therefore sufficient to show that $\Tr(e^{-\beta (\mcP-\mu N)})$ converges in $\mcH_F$. \medskip

\noindent Then,
\begin{align}
\Tr(e^{-\beta (\mcP-\mu N)}) &= \sum_{N\geq 0} \; \sum_{\{n_i\}|_N} \bra{\{n_i\}} e^{-\beta (\mcP-\mu N)} \ket{\{n_i\}} \label{NEWTr} \\
&=\sum_{N\geq 0}\; \sum_{\{n_i\}|_{N }} e^{-\beta \Lambda_{\{n_i\}_N} } 
\end{align}
where we have denoted,
\be \Lambda_{\{n_i\}_N} \equiv \sum_i (\lambda_i - \mu) n_i \,. \ee
The second sum in \eqref{NEWTr} is restricted to those microstates $\ket{\{n_i\}}$ with a fixed $N = \sum_{i} n_{\vchi_i,\valpha_i}$. This can be understood as summing over all possible ways of arranging $N $ particles into an arbitrary number of boxes, each labelled by a single mode $i = (\vchi_i,\valpha_i)$. Notice that the dominant contribution to this sum comes from the state with the minimum $\Lambda$. This is the ground state $\ket{N ,0,0,...}$, in which all $N $ particles occupy the single-particle ground state of $\mcP$ (for a fixed $\mu$) with eigenvalue $\lambda_0$. Then, 
\be \Lambda_0 = (\lambda_0 - \mu)N \ee 
characterises the $N$-particle ground state. Let us separate this contribution to rewrite the sum as,
\begin{equation}
\sum_{\{n_i\}|_{N }} e^{-\beta \Lambda_{\{n_i\}_N} } \;\;=\;\; e^{-\beta \Lambda_0} \;\;+ \sum_{\ket{\{n_i\}} \neq \ket{N ,0,...}} e^{-\beta \Lambda_{\{n_i\}_N} } \label{partSUM}
\end{equation}
where now all sub-dominant terms are $e^{-\beta \Lambda_{\{n_i\}_N} } < e^{-\beta \Lambda_0}$. Further, we can rearrange the states in the sum (on the right hand side in \eqref{partSUM}) according to increasing values of $\Lambda$, and denote these with tildes in the new sequence, to get 
\begin{equation}
\sum_{\ket{\{n_i\}} \neq \ket{N ,0,...}} e^{-\beta \Lambda_{\{n_i\}_N} } \;\; = \;\; \sum_{\widetilde{\Lambda}_{\{n_i\}_{l,N}} > \Lambda_0} e^{-\beta \widetilde{\Lambda}_{\{n_i\}_{l,N}} } \;\;\equiv \;\; \sum_{\widetilde{\Lambda}_{l,N} > \Lambda_0} e^{-\beta \widetilde{\Lambda}_{l,N} } \,.
\end{equation}
Index $l \in \{1, 2, 3, ... \}$ labels the reorganised list of $N$-particle states in ascending order of their eigenvalues, i.e. $\widetilde{\Lambda}_{l,N} \leq \widetilde{\Lambda}_{l+1,N}$, where equality refers to possible degeneracies in adjacent states in the sequence. This can be taken into account explicitly by a further rewriting, as
\be \sum_{\widetilde{\Lambda}_{l,N} > \Lambda_0} e^{-\beta \widetilde{\Lambda}_{l,N} } = \sum_{\widetilde{\Lambda}_{L,N} > \Lambda_0} \mathbf{g}_L \, e^{-\beta \widetilde{\Lambda}_{L,N} } \ee
where now, $L \in \{1, 2, 3, ... \}$ labels the level set with eigenvalue $\widetilde{\Lambda}_L$, and we have
\be \widetilde{\Lambda}_{L,N} < \widetilde{\Lambda}_{L+1,N} \,, \quad \widetilde{\Lambda}_{L,N} < \widetilde{\Lambda}_{L,N+1} \,. \ee 
$\mathbf{g}_L$ is the degree of degeneracy\footnote{Degree of degeneracy is the dimension of the eigenspace of the degenerate states, or equivalently, the multiplicity of the corresponding spectral value.} of the level $L$, which is assumed to be finite for every $L$, and such that it increases less than exponentially with increasing $\tilde{\Lambda}$. \medskip

\noindent Together, we have the double series
\be \label{DSer} \Tr(e^{-\beta (\mcP-\mu N)}) = \sum_{N \geq 0} \sum_{\widetilde{\Lambda}_{L,N} \geq \Lambda_0}  \mathbf{g}_L \, e^{-\beta \widetilde{\Lambda}_{L,N} } \ee
where now, we have allowed for degeneracy in $\Lambda_0$ (for generality), and $\widetilde{\Lambda}_{L=0} = \Lambda_0$. Each (so-called row and column) series, of the above double series, converges by ratio test,
\be 
 \lim_{L \rightarrow \infty} \frac{\mathbf{g}_{L+1}}{\mathbf{g}_L} \, e^{-\beta (\widetilde{\Lambda}_{L+1,N} - \widetilde{\Lambda}_{L,N})} < 1 \,, \quad  \lim_{N \rightarrow \infty}  e^{-\beta (\widetilde{\Lambda}_{L,N+1} - \widetilde{\Lambda}_{L,N})} < 1 \,.
\ee
Therefore, the double series in \eqref{DSer} converges absolutely \cite{ghorpade2010}. $\hfill \square$ \medskip

\noindent We remark that the above proof can be applied directly to the analogous case of canonical states of the form \eqref{posextrho}, to show that they are trace-class in a given $N$-particle Hilbert space \cite{Kotecha:2018gof}. \\ 



\noindent The trace can be evaluated straightforwardly, exactly along the lines of calculation of a grand-canonical partition function for an ideal Bose gas in standard quantum statistical mechanics. In occupation number basis, we have
\begin{align}
\Tr(e^{-\beta(\mcP - \mu N)})  &=    \sum_{\{n_i\}} e^{-\beta \sum_i(\lambda_i - \mu)n_i } \\
 &= \sum_{\{n_i\}} e^{-\beta (\lambda_0 - \mu)n_0} e^{-\beta (\lambda_1 - \mu)n_1}...  \\
&= \sum_{n_0=0}^\infty e^{-\beta (\lambda_0 - \mu)n_0} \sum_{n_1=0}^\infty e^{-\beta (\lambda_1 - \mu)n_1} ... \\
&= \prod_{\vchi,\valpha} \sum_{n_{\vchi\valpha}=0}^\infty e^{-\beta (\lambda_{\vchi\valpha} - \mu)n_{\vchi\valpha}} = \prod_{\vchi,\valpha} \frac{1}{1-e^{-\beta(\lambda_{\vchi\valpha}-\mu)}} \,.
\end{align} $\hfill \square$ \\


\noindent \textbf{Remark.} The above Proofs $1-3$ are directly applicable to the analogous case of self-adjoint extensive operators $\mcP$ with discrete spectrum that are semi-bounded on $\mcH_F$, instead of simply being positive. Being semi-bounded essentially means that the spectrum is either bounded from below or from above. In the case when $\mcP$ is bounded from below, i.e. $\exists$ a minimum eigenvalue $\lambda_0$ (be it non-negative or negative), then the corresponding grand-canonical state satisfies the above proofs for $0 < \beta < \infty$ and $\mu < \lambda_0$. An example of this case is precisely the one that is considered in the above Lemmas. On the other hand, in the case when $\mcP$ is bounded from above, i.e. $\exists$ a maximum eigenvalue $\lambda_0$ (be it non-negative or negative), then the corresponding grand-canonical state satisfies the above proofs for $-\infty < \beta < 0$ and $\mu > \lambda_0$. An example of this case is the state \eqref{RHOMOM} considered in section \ref{inttrans}, for the regularised momentum operators generating internal translations along $\mathbb{R}^n$.


\section{KMS condition and Gibbs states} \label{appkmsgibb}

We consider the question of when Gibbs density operators are the unique normal KMS states. The statements of the lemma, and of the following corollary, are only slight variations of the traditional one encountered in standard algebraic quantum statistical mechanics (see Remark 1 below). The proof is directly in line with the standard ones; specifically, we have made combined use of strategies suggested in \cite{Haag:1992hx} and \cite{Emch1980}. \\

\noindent \textbf{Lemma.} Let $\mathfrak{A}$ be a C*-algebra which is irreducible on a Hilbert space $\mcH_{\mfA}$, i.e. the commutant $\mathfrak{A}' := \{A \in \mathcal{B}(\mcH_{\mfA}) \, | \, [A,B] = 0, \; \forall  B \in \mfA \}$ is a multiple of identity, $\mathfrak{A}'= \mathbb{C}I \equiv \{ cI \,|\, c\in \mathbb{C}\}$. Let $\alpha_t$ be a 1-parameter group of *-automorphisms of $\mfA$ which are implemented on $\mcH_{\mfA}$ by a group of unitary operators $U(t) = e^{i\mcG t}$, where $t\in \mathbb{R}$ and $\mcG$ is a self-adjoint operator on $\mcH_{\mfA}$. Consider a normal algebraic state $\omega_\rho$ on $\mfA$, i.e. $\omega_\rho[A] = \Tr(\rho A)$ (for all $A \in \mfA$), where $\rho$ is a density operator on $\mcH_{\mfA}$. Then, $\omega_\rho$ is an $\alpha$-KMS state at value $\beta$, if and only if the corresponding density operator is of the Gibbs form, 
\be \rho = c \, e^{-\beta \mcG} \ee
where, $c \in \mathbb{C} \backslash\{0\}$ and $\beta \in \mathbb{R}_{>0}$. \medskip

\noindent \textbf{Proof.} Since $\omega_\rho$ is a KMS state, by definition (see page \pageref{FNstate}) we have a complex function $F_{AB}(z)$, for every $A,B\in \mfA$, which satisfies the following boundary conditions,
\begin{align}
F_{AB}(t) &= \Tr(\rho \, A \, e^{i\mcG t}B e^{-i\mcG t})  \label{BCkms1} \\
F_{AB}(t + i\beta) &= \Tr(\rho \, e^{i\mcG t}B e^{-i\mcG t} \, A) \label{BCkms2}
\end{align}
where $t = \text{Re}(z)$. Further by definition, $F_{AB}$ is analytic on $\mfI = \{z \in \mathbb{C} \; | \; 0 < \text{Im}(z) < \beta \}$, and continuous and bounded on its closure $\bar{\mfI}$, for a given $0 < \beta < \infty$. \medskip

\noindent Consider $A = I$. We can do this because the above conditions must hold for every pair of elements in $\mfA$, thus also for this simple case. Then, we have an equality 
\be F_{IB}(t) = F_{IB}(t+i\beta) \ee
for all $t \in \mathbb{R}$ and $B \in \mfA$. Since $F_{IB}$ is analytic on $\mfI$, we can continuously move to another point, say $z = (t+a)+ib \in \mfI$, where the equality will still hold, i.e. $F_{IB}(z) = F_{IB}(z+i\beta)$. By analytic continuation, we see that $F_{IB}$ is analytic on full $\mathbb{C}$, and periodic along the imaginary axis, with period $\beta$. Further recall that, $F_{IB}$ is bounded. Therefore, $F_{IB}$ is bounded and analytic on $\mathbb{C}$, and by Liouville's theorem, it is constant. That is,
\be  \Tr(\rho \, e^{i\mcG t}B e^{-i\mcG t}) = \text{constant} \,. \ee
Since this is true for all $t$ and $B$, we have that the state itself is invariant, i.e.
\be \label{RINV} e^{i\mcG t} \rho e^{-i\mcG t} = \rho  \ee
as expected. In this simple setting, we have basically recovered the known result that any algebraic KMS state, say $\omega_{\text{KMS}}$, is stationary with respect to its defining automorphism group, i.e. $\omega_{\text{KMS}}[\alpha A] = \omega_{\text{KMS}}[A]$ (for all $A$). Result \eqref{RINV} will be used in the following. \medskip

\noindent Now for any $A,B \in \mfA$, let us consider the boundary conditions \eqref{BCkms2} and \eqref{BCkms1} as follows. For \eqref{BCkms2}, we have
\begin{align}
F_{AB}(t + i\beta) &= \Tr(e^{i\mcG t}e^{-i\mcG t} \, \rho \, e^{i\mcG t}B e^{-i\mcG t} \, A) \\
&= \Tr(e^{i\mcG t} \, \rho \, B e^{-i\mcG t} \, A) \\
&= \Tr(\rho \, B \, e^{-i\mcG t} A e^{i\mcG t}) \label{Rzero}
\end{align}
using \eqref{RINV}, and cyclicity of trace. For \eqref{BCkms1}, by continuity we get
\begin{align}  
F_{AB}(t) \to F_{AB}(t + i\beta) &= \Tr(\rho \, A \, e^{i\mcG (t + i\beta)}B e^{-i\mcG (t + i\beta)})  \\
&= \Tr(\rho \, A \, e^{i\mcG t}\,e^{-\beta \mcG}\, B \, e^{-i\mcG t} \, e^{\beta \mcG}) \\
&= \Tr(e^{-\beta \mcG}\, B \, e^{\beta \mcG} \, e^{-i\mcG t} \,\rho \, A \, e^{i\mcG t}) \\
&= \Tr(e^{-\beta \mcG}\, B \, e^{\beta \mcG} \, \rho \, e^{-i\mcG t} A e^{i\mcG t})  \label{Lzero}
\end{align}
again using \eqref{RINV}, and cyclicity of trace. Then from \eqref{Lzero} and \eqref{Rzero}, we have
\be \Tr(e^{-\beta \mcG}\, B \, e^{\beta \mcG} \, \rho \, e^{-i\mcG t} A e^{i\mcG t}) = \Tr(\rho \, B \, e^{-i\mcG t} A e^{i\mcG t})  \ee
which, we recall, holds for every $A \in \mfA$, $t\in \mathbb{R}$, by definition of a KMS state. Therefore,
\begin{align} 
e^{-\beta \mcG} B \, e^{\beta \mcG} \rho = \rho \, B \quad &\Rightarrow \, B \, e^{\beta \mcG} \rho =  e^{\beta \mcG} \rho \, B \\
&\Rightarrow \, [e^{\beta \mcG} \rho, B] = 0 \quad (\forall B \in \mfA) \\
&\Rightarrow \, e^{\beta \mcG} \rho \in \mfA'  \\
&\Rightarrow \, e^{\beta \mcG} \rho = cI 
\end{align} 
where the last step is due to irreducibility of $\mfA$ on $\mcH_{\mfA}$. Validity of the converse statement, that a Gibbs state satisfies the KMS condition, can be found detailed in numerous standard texts (see, for example \cite{Haag:1992hx,Emch1980,Strocchi:2008gsa}, or the original papers \cite{Kubo:1957mj,Martin:1959jp,Haag:1967sg}).  $\hfill \square$ \\

\noindent \textbf{Remark 1.} In standard quantum statistical mechanics for material systems on spacetime, a commonly encountered example of the above is the case of finite systems, e.g. a box of gas \cite{Haag:1992hx,Emch1980,Strocchi:2008gsa}. In this case, $\mathfrak{A} = \mathcal{B}(\mathcal{H})$ is the algebra of all bounded linear operators on some Hilbert space, which is naturally irreducible. Then, given the unitary group of time translations $U(t)=e^{iHt}$, we have that canonical Gibbs states $e^{-\beta H}$ are the unique normal KMS states, according to the above proof. This can be extended directly for grand-canonical states, by considering instead a generator $\tilde{H} = H - \mu N$ \cite{Haag:1992hx,Strocchi:2008gsa}. \\

\noindent \textbf{Remark 2.} There are two main features of the system that are vital for this proof: existence of a 1-parameter group of unitary operators $U(t)$ which is strongly continuous (thus, associated with a self-adjoint generator $\mcG$, by Stone's theorem); and, irreducibility of the algebra $\mfA$ on a given Hilbert space. As long as any algebraic system has these two features, the above proof will follow through. Therefore, we have the following: \\

\noindent \textbf{Corollary.} Let $\mathcal{A}$ be a C*-algebra, and $\pi$ be an irreducible representation of $\mathcal{A}$, with GNS Hilbert space $\mcH_\pi$. By irreducibility we have, $\pi(\mathcal{A})' = \mathbb{C}I$. Let $\alpha_t$ be a 1-parameter group of automorphisms of $\mathcal{A}$, which are implemented by a strongly continuous group of unitary operators $U(t) = e^{i\mcG t}$ on $\mathcal{H}_\pi$. Then, the unique normal KMS state, with respect to $U(t)$ at value $\beta$, on the algebra $\pi(\mathcal{A})$ is a Gibbs state, of the form $e^{-\beta \mcG}$.  



\section{Strong continuity of map $U_X$} \label{appcontgrp}

\noindent \textbf{Lemma.} Let $\mathbb{G}$ be a Lie group with Lie algebra $\mathfrak{G}$, $\mathfrak{H}$ be a Hilbert space, and $\mathcal{U}(\mathfrak{H})$ be the group of unitary operators on $\mathfrak{H}$, as considered in section \ref{inttrans}. Given a continuous map $g_X: \mathbb{R} \to \mathbb{G} ,\, t \to g_X(t)$, and a strongly continuous map $U: \mathbb{G} \to \mathcal{U}(\mathfrak{H}) ,\, g \mapsto U(g)$, then $U_X := U \circ g_X : t \mapsto U_X(t)$ is strongly continuous in $\mathfrak{H}$, i.e.
\be || (U_X(t_1) - U_X(t_2)) \psi || \to 0 \, \quad \text{as} \; t_1 \to t_2 \ee
for a given $X\in \mathfrak{G}$, any $t_1,t_2 \in \mathbb{R}$, and all $\psi \in \mathfrak{H}$. \cite{Kotecha:2018gof} \medskip

\noindent \textbf{Proof.} Recall that, by strong continuity of $U$ we have 
\be \label{STRU} ||(U(g_1) - U(g_2))\psi|| \to 0 \, \quad \text{as} \; g_1 \to g_2 \ee
for any $g_1,g_2 \in \mathbb{G}$ and all $\psi \in \mathfrak{H}$. Then, for any $t_1,t_2 \in \mathbb{R}$ and all $\psi \in \mathfrak{H}$, we have
\begin{align}
||(U_X(t_1) - U_X(t_2))\psi|| &= ||(U(g_X(t_1)) - U(g_X(t_2)))\psi|| \\
&= ||(U(g_1) - U(g_2))\psi||
\end{align}
where $g_1 \equiv g_X(t_1)$ and $g_1 \equiv g_X(t_2)$ are arbitrary elements on the curve $g_X(t) \in \mathbb{G}$. Now, by continuity of $g_X$, we have $g_1 \to g_2$ as $t_1 \to t_2$. Then, using strong continuity of $U$ in \eqref{STRU}, we have
\be
||(U(g_1) - U(g_2))\psi|| \to 0  \,,\quad \text{as} \; t_1 \to t_2 \,.
\ee   $\hfill \square$


\section{Normalisation of Gibbs state for closure condition} \label{AppClosure}

\noindent \textbf{Lemma.} The partition function for the generalised Gibbs state $\rho_\beta$ (equation \eqref{closuregibbs}) on $\Gamma_{\{A_I\}}$ with respect to the closure condition, is given by
\be Z_\beta = \prod_{I=1}^4 \frac{4\pi A_I}{||\beta||} \sinh (A_I ||\beta||)  \ee
as in equation \eqref{closZb}. \medskip

\noindent \textbf{Proof.} This proof follows the strategy presented in \cite{e18100370}, for the analogous case of action of 3d rotations on a single sphere. Recall that, $\beta \in \su(2)$, $\Gamma_{\{A_I\}} = S^2_{A_1} \times \dots \times S^2_{A_4} \ni m$, and $A_I \in \mathbb{R}_{>0}$ ; $J(m) := \sum_{I=1}^4 X_I \in \su(2)^*$ is a smooth dual algebra-valued function on $\Gamma_{\{A_I\}}$, with $||X_I ||=A_I$; and, $\beta.J(m)$ is a smooth real-valued function on $\Gamma_{\{A_I\}}$. Then, the partition function is
\begin{align} 
Z_{\beta} &= \int_{\Gamma_{\{A_I\}}} d\lambda \; e^{-{\beta \cdot J}}  \\
&=  \int_{S^2_{A_1}} \dots \int_{S^2_{A_4}} [d\mu]^4 \; e^{-\sum_{I=1}^4 \beta . X_I} \\
&= \prod_{I=1}^4 \int_{S^2_{A_I}} d\mu \; e^{-\beta . X_I}
\end{align}
where $d\mu$ is an area element on a single 2-sphere with a fixed radius $A_I$. \medskip

\noindent We can naturally embed this system in Euclidean $\mathbb{R}^3$, by identifying both $\su(2)$ and $\su(2)^*$ with $\mathbb{R}^3,$\footnote{There exists a canonical isomorphism between 3d vectors and rotation matrices (i.e. adjoint representation of $\su(2)$ or $\mathfrak{so}(3)$) given by: $\mathbb{R}^3 \ni \vec{b} = (b_1,b_2,b_3) \mapsto \hat{b} = \sum_{a=1}^3 b_a \hat{L}_a \in \su(2)$, where $\h{L}_a$ are matrix generators of 3d rotations. Equivalently, in the standard cartesian basis, the isomorphism is: $\vec{e}_x,\vec{e}_y,\vec{e}_z \mapsto \hat{L}_1,\hat{L}_2,\hat{L}_3$. The Lie bracket structure on $\mathbb{R}^3$ is given by the standard cross product: $[\hat{b}_1,\hat{b}_2] = \widehat{\vec{b}_1 \wedge \vec{b}_2}$ .} and taking the surface measure on a single sphere as that induced by $\mathbb{R}^3$. Then, for any $b \in \su(2), x \in \su(2)^*$, the inner product, $b.x = \vec{b}.\vec{x}$, is the usual dot product for 3d vectors. Now, let $\beta \equiv \vec{\beta}$ be an arbitrary non-zero vector in $\su(2) \cong \mathbb{R}^3$. Further, let us choose an orthonormal cartesian frame $(\vec{e}_x,\vec{e}_y,\vec{e}_z)$ in $\mathbb{R}^3$, such that
\be \vec{\beta} = ||\beta|| \vec{e}_z \ee
where $ || \beta || \in \mathbb{R}_{>0}$. Recall that by definition, a point $m = (X_I) \in \Gamma_{\{A_I\}}$ identifies four points (or equivalently, the associated vectors) $X_I \in S^2_{A_I}$, one each on the individual 2-spheres. Any such vector $\vec{X}_I$ can be written in spherical coordinates $(\theta,\phi)$, for fixed radius $A_I$, as
\be \vec{X}_I = A_I(\sin \theta \cos \phi \, \vec{e}_x + \sin \theta \sin \phi \, \vec{e}_y + \cos \theta \, \vec{e}_z )\ee
where, $\theta \in [0,\pi], \phi \in [0,2\pi]$ as per common convention. The standard area element in this parametrization is 
\be d\mu = A_I^2 \sin \theta \, d\theta d\phi \,. \ee
Then, the contribution to $Z_\beta$ from each integral with fixed $I$ is given by,
\begin{align}
 \int_{S^2_{A}} d\mu \; e^{-\vec{\beta} . \vec{X}} &= A^2 \int_{0}^{2\pi} d\phi \int_{0}^{\pi} d\theta \sin\theta \; e^{- A||\beta|| \cos \theta}  \\
 &= \frac{2\pi A^2}{A||\beta||} \int_{-A||\beta||}^{A||\beta||} du \; e^{u} \\[10pt]
 &= \frac{4\pi A}{||\beta||} \left( \frac{e^{A|| \beta ||} - e^{-A|| \beta ||}}{2} \right) \\[10pt]
 &= \frac{4\pi A}{||\beta||} \sinh (A||\beta||)
\end{align}
where, we have substituted $u := -A||\beta|| \cos \theta$.   $\hfill \square$

\end{subappendices}

\chapter{Conclusions}
\label{CONC}

\begin{quotation}
\begin{center} Knowledge is love and light and vision. ---\emph{Helen Keller} \end{center}
\end{quotation}


\vspace{3mm}

\noindent In this thesis, we have discussed: aspects of a generalised framework for equilibrium statistical mechanics in the context of background independent systems; and, thermal aspects of a candidate quantum spacetime composed of many combinatorial and algebraic quanta, utilising their field theoretic formulation of group field theory. Specifically, we have focussed on generalisation of Gibbs states. \cite{Kotecha:2018gof,Chirco:2018fns,Chirco:2019kez,Kotecha:2019vvn,Assanioussi:2019ouq,Assanioussi:2020hwf}


\section*{Towards generalised equilibrium statistical mechanics}

We have presented aspects of a potential extension of equilibrium statistical mechanics for background independent systems with an arbitrarily large but finite number of degrees of freedom. In particular, we have focused on the definition of equilibrium states, based on a collection of results and insights from studies of constrained systems on spacetime and discrete quantum gravitational systems devoid of standard spacetime-related structures. While various proposals for a generalised notion of statistical equilibrium have been summarised (sections \ref{characG} and \ref{past}), one in particular, based on the constrained maximisation of information entropy has been stressed upon (section \ref{maxent}). As we have detailed, this characterisation is in the spirit of Jaynes' method, wherein the constraints are average values of a set of macroscopic observables $\{\braket{\mcO_a} = U_a\}$ that an observer has access to, and maximising the entropy under these constraints amounts to finding the least-biased distribution over the microscopic states such that the statistical averages of the same $\mcO_a$ coincide with their given macroscopic values $U_a$. The resultant state is a generalised Gibbs density function or operator (equations \eqref{genstate1} and \eqref{genstate2} respectively), characterised by a set of several observables and a conjugate multivariable temperature. \

Further, we have discussed how this notion of equilibrium in a generalised Gibbs state, and its associated stationarity with respect to the modular flow, is democratic, in that it does not require preferring any one of these observables as special over the others (section \ref{modstat}). We have also investigated preliminary aspects of a generalised thermodynamics as directly implied by these states, including defining the basic thermodynamic potentials (section \ref{TDPot}), considering the issue of deriving a single common temperature (section \ref{commtemp}), and discussing generalised zeroth and first laws (section \ref{zflaws}). 

We have argued in favour of the potential of this thermodynamical characterisation based on the maximum entropy principle, by highlighting its many unique and valuable features (section \ref{rem}). For instance, this characterisation is comprehensive, accommodating all past proposals for defining Gibbs states. It is also inherently observer-dependent, being defined using the observed macrostate $\{\braket{\mcO_a} = U_a\}$. Further, this method does not require a pre-defined automorphism to define equilibrium with respect to, unlike the KMS characterisation. Although, once a state has been defined, one can (if one wants) extract its modular flow with respect to which it will satisfy the KMS condition (section \ref{modstat}). Therefore, this proposal could be especially useful in quantum gravitational contexts (chapter \ref{TGFT}), where we may be interested in geometric quantities such as area and volume (section \ref{posgibbs}) which may not necessarily be generators of some automorphism of a system a priori.   \

An important extension of our considerations, for future work, is a suitable inclusion of constrained dynamics and its consequences, before any deparametrization. For example, the considerations in section \ref{congibb}, associated with classical constraint functions, require a more complete understanding as to their interpretation in terms of effective constraints and their applications \cite{Bojowald:2009zzc,Bojowald:2009jj,Bojowald:2010xp,Bojowald:2010qw}. Aspects of generalised thermodynamics also require further development. For instance in the first law as presented in section \ref{zflaws}, the additional possible work contributions need to be identified and understood, particularly in the context of background independence, along with the interpretation of generalised heat terms \cite{souriau1,e18100370,Chirco:2019tig,melechirco}. It would also be interesting to understand better the roles of and relation to (quantum) information, entropy and (macroscopic) observers in these contexts \cite{Hoehn:2017gst,ZeilingerFound,Brukner2003,Jaynes1992,fuchs2002quantum,Rovelli:2015dha,bruknerQCaus,PhysRevA.75.032110,Spekkens2016}, and the relation to quantum reference frames \cite{GiacominiQMCO,Vanrietvelde20,delaHamette:2020dyi,Krumm:2020fws,Hohn:2019cfk,Hoehn:2021wet,Bartlett:2006tzx}. \

Such investigations into generalised statistical and thermodynamical aspects may benefit from working with some specific, physically motivated example, for example stationary black holes. For instance, the thermodynamical characterisation could be applied in a spacetime setting, with respect to the mass, charge and angular momentum observables, like we alluded to in the Introduction. In addition to clarifying some of the aspects that we mentioned above, such a setting could further help unfold the physical mechanism for the selection of a single common temperature, thus also of physical time and energy, starting from a generalised Gibbs measure where none is preferred a priori. This may also have an illuminating interplay with quantum reference frames in spacetime \cite{Giacomini:2021gei,Smith:2019imm,Hohn:2019cfk,Castro-Ruiz:2019nnl}.


\section*{Towards thermal quantum spacetime} 

\subsection*{Many-body formulation and group field theory}

We have considered equilibrium statistical mechanical aspects of a candidate quantum gravitational system, composed of many quanta of geometry. The choice of these quanta is inspired directly from boundary structures in various discrete approaches including loop quantum gravity, spin foams and simplicial gravity. They are the combinatorial building blocks (i.e. boundary patches) of graphs, labelled with algebraic data (usually of group $SU(2)$) encoding discrete geometric information (sections \ref{atomkin} and \ref{bosgft}), e.g. labelled 4-valent nodes or tetrahedra in 4d models. Their many-body dynamics is dictated by non-local interaction vertices (sections \ref{intdyn} and \ref{effgft}), e.g. 4-simplices in 4d models, resulting in a discrete quantum spacetime, e.g. 4d simplicial complex. Statistical states can then be defined on a multi-particle state space e.g. of an arbitrarily large but finite number of tetrahedra. Then, generalised Gibbs states can be defined using the thermodynamical characterisation in general, or if a suitable 1-parameter group of automorphism exists then equivalently using the dynamical KMS condition characterisation (sections \ref{GeneqmS} and \ref{gengibb}). \

In particular, we have shown that a coarse-graining (using coherent states) of a class of generalised Gibbs states of a system of such quanta of geometry, with respect to the dynamics-encoding kinetic and vertex operators, naturally gives rise to covariant group field theories (section \ref{effgft}). In this way, we have interpreted a group field theory as an effective statistical field theory, extracted from the underlying statistical quantum gravitational system, and have thus provided a statistical basis for the standard understanding of these quanta as being excitations of group fields. \

In this thesis, we have considered complex-valued scalar group fields, defined over a domain space comprised of locally compact, connected and unimodular Lie groups. In particular, we have considered domain manifolds of the general form $G^d \times \mathbb{R}^n$, with integers $d \geq 1$ and $n \geq 0$ (sections \ref{atomkin} and \ref{bosgft}). In some specific examples, we have chosen $G = SU(2)$, $d=4$ and $n=1$, corresponding to some common choices for 4-dim GFT models, minimally coupled to a single real-valued scalar matter field (section \ref{bosgft}). The dynamics of group fields is encoded in an action function (or from a statistical standpoint, an effective Hamiltonian or a Landau-Ginzburg free energy function), with a non-local interaction term defining the bulk vertex of the 2-complex (equivalently, of the GFT Feynman diagram), and a local kinetic term defining the propagator associated with bulk bondings between these vertices (section \ref{intdyn}).   \

The many-body perspective has been our main technical strategy for the development and investigation of thermal equilibrium aspects of group field theory, in a fully background independent context for the fundamental candidate building blocks of quantum spacetime, and within the full theory as opposed to special approximations. Specifically, this perspective has offered advantages at two levels. First, the suggested formal description of spacetime as a many-body quantum system has allowed us to handle these issues within a mathematical formalism that maintains close analogies with that used for more standard physical systems. This, in a way, has permitted us to move forward without having fully solved all the conceptual issues, especially surrounding background independence, implicated in the problem. Second, while GFTs are background independent from the point of view of spacetime physics (in the sense that spacetime itself has to be reconstructed in most of its features), their mathematical definition as field theories on Lie groups has allowed us to work with the background structures of the group manifold playing technically a very similar role to what spacetime structures play in usual field theories, e.g. for condensed matter systems. \

Along this line, we have first established the ground work for the system's classical and quantum kinematics, to later consider its statistical aspects. Specifically, we have described a Weyl algebraic formulation for the choice of bosonic quanta (section \ref{weylGFT}), presented the required details of the Fock representation associated with a degenerate vacuum (sections \ref{fockspace} and \ref{usbas}), and constructed groups of unitarily implementable translation *-automorphisms (section \ref{AUT}) to be used later for defining structural equilibrium states. We have also outlined a procedure to deparametrize an originally constrained many-body system, to define from it a canonical Hamiltonian system equipped with a good clock variable (section \ref{depgft}).

\subsection*{Generalised Gibbs states}

Based on the above many-body framework, we have constructed examples of statistical Gibbs states for discrete quantum geometries composed of classical and quantum tetrahedra, and in general polyhedra (section \ref{gengibb}). \

As a first example of applying the thermodynamical characterisation, we have presented a class of Gibbs density operators generated by extensive, positive (more generally, semi-bounded either from above or below, see the Remark in appendix \ref{posapp2}) operators defined on the Fock Hilbert space of the degenerate vacuum (section \ref{genposgibbs}). As a special case, we have discussed a state associated with a geometric volume operator (section \ref{volposgibbs}). Such a state can be understood as describing a quantum gravitational state underlying a region of space with fixed macroscopic (average) volume. We have shown that a direct consequence of a system being in such a state, is the occurrence of Bose-Einstein condensation to the single-particle ground state of the volume operator, much like the case of non-relativistic Bose gases characterised by a free Hamiltonian. We have thus presented a model-independent, statistical mechanism for generating a low spin phase (e.g. spin-1/2, when neglecting the degenerate spin-0 case, for $SU(2)$ data), starting from a thermal state. Such phases are encountered often in studies related to group field theory and loop quantum gravity. \

For further work along this line, it would be interesting to study quantum black hole states of these degrees of freedom, with thermality potentially associated with area operators \cite{PhysRevD.55.3505,Oriti:2018qty,PhysRevLett.116.211301,DiazPolo:2011np,Perez:2017cmj}. In general, as we have stressed before, it is important to be able to identify suitable observables to characterise an equilibrium state of physically relevant cases with. Further investigating thermodynamics of quantum gravitational systems would benefit from confrontation with studies of thermodynamics of spacetime \cite{Jacobson:1995ab,Bardeen:1973gs,Padmanabhan:2009vy,Israel:1976ur}. For this we may need to consider the quantum nature of the degrees of freedom, and use insights from the field of quantum thermodynamics \cite{doi:10.1080/00107514.2016.1201896}, which itself has interesting links to quantum information theory \cite{Goold_2016}. \

Then, we have considered the KMS condition characterisation for generalised Gibbs states. Along the lines of arguments in standard algebraic quantum statistical mechanics, we have clarified that the unique normal KMS states in an algebraic system (which is not necessarily GFT) are Gibbs states, when the algebra of observables is irreducible on the given Hilbert space and the system is equipped with a strongly continuous 1-parameter group of unitary transformations (appendix \ref{appkmsgibb}). Subsequently, we have identified the same structures in the present GFT system, for the construction of (unique KMS) Gibbs states. \

Specifically, we have constructed two classes of Gibbs states associated with momentum operators (section \ref{momgibbs}), which satisfy the KMS condition with respect to 1-parameter groups of transformations defined previously in sections \ref{AUT} and \ref{depgft}. In the first (section \ref{inttrans}), we have considered 1-parameter groups of translations on the base manifold $G^d \times \mathbb{R}^n$, and have constructed structural Gibbs states associated with their corresponding self-adjoint momentum operators. These states encode equilibrium with respect to internal translations along the base manifold. Since the group manifold is in general curved, then naturally the corresponding notion of thermality depends on the trajectory used to define it, just like in the case of Rindler trajectories on Minkowski spacetime. For the second class of states (section \ref{physeqm}), we recall that the primary reason to couple scalar fields, by extending the base manifold of group fields by $\mathbb{R}$ degrees of freedom, was to subsequently use them to define relational clock reference frames. It was then natural to seek Gibbs states, still generated by the momentum of the scalar field, but within a deparametrized system thus encoding relational dynamics. With this in mind, we have presented relational physical Gibbs states defined with respect to clock Hamiltonians, based on the result of deparametrization as detailed earlier in section \ref{depgft}. \

Lastly, we have considered classical equilibrium configurations of a system of many tetrahedra, using the thermodynamical characterisation for constraint functions. The main idea here is that the imposition of constraints can be understood from a statistical standpoint, either being satisfied strongly (i.e. exactly, $C=0$) via a microcanonical distribution on the extended unconstrained phase space, or effectively (i.e. on average, $\langle C \rangle = 0$) via a canonical Gibbs state (section \ref{congibb}). We have considered two examples for preliminary investigations. However, further work is required to understand physical consequences of our examples in quantum gravity; and to understand better the implementation and consequences of effective constraints, $\braket{C}$, in quantum gravity \cite{Bojowald:2009zzc,Bojowald:2009jj,Bojowald:2010xp,Bojowald:2010qw}. \

As a first example, we have considered the simple case of the classical closure condition associated with a single tetrahedron (section \ref{entclosure}). The corresponding Gibbs distribution then describes a tetrahedron geometry fluctuating in terms of its closure; and is characterised by a vector-valued temperature $\beta \in \su(2)$. It is an example of Souriau's definition of Gibbs states for Lie group actions \cite{e18100370,souriau1}, here associated with the closure constraint generating a diagonal $SU(2)$ action. We have left further investigation of the application of this state in simplicial gravity to future work. Another interesting extension would be to explore aspects of its Lie group thermodynamics, in line with the studies stemming from Souriau's generalisation of Gibbs states \cite{e18100370,souriau1,Chirco:2019tig,melechirco}. \

As another example with motivations rooted more directly in simplicial gravity, we have considered formal Gibbs distributions in a system of many tetrahedra, with respect to area-matching gluing constraints between adjacent triangular faces. This has produced fluctuating twisted geometric configurations for connected simplicial complexes formed by the same tetrahedra (section \ref{glucondgibb}). This line of investigation can be used to explore specific examples of simplicial gravity (or group field theory) models with direct or stronger geometric interpretation, and thus of greater interest for quantum gravity. For instance, one could consider the state space of geometric (in the sense of metric) tetrahedra and utilise a generalised Gibbs state to define the partition function with a dynamics encoded by the Regge action \cite{Regge:2000wu,Regge:1961px}. Another interesting direction would be to define a Gibbs density implementing not only gluing constraints but also shape-matching constraints \cite{Dittrich:2008va}, or simplicity constraints, on the twisted geometry space \cite{Freidel:2010aq, Rovelli:2010km}, thus reducing to a proper Regge geometry starting from $SU(2)$ holonomy-flux data.


\subsection*{Thermofield double vacua and inequivalent representations}

In section \ref{threpcond}, we have constructed finite temperature, equilibrium phases associated with a class of generalised Gibbs states in group field theory, based on their non-perturbative thermal vacua. For this, we have utilised tools from the formalism of thermofield dynamics (section \ref{tfd}). The vacua are squeezed states encoding entanglement of quantum geometric data (section \ref{finite}), and are unitarily inequivalent to the class of degenerate vacua (section \ref{zero}). Entanglement is expected to be a characteristic property of a physical quantum description of spacetime in general \cite{Bombelli:1986rw,Srednicki:1993im,Ryu:2006bv,VanRaamsdonk:2010pw,Hoehn:2017gst,VanRaamsdonk:2016exw,Marolf:2017jkr,Bianchi:2012ev,Baytas:2018wjd,Livine:2017fgq,Chirco:2017xjb,Chirco:2019dlx,Chirco:2017wgl,Chirco:2017vhs,Colafranceschi:2020ern}. This setup also opens the door to using such techniques in discrete quantum gravity, thus facilitating exploration of the phase structure of quantum gravity models characterised by generalised thermodynamic parameters $\beta_a$; and, complementing renormalization investigations in group field theory \cite{Carrozza:2016vsq,Carrozza:2017vkz,Benedetti:2014qsa} and possibly other related approaches \cite{Dittrich:2014mxa, Bahr:2014qza, Bahr:2016hwc}. \

Further, we have introduced coherent thermal states which, in addition to carrying statistical fluctuations in a given set of observables, are also condensates of quantum geometry (section \ref{CTS}). Zero temperature coherent states in group field theory have been used to obtain an effective description of flat, homogeneous and isotropic cosmology (flat FLRW), where certain quantum corrections arise naturally and generate a dynamical modification with respect to classical gravity, preventing the occurrence of a big bang singularity along with cyclic solutions in general \cite{Oriti:2016qtz,Oriti:2016acw,Gielen:2016dss,Pithis:2019tvp}. Encouraged by these results, the introduction of statistical condensates, like coherent thermal states, may bring further progress to the GFT condensate cosmology program by offering a tangible and controllable way of incorporating perturbations in relevant observables. Such considerations could be valuable say for understanding the quantum gravitational origin of structure formation. They could also lead to modifications during early times in the previously studied homogeneous and isotropic flat cosmology models in GFT, such as altering the inflation rate \cite{Oriti:2016qtz,deCesare:2016rsf,Assanioussi:2020hwf}, as considered later in section \ref{GFTCC}. \

The thermal vacua constructed here (section \ref{finite}) are thermofield double states \cite{Takahasi:1974zn,Israel:1976ur,Unruh:1976db,Bisognano:1976za,Hartle:1976tp,Sewell:1982zz,Maldacena:2001kr}. This setup could thus also be useful for the study of quantum black holes. In group field theory for example, black holes have been modelled as generalised condensates \cite{Oriti:2018qty,PhysRevLett.116.211301}, which must also possess related thermal properties. Thermal coherent states may then provide just the right type of technical structure, in order to study the statistical and thermal aspects of the corresponding quantum black holes \cite{Perez:2017cmj,DiazPolo:2011np,Oriti:2018qty,PhysRevLett.116.211301}. \

It would also be interesting to understand better the connection of these vacua with similar works in loop quantum gravity concerning kinematical entanglement between intertwiners in spin networks, and related further to discrete vector geometries \cite{Bianchi:2016tmw,Livine:2017fgq,Baytas:2018wjd}, especially since the squeezed vacua constructed here essentially encode entanglement between gauge-invariant spin network nodes (i.e. intertwiners), but at a field theory level. Moreover, our construction can be extended even further to more general two-mode squeezed vacua. For instance, condensates of correlated quanta, like dipole condensates \cite{Gielen:2016dss}, may be straightforwardly constructed and studied in this setup. Considering correlations between different modes of the quanta, which encode quantum geometric data, might also make comparisons with the studies in LQG mentioned above \cite{Bianchi:2016tmw,Livine:2017fgq,Baytas:2018wjd} more direct. \

 Finally, by providing quite a straightforward handle on collective, quasi-particle modes in discrete quantum gravity, while still allowing for access to different inequivalent representations, this framework may bring closer the studies of microscopic theories of quantum gravity and analogue gravity models \cite{Barcelo:2005fc}.


\subsection*{Condensate cosmology with volume fluctuations}

As a preliminary application in a more physically relevant setting, we have studied some implications of the presence of statistical volume fluctuations in the context of group field theory by using coherent thermal states for condensate cosmology (section \ref{GFTCC}). In the GFT condensate cosmology program, a quantum gravitational phase of the universe is modelled as a condensate. Along these lines, we have considered a thermal condensate (section \ref{vgibbs}), consisting of: a pure condensate representing an effective macroscopic homogeneous spacetime, like in previous studies \cite{Oriti:2016qtz,Oriti:2016acw,Gielen:2016dss,Pithis:2019tvp}; and, a static thermal cloud over the pure condensate part, encoding statistical fluctuations of quantum geometry.

The presented model recovers cosmological dynamics of a flat FLRW universe with a minimally coupled scalar field at late times (section \ref{lateev}), when the condensate dominates the thermal cloud. While at early times (section \ref{earlyev}), when the thermal part dominates the condensate, the model displays quantum and statistical corrections, the latter being the primary difference with respect to earlier works based on pure (zero temperature, or non-thermal) condensates. In particular, we have shown that the singularity is generically resolved with a bounce between a contracting and an expanding phase, and that there exists an early phase of accelerated expansion (section \ref{earlyev}). The expansion phase is characterised by an increased number of e-folds compared to those achieved in the previous zero temperature analysis of the same class of free GFT models. This increase in the number of e-folds, obtained in absence of interactions, is attributed to the presence of the thermal cloud. This is in contrast to previous conclusions (see \cite{deCesare:2016rsf} and related works) that an increase in the number of e-folds necessarily requires non-zero interactions. However, the maximum number of e-folds achieved here ($\approx 0.3$) are still negligible. Thus, considering a dynamical thermal cloud and non-trivial interactions would be expected to result in a sufficiently long period of this geometric inflation. These would be valuable extensions of the present work.   \

Consideration of a dynamical (non-static) thermal cloud would allow for further investigations of consequences of the presence of a thermal cloud on the effective physics of the system, even in free models. Subsequently, it would naturally be interesting to consider an interacting model in the presence of thermal fluctuations. In fact, these aspects of having a dynamical thermal cloud and an overall interacting theory are intimately related, as discussed briefly in section \ref{disc}. In general, one could systematically extend the previous studies in GFT condensate cosmology \cite{Pithis:2019tvp,Oriti:2016acw,Gielen:2016dss} to the case with thermal fluctuations using our setup, and investigate various aspects including dynamical analysis of fluctuations, perturbations, anistropies and inhomogeneities. \

For our analysis, we have introduced a suitable generalisation of relational clock frames in GFT, by considering clock functions $t(\phi)$, implemented as smearing functions (section \ref{clock}). Consequently, we have formulated the effective equations of motion and the dynamical quantities as functionals of $t$ (section \ref{eff2}). In comparison with past works where relational quantities are functions of the coordinate $\phi$, the clock frames introduced here resolve divergences associated with coincidence limits $\phi_1 \to \phi_2$. A more complete understanding of relational frames in GFT and their precise occurrence from a physical mechanism of deparametrization is left to future work. \

Lastly, we have understood $\beta$ as a statistical parameter that controls the extent of depletion of the condensate into the thermal cloud, and overall the strength of statistical fluctuations of observables in the system. The question remains whether it also admits a geometrical interpretation. Taking guidance from classical general relativity, we know that spatial volume generates a dynamical evolution in constant mean curvature foliations, wherein the temporal evolution is given by the so-called York time parameter. Constant York time slices are thus constant extrinsic curvature scalar (mean curvature) slices, and the two quantities are proportional to each other. In this case, one could attempt to understand $\beta$ as the periodicity in York time, equivalently in scalar extrinsic curvature (both of which are conjugates to the spatial volume). In particular for homogeneous and isotropic spacetimes, York time is further proportional to the Hubble parameter \cite{Roser:2014foa}. A detailed investigation of such aspects and their implications would be interesting for future work, also for studying thermodynamical aspects of the present quantum gravitational system.


\section*{Final remarks}

We have asked the question: can we characterise a macroscopic system as being in equilibrium without there being a notion of time? And we have tentatively answered: yes; if the system has maximum total entropy, compatible with its macrostate (e.g. average volume, energy, etc.), then we can say that it is in equilibrium. In traditional statistical thermodynamics, this is a direct consequence of the second law. Thus, the validity of the second law ensures the validity of this characterisation for statistical equilibrium. But in this thesis, in the context of background independent systems, we have understood this ``thermodynamical characterisation'' as being more fundamental, coinciding with the understanding of the quality of entropy in a macroscopic dynamical system as being more fundamental than an evolution parameter (i.e. time), the latter being a feature only of the special class of deparametrized (or deparametrizable) constrained systems. We are accustomed to understanding (and subsequently treating) statistical equilibrium as being synonymous with the property of stationarity in time\footnote{Or, with the property of satisfying Kubo's correlation functions, i.e. the KMS boundary conditions, with respect to the given time evolution.}, often to the extent that the property of maximal entropy is understood as simply being an additional feature of an equilibrium configuration that is not fundamental to its definition. However in the context of dynamically constrained (or time reparametrization-invariant) systems, wherein a preferred, external, global choice of a time variable is absent, the latter property of maximal entropy may be more definitive. 

The justification of this suggestion must naturally lie in its consequences for physical systems, both gravitational and non-gravitational. Since we have arrived at this suggestion in our attempts to define Gibbs states (and therefore, also understand their construction procedures) within the context of the present discrete quantum gravitational system, our investigations are also limited accordingly. In particular, much like in almost any approach to quantum gravity, our investigations largely face a disconnect from established spacetime physics (even though we have made preliminary attempts at connecting with a candidate macroscopic phase and cosmology, in sections \ref{threpcond} and \ref{GFTCC}). This disconnect is especially stark in more radical approaches that are not based on standard spacetime structures, and is tied with the difficult open problem of emergence of spacetime; in this thesis, we have not directly tackled the problem of emergence, however our results may provide some useful tools to address it\footnote{A quantum statistical framework can indeed be used, starting from a given quantum gravitational model, to extract and analyse the collective behaviour of the underlying degrees of freedom. It is at this coarse-grained level of description that we expect continuum spacetime and geometry to emerge \cite{Oriti:2018dsg,Oriti:2013jga,Oriti:2018tym}.}. Therefore, the following two broad but important directions call for further work: physical consequences of generalised statistical equilibrium in quantum gravity and connection with known spacetime phenomena; and, investigation of generalised statistical equilibrium in constrained systems on spacetime\footnote{We note that this latter case has been studied in some works, e.g. \cite{e18100370,souriau1,Montesinos:2000zi,Chirco:2019tig,melechirco}.}.

Moreover, these tasks of formulating a framework for background independent statistical mechanics, applicable also to gravity (since like any other dynamical field, gravitational field must also undergo thermal fluctuations), and that of investigating the statistical mechanics of quanta of geometry are formally different. We have dealt with the latter\footnote{For efforts toward the former, see for example \cite{Chirco:2019tig,melechirco,Rovelli:1993ys,Connes:1994hv,Rovelli:2012nv,Chirco:2013zwa,Rovelli:2010mv,Montesinos:2000zi,Chirco:2016wcs}.}. The conceptual challenges, like timelessness, that are encountered when considering the statistical mechanics of general relativistic spacetime, and of pre-geometric quanta underlying a spacetime (as defined in some quantum gravity framework) are similar. But they are two separate issues, even if expected to be related eventually\footnote{Even then, to uncover this explicit relation, and see the interplay with spacetime physics, would require tackling the issue of emergence more directly.}; and, developments in one can lead to a more refined understanding of the other. 

Finally, we recognise that insights from statistical mechanics and thermodynamics were crucial for the birth of the quantum hypothesis by Planck\footnote{For interesting discussions, see for instance \cite{kuhn1987,planck2012eight}.}, and thus of quantum theory. They may prove to be crucial also in our quest for understanding the fundamental nature of gravity.

\backmatter

\chapterstyle{default} 



\bibliographystyle{unsrt}

\bibliography{refthesis} 


\printindex 


\end{document}